\documentstyle[12pt]{book}
\newcommand{\be}{\begin{equation}}
\newcommand{\ee}{\end{equation}}
\newcommand{\beq}{\begin{eqnarray}}
\newcommand{\eeq}{\end{eqnarray}}

\def\NPB#1#2#3{{\it Nucl.~Phys.} {\bf{B#1}} (19#2) #3}
\def\PLB#1#2#3{{\it Phys.~Lett.} {\bf{B#1}} (19#2) #3}
\def\PRD#1#2#3{{\it Phys.~Rev.} {\bf{D#1}} (19#2) #3}
\def\PRL#1#2#3{{\it Phys.~Rev.~Lett.} {\bf{#1}} (19#2) #3}
\def\CMP#1#2#3{{\it Comm.~Math.~Phys.} {\bf{#1}} (19#2) #3}
\def\JGP#1#2#3{{\it J.~Geom.~Phys.} {\bf{#1}} (19#2) #3}
\def\IJMP#1#2#3{{\it Int.~Jour.~Mod.~Phys.} {\bf{#1}} (19#2) #3}
\def\CQG#1#2#3{{\it Class.~Quant.~Grav.} {\bf{#1}} (19#2) #3}
\def\ZPC#1#2#3{{\it Z.~Phys.} {\bf C#1} (19#2) #3}
\def\PTP#1#2#3{{\it Prog.~Theor.~Phys.} {\bf#1}  (19#2) #3}
\def\MPLA#1#2#3{{\it Mod.~Phys.~Lett.} {\bf#1} (19#2) #3}
\def\PR#1#2#3{{\it Phys.~Rep.} {\bf#1} (19#2) #3}
\def\AP#1#2#3{{\it Ann.~Phys.} {\bf#1} (19#2) #3}
\def\RMP#1#2#3{{\it Rev.~Mod.~Phys.} {\bf#1} (19#2) #3}
\def\HPA#1#2#3{{\it Helv.~Phys.~Acta} {\bf#1} (19#2) #3}
\def\JETPL#1#2#3{{\it JETP~Lett.} {\bf#1} (19#2) #3}

\begin{document}
\def\part{\partial}
\def\btd{\bigtriangledown}
\def\ap{\a^{'}}
\def\ww{\wedge}
\def\ran{\rangle}
\def\lan{\langle}
\def\bra{\rangle}
\def\ket{\langle}
\def\rar{\rightarrow}
\def\lrar{\leftrightarrow}
\def\ra{\rightarrow}
\def\lra{\leftrightarrow}
\def\a{\alpha}
\def\b{\beta}
\def\g{\gamma}
\def\G{\Gamma}
\def\D{\Delta}
\def\d{\delta}
\def\e{\epsilon}
\def\z{\zeta}
\def\h{\eta}
\def\th{\theta}
\def\k{\kappa}
\def\l{\lambda}
\def\L{\Lambda}
\def\m{\mu}
\def\n{\nu}
\def\x{\xi}
\def\X{\Xi}
\def\p{\pi}
\def\P{\Pi}
\def\r{\rho}
\def\s{\sigma}
\def\S{\Sigma}
\def\t{\tau}
\def\f{\phi}
\def\F{\Phi}
\def\c{\chi}
\def\w{\omega}
\def\W{\Omega}
\def\de{\partial}
\def\bra{\langle}
\def\ra{\rightarrow}
\def\ket{\rangle}
\hyphenation{ne-ga-ti-va}
\hyphenation{mo-del-li}
\hyphenation{si-ste-ma-ti-ca}
\hyphenation{pro-ble-ma}
\hyphenation{mo-du-li}
\hyphenation{in-va-ri-an-za}
\hyphenation{ri-nor-ma-liz-za-zio-ne}
\hyphenation{e-ner-gi-a}
%








%
\begin{titlepage}
\begin{center}
\hfill  \quad ROM2F-98/43 \\
\vskip 1cm

{\large \bf  Universit\`a degli Studi di L'Aquila }  \\ \vskip 0.5cm
{ \bf Facolt\`a di Scienze MM.FF.NN.} \\ \vskip 1.5cm
{\large Tesi di Laurea in Fisica} \\ \vskip 2cm
{\Large \bf Aspetti non perturbativi della Teoria delle Stringhe } \\
\vskip 3cm
\end{center}

Candidato \hfill\hfill Relatore interno      
\vskip 0.3cm
Giuseppe D'Appollonio \hfill Prof. Aurelio Grillo
\vskip 0.1cm
Matr. n.125265 \hfill\hfill
\vskip 0.8cm
\hfill\hfill Relatore esterno
\vskip 0.3cm
\hfill\hfill Prof. Augusto Sagnotti

\vspace{1.5cm}

\noindent
Unabridged version of the Thesis 
presented to the University of L'
Aquila, in partial fulfillment of the requirements for the ``Laurea''
degree in Physics, October 1998. 
Work carried out at the University
of L'Aquila and at the University of Rome ``Tor Vergata''. In Italian.

\vskip 3.5cm
\begin{center}
Anno Accademico 1997-98
\end{center}
\vfill
\end{titlepage}

\pagenumbering{arabic}
\chapter*{Prefazione}
\addcontentsline{toc}{chapter}{Prefazione}
\markboth{} {}

Il lavoro di Tesi \`e stato svolto presso il Dipartimento di Fisica dell' Universit\`a
di L'Aquila e il Dipartimento di Fisica della II 
Universit\`a di Roma ''Tor Vergata". Vorrei ringraziare anzitutto
il Prof. Augusto Sagnotti, per avermi introdotto 
alla teoria delle superstringhe e 
per avermi guidato nello studio di un soggetto cos\`\i \, vasto fino a 
rendermi capace di dare qualche contributo originale. Vorrei ringraziare
anche l'Ecole Polytechnique per avermi assegnato una borsa di studio 
che ha reso possibile un lungo soggiorno a Parigi durante il quale
questo lavoro \`e stato terminato, e soprattutto
il Prof. Ignatios Antoniadis e il Dott. Emilian Dudas
per avermi dato l'opportunit\`a di collaborare con 
loro all' Ecole Polytechnique. 
Vorrei ringraziare infine 
il Prof. Aurelio Grillo per l'interesse mostrato verso 
questa Tesi. Riservo il ringraziamento pi\`u sincero per i miei genitori
ai quali vorrei dedicare questa Tesi.  

\vskip 1cm

L'Aquila, ottobre 1998   \hspace{5cm}  Giuseppe D'Appollonio

\tableofcontents
\markboth{}{}

\chapter{Introduzione}
\markboth{} {}

L'argomento centrale di questa Tesi \`e la rottura spontanea della 
supersimmetria in modelli di stringhe aperte.
I prossimi tre capitoli contengono una descrizione dettagliata
della formulazione perturbativa
della teoria di stringhe, una teoria quantistica 
consistente delle interazioni di gauge e gravitazionali;
il quinto capitolo descrive i recenti risultati sulle dualit\`a di stringa, 
generalizzazioni della dualit\`a elettromagnetica, che legano i diversi modelli 
tra loro; il sesto capitolo, infine, descrive meccanismi di rottura spontanea 
della supersimmetria in teoria di stringhe aperte.

L'idea fondamentale alla base della teoria di stringhe 
\`e estremamente semplice: le 
eccitazioni puntiformi della teoria dei campi vengono sostituite
con eccitazioni unidimensionali che interagiscono in modo essenzialmente 
geometrico attraverso la fusione e la divisione delle loro superfici di 
universo. 
Il formalismo funzionale permette di 
descrivere la propagazione di una stringa in modo analogo alla 
propagazione di una particella, sostituendo alla somma sulle
linee d'universo una somma sulle superfici 
che collegano le configurazioni iniziali e finali della stringa;
si \`e quindi condotti 
in modo naturale a studiare attentamente teorie di campo 
bidimensionali. 
Nel caso della stringa bosonica, ad esempio, 
la teoria bidimensionale \`e invariante conforme a livello classico 
e contiene solo 
i campi che descrivono l'immersione della superficie nello spaziotempo.  
L'invarianza conforme gioca un ruolo essenziale 
nella quantizzazione perch\`e 
elimina dallo spazio di Hilbert tutti gli stati di norma negativa e deve 
essere quindi preservata dalle correzioni quantistiche. 
L'assenza di anomalie conformi \`e 
una delle prime condizioni di consistenza riconosciute in modo chiaro e 
poich\`e \`e legata ad una quantit\`a, detta 
carica centrale, che misura il contenuto operatoriale della 
teoria, pone restrizioni sulla dimensione 
dello spaziotempo. Nel caso della stringa bosonica la dimensione critica \`e
$D=26$. 

Consideriamo ora le eccitazioni della stringa. Esse
formano una torre di stati di massa e spin via via crescente; la loro massa 
\`e dell'ordine della massa di Planck e questo indica chiaramente che esse,
essenziali per il comportamento soffice ad alte energie delle ampiezze di 
stringa, rivestono solo 
un ruolo marginale nella descrizione della fisica di bassa energia,
dominata dai modi di vibrazione non massivi che sono
uno scalare $\f$, detto dilatone, una 2-forma antisimmetrica
$B_{\m\n}$ e un campo simmetrico a traccia nulla $g_{\m\n}$, che viene 
identificato con il tensore metrico. 
La presenza della corrispondente 
particella di spin 2 nello spettro della stringa ha motivato 
lo studio 
della teoria come una teoria quantistica della gravit\`a.

La stringa bosonica non \`e sufficiente per fornire una descrizione 
consistente di tutte le interazioni. Tra i suoi modi di vibrazione 
figura infatti un tachione e il vuoto risulta instabile; 
inoltre essa non contiene eccitazioni fermioniche. 
Questi problemi sono stati risolti con l'introduzione della superstringa. La 
teoria sulla superficie di universo \`e in questo caso superconforme e accanto 
alle coordinate bosoniche figurano dei campi fermionici anticommutanti. La 
dimensione critica per la superstringa \`e $D=10$ e lo spettro di bassa energia 
riproduce quello delle teorie di supergravit\`a in dieci dimensioni. 
Da questo punto di vista le teorie di stringhe 
rappresentano una regolarizzazione quantistica consistente 
delle supergravit\`a, che sono teorie di campo non rinormalizzabili.
Esistono 
in effetti cinque tipi di stringhe supersimmetriche 
\footnotemark \footnotetext{La supersimmetria \`e una simmetria continua che 
ruota campi bosonici in campi fermionici.}
in dieci dimensioni. In 
particolare la superstringa di Tipo IIA ha come 
limite di bassa energia la supergravit\`a non chirale di Tipo IIA, mentre la 
superstringa di Tipo IIB ha come limite di bassa energia la supergravit\`a 
chirale di Tipo IIB. 
La Tipo IIA e la Tipo IIB sono teorie di stringhe chiuse e orientate. \`E 
possibile ottenere anche una teoria di stringhe chiuse non orientate e di 
stringhe aperte, detta teoria di Tipo I, che ha come limite di bassa energia la 
supergravit\`a con una carica di supersimmetria in dieci dimensioni accoppiata 
a una teoria di super Yang-Mills con gruppo di gauge $SO(32)$.
L'ultimo tipo di stringa consistente in dieci dimensioni \`e la stringa 
eterotica che si ottiene combinando i modi sinistri di una superstringa e i 
modi destri di una stringa bosonica. Esistono due modelli eterotici in dieci 
dimensioni, uno con gruppo di gauge $SO(32)$ (che indichiamo per brevit\`a con 
HO) ed uno con gruppo di gauge
$E_8 \times E_8$ (che indichiamo con HE).
\`E interessante notare l'origine profondamente diversa
dei campi di gauge che compaiono 
nei modelli di stringhe aperte e nei modelli eterotici. Nel primo caso
i campi di gauge derivano dal settore aperto e le 
cariche sono assegnate agli estremi 
delle stringhe; nel secondo caso sono le sedici coordinate bosoniche 
destre, 
che non possono essere combinate con corrispondenti coordinate bosoniche 
sinistre, 
a dare origine ad una distribuzione di carica sulle stringhe chiuse
che si riflette nell'invarianza di gauge della teoria 
spaziotemporale.

Dalla discussione nel secondo capitolo emerger\`a che 
una teoria di stringhe perturbativa \`e essenzialmente una teoria 
conforme definita in modo consistente su superfici di Riemann di
genere arbitrario. La 
consistenza della teoria conforme si riflette in una serie di importanti 
propriet\`a della teoria spaziotemporale come la relazione tra spin e 
statistica e l'assenza di anomalie. 
A tutt'oggi \`e estremamente sorprendente che un'idea semplice come 
l'espansione in superfici sia tanto potente da descrivere teorie 
quantistiche che unificano interazioni di gauge e 
gravitazionali.

Il numero di teorie conformi adatte a definire un vuoto perturbativo
\`e enorme e questo rappresenta il problema pi\`u grave della teoria di 
stringhe. La grande variet\`a di modelli che si possono ottenere emerge in modo 
evidente quando si considerano le compattificazioni della teoria, ovvero si 
studiano vuoti nei quali un certo numero di dimensioni formano una variet\`a di 
volume molto
ridotto e risultano invisibili a bassa energia. Il vuoto \`e stabile se 
la teoria possiede almeno una supersimmetria spaziotemporale, e questa 
condizione implica che le variet\`a geometriche rilevanti sono delle variet\`a 
complesse note come spazi di Calabi e Yau. Nel terzo capitolo descriviamo le 
varie classi di compattificazioni ed i metodi sviluppati per calcolare le 
propriet\`a basilari
della fisica di bassa energia. Qui sottolineiamo come
lo studio delle compattificazioni della teoria di stringhe 
abbia rivelato una 
relazione profonda tra concetti algebrici e geometrici ed abbia 
messo in evidenza il 
modo sottile con cui le stringhe vedono la geometria dello spaziotempo. 
L'esempio pi\`u semplice a questo proposito \`e la T dualit\`a; quando la 
teoria viene compattificata su un cerchio, gli stati vengono classificati 
non solo da un momento quantizzato, come in una teoria di campo usuale, ma 
anche da un numero di avvolgimento, caratteristico di un oggetto esteso, che 
pu\`o avvolgersi attorno ad una direzione compatta. Quando si considera 
il limite in cui il raggio del cerchio va a zero, tutti i modi interni di una 
teoria di campo si disaccoppiano perch\`e divengono infinitamente massivi. In 
teoria di stringhe, invece, il disaccoppiarsi dei modi di momento coincide con 
l'annullarsi della massa degli infiniti stati di avvolgimento. Vedremo che 
questo ha come
conseguenza l'equivalenza per una teoria di stringhe tra la
propagazione su un cerchio di raggio $R$ e su un cerchio 
di raggio $\frac{\ap}{R}$ \footnotemark \footnotetext{ $\ap$ \`e 
una costante che 
fissa la tensione della stringa.}.
Questo tipo di equivalenze possono essere estese a tutte le compattificazioni 
toroidali e danno origine 
a gruppi discreti di T dualit\`a; i gruppi continui corrispondenti sono
simmetrie delle teorie effettive di supergravit\`a.
La T dualit\`a riduce anche il numero di teorie indipendenti in dieci 
dimensioni; \`e infatti facile verificare che il limite della teoria di Tipo 
IIB compattificata su un cerchio per $R \rar 0$ \`e la Tipo IIA in 
dieci dimensioni e viceversa. Allo stesso modo la teoria HE e la teoria HO sono 
T duali.

Ogni compattificazione \`e caratterizzata da un numero pi\`u o meno grande 
di moduli, deformazioni marginali della teoria conforme che si manifestano 
nella teoria di bassa energia come particelle scalari con accoppiamenti 
puramente derivativi. I moduli riflettono le possibili deformazioni 
della teoria e variandoli in modo continuo si pu\`o passare da un vuoto ad un 
altro; questo mostra in modo chiaro che gi\`a a livello perturbativo gran parte 
delle teorie che si ottengono compattificando i modelli in dieci dimensioni 
sono connesse tra loro. Il problema di come la teoria fissi il valore dei
moduli
selezionando un particolare stato di vuoto non \`e ancora stato risolto, 
e probabilmente non lo sar\`a fino a quando non avremo a disposizione una 
formulazione non perturbativa.

Finora abbiamo parlato di compattificazioni di stringhe chiuse; la costruzione 
sistematica di modelli di stringhe aperte \`e pi\`u delicata perch\`e la serie 
perturbativa coinvolge superfici con bordi e superfici non orientate e le 
condizioni di consistenza per teorie conformi su questo tipo di
superfici sono molto pi\`u sottili. La soluzione di questo problema, descritta 
in dettaglio nel quarto capitolo, 
segue da 
un'osservazione che collega in modo molto elegante la teoria di Tipo IIB e la 
teoria di Tipo I. La stringa di 
Tipo IIB \`e simmetrica rispetto allo scambio di 
modi destri e modi sinistri ed \`e quindi possibile considerare la teoria che 
si ottiene quozientando la $IIB$ con $\W$, 
dove $\W$ \`e l'operazione di parit\`a 
sulla  superficie di universo. Il quoziente della teoria si ottiene proiettando 
con $\W$ lo spettro chiuso, e d\`a origine ad uno spettro chiuso non 
orientato.
La consistenza della teoria richiede allora l'introduzione 
di settori $\it{twistati}$ con $\W$, che 
corrispondono alle stringhe aperte. Lo sviluppo sistematico di questa idea ha 
permesso di studiare le teorie conformi su superfici con bordi e non 
orientate e di costruire grandi classi di modelli di stringhe aperte.
L'importanza di questi modelli \`e duplice; da una parte descrivono il limite 
di forte accoppiamento di modelli eterotici, dall'altra presentano nel loro 
spettro perturbativo degli oggetti, noti come D-brane, che sono stati 
fondamentali per la comprensione delle propriet\`a non perturbative della 
teoria. 

Le D brane, descritte sempre nel quarto capitolo, corrispondono a
stringhe aperte con condizioni al bordo di Dirichlet in $p$ direzioni; 
\`e possibile 
mostrare che queste teorie conformi dal punto di vista spaziotemporale 
rappresentano 
degli oggetti estesi p-dimensionali, carichi rispetto ai campi del 
settore di Ramond della teoria di stringhe e invarianti rispetto ad 
un certo numero di cariche di supersimmetria.
Data la semplicit\`a della loro descrizione, le D-brane si sono rivelate 
essenziali per 
stabilire in modo preciso relazioni di dualit\`a tra le varie teorie di 
stringhe e sono delle sonde della struttura su piccola scala dello spaziotempo 
pi\`u sensibili delle stringhe stesse.

Come abbiamo gi\`a accennato,
la principale limitazione della teoria di stringhe
deriva dalla sua origine: nata come 
generalizzazione di una serie perturbativa, ancora oggi non \`e chiaro quali 
siano i suoi principi fondamentali. 
Un progresso notevole per una comprensione completa della teoria \`e stato 
fatto grazie alla scoperta delle dualit\`a, congetture riguardanti
l'equivalenza di due o pi\`u teorie di stringhe che,
a livello perturbativo, appaiono distinte.
La formulazione di queste congetture, che danno
finalmente concretezza all'idea che tutte le teorie di stringhe non siano altro 
che espansioni asintotiche di un'unica teoria fondamentale,
\`e stata motivata da particolari simmetrie delle loro
lagrangiane effettive.
Abbiamo gi\`a osservato che le T dualit\`a si riflettono in simmetrie continue 
delle corrispondenti teorie di bassa energia;
le supergravit\`a sono caratterizzate 
anche da gruppi di simmetrie nascoste, come il gruppo non compatto
$E_{7,7}$ della supergravit\`a con $N=8$. \`E naturale pensare che una 
versione discreta di questi gruppi sia una simmetria della teoria di stringhe 
completa. Se questo \`e vero, emerge una struttura perfettamente consistente ed 
unificata delle varie teorie di stringhe che coinvolge anche la supergravit\`a 
in undici dimensioni. In effetti si crede che esista una teoria quantistica in 
undici dimensioni, chiamata provvisoriamente M teoria, nella quale le stringhe 
non rivestono un ruolo privilegiato al di fuori di particolari punti dello 
spazio dei moduli dove un'espansione in superfici \`e possibile.
In particolare, compattificando la M teoria su un cerchio si ottiene la Tipo 
IIA, e compattificando su un intervallo si ottiene la HE; i due fattori del
gruppo di gauge sono in questo caso vincolati agli estremi dell'intervallo, 
ovvero alle due estremit\`a dell'universo. In dieci dimensioni abbiamo altre 
due relazioni importanti: la Tipo I e la HO
sono S duali, ovvero l'una rappresenta il limite di forte accoppiamento 
dell'altra mentre la Tipo IIB ha un gruppo di dualit\`a non perturbativo
$SL(2,Z)$.

Le dualit\`a legano in generale 
stati perturbativi in un modello a stati non 
perturbativi in un altro; gli stati non perturbativi si presentano 
tipicamente nello spettro di bassa energia come oggetti estesi p-dimensionali, 
detti p-brane, e quando sono carichi rispetto alle forme antisimmetriche 
provenienti dal settore di Ramond hanno, come abbiamo accennato,  
una semplice descrizione come teorie conformi grazie alle D-brane. 
Questo consente un maggiore controllo delle loro 
propriet\`a e rende possibile  
una verifica precisa delle relazioni di dualit\`a, almeno per 
quanto riguarda lo spettro. La verifica dell'equivalenza a 
livello delle 
interazioni \`e agevole solo quando teoremi di non rinormalizzazione 
permettono di confrontare risultati perturbativi in una teoria con risultati 
non perturbativi in un altra ed \`e quindi cruciale l'invarianza 
rispetto ad un certo numero di cariche di supersimmetria.

Se si tiene conto degli stati non perturbativi \`e 
possibile unificare quasi totalmente lo 
spazio dei vuoti della teoria di stringhe. Punti nei quali la teoria diviene 
singolare e non consente apparentemente una transizione tra due distinti spazi 
dei moduli, si rivelano in realt\`a punti nei quali la descrizione 
perturbativa viene meno in quanto nuovi stati, associati a solitoni di massa 
nulla, compaiono nello spettro di bassa 
energia; 
una volta che questi stati vengono identificati ed il loro contributo viene
preso in considerazione, la teoria di stringhe risulta perfettamente 
consistente. 

Abbiamo descritto i vuoti perturbativi della teoria e le 
relazioni perturbative e 
non perturbative tra di essi; essenziale in ogni caso \`e l'esistenza di
un certo numero di supersimmetrie spaziotemporali. \`E la supersimmetria 
che assicura la stabilit\`a perturbativa del vuoto
ed \`e ancora la supersimmetria che permette di 
effettuare delle verifiche quantitative delle dualit\`a. 
Perch\`e la teoria di stringhe possa fornire una descrizione completa della 
natura, bisogna tuttavia affrontare il problema della rottura della 
supersimmetria. Questo problema viene discusso nel sesto capitolo.
Uno dei primi meccanismi proposti in teoria di stringhe, la 
condensazione dei gaugini, \`e stato formulato solo a livello della teoria 
effettiva di bassa energia e questo non offre un grande controllo analitico 
sulle propriet\`a del modello. Il metodo pi\`u diretto consiste nel costruire 
teorie conformi che presentano rottura spontanea della 
supersimmetria; la natura spontanea della rottura, e quindi l'esistenza di 
limiti nello spazio dei moduli nei quali la supersimmetria viene ripristinata, 
rendono plausibile che le propriet\`a di stabilit\`a della teoria originale non 
vengano del tutto compromesse e pongono vincoli sull'azione 
effettiva, che pu\`o essere determinata in modo pi\`u agevole.

Modelli con rottura spontanea della supersimmetria sono stati costruiti
estendendo alla teoria di stringhe un meccanismo introdotto in teoria dei campi 
da Scherk e Schwarz. Questo meccanismo \`e una riduzione dimensionale 
generalizzata in cui i campi mantengono una opportuna dipendenza 
dalle coordinate 
interne, fissata in modo preciso da una simmetria della teoria. In questo modo 
\`e possibile rompere spontaneamente tanto le simmetrie di gauge quanto 
la supersimmetria. 

In teoria di stringhe il meccanismo pu\`o essere 
interpretato come una deformazione del reticolo di un toro
sul quale \`e compattificata la 
teoria. 
Il problema, quando esso viene
utilizzato in teorie di 
stringhe chiuse, \`e che la scala di rottura della supersimmetria \`e 
inversamente proporzionale al volume della variet\`a interna. Per avere 
rottura spontanea nella regione del $TeV$, bisogna quindi utilizzare variet\`a 
interne molto estese la cui presenza, in modelli generici, non \`e compatibile 
con la fenomenologia.
 
Il meccanismo \`e stato recentemente esteso alla teoria di stringhe aperte e 
presenta molte caratteristiche affascinanti, sia dal punto di vista 
fenomenologico, sia dal punto di vista teorico. Come accennato in precedenza, 
modelli di stringhe aperte contengono naturalmente nel loro spettro 
perturbativo delle D-brane; quando la direzione che si utilizza per 
implementare il meccanismo di Scherk-Schwarz \`e ortogonale al volume della 
brana, lo spettro di bassa energia degli stati che vivono su di essa resta 
supersimmetrico. La massa del gravitino \`e quindi sempre
proporzionale al volume della variet\`a interna, ma la massa del gaugino resta 
nulla a livello ad albero; la rottura della supersimmetria sulla brana
viene mediata da 
interazioni gravitazionali, soppresse dalla massa di Planck $m_p$ e risulta
quindi
effettivamente $m_{1/2} \sim O(m^2_{3/2}/m_p)$. Dal punto di vista teorico 
questi vuoti sono di particolare interesse in quanto 
duali di vuoti della M teoria nei quali la 
supersimmetria \`e rotta spontaneamente utilizzando il meccanismo di Scherk e 
Schwarz lungo l'undicesima dimensione. Essi sono anche esempi di nuove 
compattificazioni della teoria in cui la gravit\`a vive nell'intero spazio 
10-dimesionale mentre campi di gauge ed altri campi di materia vivono su una 
brana di dimensione inferiore, in particolare su una 3-brana che pu\`o 
riprodurre il nostro spazio quadridimensionale. I campi che vivono sulla brana 
avvertono le altre dimensioni solo attraverso le interazioni gravitazionali e 
questo consente di avere dimensioni interne anche dell'ordine di un millimetro, 
dato che gli unici vincoli sperimentali provengono dalle verifiche della legge 
dell'inverso del quadrato nella gravit\`a Newtoniana.
Sono stati costruiti modelli in nove 
dimensioni che mettono in luce le caratteristiche essenziali del meccansimo e 
un modello in sei dimensioni, per chiarire il comportamento della teoria quando 
nello spettro sono presenti sia D9 che D5 brane. Infine sono stati costruiti
modelli in quattro dimensioni con rottura parziale della supersimmetria e 
stiamo attualmente studiando modelli che rompono spontaneamente 
$N=1$ ad $N=0$ dando uno spettro chirale.


\def\NPB#1#2#3{{\it Nucl.~Phys.} {\bf{B#1}} (19#2) #3}
\def\PLB#1#2#3{{\it Phys.~Lett.} {\bf{B#1}} (19#2) #3}
\def\PRD#1#2#3{{\it Phys.~Rev.} {\bf{D#1}} (19#2) #3}
\def\PRL#1#2#3{{\it Phys.~Rev.~Lett.} {\bf{#1}} (19#2) #3}
\def\CMP#1#2#3{{\it Comm.~Math.~Phys.} {\bf{#1}} (19#2) #3}
\def\JGP#1#2#3{{\it J.~Geom.~Phys.} {\bf{#1}} (19#2) #3}
\def\IJMP#1#2#3{{\it Int.~Jour.~Mod.~Phys.} {\bf{#1}} (19#2) #3}
\def\CQG#1#2#3{{\it Class.~Quant.~Grav.} {\bf{#1}} (19#2) #3}
\def\ZPC#1#2#3{{\it Z.~Phys.} {\bf C#1} (19#2) #3}
\def\PTP#1#2#3{{\it Prog.~Theor.~Phys.} {\bf#1}  (19#2) #3}
\def\MPLA#1#2#3{{\it Mod.~Phys.~Lett.} {\bf#1} (19#2) #3}
\def\PR#1#2#3{{\it Phys.~Rep.} {\bf#1} (19#2) #3}
\def\AP#1#2#3{{\it Ann.~Phys.} {\bf#1} (19#2) #3}
\def\RMP#1#2#3{{\it Rev.~Mod.~Phys.} {\bf#1} (19#2) #3}
\def\HPA#1#2#3{{\it Helv.~Phys.~Acta} {\bf#1} (19#2) #3}
\def\JETPL#1#2#3{{\it JETP~Lett.} {\bf#1} (19#2) #3}


\chapter{Formulazione perturbativa della teoria di stringhe}
\markboth{} {}

In questo capitolo consideriamo la pi\`u semplice azione 
classica per descrivere il moto di un oggetto esteso nello spazio-tempo.
Vedremo 
come  nel caso delle stringhe si sia  portati in modo naturale a studiare
teorie di campo conformi, come la teoria possa essere quantizzata e
descriveremo lo spettro dei vari modelli. Passeremo poi ad illustrare 
le interazioni tra questi stati e le azioni 
effettive per i modi di vibrazione pi\`u leggeri che 
dominano la dinamica a bassa energia. Nel corso della discussione emergeranno 
le regole essenziali di consistenza per questo tipo di modelli.


\section{Stringa bosonica}
\markboth{} {}

Per un oggetto con p=d-1 dimensioni spaziali (p-brana) che si propaga in uno 
spazio tempo D-dimensionale con metrica $ \h_{\m\n}=(-+...+) $ consideriamo 
l'azione di Nambu-Goto
\be
S= -T\int{d^d\x\sqrt{-g}}   \ ,
\ee
dove $ g = det g_{\a\b} $ e
$ g_{\a\b} $ \`e la metrica indotta sul volume della p-brana 
\be
 g_{\a\b}=\h_ {\m\n}\partial_\a{X^\m}\partial_\b{X^\n}    \ .
\ee
Possiamo anche scrivere un'azione classicamente equivalente considerando 
$ g_{\a\b} $ 
una variabile indipendente
\be
S= T\int{d^d\x( \frac{1}{2}\sqrt{-g}g^{\a\b}\partial_\a{X^\m}\partial_\b{X^\n}\h
_{\m\n}+\l \sqrt{-g}} )   \  .
\label{ng}
\ee
L'equazione del moto per $ g_{\a\b} $ \`e infatti
\be
\frac{\delta S}{\delta g^{\a\b}} =  \frac{T}{4}( g_{\a\b}g^{\r\s}
\partial_\r{X^\m}
\partial_\s{X^\n}\h_{\m\n} \nonumber \\
 -2\partial_\a{X^\m}\partial_\b{X^\n}\h_{\m\n} )-\frac{\l}{2}g_{\a\b} = 0  
\quad  ,
\ee
e prendendo la traccia
\be
(d-2)Tg^{\a\b}\partial_\a{X^\m}\partial_\b{X^\n}\h_{\m\n}=2\l d  \   .
\ee
Si vede, quindi, che in dimensione diversa da due e con
 $ \l=T\frac{(d-2)}{2} $ ,  $  g_{\a\b} $ 
corrisponde alla metrica indotta, mentre, per $d=2$,     $\l=0$ e $ g_{\a\b} $ 
\`e solo conformemente equivalente alla metrica indotta. Nel caso $ d=2 $, che 
corrisponde alla stringa, l'azione (\ref{ng}), oltre ad essere 
invariante rispetto a trasformazioni di Poincar\`e
\be
\d X^{\m} = \w^{\m}{}_{\n} X^\n + a^{\m} \ ,  \hspace{2cm}    \d g_{\a\b} = 0 \ ,
\label{sng1}
\ee 
e rispetto a diffeomorfismi
\be
\d g_{\a\b}= \btd_\a \x_\b+\btd_\b \x_\a  \ , \hspace{2cm} \d X^\m= \x^\a 
 \partial_\a{X^\m}   \  ,
\label{sng2}
\ee
\`e anche invariante rispetto a trasformazioni di Weyl
\be
\d X^\m = 0 \ ,  \hspace{2cm} \d g_{\a\b}=2\L(\x)g_{\a\b}   \  .
\label{sng3}
\ee
L'invarianza di Weyl della teoria bidimensionale \`e essenziale per la 
costruzione di una teoria quantistica consistente; come vedremo le teorie di 
campo conformi (CFT) rappresentano soluzioni "classiche" della teoria di 
stringhe, e si riscontra una relazione profonda tra propriet\`a bidimensionali 
 (invarianza conforme, invarianza modulare, algebre di 
correnti) e propriet\`a spaziotemporali (assenza di divergenze,
 cancellazione delle anomalie, supersimmetria). \`E questa caratteristica che 
seleziona le stringhe tra gli altri oggetti estesi; in due dimensioni si ha un 
grande controllo sulla teoria quantistica e si ottiene uno spettro discreto, 
mentre non \`e noto come quantizzare le altre brane. Inoltre si pu\`o avere 
uno spettro continuo, come \`e stato dimostrato per la membrana della 
supergravit\`a in undici dimensioni. Torneremo comunque a discutere queste 
brane, essenziali nella teoria, quando parleremo delle dualit\`a.

Concentriamo quindi la nostra attenzione sulle stringhe; in $ d=2 $ l'azione 
(\ref{ng}) diviene
\be
S= - \frac{T}{2}\int{d^d\x 
 \sqrt{-g}g^{\a\b}\partial_\a{X^\m}\partial_\b{X^\n}\h_{\m\n}  }   \   ,
\ee
e variando S rispetto alla metrica si ottiene il tensore energia-impulso
\be
T_{\a\b} = - \frac{2}{T}\frac{1}{\sqrt{-g}}\frac{\d S}{\d g^{\a\b}} = 
\partial_\a{X}\partial_\b{X} - \frac{1}{2} g_{\a\b}g^{\r\s}\partial_\r{X}
\partial_\s{X}   \  .
\label{vin}
\ee
Variando S rispetto a $ X^{\m} $ otteniamo le equazioni del moto
\be
\partial_{\a}{(\sqrt{-g}g^{\a\b}\partial_{\b}{X^\m})}=0  \  ,
\ee
e le simmetrie (\ref{sng1}) - (\ref{sng3}) danno origine alle correnti
\be
P^{\a}{}_{\m}= -T\sqrt{-g}g^{\a\b}\partial_{\b}{X_\m}   \ ,
\label{gen1}
\ee
\be
J^{\a}{}_{\m\n} = -T\sqrt{-g}g^{\a\b}(X_{\m}\partial_{\b}{X_{\n}}-
X_{\n}\partial_{\b}{X_{\m}})   \  ,
\label{gen2}
\ee
e quindi alle corrispondenti cariche conservate $ P_{\m} $ e $ J_{\m\n} $. 
Osserviamo che su 
una superficie \`e sempre possibile scegliere un sistema di coordinate 
locali, dette coordinate $ \it{isoterme} $, nelle quali la metrica \`e 
conformemente piatta
\be
g_{\a\b}(\x)=e^{2\L (\x)}\h_{\a\b}  \  .
\ee
Sono stati sviluppati diversi metodi per quantizzare la teoria. Un primo metodo 
\`e la $ \bf{quantizzazione} $  $ \bf{ canonica} $
$ \bf{ covariante} $.

Nel gauge conforme l'azione diviene
\be
S = \frac{T}{2} \int{d \s d \t ( \dot{X}^2-X^{'2} )}   \ ,
\ee
dove $\t \in R$ e $ \s \in [0,2 \p]$ parametrizzano il worldsheet;
l'equazione del moto \`e l'equazione delle onde in due dimensioni. Se
scegliamo per il campo $ X^\m $ condizioni al bordo periodiche 
\be
X^\m (\t,\s + 2\p)=X^\m (\t,\s)   \  ,
\ee
abbiamo una teoria di stringhe chiuse ; se imponiamo condizioni al bordo di 
Neumann
\be
\partial_{\s}{X^\m}(\t,0)=\partial_{\s}{X^\m}(\t,\p)=0  \  ,
\ee
abbiamo una teoria di stringhe aperte. Discuteremo condizioni al contorno di 
Dirichlet nel terzo capitolo. Nel caso della stringa chiusa la soluzione si 
pu\`o scrivere come somma di un'onda progressiva e di un onda regressiva
 $ X^\m (\t,\s)=X^\m_{L}(\t + \s)+X^\m_{R}(\t-\s) $ e l'espansione
in modi normali \`e la seguente
\be
X^\m_{L} (\t + \s) = \frac{x^\m}{2}+\frac{p^\m}{2\p T}(\t+\s)+
\frac{i}{\sqrt{4\p T}}\sum_{k\neq 0}{ \frac{\bar{\a_k}}{k}e^{-ik(\t + \s)}} \ ,
\ee
\be
X^\m_{R} (\t - \s) = \frac{x^\m}{2}+\frac{p^\m}{2\p T}(\t-\s)+
\frac{i}{\sqrt{4\p T}}\sum_{k\neq 0}{ \frac{\a_k}{k}e^{-ik(\t - \s)}} \ ,
\ee
mentre per la stringa aperta abbiamo
\be
X^\m (\t, \s) = {x^\m} +\frac{p^\m \t}{\p T}+
\frac{i}{\sqrt{\p T}}\sum_{k\neq 0}{ \frac{\a_k}{k}e^{-ik\t}cos(k\s)}  \ .
\ee
Se introduciamo delle coordinate di cono di luce sul worldsheet
\be
\x_+ = \t + \s \ ,  \hspace{3cm} \x_- = \t - \s  \ ,
\label{lc}
\ee
le componenti del tensore energia-impulso sono
\be
T_{++}= \frac{1}{8}( \dot{X}+X^{'})^{2}  \ , \hspace{1cm}
T_{--}= \frac{1}{8}( \dot{X}-X^{'})^{2}  \ , \hspace{1cm}
T_{+-}=T_{-+}=0   \ ,
\ee 
e i vincoli (\ref{vin}) diventano $ (\dot{X} \pm X^{'})^2 =0 $. Possiamo 
esprimere questi vincoli in modo equivalente introducendo gli operatori di 
Virasoro, definiti
come componenti di Fourier del tensore energia-impulso. Per la stringa 
chiusa abbiamo quindi due famiglie di operatori
\be
L_n = 2T \int_{0}^{2\p}{d \s T_{--}e^{-in(\t - \s)}}  \ ,  \hspace{1cm}
\bar{L}_n = 2T \int_{0}^{2\p}{d \s T_{++}e^{in(\t + \s)}}  \ ,
\ee
e in termini degli oscillatori
\be
L_n = \frac{1}{2}\sum_m{ \a_{n-m} \a_m} \  , \hspace{1cm} 
\bar{L_n} = \frac{1}{2}\sum_m{ \bar{\a}_{n-m} \bar{\a}_m} \ .
\ee
Per la stringa aperta abbiamo un solo tipo di operatore dato da
\be
L_n = 2T \int_{0}^{\p}{d \s [T_{--}e^{-in(\t - \s)}+T_{++}e^{in(\t + \s)}]} =
      \frac{1}{2}\sum_m{ \a_{n-m} \a_m} \ ,
\ee
e in entrambi i casi quantizziamo la teoria imponendo le relazioni di 
commutazione canoniche
\be
[X^{\m}(\t,\s), \dot{X}^{\n}(\t,\s^{'})] = \frac{\h^{\m\n}}{T} \d (\s - \s^{'}) \
 .
\ee
Per la stringa chiusa abbiamo quindi
\be
[x^{\m},p^{\n}]=i \h^{\m\n}   \  ,  \hspace{1cm}
[\a^{\m}_m, \a^{\n}_n]=m\d_{m+n,0} \h^{\m\n}  \  ,  \hspace{1cm}
[\bar{\a}^{\m}_m, \bar{\a}^{\n}_n]=m\d_{m+n,0} \h^{\m\n} \ ,
\label{op}
\ee
e per la stringa aperta abbiamo le stesse espressioni salvo l'assenza dei modi 
$ \bar{\a}^{\n}_{n} $. La realit\`a di $ X^{\m} $  d\`a infine
$  ( \a^{\n}_n )^{\dag}= \a^{\n}_{-n} $.

Lo spazio di Hilbert della teoria \`e costruito partendo da uno stato 
fondamentale $ |k^{\m} \ran $ caratterizzato da
\be
p^{\m} |k^{\m} \ran = k^{\m}|k^{\m} \ran   \  ,  \hspace{1cm}
\a^{\m}_{m}|k^{\m} \ran = 0  \  ,  \hspace{1cm} m > 0  \ ,
\ee
e agendo con i modi di frequenza negativa. In questo schema di quantizzazione 
l'invarianza di Lorentz \`e manifesta; possiamo facilmente scrivere i 
generatori dell'algebra (\ref{gen1}) - (\ref{gen2}) in termini degli operatori 
in (\ref{op}) e verificare le relazioni di commutazione. Tutti gli 
operatori sono ordinati normalmente; in particolare per gli $ L_n $ con
$ n \neq 0 $  si ha  $ L_n = : L_n:, $ mentre $ L_0 $ \`e sensibile
all'ordinamento normale e  $ L_0 = :L_0: - a $, con $ a $ una costante
per il momento indeterminata. Concentriamoci sui soli operatori 
$L_n$, o se si vuole sulla stringa aperta; la discussione per gli $\bar{L}_n$ 
\`e del tutto analoga. Si pu\`o mostrare che  gli operatori $ L_n $ realizzano
un'algebra nota come algebra di Virasoro
\be
[L_n,L_m] = (n-m)L_{n+m} + \frac{D}{12}n(n^2-1) \d_{n+m,0} \ .
\label{vir}
\ee
La quantizzazione covariante presenta delle sottigliezze perch\`e lo spazio di 
Hilbert contiene stati con norma negativa. Ad esempio
\be
\lan k| \a^{\m}_{1} \a^{\n}_{-1} |k^{'} \ran = (2\p)^{D} \d (k-k^{'}) \h^{\m\n} \
 .
\ee
Per ottenere una teoria unitaria \`e necessario selezionare un sottospazio 
contenente 
solo stati con norma positiva. Questo \`e possibile imponendo che gli 
stati fisici soddisfino le condizioni
\be
L_{n} | \f \ran =0 \hspace{1cm} n > 0  \,
\ee
\be
(L_0 - a) | \f \ran = 0 \ .
\label{lo}
\ee
L'estensione centrale dell'algebra non consente di porre
$ L_{n} | \f \ran =0 $ per $ n $ negativo.
In questo modo il vincolo (\ref{vin}) viene introdotto anche nella teoria 
quantistica, e uno studio dettagliato del 
sottospazio degli stati fisici mostra che questo contiene esclusivamente
stati a norma positiva per $ D \leq 26 $ e $ a \leq 1 $.
Da eq. (\ref{lo}) possiamo ricavare la 
condizione di mass-shell ; essendo $ m^2 = -p^2 $ e definendo l'operatore 
numero $ N = \sum_{n=1}^{\infty} { \a_{-n} \a_{n} } $, abbiamo
\be
m^2= \frac{1}{\a{'}} ( N-a ) \ .
\ee
Per le stringhe chiuse si ha invece
\be
m^2= \frac{2}{\a{'}} ( N + \bar{N} -2a ) \ .
\ee
Inoltre dall'espressione analoga per $ \bar{L_0} $ si ottiene la condizione 
$ N = \bar{N} $, l'unica relazione tra i settori destro e sinistro della
teoria.
Gli stati della forma $ L_{-n}|\chi \rangle , n>0 $ sono ortogonali a tutti 
gli stati fisici e sono detti $  \it{spuri} $; quando sono anche fisici vengono 
detti $ \it{nulli} $ e sono equivalenti al vettore nullo : per questo 
motivo gli stati 
fisici sono definiti come classi di equivalenza modulo stati nulli. Per 
studiare lo spettro della teoria dobbiamo quindi identificare livello per 
livello 
stati fisici e stati nulli. Quando $ N=0 $ abbiamo un tachione; quando $ N=1 $ 
lo stato pi\`u generale \`e $ |\e k \ran = \e_{\m} \a^{\m}_{-1} |k \ran $ e la 
condizione $ L_1 |\e k \ran =0 $ d\`a  $ \e^{\m} k_{\m} = 0 $,  mentre
$ L_0 |\e k \ran =0 $ corrisponde a $ m^2 = - k^2 = \frac{1-a}{\a^{'}} $.
C' \`e uno stato spurio $ L_{-1}|k\ran = 2k \a_{-1}|k\ran $ che \`e nullo se $ 
k^2 =0 $; in questo modo per $ D=26 $ e $ a=1 $ l'equivalenza modulo stati 
nulli diviene per gli stati con $ N=1 $ l'usuale invarianza di gauge
 $ \e^\m \sim \e^\m + k^\m $. Il fatto che stati nulli siano presenti ad ogni 
livello fa pensare che l'invarianza di gauge sia solo parte di una simmetria 
molto pi\'u vasta della teoria di stringhe. In modo del tutto analogo si 
verifica che per la stringa bosonica chiusa per $ N=\bar{N}=1 $ gli stati 
$ \e_{\m\n} \a^\m_{-1} \a^\n_{-1} |k \ran $ soddisfano
$ k^\m \e_{\m\n}= n^\n \e_{\m\n} = 0 $ con $ \e_{\m\n} $ definito a meno di 
$ \e_{\m\n} \sim \e_{\m\n} + k_\m \z_\n + \z^{'}_\m k_\n $ dove 
$ k \z= k \z^{'} =0 $.

Un altro metodo, la quantizzazione nel $ \bf{cono} $
 $ \bf{di} $ $ \bf{luce}$,
presenta il vantaggio che tutti gli stati sono fisici. Introduciamo le 
coordinate $ X^{\pm} = X^0 \pm X^D $ nello spaziotempo; sul worldsheet,
fissato il gauge conforme e nelle coordinate (\ref{lc}), abbiamo ancora
la possibilit\`a di effettuare trasformazioni del tipo 
\be
\x^{'}{}_{+}= f(\x_{+})   \ ,  \hspace{3cm} \x^{'}{}_{-}= g(\x_{-}) \ .
\ee
Se sfruttiamo queste trasformazioni per porre $ X^{+} = x^{+} + \a^{'}p^{+}\t $
, \`e possibile risolvere i vincoli esprimendo $ X^{-} $
in funzione delle coordinate trasverse $ X^{i}  $ e procedere quindi 
alla quantizzazione: lo spazio di Hilbert conterr\`a manifestamente solo
stati con norma positiva,  mentre si dovr\`a verificare accuratamente
l'invarianza di Lorentz della teoria. Otteniamo infatti
\be
( \dot{X} + X^{'} )^{-} = \frac{(\dot{X} + X^{'})^{2}}{\a^{'}p^{+}} \ ,
\ee
e possiamo quindi esplicitare
\be
\a^{-}_{n} = \frac{2}{\sqrt{2\a^{'}}p^{+}}L_{n}  \ ,
\ee
e analogamente per $ \bar{\a}^{-}_{n} $ . Utilizzando $ \a^{-}_{0} = 
\frac{p^{-}}{\sqrt{\p T}} $, si ottiene la condizione di mass-shell
\be
\a^{'}m^{2} = 4( N - a )  \ .
\ee
Si pu\`o quindi verificare che gli operatori $ J^{\m\n} $ soddisfano l'algebra
di Lorentz per $ d=26 $ e $ a=1 $ ; in particolare l'unico commutatore da 
calcolare con cautela \`e $ [j^{i-},j^{j-}] $ .
\`E semplice interpretare lo spettro della teoria. Nel caso della stringa 
chiusa abbiamo uno stato fondamentale tachionico $ |k^\m \rangle $ con $ m^{2} =
 - \frac{4}{\a{'}} $, mentre gli stati 
$ \a^i_{-1} \bar{\a}^j_{-1} |k^{\m} \ran $  con
$ m^2 = 0 $ si decompongono in rappresentazioni irriducibili di $ SO(24) $
\be
\a^i_{-1} \bar{\a}^j_{-1}|k \ran = \a^{[i}_{-1} \bar{\a}^{j]}_{-1}|k \ran +
 [ \a^{(i}_{-1} \bar{\a}^{j)}_{-1}|k \ran -
 \frac{ \d^{ij}}{24} \a^k_{-1} \bar{\a}^k_{-1}|k \ran ] 
 + \frac{ \d^{ij}}{24} \a^k_{-1} \bar{\a}^k_{-1}|k \ran   \ ,
\ee
 e possono essere interpretati come un gravitone $ G_{\m\n} $ 
, un tensore antisimmetrico $ B_{\m\n} $ ed uno scalare, il dilatone,
$ \f $ . Al livello successivo gli stati possono ancora essere organizzati in 
rappresentazioni di $ SO(24) $ e risultano via via pi\`u massivi.
Per la stringa aperta la condizione di mass-shell diviene
 $ m^2 = \frac{1}{\a^`}(N-1) $, lo stato fondamentale \`e ancora una volta 
tachionico mentre gli stati $ \a^i_{-1} |k^\m \ran $ hanno massa nulla e 
trasformano come un vettore di $ SO(24) $.

Per discutere la teoria dal punto di vista dell'integrale funzionale 
effettuiamo una continuazione analitica $ \t \mapsto -i \t $ in modo da avere 
un worldsheet con segnatura euclidea $ (+,+) $. Il passaggio dalla metrica 
euclidea a quella minkowskiana \`e molto delicato nella gravit\`a 
quadridimensionale, ma nel nostro caso \`e possibile dimostrare l'equivalenza 
delle due formulazioni \cite{D'Ho}. Introducendo le coordinate complesse
$ w= \t +i \s $ , $ \bar{w} = \t + i \s $, nel gauge conforme l'azione diviene
\be
S = T \int{d^2w    \partial_w{X} \partial_{\bar{w}}{X}}  \  .
\ee
Ora la teoria \`e definita sulla striscia $ [0,2\p] \times R $, ma la 
trasformazione conforme $ z = e^w $ consente di definirla su
 $ C^* = C \backslash \{0\} $. In queste coordinate le traslazioni temporali 
corrispondono a dilatazioni e le traslazioni spaziali a rotazioni ; \`e 
quindi naturale interpretare come hamiltoniana della teoria il generatore delle 
dilatazioni. Poich\`e l'invarianza conforme \`e essenziale per la consistenza 
perturbativa della teoria di stringhe, illustriamo brevemente i principali 
aspetti delle teorie di campo conformi in due dimensioni.


\section{Teorie di campo conformi}   
\markboth{} {}

In uno spazio $ R^{(p,q)} $, dove $ (p,q) $ indica la segnatura della metrica, 
l'algebra del gruppo conforme per $ p+q \neq 2 $ \`e isomorfa a
quella di  $ O(p+1,q+1) $, mentre in due dimensioni \` e un'algebra 
infinito dimensionale. Se fissiamo una segnatura euclidea ed utilizziamo 
coordinate complesse, i generatori sono $ l_n = -z^{n+1} \partial $ e
$ \bar{l}_n = -\bar{z}^{n+1} \bar{\partial} $, con $ n \in Z $ e soddisfano le 
relazioni
\be
 [l_n,l_m]=(n-m)l_{n+m} \  ,  \hspace{0.5cm}
 [\bar{l}_n,\bar{l}_m]=(n-m)\bar{l}_{n+m} \ ,  \hspace{0.5cm}
   [l_n,\bar{l}_m]=0 \ . 
\ee
Se consideriamo  $ z $ e $ \bar{z} $ come variabili indipendenti abbiamo 
effettivamente la somma di due algebre isomorfe e la teoria si separa in due 
settori che dal punto di vista della teoria di stringhe  corrispondono
rispettivamente ai modi destri e ai modi sinistri. Il sottogruppo $ Sl(2,C) $ 
generato da
$ \{l_{-1},l_{0},l_1 \} \cup \{\bar{l}_{-1},\bar{l}_{0},\bar{l}_1 \} $
comprende solo le trasformazioni globalmente definite su $ S^2 $ e 
per questo \`e
detto gruppo conforme globale. Lo studio sistematico di teorie conformi 
quantistiche bidimensionali ha avuto inizio con l'importante articolo di 
Belavin, Polyakov e Zamolodchikov \cite{BPZ};
supponiamo di avere una 
famiglia di campi $ \{ \f_i \} $, detti campi primari, che formano un'algebra 
associativa e completa rispetto al prodotto a piccole distanze
\be
\f_i(z) \f_j(w) \sim \sum_k c_{ijk}(z,w) \f_k(w)  \ .
\label{alg}
\ee
Supponiamo inoltre che essi trasformino rispetto a  
cambiamenti di coordinate $ z \mapsto f(z) $ come
\be
\f (z,\bar{z}) \mapsto ( \partial{f} )^h
 ( \bar{\partial}f )^{\bar{h}} \f(f,\bar{f})  \  ,
\ee
dove $ (h,\bar{h}) $ sono i pesi conformi del campo. Per trasformazioni
 infinitesime $ z \mapsto z + \e (z) $ abbiamo quindi 
\be
\d_{\e,\bar{\e}} \f = [ ( h\partial \e + \e \partial ) +
( \bar{h}\bar{\partial}\bar{\e} + \bar{\e} \bar{\partial} )] \f  \  .
\label{var}
\ee
La condizione che il tensore energia impulso abbia traccia nulla si traduce,
in coordinate complesse, in $ T_{z \bar{z}} = 0 $ e le leggi di 
conservazione implicano che le due componenti non nulle, $ T_{zz} $ e
 $ T_{\bar{z} \bar{z}} $, sono rispettivamente olomorfe ed antiolomorfe. 
Abbiamo quindi  un'infinit\`a di cariche conservate
 $ Q_{ \e} = \oint{dz \frac{\e (z)}{2 \p i}T(z)} $ e possiamo riscrivere la 
$ (1.42) $ come $ \d \f = [Q, \f] $. Questi commutatori possono essere 
calcolati facilmente utilizzando la definizione di prodotto R-ordinato ed il 
teorema di Cauchy per esprimerli come integrali sul piano complesso; in 
particolare un campo primario ha la seguente OPE con il tensore energia impulso
\be
T(z)\f (w) = \frac{h \f(w) }{(z-w)^{2}} + \frac{ \partial \f (w) }{z-w} + ... \  ,
\label{tope}
\ee
dove i termini elisi sono operatori con coefficienti non singolari nel limite
 $ z \rar w $.
La forma pi\`u generale del prodotto di $ T $ con se stesso mostra come esso 
non sia un campo primario per la presenza di un polo del quarto ordine
\be
T(z)T(w) = \frac{c}{2} \frac{1}{(z-w)^4}+\frac{2T(w)}{(z-w)^2}+
\frac{\partial T(w)}{z-w} \label{tt}  \   .
\ee
La costante $ c $, detta carica centrale, \`e legata ad una anomalia 
conforme che si manifesta, quando la teoria viene definita su di una superficie 
curva, in un mancato disaccoppiamento del fattore conforme della metrica;
si ha infatti $ T^{\a}{}_{\a}=\frac{c}{96 \p^3} \sqrt{g} R $,   
dove $ R $ \`e la
curvatura della superficie. Questa legge di trasformazione
 pu\`o essere integrata dando, per trasformazioni finite $ z \mapsto f(z) $,
\be
T(z) \mapsto (\partial f)^2 T(f) + \frac{c}{12}
\left ( \frac{\partial^3 f}{\partial f} - \frac{3}{2}
 \left (\frac{\partial^2 f}{\partial f} \right )^2 \right ) \   ,
\ee
dove il termine aggiuntivo non \`e altro che la derivata di Schwartz. 
L'invarianza 
conforme pone forti vincoli sulle funzioni di correlazione: determina quelle 
a due punti 
\beq
< \f_i(z_1) \f_j(z_2) > = 0 \hspace{1cm} se \hspace{1cm} 
h_1 \ne h_2   , \nonumber \\
= \frac{c_{ij}}{r_{12}^{2h}} \hspace{1cm} se \hspace{1cm} h_1 = h_2  \  ,
\eeq
dove $r_{ij} = |z_i - z_j|$ e le costanti $c_{ij}$ vengono fissate dalla 
normalizzazione dei campi, e determina anche quelle a tre punti 
\be
< \f_i(z_1) \f_j(z_2) \f_k(z_3) > = \frac{c_{ijk}}
{r_{12}^{h_i+h_j-h_k}r_{23}^{h_j+h_k-h_i}r_{31}^{h_k+h_i-h_j}}  \  ,
\ee
a meno delle costanti $ c_{ijk} $ che dipendono dal particolare modello 
studiato.
Le funzioni a quattro o pi\`u punti dipendono invece in modo non banale da 
birapporti.
L'OPE in (\ref{alg}) si riscrive
\be
\f_i(z) \f_j(w) = \sum_k c_{ijk} (z-w)^{h_k-h_i-h_j}
(\bar{z}-\bar{w})^{\bar{h}_k-\bar{h}_i-\bar{h}_j} \f_k(w)  \  ,
\ee
e si pu\`o verificare che i coefficienti $ c_{ijk} $ sono gli stessi che 
compaiono nella funzione a tre punti.
Una teoria conforme \`e completamente specificata una volta assegnata
la sua carica centrale, il suo contenuto di campi primari e le correlazioni
tra primari (determinate dai $c_{ijk}$); in casi particolari
\`e nota la soluzione 
completa, come per i modelli minimali, le algebre affini o gli orbifolds. In 
teoria di stringhe si considera tipicamente una sottoalgebra di campi 
mutuamente locali, vale a dire campi il cui prodotto operatoriale \`e ad un 
solo
valore, in modo che le funzioni di correlazione siano ben definite su ogni 
superficie di Riemann; ulteriori restrizioni derivano dal richiedere 
che anche la dipendenza dai moduli della superficie
sia ad un solo valore.

\`E utile 
espandere i campi in serie di Laurent e calcolare le relazioni di commutazione 
tra i modi; l'informazione contenuta in questi commutatori \`e del tutto 
equivalente all'espansione in prodotto di operatori dei campi. Se per ogni 
campo 
primario scriviamo $ \f (z) = \sum_{n \in Z} \f_n z^ {-n-h} $ e la relazione 
inversa $ \f_n = \oint{ \frac{dz}{2 \p i} \f(z) z^{n+h-1} } $, la (\ref{tope})
e la (\ref{tt}) divengono 
\be
[l_n,l_m] = l_{n+m} + \frac{c}{12} n(n^2-1) \d_{n+m,0}  \  ,
\ee
\be
[l_n, \f_{m}] = [n(h-1)-m] \f_{n+m}   \  ,
\ee
L'algebra degli operatori $ l_n $ \`e detta algebra di Virasoro. Le sue 
rappresentazioni possono essere costruite partendo da un vettore di peso 
massimo $ |h \ran $ tale che
\be
l_0 |h\ran = h |h\ran  \  , \hspace{1cm} l_n |h\ran = 0 
\hspace{1cm} \forall n>0   \  .
\ee
Gli stati $ l_{-n_1}...l_{-n_{k}} |h\ran $ con $ \sum_{i=1}^k n_i = N $ sono i 
discendenti di livello $ N $;  $ |h\ran $ insieme ai suoi discendenti 
costituisce il modulo di Verma, una rappresentazione in genere riducibile 
dell'algebra. Ad ogni campo primario di peso $ h $ possiamo associare uno 
stato $ |h \ran = \f (0) |0\ran $, ed \`e facile verificare che si tratta di uno 
stato di peso massimo. I discendenti sono creati dai campi
\be
L_{-n}\f (z) = \oint{ \frac{dw}{2 \p i} w^{n-1}T(w) \f (z)}  \  ,
\ee
e si pu\`o mostrare utilizzando le identit\`a di Ward della teoria che le 
funzioni di correlazione dei campi discendenti sono determinate da quelle dei 
campi primari. Il carattere di una rappresentazione associata ad uno stato di 
peso massimo \`e dato da
\be
\chi_h(q) = tr( q^{l_0-\frac{c}{24}}) =
\frac{q^{h-\frac{c}{24}}}{\prod_{n=1}^{\infty}(1-q^n)}  \  ,
\ee
e conta il numero di discendenti ai vari livelli.
Definiamo infine l'aggiunto di un operatore:
\be
\f ^{\dag} (z,\bar{z}) = \f ( \frac{1}{\bar{z}}, \frac{1}{z} )(\bar{z})^{-2h}
(z)^{-2\bar{h}}  \  .
\ee
Imponendo che $ \lim_{z \rar 0} \f (z)|0\ran $ sia regolare si ottiene
$ \f_n |0\ran=0 $ per $ n \geq 1-h $; in modo analogo $ \lan 0| \f_n =0 $ per
$ n \leq h-1 $. In particolare il vuoto \`e annullato da $ l_{-1}$,$l_0$ e 
$l_1$ ( e dai corrispondenti operatori barrati ) e viene quindi 
detto vuoto $ SL_2C$ 
invariante.

Spesso l'algebra di Virasoro pu\`o essere realizzata partendo da un'algebra di 
simmetrie pi\`u vasta. Un caso particolarmente importante \`e quello delle 
algebre affini, caratterizzate da operatori $ j^a(z) $ di 
dimensione $ (1,0) $, con una espansione in modi $ j^a(z) = \sum_{n\in Z} j^a_n
z^{-n-1}$, che soddisfano le relazioni
\be
[j^a_m,j^b_n]= mG^{ab} \d_{m+n,0} + if^{ab}{}{}_c j^{c}_{m+n}  \  .
\label{km}
\ee

Utilizzando l'associativit\`a del prodotto di operatori si pu\`o verificare che 
$ f^{abc}=f^{ab}{}{}_dG^{dc} $ sono quantit\`a totalmente antisimmetriche e 
rappresentano le costanti di struttura di un'algebra di Lie con metrica 
invariante $ G^{ab} $. Le rappresentazioni vengono costruite partendo da stati 
primari affini $ |r_i\ran $ caratterizzati da
\be
j^a_0 |r_i\ran = i(t_r)^a_{ij}|r_j\ran  \  , \hspace{1cm} j^a_n|r_i\ran = 0 
\hspace{1cm} \forall n>0   \  .
\ee
dove le $t_r$ sono matrici in una qualche rappresentazione dell'algebra di Lie.
Gli altri stati si ottengono agendo con i modi di frequenza negativa. Una 
restrizione sulle possibili rappresentazioni dell'algebra di Lie che possono 
utilizzarsi per costruire rappresentazioni dell'algebra affine viene 
dall'unitariet\`a: solo particolari rappresentazioni sono ammesse e sono dette 
integrabili. Consideriamo dei gruppi semplici compatti e diagonalizziamo la 
metrica di Cartan-Killing ponendo $ G^{ab}=k \d^{ab} $; 
la costante $k$ \`e detta livello dell'algebra affine. 
L'operatore
\be
T(z) = \frac{1}{k+\bar{h}} : j^a(z)j^a(z) : \ ,
\ee
soddisfa l'algebra di Virasoro con una carica centrale
$ c = \frac{kD}{k+\bar{h}} $,  dove $D$ \`e la 
dimensione del gruppo e $\bar{h}$ \`e
il numero di Coxeter duale, definito come la traccia, opportunamente normalizzata, del
Casimir quadratico nell'aggiunta. Gli operatori che creano gli stati pimari affini 
sono anche primari dell'algebra di Virasoro con peso
$ h_r = \frac{c_r}{k+\bar{h}} $, dove $c_r$ \`e il Casimir quadratico della 
rappresentazione.

Consideriamo ora una serie di semplici esempi che saranno particolarmente 
utili nel 
seguito. Il primo \`e il caso di un bosone 
libero descritto dall'azione
\be
S = \frac{1}{4 \p} \int{d^2z \partial X(z) \bar{\partial} X(z)}  \  .
\ee
La funzione a due punti \`e $ \lan X(z)X(w) \ran = - ln |z-w|^2 $ che, 
limitandoci alla parte olomorfa della teoria, scriviamo come
\be
\lan X(z)X(w) \ran = - ln (z-w)  \  .
\label{cbos}
\ee
Il tensore di energia impulso \`e 
\be
T(z) = -\frac{1}{2} : (\partial X)^2 : =
 - \frac{1}{2} \lim_{w \rar z} \left [ \part{X(w)} \part{X(z)} + \frac{1}{(z-w)^2} \right ] \ ,
\ee
dove il prodotto normalmente ordinato \`e definito rimuovendo la 
singolarit\`a nel limite in cui i due operatori coincidono; utilizzando
il teorema di Wick si ottiene
\be
T(z) \part X(w) \sim \frac{\part X(w)}{(z-w)^2} + \frac{\part^2 X(w)}{z-w}
+ reg  \  ,
\ee
\be
T(z)T(w) \sim \frac{1}{2(z-w)^4} + \frac{T(w)}{(z-w)^2} +
\frac{2 \part T(w)}{z-w} + reg  \  .
\ee
Abbiamo quindi una teoria conforme con $ c=1 $ e $ \part X $ rappresenta un 
campo primario con $ h = 1 $. Anche gli operatori $ : e^{i k X(z)} :$ sono 
primari con $ h = \frac{k^2}{2} $ e funzione di correlazione
$ \lan e^{-i k X(z)}e^{i k X(z)} \ran = (z-w)^{k^2} $. In generale una 
funzione di correlazione $ \lan e^{i k_1 X(z)}...e^{i k_n X(z)} \ran $
\`e non nulla solo se $ \sum_{i=1}^n k_i = 0 $; questa legge di conservazione
diviene pi\`u chiara se si guarda alla teoria come realizzazione
di un algebra affine $ U(1) $ a livello $ k=1 $. Infatti la corrente \`e data
da $ j(z) = i \part X $, gli operatori $ e^{i k X(z)} $ hanno carica $ k $
e le correlazioni non banali corrispondono a carica totale nulla.
Il carattere \`e dato da $ \frac{q^{ \frac{k^2}{2}}}{\h (q)} $ dove
$ \h (q) = q^{\frac{1}{24}} \prod_{n=1}^{\infty} (1-q^n) $. Una semplice 
variante della teoria si ottiene considerando
\be 
S = \frac{1}{4 \p} \int{d^2z[  \partial X(z) \bar{\partial} X(z)} +
 \frac{Q}{2}\sqrt{g} R X(z)]  \  .
\ee
Il termine aggiuntivo coinvolge la curvatura scalare della superficie e non 
altera la dinamica, ma l'espressione del tensore energia-impulso  diviene
\be
T(z) = -\frac{1}{2} : (\partial X)^2 : + \frac{Q}{2} \part^2 X  \  .
\ee
In questo modo viene alterata sia la carica centrale,
che diviene  $ c = 1 + 3Q^2 $,  sia la 
dimensione degli operatori $ e^{ikX(z)} $, che diviene $ h = \frac{k(k+Q)}{2}$. 
In effetti c'\`e una anomalia nel prodotto, che diventa 
\be
T(z)j(w) \sim \frac{Q}{(z-w)^3} + \frac{j(w)}{(z-w)^2} + 
\frac{\part j(w)}{z-w} + reg  \  ,
\ee
e per questo le funzioni di correlazione sono non nulle solo se
$ \sum_{i=1}^n k_i = Q $.

Descriviamo ora un fermione libero
\be
S= -\frac{1}{8 \p} \int{ d^2z ( \psi(z) \bar{\part} \psi(z) +
\bar{\psi(z)} \part \bar{\psi(z)} )}  \  .
\ee
Dalle equazioni di moto segue che $ \psi $ \`e analitico e $ \bar{\psi} $ 
antianalitico; limitiamoci a considerare la parte analitica.
In questo caso $ \lan \psi(z) \psi(w) \ran = \frac{1}{(z-w)} $ ,
$ T(z) = -\frac{1}{2} : \psi(z) \part \psi(z) : $, e si pu\`o verificare che
$ c= \frac{1}{2} $ e che la dimensione di $ \psi $ \`e $ \frac{1}{2} $.
A questi campi anticommutanti possono essere date due diverse condizioni al 
contorno : periodiche o di Neveu-Schwarz $ \psi(z) = \psi(e^{2 \p i}z) $ e 
antiperiodiche o di Ramond  $ \psi(z) = -\psi(e^{2 \p i}z) $; entrambe le scelte danno
correnti di Noether periodiche. I modi dell'espansione
$ \psi(z) = \sum_{n} \psi_n z^{-n-\frac{1}{2}} $
sono semiinteri nel primo caso ($n \in Z+\frac{1}{2} $) e  interi nel secondo 
$ ( n \in Z ) $.
Un sistema di $N$ fermioni pu\`o essere utilizzato per realizzare un'algebra 
$ O(N)_1 $; infatti l'azione 
$  S= -\frac{1}{8 \p} \int{ d^2z ( \psi^i \bar{\part} \psi^i)} $ ha una 
simmetria globale $ O(N) $, le correnti conservate
$ J^{ij} = i :\psi^i \psi^j : $ verificano l'algebra (\ref{km}) con $ k=1$ e le 
costanti di struttura di $ O(N) $. Le rappresentazioni integrabili sono quella 
di vuoto (O), quella vettoriale (V) e le rappresentazioni spinoriali. In 
particolare, 
per N dispari , si ha una sola 
rappresentazione spinoriale (S), mentre per N pari se ne hanno due , (S) e (C).
Queste sono coniugate tra loro se $ N = 4k + 2 $, e sono autoconiugate se 
$ N = 4k $.
Se definiamo nel settore NS lo stato di vuoto 
imponendo $ \psi^i_n |0 \ran = 0 $ per $ n > 0 $, \`e facile verificare
che  $ |0\ran $ \`e un primario della rappresentazione O mentre gli stati
$ \psi^i|0 \ran $ sono primari per la rappresentazione V. Gli altri stati si 
ottengono agendo con i modi negativi delle correnti; osserviamo che questi 
contengono un numero pari di oscillatori fermionici. Quindi se introduciamo
un operatore che distingue stati con  numeri dispari e stati con  numeri 
pari di oscillatori, possiamo separare nel settore NS la rappresentazione O 
dalla V. Definendo infatti
$ (-1)^F $ con $ F = \sum_{n=\frac{1}{2}}^{\infty} \psi^i_{-n} \psi^i_n $; 
abbiamo $ \{(-1)^F, \psi^i_n\}=0 $ e $ (-1)^F|0\ran = |0 \ran $; possiamo 
costruire
\be
Tr_{NS} q^{L_0-\frac{c}{24}} = (\frac{\th_3}{\h})^{\frac{N}{2}}  , \hspace{2cm}
Tr_{NS}(-1)^F q^{L_0-\frac{c}{24}} = (\frac{\th_4}{\h})^{\frac{N}{2}} \ ,
\ee 
e quindi
\be
\chi_O = Tr_{NS}\frac{(1+(-1)^F)}{2} q^{L_0-\frac{c}{24}} = 
\frac{1}{2}[(\frac{\th_3}{\h})^{\frac{N}{2}}+(\frac{\th_4}{\h})^{\frac{N}{2}}]
\label{caro}  \  ,
\ee
\be
\chi_V = Tr_{NS}\frac{(1-(-1)^F)}{2} q^{L_0-\frac{c}{24}} =
\frac{1}{2}[(\frac{\th_3}{\h})^{\frac{N}{2}}-(\frac{\th_4}{\h})^{\frac{N}{2}}]
\label{carv} \  .
\ee
Passiamo ora al settore di Ramond; dalle relazioni di anticommutazione 
\be
\{\psi_n\psi_m\} = \d_{n+m,0}  \  ,
\ee
si vede che definendo $ \g^i = \sqrt{2} \psi^i_0 $ si realizza
un'algebra di Clifford, e quindi il vuoto \`e uno spinore di $ O(N) $ che 
indichiamo con $ |S\ran $; se
consideriamo $N$ pari, lo spinore si decompone in due 
rappresentazioni di chiralit\`a opposta $ |S\ran $ e $ |C\ran $, con
\be
 \g^{n+1}|S\ran=|S\ran  \ ,  \hspace{2cm} \g^{n+1}|C\ran= - |C\ran   \  ,
\ee
dove l'operatore di chiralit\`a \`e $ \g^{n+1} = \prod_{i=1}^{n} \g^i $.
Ci aspettiamo che agendo con i modi negativi delle correnti su questi stati del 
settore di Ramond si generino le rappresentazioni S e C; possiamo infatti 
verificare che i due stati  in $ (1.60) $ hanno la dimensione corretta data 
dalla $ (1.48) $ calcolando $ \lan S|T(z)|S\ran = \frac{h_S}{z^2} $. Se 
utilizziamo
$ \lan S|\psi^i(z)\psi^j(w)|S\ran = \d^{ij} \sqrt{\frac{z}{w}}\frac{1}{z-w}  $
si ottiene $ h_S = \frac{N}{16} $ e analogamente per $ C$. Introducendo anche 
in questo settore l'operatore
$ (-1)^F = \g^{n+1}(-1)^{\sum_{n=1}^{\infty} \psi^i_{-n}\psi^i_n} $ con
$ (-1)^F|S \ran = |S \ran $ e $ (-1)^F|C \ran = |C \ran $,  possiamo ottenere
\be
\chi_S = \chi_C = (\frac{\th_2}{\h})^{\frac{N}{2}}   \ .
\label{cars}
\ee
Come si vede, abbiamo dovuto introdurre due diversi stati di vuoto, uno per il 
settore NS ed uno per il settore R.
In realt\`a  \`e possibile dare una descrizione pi\`u 
elegante del sistema definendo gli operatori di spin, campi primari che 
trasformano nella rappresentazione spinoriale e creano il vuoto di Ramond 
agendo sul vuoto di Neveu-Schwarz. Per fare questo dobbiamo considerare una 
costruzione tipica delle teorie conformi in due dimensioni, detta 
$ \bf{bosonizzazione} $. Partiamo ancora una volta da 
$ S = -\frac{1}{8\p}\int (\psi^i \bar{\part}\psi^i ) $ con $ i=1,...,2n $ e 
definiamo 
\be
\psi^k = \frac{1}{\sqrt{2}}( \psi^{2k-1} + i\psi^{2k}) , \hspace{1cm}
\psi^{\bar{k}} = \frac{1}{\sqrt{2}}( \psi^{2k-1} - i \psi^{2k})  \   ,
\ee
con $k=1,...,n $; le funzioni di correlazione sono 
$ \lan \psi^k(z) \psi^{\bar{l}}(w) \ran = \frac{\d^{k \bar{l}}}{z-w} $.
Se $ \f^i $  sono $n$  bosoni liberi, gli operatori $ e^{\pm i \a_k \f } $, 
dove 
$ \a^i_k = \d^i_k $ \`e un vettore unitario ad $ n $ componenti nella direzione 
k-esima, descrivono la stessa teoria degli operatori $(\psi^k$ ,$ \psi^{\bar{k}})$
in quanto hanno lo stesso peso conforme, la stessa carica 
centrale e riproducono la funzione di correlazione 
\be
\lan \partial X(z) \partial X(w) \ran = - \frac{1}{(z-w)^2}  \  ;
\ee
questo basta ad 
assicurare una corrispondenza completa tra tutte le funzioni di correlazione
dei due sistemi. In effetti, per fare in modo che gli operatori
$ e^{\pm i \a_k \f } $ anticommutino per $ k \neq l $, 
\`e necessario effettuare una trasformazione di Klein
introducendo degli 
operatori $c(k)$ indipendenti da $z$ detti $ \it{cocicli} $ ed identificando 
$ \psi^k$ con $c(k)e^{\pm i \a_k \f } $. Esprimendo correnti del sistema 
fermionico in termini di bosoni, si ottiene
\be
J^{kl}=c(k)c(l)e^{i\a_{kl}\f} \ , \hspace{2cm} a^i_{kl}= \d^i_k+ \d^i_l  \ ,
\ee
\be
J^{k\bar{l}}=c(k)c(\bar{l})e^{i\a_{k\bar{l}}\f} \  , \hspace{2cm}
a^i_{k\bar{l}}= \d^i_k - \d^i_l  \ ,
\ee
\be
J^{k\bar{k}}= i \part \f^k  \  .
\ee 
\`E interessante notare che i generatori dell'algebra di Cartan sono 
realizzati dalle correnti $U(1)$ della teoria bosonica, mentre gli altri 
generatori sono esponenziali costruiti con i vettori peso della 
rappresentazione vettoriale di $O(2n)$.
Possiamo allora pensare di ottenere operatori di spin utilizzando vettori peso
della rappresentazione spinoriale; se poniamo infatti
\be
S_\a(z) = e^{i \a_{\a} \f }  \  ,
\ee
dove gli $ \a_{\a} $ sono $ 2^n$ vettori con n componenti pari
a $ \pm \frac{1}{2} $ otteniamo i due spinori di $O(2n)$ :
S se il numero di segni $+$ in $ \a_{\a}$ \`e pari e C se \`e dispari. Il peso 
conforme \`e $ h = \frac{n}{8} = \frac{N}{16} $ e i prodotti rilevanti sono
\be
J^{kl}(z)S_{\a} (w) = \frac{1}{z-w} \frac{1}{4}[\g^k , \g^l]_{\a\b}S_\b (w)  \ ,
\ee
\be
S_\a (z)S_\b (w) = \frac{ \d_{\a\b}}{(z-w)^{\frac{N}{8}}}+
\g^i_{\a\b} \frac{\psi^i}{(z-w)^{\frac{N}{8}-\frac{1}{2}}}+
\frac{i}{2}[ \g^i,\g^j]_{\a\b} \frac{J^{ij}}{(z-w)^{\frac{N}{8}-1}}+...  \ ,
\ee
\be
\psi^i(z)S_\a(w) = \frac{1}{\sqrt{z-w}} (\g^i)^{\a\b}S_\b (w) \  .
\ee
L'ultimo prodotto \`e consistente con l'idea che gli operatori di spin, 
interpolando tra il settore NS ed il settore R e quindi cambiando condizioni al 
contorno periodiche in antiperiodiche, introducano un taglio nel piano 
complesso.
 
Questa costruzione di un'algebra affine a livello $ k=1 $ in termini di bosoni 
liberi pu\`o essere generalizzata ad ogni algebra G semplicemente allacciata
$ ( A_n , D_n , E_6 , E_7 , E_8 ) $ ed \`e nota come costruzione di 
Halpern-Frenkel-Kac-Segal. 
Scegliamo come generatori di G la base di Cartan-Weyl
e indichiamo con $ \L_R $ il reticolo delle radici, che hanno tutte la stessa
lunghezza fissata per convenzione pari a $\sqrt{2}$.
Se $r=rank(G) $ consideriamo un sistema di $r$ bosoni e 
poniamo $ H^i= i \part X^i $ per i generatori della sottoalgebra di Cartan;
associando inoltre ad ogni $\l \in \L_R $ un operatore
$ J^{\l}(z) = e^{i \l X(z)}c_{\l}(z) $. I $ c_{\l} $ sono dei cocicli con la 
propriet\`a che $ c_\l c_\m = (-1)^{\l \cdot \m} c_\m c_\l $ e 
$ c_\l c_\m = \e ( \l ,\m ) c_{\l+\m} $, con $ \e= \pm 1$.
Utilizzando
\beq
:e^{i \l X(z)}::e^{i \m X(w)}: =
(z-w)^{\l \cdot \m}:e^{i \l X(z) +i \m X(w)}:=  \nonumber \\
(z-w)^{\l \cdot \m}:e^{i (\l + \m)}(1+i(z-w) \l \part X(z) + O(z-w)): \   ,
\eeq
si pu\`o verificare che i modi di questi operatori soddisfano le regole di
commutazione dell'algebra affine: 
\be
[H^i_m,H^j_n] = m \d^{ij} \d_{m+n,0}  \  , \hspace{1cm}
[H^i_m,J^\l_n] = \l^i J^\l_{m+n} \  ,
\ee
\beq
[J^\l_m,J^\m_n] & = & 0 \hspace{3.9cm} {\it se~ }  \l + \m \ni \l_R  \ ,  
\nonumber \\
  & = & \e( \l , \m ) J^{\l + \m}_{m+n} \hspace{2cm} {\it se~ }  
\l + \m \in \L_R  \  ,  \nonumber \\
 & = & \l H_{m+n} + m \d_{m+n,0} \hspace{0.9cm} {\it se~ }  \l + \m =0  \  .
\eeq

Si possono introdurre operatori nelle varie rappresentazioni dell'algebra 
semplicemente utilizzando esponenziali con vettori appartenenti al reticolo dei 
pesi $ \L_W $, corrispondenti alla rappresentazione desiderata. Il reticolo 
dei pesi si suddivide in classi di coniugio rispetto a $ \L_R $ e bisogna 
scegliere queste classi in modo che si chiudano rispetto all'addizione per 
avere una OPE consistente; in particolare, queste possibili scelte delle classi 
di coniugio sono in corrispondenza con i sottogruppi del centro del gruppo.
Il carattere della rappresentazione corrispondente al sotto reticolo (l)
\`e 
\be
\chi^G_{(l)} = tr_{(l)} q^{L_0-\frac{c}{24}} =
\frac{1}{( \h (q))^d}\sum_{ \l \in (l) } q^{ \frac{ \l^2}{2} }  \ ,
\ee
dove $ q = e^{2 \p i \t } $. \`E facile verificare, e verr\`a fatto in modo
esplicito in seguito, che il carattere \`e invariante rispetto a trasformazioni
modulari sul parametro $ \t $ se il reticolo scelto \`e pari e autoduale.

Un'altra teoria conforme che si riveler\`a importante nel seguito comprende
una coppia di campi $ (b,c) $ di dimensioni $ j $ e $ 1-j $. 
Consideriamo sia il caso bosonico che quello fermionico, distinguendoli con un 
parametro $ \e = \mp $. L'azione \`e
\be
S = \frac{1}{ \p} \int d^2z b(z) \bar{\part} c(z)   \ ,
\ee
e $  \lan b(z)c(w) \ran  = \frac{ \e}{z-w} $. Come negli esempi precedenti,
possiamo calcolare il tensore di energia impulso
$ T(z) = -j b \part c + (1-j) \part b c $ e la carica centrale 
$ c = \e ( 1 - 3Q^2) $ con $ Q = \e (1-2j) $. Il sistema ha una simmetria 
$U(1) $ 
\be
c \mapsto e^{-i \th}c  ,  \hspace{2cm} b \mapsto e^{i \th}b  \  ,
\ee
con corrente $ j = - :bc: $. Questa corrente ha un' anomalia nel prodotto con 
$T$
\be
T(z)j(w) = \frac{Q}{(z-w)^3} + \frac{j(w)}{(z-w)^2}+\frac{\part j(w)}{z-w}  \ ,
\ee
che in termini di modi d\`a
\be
[L_m,j_n] = -nj_{m+n} + \frac{Q}{2}m(m+1) \d_{m+n,0}  \  .
\ee
Le funzioni di correlazione sono non nulle solo quando violano la carica 
associata con $j$ di una quantit\`a $Q$. Questa anomalia \`e 
legata alla presenza di zero modi degli operatori $ (b,c) $ ; in particolare 
la differenza tra gli zero modi di c e quelli di b \`e pari a
$ - \frac{\e}{2}Q \chi $ dove $ \chi = 2 - 2g $ \`e la caratteristica di Eulero 
della superficie sulla quale si considera definita la teoria.


\section{Integrale di Polyakov}
\markboth{} {}

In coordinate complesse \`e molto semplice descrivere la geometria di una
superficie; le uniche componenti non nulle della connessione di Christoffel 
sono
\be
\G^z_{zz} = \part \f \   , \hspace{2cm} \G^{\bar{z}}_{\bar{z}\bar{z}} =
\bar{\part} \f  \ .
\ee
Inoltre la curvatura scalare \`e $ R= -4 e^{- \f} \part \bar{ \part} \f $; 
definiamo la derivata covariante per tensori con $n$ indici olomorfi
\be
\btd^z_n T^{z...z} = g^{z \bar{z}} \bar{\part} T^{z...z} \ , \hspace{1cm}
\btd^n_z T^{z...z} = ( \part + n \part \f)T^{z...z}  \  .
\ee
Rispetto al prodotto tra due tensori di tipo $n$,
\be
 (T_1.T_2) = \int dz \sqrt{g} (g_{z \bar{z}})^n \bar{T_1}(z) T_2(z)  \  ,
\ee
l'aggiunto di $ \btd^z_n$ \`e $ - \btd^{n+1}_z$,  e analogamente 
$ ( \btd^n_z)^{ \dag} = - \btd^z_{n-1} $. Il teorema di Riemann-Roch fornisce 
l'indice dell'operatore $ \btd^z_n$, definito come differenza tra il numero di zero modi
dell'operatore stesso e del suo aggiunto
\be
ind \btd^z_n = (2n-1)(g-1)  \  .
\ee
Vediamo come definire l'integrazione su tutte le possibili metriche 
bidimensionali; la variazione della metrica dovuta ad una riparametrizzazione
infinitesima $ \x $ e ad un riscalamento di Weyl $ g \mapsto e^{2 \w}g $
\`e data da
\be
\d g_{\a\b} = \btd_\a \x_\b + \btd_\b \x_\a + 2\w g_{\a\b} =
(P \x )_{\a\b} + 2\bar{ \w} g_{\a\b}   \  ,
\ee
dove $ \bar{ \w} = \w + \frac{1}{2} \btd_\g \x^\g $ e 
$ (P \x )_{\a\b} = \btd_\a \x_\b + \btd_\b \x_\a - g_{\a\b}( \btd_\g \x^\g ) $.
L'operatore P trasforma vettori in tensori simmetrici a traccia nulla : vettori 
nel nucleo di P equivalgono a riscalamenti di Weyl e sono detti
vettori di Killing conformi; questa riscrittura evidenzia quindi 
una sovrapposizione tra gruppo dei diffeomorfismi e 
riscalamenti di Weyl,sovrapposizione che si dovr\`a tenere presente  nella 
definizione della misura, come bisogner\`a considerare
l'esistenza di deformazioni della metrica non  connesse 
all'identit\`a e pertanto non esprimibili in termini di campi vettoriali.

Possiamo rappresentare il determinante dell'operatore $ PP^{\dag} $ 
introducendo dei ghosts $ c^{ \a} $ e degli antighosts $ b_{\a\b} $
\be
\sqrt{det(PP^{\dag})} = 
\int DcDb e^{\frac{i}{ \p}\int \sqrt{g} g^{\a\b}b_{\a\b} \btd_\b c^\a }  \ .
\ee
Nel gauge conforme l'azione diviene
\be
S_g = \frac{1}{2 \p} \int [ b_{zz} \btd_{\bar{z}} c^z 
+ b_{\bar{z} \bar{z}} \btd_{z} c^{\bar{z}}]  \  ,
\ee
e corrisponde al sistema $ (b,c) $ con $ j=2 $ considerato 
precedentemente, che rappresenta quindi
una parte comune a tutte le teorie di 
stringhe in quanto legata all'invarianza per riparametrizzazioni della 
teoria bidimensionale. L'indipendenza dal fattore conforme della metrica \`e 
una questione pi\`u sottile; classicamente il campo $ \f $ si disaccoppia dalla 
teoria ma effetti dovuti alla regolarizzazione della teoria quantistica 
introducono una dipendenza da questo campo nota come anomalia di Weyl. Il punto 
cruciale \`e che questa dipendenza \`e completamente determinata da una 
costante $c$ che \`e proprio la carica centrale della teoria conforme. Un modo 
semplice per determinare la forma di questa dipendenza \`e il seguente.
Se come al solito definiamo 
\be
Z = \int DXDg e^{-S} = e^{-W}  \  ,
\ee
variando rispetto alla metrica si pu\`o verificare che 
\be
\btd_z ( \frac{1}{\sqrt{g}}\frac{ \d W}{ \d \f }) =
\btd^z ( \frac{1}{\sqrt{g}}\frac{ \d W}{ \d g^{zz} }) \    .
\ee
Poich\`e la misura $ | \d X |^2 = \int \sqrt{g} \d X^\m \d X^\n G_{\m\n} $ non 
varia al primo ordine con $ g^{zz} $ il secondo membro pu\`o essere 
identificato con la derivata del valore d'aspettazione di $ T_{zz} $. Il primo 
membro in generale \`e diverso da zero e quindi $ T_{zz} $ non \`e analitico; 
la sua forma \`e tuttavia fortemente vincolata : \`e un tensore di tipo $ t_z $ 
e, se si assume che sia un funzionale locale della metrica, l'unica 
possibilit\`a \`e una derivata dello scalare di Ricci 
\be
\btd^z ( \frac{1}{\sqrt{g}}\frac{ \d W}{ \d g^{zz} })=
\frac{c}{48 \p } \btd_z R   \ .
\ee
Se integriamo questa espressione e poniamo $ g = e^\f \hat{g} $, con $\hat{g} $ 
metrica di riferimento, otteniamo la dipendenza di $W$ da $ \f $  :
\be
W = \frac{c}{96 \p} \int [ \sqrt{g}g^{\a\b} \part_\a \f \part_\b \f +
\frac{ \m^2}{2} e^{ \f}]  \  .
\ee
Ora $ \btd_z <T_{zz}> \neq 0 $, ma se  definiamo
\be
 \tilde{T} = T + \frac{c}{48 \p} ( 2 \part^2 \f - ( \part \f)^2 )  \  ,
\ee
abbiamo un tensore conservato; si pu\`o verificare che
\be
<\tilde{T}(z) \tilde{T}(w)> = \frac{c}{2(z-w)^4} +\frac{<\tilde{T}>}{(z-w)^2} + 
\frac{\part <\tilde{T}>}{z-w}  \  .
\ee
L'anomalia conforme si manifesta quindi in una deformazione dell'algebra 
classica di Virasoro attraverso il termine centrale legato a $c$; ha pertanto 
una semplice forma locale ed \`e possibile ridefinire il tensore di 
energia-impulso 
in modo che il suo valore di aspettazione si comporti in modo analitico.
L'azione 
\be
S = \frac{c}{96 \p} \int \sqrt{g}[ g^{\a\b} \part_\a \f \part_\b \f + 2R \f ] \ ,
\ee
\`e nota come azione di Liouville; la presenza della simmetria
\be
\hat{g} \mapsto e^{2 \w} \hat{g}  , \hspace{2cm} \f \mapsto \f - 2 \w  \ ,
\ee
permette di pensare di avere ancora una teoria conforme, 
solo con un campo ulteriore $ \f $.
Vediamo in questo modo che imponendo l'assenza di anomalie conformi la carica 
centrale della teoria sul worldsheet viene 
fissata in teoria di stringhe ad un valore critico:
sappiamo che la carica centrale dei ghosts $ (b,c) $ \`e $ c= -26 $ e quindi 
per avere una teoria con $ c=0 $ la teoria conforme definita sul worldsheet 
deve avere $ c= 26 $. In particolare, se si considera un sistema di bosoni 
liberi questo d\`a una dimensione critica per la propagazione della stringa 
bosonica pari a $ d= 26 $, in modo consistente con quanto ottenuto con gli 
altri metodi di quantizzazione.


\section{Quantizzazione BRST}
\markboth{} {}

Fissando il gauge abbiamo ridotto l'integrazione sulle metriche allo spazio dei 
moduli di superfici di Riemann, uno spazio di dimensione finita, ma 
abbiamo anche introdotto dei nuovi stati, precisamente quelli associati con i 
ghosts delle riparametrizzazioni $ b $ e $ c$. Il modo pi\`u elegante per 
descrivere gli stati fisici \`e tramite il formalismo BRST \cite{FMS}. 
Costruiamo 
l'operatore 
\be
Q = \int \frac{dz}{2 \p i} :c(z) \left (T^X + \frac{T^g}{2} \right ): =
    \int \frac{dz}{2 \p i} j_{BRST}(z) \   ,
\ee
dove $ T^g $ indica il tensore energia impulso dei ghosts. $ T^X $ 
indica il tensore di energia-impulso  
delle coordinate bosoniche ma pi\`u in generale potrebbe essere associato ad 
una arbitraria teoria conforme definita sul worldsheet.
Questo operatore \`e hermitiano e conservato; si pu\`o verificare che \`e 
nilpotente precisamente quando la carica centrale della teoria descritta da
$ T^X$ \`e 26 e questo \`e un altro metodo per determinare 
la dimensione critica. A questo punto possiamo definire gli stati fisici come 
classi della coomologia di Q. \`E opportuno bosonizzare il sistema dei ghosts 
con un campo $ \f $ con tensore di energia impulso
$ T^{\f} = \frac{1}{2} ( \part \f)^2 + \frac{3}{2} \part^2 \f $ ponendo
$ b = e^{- \f} $ e $ c = e^{ \f} $. In questo modo
$ Q = :e^{ \f} ( T^X + T^{\f} ): $. Un operatore di vertice V crea
stati fisici se  $ [Q,V] = 0 $. Se consideriamo il 
caso di un campo primario $ v $  che non contiene ghosts 
\be
Qv(z) = h_v(\part e^\f )v(z) + e^\f \part v(z)  \  ,
\ee
vediamo che per la stringa aperta \`e possibile costruire un 
operatore di vertice per uno stato
fisico integrando $v$ lungo l'asse reale quando $h_v=1$;
l'operatore in questione \`e quindi  $V = \int dz v(z) $. In
modo analogo, per la stringa chiusa gli operatori vertice sono
integrali sul piano complesso di operatori di dimensione $(1,1)$ :
$ V = \int dz d\bar{z} v(z,\bar{z}) $. Consideriamo degli esempi; per la
stringa aperta lo stato crato da $ e^{ipX(z)} $ \`e fisico quando $ p^2 = 2 $ e 
corrisponde al tachione, mentre lo stato $ \e_\m \part X^\m e^{ipX(z)} $
\`e fisico quando $ p^2 =0 $ e $ \e \cdot p = 0 $, precisamente le 
condizioni per le polarizzazioni fisiche di una particella di spin 1 a massa
nulla; per la stringa chiusa $ e^{ipX(z,\bar{z})} $ corrisponde ancora al
tachione mentre  $ \e_{\m\n} \part X^\m \bar{\part} X^\n e^{ipX} $ \`e fisico 
quando $ p^2 = 0 $ e $ \e_{\m\n}p^\m = \e_{\m\n}p^\n =0 $. Queste sono le 
condizioni per le polarizzazioni fisiche di una particella di spin 2 a massa
nulla, un tensore antisimmetrico e uno scalare.

\section{Superstringa}
\markboth{}{}

I due problemi della stringa bosonica, la presenza di un tachione e l'assenza 
di fermioni, possono essere risolti modificando la teoria bidimensionale in 
modo da renderla superconforme. L'idea iniziale \`e quella di introdurre dei 
campi anticommutanti $ \psi^\m ( \t , \s ) $  e di eliminare gli stati non fisici 
creati da $ \psi^0 $ utilizzando una nuova simmetria; dato che la teoria 
contiene ora sia gradi di libert\`a bosonici che gradi di libert\`a fermionici, 
\`e naturale pensare di utilizzare la supersimmetria. 
Scriviamo quindi l'azione che descrive la supergravit\`a in due dimensioni 
accoppiata con campi scalari $ X^\m $ e fermionici $ \psi^\m $ \cite{BDH}
\beq
S= - \frac{1}{4 \p \a^{'}} \int d^2 \x \sqrt{-g} [ g^{\a\b} \part_\a X^\m
\part_\b X_\m + i \bar{\psi}^\m \g^\a \part_\a \psi_\m + \nonumber \\
2( \bar{\chi}_\b \g^\a \g^\b \psi^\m)
( \part_\a X_\m - \frac{i}{4}\chi_\a \psi_\m)]  \  ,
\eeq
dove $ \chi $ \`e il gravitino e le matrici di Dirac in due dimensioni sono
$ \g^0 = \s^2 $ e $ \g^1 = i \s^1 $. L'azione $ (1.81) $ ha due cariche di 
supersimmetria che sono spinori di Weyl con chiralit\`a opposta. Sfruttando le 
simmetrie possiamo fissare
l'analogo del gauge conforme, che  per la teoria supersimmetrica \`e
\be
g_{\a\b} = e^{2 \L}g_{\a\b} \ ,  \hspace{2cm} \chi_\a = \g_\a \z  \  ,
\ee
dove  $ \z $ \`e uno spinore. 
Il metodo pi\`u elegante di quantizzazione \`e ancora quello BRST, che presenta
anche il vantaggio di poter scrivere semplicemente gli operatori di vertice per i
fermioni. Iniziamo dall'analisi nel cono di luce
definito imponendo
\be
X^+ = x^+ + \a^{'} p^+ \t \ , \hspace{2cm} \psi^+=\bar{\psi}^+=0 \  . \label{cl}
\ee
Sugli stati di vuoto agiamo solo con gli oscillatori nelle direzioni trasverse, 
e imponiamo solo i vincoli dovuti agli zero modi del tensore energia impulso e 
della supercorrente, che nel gauge ( \ref{cl})  \`e $ G = i \psi^\m \part X_\m $.
Oltre ai settori destro e sinistro dobbiamo anche distinguere un settore NS ed 
un settore R. Concentriamoci sui modi destri; nel settore NS la richiesta di 
invarianza di Lorentz della teoria fissa il valore della costante di 
ordinamento normale in $L_0$ e la dimensione dello spazio tempo e il risultato \`e 
$ a = \frac{1}{2} $ , $ d= 10 $. Lo stato fondamentale \`e un tachione 
$ |p\ran $, e il livello successivo contiene gli stati di massa nulla 
$ \psi^i |p\ran $, che formano un vettore di $SO(8)$. Nel settore R il vuoto \`e 
uno spinore di $O(8)$ che si decompone in uno spinore $ |S\ran $ con
chiralit\`a positiva e nel suo coniugato $ |C\ran $ , entrambi a massa nulla.
\`E possibile eliminare il tachione e contemporaneamente ottenere uno spettro 
supersimmetrico proiettando su stati con numero fermionico definito \cite{GSO}.
La proiezione va effettuata sia sul settore destro che su quello sinistro della 
teoria; la scelta di quale spinore mantenere nello spettro nel settore destro 
di Ramond \'e arbitraria, ma una volta che questa \'e stata fatta possiamo 
distinguere due teorie a seconda che  il settore sinistro venga proiettato 
nello 
stesso modo o nel modo opposto. Se nel settore R destro si sceglie 
$ (-1)^F|\bar{S}\ran = |\bar{S}\ran $, si ottiene la superstringa di tipo $IIA$ 
se nel settore sinistro si pone $ (-1)^F|C\ran = |C\ran $ e la superstringa di 
tipo $IIB$ se $ (-1)^F|S\ran = |S\ran $. Otteniamo quindi quattro settori:

(NS,NS) : contiene stati della forma 
$ \psi^i_{-1/2} \bar{ \psi}^j_{-1/2} |p\ran $, che corrispondono al dilatone, al 
tensore antisimmetrico e al gravitone.

(NS,R) : contiene stati della forma $ \psi^i_{-1/2} |p \bar{S}\ran $ che 
corrispondono ad un fermione e ad un gravitino.

(R,NS) : nella $IIB$ contiene $ \bar{\psi}^i_{-1/2} |p S\ran $ e nella $IIA$
$ \bar{\psi}^i_{-1/2} |p C\ran $.

(R,R)  : contiene dei bispinori che possono essere decomposti in forme 
antisimmetriche. Per la $IIB$ si ha $ |S \bar{S}\ran $ e per la $IIA$ 
$ |C \bar{S}\ran $.

Bisogna imporre sugli stati il vincolo di accordo tra i livelli 
$ L_0 = \bar{L}_0 $; inoltre nel settore (NS,NS) il vincolo 
$ L_0 - \frac{1}{2}=0 $ d\`a l'equazione di Klein-Gordon; nel settore (R,NS
) $ G_0=0 $ d\`a l'equazione di Dirac ed implica $ L_0 = G_0^2 =0 $, e 
analogamente in (NS,R); nel settore (R,R) $ G_0 = \bar{G}_0= 0 $, che equivalgono 
alle equazioni del moto per le forme antisimmetriche ed alle identit\`a  di 
Bianchi corrispondenti $ dF = d*F = 0 $.

Possiamo studiare la $(1.91)$ anche usando tecniche di teoria conforme. 
Nel gauge conforme e in coordinate complesse l'azione diviene
\be
S = \int d^2z ( \part X^\m \bar{\part} X_\m + \psi^\m \bar{\part} \psi_\m +
\bar{\psi}^\m \part \bar{\psi}_\m )  \  ,
\ee
e rappresenta il nostro primo esempio di una teoria con simmetria superconforme
. L'algebra corrispondente \`e un'estensione dell'algebra conforme contenente 
un nuovo generatore $ G(z) = \sum_{r \in Z} G_r z^{-r- \frac{3}{2}} $,
un campo primario di dimensione $ \frac{3}{2} $; le nuove relazioni di 
commutazione sono
\be
\{G_r,G_s \} = \frac{\hat{c}}{2} ( r^2 - \frac{1}{4} ) \d_{r+s,0} + 2L_{r+s} \  ,
\hspace{1cm} [L_m,G_r] = ( \frac{m}{2} -r ) G_{m+r}  \  .
\ee
Nel nostro caso $ G = \part X^\m \psi_\m $, ed ha le stesse propriet\`a
dei fermioni. Altre generalizzazioni naturali si hanno considerando algebre di 
supersimmetria estese : per $N=2$ si hanno due supercorrenti $ G^{\pm} $ ed una 
corrente $U(1)$ $ J$,  per $N=4$ quattro supercorrenti e tre correnti che 
realizzano $ SU(2) $ al livello $k$. Descriveremo queste algebre in maggiore 
dettaglio successivamente.

Per la determinazione della dimensione critica si procede come per la stringa 
bosonica;
i ghosts superconformi hanno una carica centrale $c=11$ e tenendo conto che 
ogni supercoordinata d\`a a $c$ un contributo pari a $ \frac{3}{2} $, si vede 
che per la superstringa la dimensione critica \`e $d = 10$.

L'operatore BRST si esprime in termini dei tensori energia-impulso e delle 
supercorrenti dei campi $ X $, $ \psi $ e dei ghosts come
\be
Q= \int \frac{d^2z}{2 \p i} \left [ c \left ( T + \frac{T^g}{2} \right ) - \g 
\left ( G + \frac{G^g}{2} \right ) \right ]  \ .
\ee
Rispetto alla stringa bosonica, uno stato fisico deve soddisfare la nuova 
condizione $ G_{-\frac{1}{2}} | \psi \ran = 0 $. Dato che $ G_{-\frac{1}{2}} $ 
pu\`o essre identificato con il generatore di trasformazioni di supersimmetria 
globali sul worldsheet questo equivale a richiedere che l'operatore di vertice 
sia supersimmetrico; ad esempio per il tachione abbiamo 
$ V_T(k) = i k \psi e^{ikX} $ e per il vettore
$ V_V(k) = \z_\m ( \part X^\m + ik \psi \psi^\m ) e^{ikX} $, con 
$ \z \cdot k =0 $ e $ k^2 = 0 $.
Per descrivere gli operatori di vertice che coinvolgono il settore di Ramond 
\`e conveniente bosonizzare i fermioni, come descritto 
precedentemente; l'operatore 
$ S_\a (z) = e^{i a_\a \f} $ ha dimensione $ h = \frac{5}{8} $, e per ottenere 
un operatore con $ h =1 $ dobbiamo tenere conto dei superghosts. La 
bosonizzazione \`e in questo caso pi\`u sottile : dobbiamo introdurre uno 
scalare $ \f $, con $ T = - \frac{1}{2} ( \part \f )^2 - \part^2 \f $, e due 
fermioni $ ( \x , \h ) $ di dimensioni $ (0,1) $. La corrispondenza \`e
$ \b = \part \x e^{- \f} $ e $ \g = \h e^{\f} $ ; l'operatore di spin per
questo sistema \`e $ e^{- \frac{\f}{2}} $, e l'operatore 
di vertice completo \`e
\be
V_{-1/2} = \bar{u}^\a S_\a e^{ikX} e^{ - \frac{\f}{2}}  \  ,
\label{fver}
\ee
con $ ku = 0 $. C'\'e una sottigliezza \cite{FMS} dovuta alla carica di 
ghost portata dall'operatore $V$, che \`e stata indicata con un pedice; i 
ghosts superconformi sono dei bosoni con una azione del primo ordine nelle 
derivate, e questo ha come conseguenza che lo spettro non \`e inferiormente 
limitato. Gli operatori $ \b_k$ , $\g_k$ agiscono su spazi caratterizzati da 
diversi valori di questa carica. Gli elementi di matrice della teoria 
conforme violano la conservazione della carica per una quantit\`a fissata e 
quindi, a seconda dell' ampiezza che si deve calcolare e del genere della 
superficie sulla quale questa viene calcolata, bisogna 
ricorrere ad operatori di 
vertice con diverse cariche di ghost. Questi operatori possono essere 
costruiti sfruttando il fatto che $ \x_0 $ non compare nelle formule di 
bosonizzazione, e quindi $ V_q = [Q, \x V_{q-1}] $ \`e BRST invariante ma non 
nullo. Questa operazione \`e detta $ \bf{picture}$ $ \bf{changing} $ e si
pu\`o dimostrare che, nelle funzioni di correlazione a genere $g \le 1$, 
muovere l'operatore che 
realizza questa trasformazione da un vertice all'altro non modifica
il risultato. Ad esempio, trasformando il vertice (\ref{fver}) si ottiene
\be
V_{1/2} = \bar{u}^\a [ e^{ \frac{\f}{2}}( \part X^\m +
 \frac{i}{4} k \psi \psi^\m ) \g^\m_{\a\b} S_\b +
 e^{ \frac{3}{2} \f } \h bS_\a]c e^{ikX}   \  .
\ee
Uno stato annullato da tutti gli oscillatori con $k>0$ \`e $ | \W \ran = 
c(0) e^{- \f(0)} |0\ran $; $ | \W \ran $ e $ |0\ran $ appartengono a spazi di 
Hilbert distinti. Gli operatori riportati precedentemente per il tachione ed il 
vettore sono nella picture $ -2 $; nella $ -1 $ si ha 
$  V_T = e^{ ikX } e^\s e^{ - \f} $ e 
$ V_V= \z \cdot \psi e^{ikX} e^\s e^{-\f} $.
Possiamo scrivere facilmente la funzione di partizione della teoria; nel 
calcolo i ghosts eliminano precisamente il contributo di due supercoordinate, 
e ricordando le (\ref{caro}) (\ref{carv}) (\ref{cars}) si ottiene
\beq
Z_{IIA} = \int \frac{d^2 \t}{ \t^2_2} \frac{1}{( \sqrt{ \t_2} \h \bar{\h})}
( \chi_V - \chi_C ) ( \bar{\chi}_V - \bar{\chi}_S )= \nonumber \\
\int \frac{d^2 \t}{ \t^2_2} \frac{1}{( \sqrt{ \t_2} \h \bar{\h})} 
\left | \frac{ \th^4_3 - \th^4_4 - \th^4_2 }{ \h^4} \right |^2=0   \ ,
\eeq
\beq
Z_{IIB} = \int \frac{d^2 \t}{ \t^2_2} \frac{1}{( \sqrt{ \t_2} \h \bar{\h})}
( \chi_V - \chi_S ) ( \bar{\chi}_V - \bar{\chi}_S )= \nonumber \\
\int \frac{d^2 \t}{ \t^2_2} \frac{1}{( \sqrt{ \t_2} \h \bar{\h})} 
\left | \frac{ \th^4_3 - \th^4_4 - \th^4_2 }{ \h^4} \right |^2=0  \  .
\eeq
L'energia di vuoto \`e quindi nulla ad un loop per le superstringhe, 
come era d'altronde da aspettarsi data la supersimmetria dello spettro. La 
proiezione effettuata per rendere la teoria consistente ha una chiara 
interpretazione geometrica legata all'invarianza modulare, vale a dire 
l'invarianza rispetto a diffeomorfismi della superficie non connessi
all'identit\`a. Queste trasformazioni, in particolare, scambiano tra loro le 
strutture di spin; ora  ogni volta che si 
introducono fermioni, si fissa con le loro condizioni al contorno una struttura 
di spin, e di conseguenza  bisogna sceglierne combinazioni tali da
risultare invarianti rispetto a trasformazioni modulari. Da 
questo punto di vista, la proiezione GSO consiste nell'includere 
i contributi di tutte le 
strutture,  sommando indipendentemente sui settori destro e sinistro.

Un operatore spaziotemporale di supersimmetria pu\`o essere identificato con 
il vertice per i fermioni nel settore di Ramond a momento nullo
\be
Q_\a = \int \frac{dz}{2 \p i} S_\a e^{- \frac{\f}{2}} \   .
\ee
L'unica sottigliezza nel verificare l'algebra di supersimmetria \`e che il 
doppio commutatore d\`a un operatore di vertice che si trova in una picture 
differente dall'originale; inoltre la relazione $ \{Q,Q\} \sim P $ \`e 
soddisfatta da $ \{ Q_{-1/2}, Q_{1/2} \} $, e questi operatori possono essere 
identificati nel calcolo di ampiezze, ovvero l'algebra \`e realizzata 
on-shell. Argomenti euristici permettono di arguire l'annullarsi dei 
tadpoles e della costante cosmologica ad ogni ordine in teoria delle 
perturbazioni \cite{Mar}. Ancora oggi, tuttavia, esistono difficolt\`a 
tecniche e di principio nel definire ampiezze a molti loops per la 
superstringa, che risultano ambigue a meno di derivate totali sullo
spazio dei moduli. 

Gli stati RR si possono decomporre in una somma di forme antisimmetriche 
\be
F_{\a\b} = \sum \frac{1^k}{k!} F_{ \m_1 ... \m_k} ( \G^{\m_1...\m_k})_{\a\b} \ ,
\ee
con $ k = 0 , 2 , 4 $ per la $IIA$ e $ k = 1 , 3 , 5 $ per la $IIB$ dove la 5-
forma \`e autoduale. Grazie alle equazioni del moto ed all'identit\`a di 
Bianchi, questi tensori possono essere espressi localmente in termini di 
potenziali; per la $IIA$ abbiamo $ C^\m$ , $C^{\m\n\r}$ ed una 0-forma che
non si propaga \footnotemark \footnotetext{Questa 0-forma corrisponde ad un
parametro libero che consente di deformare la supergravit\`a di Tipo IIA
come in \cite{Rom}}, per 
la $IIB$ $C$ , $C^{\m\n}$ , $C^{\m\n\r\s} $. I campi 
provenienti dal settore di Ramond sono peculiari : nessuno stato 
appartenente allo spettro perturbativo della teoria \`e carico rispetto a
questi campi.

\section{Interazioni}
\markboth{} {}

Come abbiamo gi\`a osservato, la formulazione attuale della teoria di stringhe 
consiste in una serie di regole che definiscono l'espansione perturbativa di 
una teoria i cui principi primi non sono ancora stati identificati;
quindi \`e come se conoscessimo solo i diagrammi di Feynman senza
avere una formulazione (formalmente) non perturbativa della 
teoria quantistica dei campi dalla quale dedurli.
La serie perturbativa della teoria di stringhe \`e una diretta 
generalizzazione della somma sui cammini ad una somma sulle 
superfici bidimensionali. 
Le superfici chiuse e orientate, rilevanti per i modelli di stringa chiusa,
sono classificate topologicamente dal genere 
$g$, che conta il numero di "loops"; ad ogni ordine 
dell'espansione abbiamo quindi un unico contributo. La serie perturbativa per 
teorie di stringhe aperte contiene anche superfici con bordi e 
superfici non orientabili, 
e verr\`a discussa in maggiore dettaglio nel terzo capitolo.

La possibilit\`a di definire in modo consistente uno spazio di Hilbert di stati
\`e legata, come abbiamo visto, all'invarianza conforme della teoria 
bidimensionale. La teoria di stringhe completa richiede tuttavia che la 
teoria conforme possa essere definita su superfici di genere arbitrario, e
questo impone un vincolo molto importante, 
noto come $\bf{invarianza~modulare}$. Infatti, una volta verificato che una 
teoria \`e localmente invariante conforme, bisogna anche tenere conto delle
trasformazioni non connesse all'identit\`a che conservano la classe 
conforme della superficie su cui essa \`e definita. Abbiamo gi\`a 
accennato allo spazio dei moduli delle superfici di Riemann, uno spazio finito 
dimensionale che parametrizza le diverse strutture complesse; l'invarianza 
modulare della teoria conforme a genere $g$ implica che la funzione di 
partizione \`e una funzione ad un solo valore su questo spazio dei moduli. Ad 
esempio, le superfici di genere $g =1$ sono parametrizzate da un modulo 
$\t \in H$ dove $H$ \`e il semipiano complesso superiore; l'azione proiettiva 
di $SL_2(Z)$ su $\t$,
\be
\t \mapsto \frac{a\t + b}{c\t + d}  \  ,  \hspace{1cm}  
\pmatrix{a & b\cr c & d} \in SL_2(Z)  \ ,
\ee
lascia invariata la classe conforme della superficie. In questo caso lo spazio 
dei moduli \`e rappresentato da un dominio fondamentale per l'azione del gruppo 
modulare su $H$, ad esempio 
$F = \{ \t_1 + i \t_2 \in H : -\frac{1}{2} \leq \t_1 \leq \frac{1}{2} \ ,
| \t|^2 > 1 \}$. Dato che $SL_2(Z)$ \`e generato da
\be
S : \t \mapsto - \frac{1}{\t} \  ,  \hspace{1cm}  T : \t \mapsto \t +1  \ ,
\label{gene}
\ee
per verificare se una teoria conforme \`e invariante modulare a 
genere uno basta verificare che la funzione di partizione sia invariante 
rispetto alle due trasformazioni in (\ref{gene}).
L'estensione di questi risultati a genere arbitrario 
\`e stata considerata in \cite{spin2}.

L'introduzione di interazioni ci permette di sottolineare ancora una volta il 
ruolo cruciale dell'invarianza conforme per la teoria di stringhe; una funzione 
di correlazione ad N punti \`e definita dall'integrale
\be
\sum_{g=o}^{\infty} \int_{\{\psi_i\}_{i=1}^n} DX \int Dh e^{-S}  \  ,
\ee
dove i $\psi_i$ sono funzionali dell'immagine tramite $X(\s,\t)$ dei bordi del 
worldsheet che rappresenta l'interazione, $\psi_i[X(\s)]$; a ciascun bordo 
corrisponde quindi una funzione complessa, proprio come la funzione d'onda 
associa un 
numero ad ogni punto nello spaziotempo. 

I termini dell'espansione sono associati a superfici 
di genere $g$ con N punture, in quanto un tubo lungo e sottile viene 
trasformato da $z= e^w$ in un disco con un piccolo buco e, nel limite in cui il 
tubo diviene infinito, in un disco con il centro rimosso. In altri termini, i 
funzionali d'onda che contengono i numeri quantici degli stati esterni di 
stringa, divengono operatori locali, $\it{operatori}$ $\it{di}$
$\it{vertice}$. L'ampiezza di scattering si scrive allora:
\be
A(1,...,N) = \sum_{g=0}^{\infty} \int DX \int Dh [V_1(k_1)...V_N(k_N)]e^{-S}  \ ,
\ee
dove $V_i(k_i)$ \`e l'integrale sul worldsheet dell'operatore di vertice, dato 
che l'emissione e l'assorbimento possono avvenire in ogni punto della 
superficie.

Vertici integrati della forma $\int d \t d\s V_i(\t,\s;k_i)$ invarianti di 
scala richiedono che $V_i$  abbia dimensione conforme due : 
$V_i(\l \xi) = \l^{-2}V_i(\xi)$. In teoria dei campi, per identificare il 
valore di aspettazione nel vuoto dei campi, si minimizza il potenziale 
effettivo; si definiscono quindi le fluttuazioni 
$\tilde{\f} = \f - \lan \f \ran$
e si procede al calcolo perturbativo. Funzioni ad un punto non nulle per
$\tilde{\f}$, segnalano che non si \`e identificato il vero vuoto della teoria, 
e pertanto le correzioni quantistiche determinano delle variazioni di 
questi valori 
di aspettazione. In teoria di stringhe l'analogo dei $\lan \f_i \ran$ sono le
funzioni ad un punto degli operatori vertice $\lan V_i \ran$. \`E facile vedere 
che l'invarianza conforme della teoria definita sul worldsheet assicura che 
abbiamo un vuoto classico; se infatti trasliamo la funzione ad un punto 
$\lan V(z) \ran$ in $\lan V(0) \ran$, riscalando secondo 
$\lan V(0) = \l^{-2} \lan V(0) \ran $ vediamo che $\lan V(0) \ran =0$. In 
modo equivalente questo risultato pu\`o essere visto come una consegunza della 
divisione per il volume dei CKV; per la sfera si tratta di $SL_2(R)$, un gruppo 
non compatto e quindi con volume infinito: otteniamo allora
un risultato nullo, perch\`e  una funzione ad un punto non contiene 
integrazioni sulla posizione degli operatori vertice in grado di  
compensare la divergenza nel denominatore.

Bisogna sempre tenere presente che in teoria di stringhe si hanno due 
espansioni; la vera espansione quantomeccanica \`e l'espansione nel genere $g$, 
che coinvolge i contributi di superfici con diversa topologia;  
la teoria 
sul worldsheet \`e a sua volta una teoria quantistica bidimensionale, che pu\`o 
essere valutata perturbativamente con parametro d'espansione
$\frac{\ap}{R^2}$, dove $R$ 
denota una dimensione tipica dello spazio in cui la stringa si propaga.

Per il calcolo delle ampiezze ad albero \`e stato sviluppato un comodo 
formalismo operatoriale nel cono di luce,
che consiste essenzialmente in manipolazioni di stati 
coerenti di oscillatori armonici; il formalismo pu\`o essere esteso ad un loop, 
ma diviene inadatto per calcoli su superfici di genere superiore, ed \`e 
comunque poco maneggevole quando si ha a che fare con correlatori che 
coinvolgono campi di spin o di twist, ovvero quando si studiano 
ampiezze per superstringhe o su orbifold.

Il formalismo covariante \cite{FMS} \`e senza dubbio pi\`u elegante e 
permette di semplificare i calcoli; rende inoltre pi\`u trasparenti 
alcune propriet\`a generali della serie perturbativa. Illustriamolo brevemente 
per la stringa bosonica.
Come abbiamo gi\`a notato il sistema $(b,c)$ ha una corrente anomala legata 
alla presenza di zero modi; ad esempio sulla sfera $c$ ha tre zero modi che 
corrispondono a $ 1,z,z^2 $. Una funzione di correlazione non nulla deve quindi
contenere tre inserzioni di $c$. Inoltre, lo stato $| \W \ran = c_1|0 \ran $
\`e diverso da zero ed \`e uno stato annullato da tutti gli operatori di 
abbassamento e con valore minimo di $L_0$; $| \W \ran $ rappresenta il 
vero stato fondamentale della teoria, e siamo portati a considerare stati 
fisici 
creati a partire da $ | \W \ran $ studiando operatori di vertice della forma
$ c(z)V(z) $, con V che non coinvolge ghosts. Le condizioni che lo stato 
$ | \psi \ran = c(0)V(0)|0\ran $ sia fisico non sono altro che 
\be
( L_0 -1 )V(0) |0 \ran = 0 \hspace{1cm}
 L_n V(0)|0 \ran = 0 \hspace{1cm} \forall n>0  \ .
\ee
L'operatore V deve essere quindi primario con peso $h_V=1$. Calcoliamo ora una 
ampiezza in teoria di stringhe aperte, l'ampiezza di Veneziano che descrive lo 
scattering di quattro tachioni:
\beq
& & A_V = \lan 0|c(z_1)e^{ip_1X(z_1)}c(z_2)e^{ip_2X(z_2)}
\int dz_3e^{ip_3X(z_3)}c(z_4)e^{ip_4X(z_4)}|0\ran   \nonumber  \\
& & = \int_0^1 dx (1-x)^{p_2 \cdot p_3}x^{p_3 \cdot p_4} \  .  
\eeq
Nel valutare questa espressione abbiamo posto $ z_1= \infty$ , $ z_2 = 1 $ , 
$ z_3 =x $ e $ z_4 =0$.
Se avessimo calcolato questa ampiezza utilizzando tutti vertici integrati 
avremmo ottenuto l'integrale di un'espressione invariante sotto l'azione di 
$SL_2 (R)$ e per rimuovere questa ridondanza avremmo dovuto sostituire 
all'integrazione su tre delle coordinate delle inserzioni l'integrazione sui 
parametri del gruppo calcolando lo Jacobiano appropriato 
\be
\int \frac{ dz_1dz_2dz_3dz_4}{Vol(SL_2R)} \mapsto
\int dz_3(z_1-z_2)(z_1-z_4)(z_2-z_4)    \  .
\ee
La presenza dei ghosts produce automaticamente la giusta misura 
d'integrazione perch\`e
\be
\lan0|c(z_1)c(z_2)c(z_4)|0\ran = (z_1-z_2)(z_1-z_4)(z_2-z_4)    \  .
\ee
In generale sono proprio i ghosts, data la loro intima connessione con la 
geometria del worldsheet, ed in particolare con moduli e vettori di Killing 
conformi, a produrre la misura d'integrazione opportuna sullo spazio dei moduli 
delle superfici di Riemann per ogni genere e per ogni numero 
d'inserzioni esterne.

In teoria di stringhe chiuse, l'ampiezza a quattro tachioni \`e detta ampiezza 
di Shapiro e Virasoro, e pu\`o essere calcolata in modo del tutto analogo
\be
A_{SV} = \int d^2z |1-z|^{2p_2 \cdot p_3}|z|^{2p_3 \cdot p_4}  \  .
\ee
\`E interessante osservare che da questo punto di vista il vincolo
$ L_0 = \bar{L_0} $ viene imposto sugli stati fisici 
dall'integrazione su tutto il piano complesso delle inserzioni degli operatori 
di vertice.

Un metodo molto usato per il calcolo di funzioni di correlazione fa uso del 
tensore di energia-impulso; se si conosce il valore di aspettazione di $T(z)$ 
in presenza di un operatore vertice, \`e possibile ricavare un'equazione del 
primo ordine per le funzioni di correlazione di questi operatori. 
Consideriamo ad esempio
il caso di un correlatore di
operatori di spin; sul toro l'operatore di spin per $2N$ fermioni 
pu\`o essere costruito come prodotto di $N$ operatori $SO(2)$, possiamo 
limitarci a calcolare funzioni di correlazione a $n$ punti sul toro per 
$S^{\pm}$, dove $S^{\pm} \sim e^{\pm \frac{i}{2} \f}$ con $\psi \sim e^{i \f}$.
Abbiamo $T(z) = \lim_{z \rar w}[\frac{1}{2} \part \bar{\psi}(z) \psi(w) -
\frac{1}{2}\bar{\psi}(z) \part \psi (w) + \frac{1}{(z-w)^2}]$, e se 
indichiamo con 
$\n = 1,2,3,4$ le strutture di spin $(P,P)$ , $(A,P)$ , $(A,A)$ e $(P,A)$  \ , 
possiamo calcolare
\beq
& & G_{\n}(z,w;z_1,z_2) = \frac{\lan \bar{\psi}(z)\psi(w) 
S^+(z_1)S^-(z_2) \ran_{\n}}{\lan S^+(z_1)S^-(z_2) \ran_{\n}}  \nonumber \\
& & = \frac{\th_1^{'}(0)}{\th_{\n}(\frac{z_2-z_1}{2})}
(\frac{\th_1(z_1-z_2)\th_1(w-z_1)}{\th_1(z-z_1)\th_1(w-z_2)})^{1/2}
\frac{\th_{\n}(z-w+\frac{z_2-z_1}{2})}{\th_1(z-w)}   \   .
\eeq
Questa funzione \`e completamente fissata dalle condizioni di periodicit\`a 
attorno ai cicli di omologia del toro e dalle singolarit\`a nei vari limiti in 
cui due variabili coincidono. A questo punto
\be
A(z,z_1,z_2) = 
\frac{\lan T(z)S^+(z_1)S^-(z_2) \ran}{\lan S^+(z_1)S^-(z_2) \ran} = 
\lim_{z \rar w}[ \frac{1}{2} \part_z G - \frac{1}{2} \part_w G + 
\frac{1}{(z-w)^2}] \   ;
\ee
dato che per $z \sim z_2$ si ha $A \sim \frac{h_S}{(z-z_2)^2} + 
\frac{1}{z-z_2} \part_{z_2} ln \lan S^+(z_1)S^-(z_z) \ran$, otteniamo per 
$\lan S^+(z_1)S^-(z_z) \ran$ l'equazione differenziale
\be
\part_{z_2} ln \lan S^+(z_1)S^-(z_z) \ran =
\frac{1}{2} \frac{\th^{'}_1(\frac{z_2-z_1}{2})}{\th_1(\frac{z_2-z_1}{2})}
-\frac{1}{4} \frac{\th^{'}_1(z_2-z_1)}{\th(z_2-z_1)}   \   ,
\ee
che risolta d\`a infine
\be
\lan S^+(z_1)S^-(z_z) \ran_{\n} = K_{\n}[\th_1(z_2-z_1)]^{-1/4}
\th_{\n}(\frac{z_2-z_1}{2})  \   ,
\ee
dove $K_\n$ \`e una costante che pu\`o essere fissata utilizzando la
fattorizzazione delle ampiezze.
In modo analogo \`e possibile determinare le funzioni di correlazione per i 
ghosts e quindi, combinando opportunamente questi risultati, funzioni di 
correlazione per operatori vertice bosonici e fermionici nelle pictures 
opportune.

\section{Azioni effettive}
\markboth{} {}

Vediamo ora come l'invarianza conforme della teoria bidimensionale che 
descrive la 
propagazione di una stringa in presenza di campi di background 
determina le equazioni del moto \cite{cb1}. 
L'azione accoppia questi campi di background 
proprio agli operatori di vertice che ne descrivono l'emissione:
\beq
S= \frac{1}{4\p \a^{`}} \int{ d^2 \x [\sqrt{g} g^{\a\b} G_{\m\n} 
   \partial_{\a}{X^\m} \partial_{\b}{X^\n} +}  \nonumber \\
   \e^{\a\b} B_{\m\n}
   \partial_{\a}{X^\n} \partial_{\b}{X^\n} + \frac{\a^{'}}{2}\sqrt{g}R \f]  \  .
\label{sig}
\eeq
Utilizziamo il metodo del campo di background, scrivendo
\be
X^\m ( \x ) = X^\m_0( \x ) + \p^\m( \x )  \ ,
\ee
ed effettuando l'integrazione funzionale solo sulle fluttuazioni $ \p^\m$; 
perch\`e il metodo sia manifestamente covariante dal punto di vista 
spaziotemporale introduciamo il sistema di coordinate normali di Riemann. 
Consideriamo  la geodetica $ \l(t) $ che collega i punti $X^\m_0$ 
e $X^\m_0 + \p^\m$ ed esprimiamola in termini del vettore tangente a $\l$ per 
$t=0$, che indichiamo con $\h$
\be
\l^\m (t) = X^\m_0 + \h^\m t - \frac{1}{2} \G^\m_{\s_1\s_2} \h^{\s_1} \h^{\s_2} 
t^2 - \frac{1}{3} \G^\m_{\s_1 \s_2 \s_3} \h^{\s_1} \h^{\s_2} \h^{\s_3} t^3+  ... \
,
\ee
Per $t=1$ abbiamo un cambiamento di coordinate in un intorno di $X^\m_0$ dal 
campo $ \p$ al sistema di coordinate $ \h$ che \`e appunto il sistema di 
coordinate normali
\be
\p^\m = \h^\m - \frac{1}{2} \G^\m_{\s_1\s_2} \h^{\s_1} \h^{\s_2} t^2 + ...  \ ,
\ee
In questo sistema i simboli di Christoffel si annullano, come i 
simboli $ \G^\m_{\s_1\s_2...\s_n}$ di ordine superiore, e l'espansione di un 
tensore \`e quindi molto pi\`u semplice. Otteniamo ad esempio
\be
G_{\m\n}(X_0 + \p) = G_{\m\n}(X_0) + \frac{1}{3}R_{\m\l\s\n} \h^\l \h^\s + ... \  ,
\ee
Espandiamo nello stesso modo $\part_a(X^\m_0 + \p^\m)$, $B_{\m\n}$ e $\f$; 
introduciamo anche un vielbein $e^i_\m (X_0)$ con 
$e^i_\m(X_0) e^j_\n (X_0) \d_{ij} = G_{\m\n}(X_0)$, 
in modo da diagonalizzare il 
termine cinetico per gli $\h^i = e^i_\m (X_0) \h^\m$. A questo punto possiamo 
espandere l'azione (\ref{sig}), trascurando i termini lineari in $\h$ e 
mantenendo quelli con al pi\`u due derivate dei campi di background ( 
l'espansione perturbativa funziona quindi per campi lentamente variabili su 
scale dell'ordine di $\sqrt{\ap} $), ottenendo
\beq
& & S(X_0 + \p) = S(X_0) + \frac{1}{4 \p \ap} \int d^2 \x \sqrt{\g} \g^{ab}
[G_{\m\n} \btd_a \h^\m \btd_b \h^\n + 
R_{\m\l\s\n} \part_a X^\m_0 \part_b X^\n_0 \h^\l \h^\s] + \nonumber \\
& & \frac{1}{3 \p \ap} \int d^2 \x \sqrt{\g} \g^{ab}
R_{\m\l\s\n} \part_a X^\m_0 \h^\l \h^\s \btd_b \h^\n + 
\frac{1}{12 \p \ap} \int d^2 \x \sqrt{\g} \g^{ab}
R_{\m\l\s\n} \h^\l \h^\s \btd_a \h^\m \btd_b \h^\n + \nonumber \\
& & \frac{1}{4 \p \ap} \int d^2 \x \e^{ab}
[H_{\m jk} \part_a X^\m_0 \btd_b \h^i \h^j + 
\frac{1}{2} \btd_i H_{\m\n j} \part_a X^\m_0 \part_b X^\n_0 \h^i \h^j ] +
\nonumber \\
& & \frac{1}{12 \p \ap} \int d^2 \x \e^{ab}
H_{ijk} \h^i \btd_a \h^j \btd_b \h^k + 
\frac{1}{8 \p } \int d^2 \x \sqrt{\g} R \btd_i \btd_j \f \h^i \h^j + ...  \ .
\eeq

La strategia per determinare l'anomalia di Weyl \`e la seguente; calcoliamo 
perturbativamente $<T_{zz}>$ ed utilizziamo l'equazione di conservazione 
$\btd^{\bar{z}}<T_{\bar{z}z}> + \btd^z<T_{zz}> = 0 $ per dedurre 
$<T_{\bar{z}z}>$. Il calcolo su un worldsheet piatto \`e particolarmente 
semplice; l'equazione di conservazione diviene $ q_+<T_{-+}>+q_-<T_{++}> = 0$
e bisogna calcolare solo tre diagrammi; ad esempio uno solo coinvolge la 
curvatura ed \`e
\be
\int \frac{d^2k}{2 \p} \frac{k_+(k_++q_+)}{k^2(k+q)^2}
R_{\m\n}\part_a X^\m_0 \part^a X^\n_0 (q) \  ,
\ee
e poich\`e 
\be
\int \frac{d^2k}{2 \p} \frac{k_+(k_++q_+)}{k^2(k+q)^2} = -\frac{q_+}{4q_-} \  ,
\ee
si ottiene
\be
<T_{-+}>( \x ) = \frac{1}{4}R_{\m\n}(X_0)\part_a X^\m_0 \part^a X^\n_0 ( \x ) \ .
\ee

L'inclusione del dilatone, che compare 
accoppiato alla curvatura scalare, \`e pi\`u laboriosa. 
Il suo contributo a $\b^G_{\m\n}$ e 
$\b^B_{\m\n}$   pu\`o essere 
determinato facilmente se si osserva che esso compare con una potenza di $\ap$
in pi\`u rispetto a $B_{\m\n}$ e $G_{\m\n}$ e quindi il calcolo da effettuare 
\`e classico; inoltre, mentre
l'accoppiamento del dilatone si annulla su un worldsheet piatto, non si annulla 
la sua variazione rispetto alla metrica, che altera il tensore di 
energia-impulso. C'\`e infatti un termine aggiuntivo :
\be
T^{\f}_{-+} = \Box \f (X( \x ))  \ .
\ee
Possiamo calcolare questo contributo a livello classico e cancellarlo 
con il contributo ad un loop del gravitone e del tensore antisimmetrico,
che sono dello stesso ordine in $ \ap$. In questo modo otteniamo
\be
\b^G_{\m\n} = R_{\m\n} - \frac{1}{4}H^2_{\m\n} + 2\btd_\m \btd_\n \f  \ ,
\ee
\be
\b^B_{\m\n} = \frac{1}{2} \btd^\l H_{\l\m\n} - \btd^\l \f H_{\l\m\n}  \ .
\ee
Anche per determinare $\b^{\f}$ possiamo evitare calcoli su spazi curvi 
studiando invece funzioni a due punti del tensore energia impulso su uno 
spazio piatto.
Dopo aver valutato alcuni diagrammi a due loops si ottiene
\be
\b^{\f} = \frac{D}{6} + \frac{\ap}{2}[ -R + \frac{1}{12}H^2 + 4(\btd \f)^2
-4 \btd^2 \f]   \ .
\ee

Perch\`e il modello $\s$ sia invariante di Weyl dobbiamo porre a zero tutti i 
coefficienti dell'anomalia : $\b^G_{\m\n} = \b^B_{\m\n} = \b^{\f} = 0$; 
un'importante condizione di consistenza di questo metodo \`e che quando 
$\b^G_{\m\n}$ e $\b^B_{\m\n} $ vengono posti a zero, si pu\`o mostrare che 
$\b^{\f}$, che viene interpretato come carica centrale della teoria conforme, 
\`e costante. Resta da 
vedere se i coefficienti dell'anomalia possono essere annullati 
contemporaneamente; questo \`e possibile proprio perch\`e queste equazioni sono 
derivabili dall'azione 
\be
S = \int d^{D}X \sqrt{G} e^{-2 \f} [ R + 4 ( \btd \f)^2 - \frac{1}{12}H^2]  \ ,
\ee
ovvero, riscalando la metrica $ G = g e^{\frac{4}{D-2} \f}$
\be
S = \int d^{D}X \sqrt{g}[ R -\frac{4}{D-2} ( \btd \f)^2 - \frac{1}{12}
e^{-\frac{8 \f}{D-2}} H^2]   \ .
\ee
Questa non \`e altro che l'azione di Einstein con termini cinetici per il 
dilatone e per la 2-forma antisimmetrica. Si vede cos\`\i \, che richiedere 
l'invarianza conforme del modello $\s$ equivale ad imporre determinate 
equazioni per i campi di background, e queste equazioni sono 
quelle della relativit\`a generale accoppiata con gli altri campi 
dello spettro a massa nulla della stringa.
Il legame diretto con la teoria di stringhe si vede se si utilizza questa 
azione per calcolare ampiezze di scattering ad albero; si pu\`o verificare che 
esse riproducono i risultati di stringa espandendo questi ultimi in potenze di 
$\ap$. 

Questo \`e in realt\`a  
un altro metodo  per determinare le azioni 
effettive.
Noto lo spettro, abbiamo parte dell'informazione necessaria per costruire 
un'azione effettiva; bisogna tuttavia specificare anche le interazioni e per 
fare questo si calcolano ampiezze in teoria di stringhe e si fissano gli 
accoppiamenti della lagrangiana in modo da riprodurre, ordine per ordine in 
$ \ap$, la stessa ampiezza. L'azione effettiva contiene per costruzione solo i 
modi non massivi e quindi lo scambio di modi massivi si manifesta 
indirettamente attraverso accoppiamenti non rinormalizzabili.
Per i modelli supersimmetrici in dieci dimensioni, e pi\`u in generale 
per teorie 
effettive  con almeno sedici supercariche, esiste un'unica lagrangiana del 
secondo ordine nelle derivate consistente con tutte le simmetrie della teoria e 
quindi, in questi casi, l'azione effettiva pu\`o essere scritta senza fare 
calcoli. Descriviamo in maggiore dettaglio questo metodo e come si ottengono 
termini superiori nell'espansione in derivate \cite{g1,g2}.

Noto lo spettro, costruiamo $L_{2pt}$, la lagrangiana libera che descrive 
le particelle non massive della teoria; aggiungiamo quindi dei termini cubici, 
che tengono conto dei loro accoppiamenti a tre punti in modo da riprodurre il 
vertice della teoria di stringhe, ottenendo $L_{3pt}$. Consideriamo quindi 
un'ampiezza a quattro punti; l'unitariet\`a garantisce che i poli non massivi 
sono quelli generati dai grafici ad albero di $L_{3pt}$, 
mentre gli altri contributi, dovuti allo scambio di particelle massive,
non sono singolari per piccoli momenti esterni ed ammettono quindi 
un'espansione in potenze di $\ap p^2$. Ogni termine di questa espansione pu\`o 
essere riprodotto da un operatore locale che includiamo nella lagrangiana 
costruendo $L_{4pt}$. In altri termini, costruiamo un'ampiezza a quattro punti 
in teoria di stringa $A^4_{str}$ e un'ampiezza in teoria di campo $A^4_{ft}$ 
utilizzando $L_{2pt} + L_{3pt}$; per unitariet\`a la differenza 
$A^4_{str} - A^4_{ft}$ non contiene poli per $p^2 \sim 0$ e possiamo 
espanderla ottenendo i nuovi termini di contatto.
La procedura pu\`o essere iterata costruendo ogni $L_{Npt}$.
Questa lagrangiana, come quella che si ottiene dalle funzioni $\b$, \`e 
determinata solo a meno di una trasformazione non singolare dei campi, in 
quanto le ampiezze di scattering non ne vengono alterate. Se infatti 
$ \f^{'} \mapsto \f( \f^{'})$, 
$\frac{\d L[\f]}{\d \f_i} = \frac{\d L^{'}[\f^{'}(\f)]}{\d \f_i} =
\sum_j \frac{\d L^{'}}{\d \f^{'}_j} \frac{\d \f^{'}_j}{\d \f_i} $; allo stesso 
modo $ \b_i( \f) = \L \frac{ \part \f_i( \f^{'})}{\part \L} =
\b^{'}_j ( \f^{'})\frac{\part \f_i}{\part \f^{'}_j}$. Fissato il modello $\s$, 
entrambe le condizioni devono essere soddisfatte e perch\`e il sistema
$\b_i = 0 $ , $ \frac{\d L}{\d \f_i} =0$ non sia sovradeterminato, le due 
equazioni devono coincidere a meno di ridefinizioni dei campi, come \`e stato 
esplicitamente verificato per i primi termini della serie.

Consideriamo ora il settore bosonico delle teorie di Tipo II, che contengono
$h_{mn}$ , $B_{mn}$ e $\f$. La presenza di una particella priva di massa
di spin due e 
l'invarianza di Lorentz della teoria assicurano che l'azione effettiva per $h$ 
\`e l'azione di Einstein, a meno di correzioni di ordine superiore nelle
derivate; per l'invarianza di 
gauge della 2-forma  $ B_{mn} \mapsto B_{mn} + \part_{[m}\e_{n]}$, 
solo la combinazione gauge invariante $H = dB = \part_{[c}B_{ab]}$ pu\`o 
comparire nella lagrangiana. Infine uno shift del dilatone modifica la 
costante che moltiplica l'azione classica di stringa; questo indica che, 
dopo una trasformazione conforme, il dilatone si accoppia tramite
$e^{c (1+w) \f}$, dove $w$ \`e il peso conforme del termine nella lagrangiana e 
$c$ una costante. Complessivamente:
\be
L_{2pt} = \frac{1}{2 \k^2} R - \frac{1}{6}e^{-2c \f}H^2 - \frac{1}{2}(\btd \f)^2   \ ,
\ee
con $\k^2 = 8 \p G$. L'ampiezza per tre vettori di momento $k_i$
e polarizzazione $\r_i$ in teoria di stringhe aperte \`e:
\be
A_3^o = \r_1K_2 \r_2 \r_3 + \r_2K_3 \r_3 \r_1 + \r_3K_1 \r_1 \r_2  \  .
\ee
Ponendo inoltre $\t_{mn} = \r_m \otimes \r_n$, l'ampiezza per stringhe chiuse \`e:
\beq
& & A_3 = 4g^2 \tilde{A}^o_3 \otimes A^o_3 = \nonumber \\
& & g[(k_2 \t_1 k_2)tr(\t_2 \t^T_3) +  (k_3 \t_2 \t_3^T \t_1 k_2) + 
(k_1 \t_3 \t_2^T \t_1 k_2)  + perm. cicl. ]  \   .
\label{a3}
\eeq
$A_3$ contiene solo termini con due derivate, e questo implica che non sono 
presenti nuove interazioni; possiamo comunque legare $\k$ e $c$ a $g$ e $\ap$. 
Con $\t_{mn} = h_{mn}$, simmetrico e a traccia nulla, 
otteniamo da (\ref{a3}) l'ampiezza per tre gravitoni:
\beq
A_{hhh} = g[(k_2 \h_1 k_2)tr(\h_2 \h_3) + & (k_3 \h_2 \h_3 \h_1 k_2) + &
(k_1 \h_3 \h_2 \h_1 k_2)  + perm. cicl. ]    \ .
\eeq
Questa ampiezza \`e ottenibile da
\be
L_{hhh} = -\frac{g}{2}(h^{ab}h^{cd}\part_a\part_b h_{cd} +2 \part^d h_ab 
\part^a h^{bc} h_{cd})   \   ,
\ee
mentre espandendo l'azione di Einstein con $g_{mn} = \h_{mn} + 2 \k h_{mn}$ si 
ottiene il termine cubico :
\be
\frac{1}{2 \k^2} \sqrt{-g}R = - \k(h^{ab}h^{cd}\part_a\part_b h_{cd} +
2 \part^d h_ab \part^a h^{bc} h_{cd})   \   .
\ee
Abbiamo quindi 
\be
2 \k = g (2 \ap)^2  \  .
\ee
Analogamente, si pu\`o determinare $c$ calcolando $A_{BBD}$ e si ottiene 
$ c = \frac{\k}{\sqrt{2}}$.

Passiamo ora all'ampiezza a quattro punti, che per le stringhe aperte \`e
\be
A_4^o = - \frac{\G(-s/2) \G(-t/2)}{\G(1+u/2)} t_{mnpqrsuv}
f_1^{mn}f_2^{pq}f_3^{rs}f_4^{uv}  \  ,
\ee 
e per le stringhe chiuse
\beq
A_4 = 2g^2 \tilde{A}_4^o \otimes A^o_4 
= -2g^2  \frac{\G(-s/8) \G(-t/8) \G(-u/8)}{ \G(1+s/8) \G(1+t/8) \G(1+u/8)} 
\nonumber \\
\frac{1}{16} [t_{abcdefgh} \tilde{f}_1^{ab} \tilde{f}_2^{cd} 
\tilde{f}_3^{ef} \tilde{f}_4^{gh}]
\otimes
\frac{1}{16}[t_{mnpqrsuv}f_1^{mn}f_2^{pq}f_3^{rs}f_4^{uv}] \  ,
\eeq
dove $s$ , $t$ , $u$ sono gli usuali invarianti cinematici, 
$f_i^{mn} = k_i^m \r_i^n - k_i^n \r_i^m$ e
\be
t_{mnpqrsuv}f_1^{mn}f_2^{pq}f_3^{rs}f_4^{uv} =
\frac{1}{2} [tr(f_1f_2f_3f_4) - \frac{1}{4}tr(f_1f_3)tr(f_2f_4) + 
perm. cicl.]  \ .
\ee
Per ottenere termini al pi\`u quartici nei momenti possiamo espandere
\be
\frac{\G(-s/8) \G(-t/8) \G(-u/8)}{ \G(1+s/8) \G(1+t/8) \G(1+u/8)} \sim
- \frac{2^9}{stu} - 2 \z(3) \ . 
\ee
Il primo termine corrisponde ad interazioni gi\`a 
presenti in $L_{3pt}$; il secondo termine d\`a origine a nuovi accoppiamenti 
che, essendo $\z(3)$ un numero trascendente, non possono essere determinati
semplicemente con considerazioni di supersimmetria. 
Per il gravitone, $f^{ab} \otimes f_{cd}$ corrisponde al tensore di Riemann 
linearizzato: $R^{ab}{}_c{}_d \sim - \k\part^{[a}\part_{[c}h^{b]}_{d]} =
\k k^{[a}k_{[c} \tilde{\r}^{b]}\otimes \r_{d]} =
\k f^{ab} \otimes f_{cd}$ e, di conseguenza,
\be
L_{\z(3) gr} = \frac{\z(3)}{3 \cdot 2^7 \k^2} e^{-\frac{3}{\sqrt{2}} \k \f}
t^{abcdefgh}t^{mnpqrsuv}R_{abmn}R_{cdpq}R_{efrs}R_{ghuv}   .
\ee
In modo analogo, per la 2-forma 
$f_{ab} \otimes f_{cd} \sim \btd_{[a}H_{b]cd}$, e per il dilatone
$f_{ab} \otimes f^{cd} \sim - \frac{1}{\sqrt{8}} 
\h_{[a}^{[c}\btd_{b]} \btd^{d]} \f$; se si definisce
\be
\bar{R}_{ab}{}^c{}^d = R_{ab}{}^c{}^d + \k e^{-\frac{1}{\sqrt{2}}\k D}
\btd_{[a} H_{b]}{}^c{}^D - \frac{1}{\sqrt{8}} \k 
\h_{[a}{}^{[c} \btd_{b]} \btd^{d]} \f  \ ,
\ee
si pu\`o scrivere la lagrangiana completa descrivente gli accoppiamenti 
quartici a meno di termini $O(p^6)$:
\beq
L_{4pt} & = & \frac{1}{2 \k^2} R - \frac{1}{6} e^{- \sqrt{2} \k \f} H^2 
- \frac{1}{2}(\btd \f)^2  
+ \frac{\z(3)}{2^8 \k^2} e^{-\frac{3}{\sqrt{2}} \k D}
[\bar{R}_{abmn}\bar{R}^{bcnp}\bar{R}_{cd}{}^{qm}\bar{R}^{da}{}_{pq} \nonumber \\
& + & \frac{1}{2}\bar{R}_{abmn}\bar{R}^{bcnp}\bar{R}_{cdpq}
\bar{R}^{daqm} - 
\frac{1}{2}\bar{R}_{abmn}\bar{R}^{bcmn}\bar{R}_{cdpq}\bar{R}^{dapq}  \nonumber \\
& - & \frac{1}{4} \bar{R}_{abmn}\bar{R}^{bcpq}\bar{R}_{cd}{}^{mn}
\bar{R}^{da}{}_{pq} 
+ \frac{1}{16}\bar{R}_{abmn}\bar{R}^{bapq}
\bar{R}_{cdmn}\bar{R}^{dcpq} \nonumber \\ 
&+& \frac{1}{32}\bar{R}_{abmn}\bar{R}^{bamn}\bar{R}_{cdpq}\bar{R}^{dcpq}] \ .
\eeq

La stringa di Tipo I e la stringa eterotica hanno come limite di bassa energia
la supergravit\`a con $N=1$ in dieci dimensioni accoppiata a SYM;
gi\`a dalla funzione a tre punti, in questo caso, si ottengono importanti 
modifiche della teoria di supergravit\`a. Includendo 
l'azione di Yang-Mills per i campi di gauge abbiamo
\be
L_{2pt}= \frac{1}{2 \k^2} R - \frac{1}{6} e^{- \sqrt{2} \k D} H^2 
- \frac{1}{2}(\btd D)^2  - \frac{1}{4}e^{-cD}F^{2}  \  .
\ee
In particolare, per la stringa eterotica le ampiezze a tre punti 
\beq
A_{ggg} &=& g[(k_2 \t_1 k_2)tr(\t_2 \t_3) + (k_3 \t_2 \t_3 \t_1 k_2) + 
(k_1 \t_3 \t_2 \t_1 k_2)   \nonumber \\ 
&+& \frac{1}{4}(k_2\t_1k_2)(k_3\t_2\t_3^Tk_1) + perm. cicl. ]   \ ,  \nonumber \\
A_{AAg} &=& -\frac{g}{2}[Tr(A_1A_2)(k_1\t_3k_1)+Tr(A_1^aA_2^b)
(k_{2a}k_1^c\t_{3cb}+k_{3b}k_1^c\t_{3ca})] \ ,  \nonumber \\
A_{AAA} &=& -g[Tr([A^m_1,A_{2m}]A^n_3)k_{1n} + perm.cicl.]  \ ,
\eeq
portano a ridefinire
\be
H_{abc} = \part_{[c}B_{ab]} - \frac{\k}{4}\w^{CS}_Y + 
\frac{\k}{4g_Y^2}\w^{CS}_L   \  ,
\label{hb}
\ee
dove $g_Y$ \`e la costante d'accoppiamento di gauge e dove
\beq
\w^{CS}_Y = A \ww F - \frac{1}{3} A \ww A \ww A  \ , & &
\w^{CS}_L = \w \ww R -\frac{1}{3} \w \ww \w \ww \w  \    ,
\eeq
sono le forme di Chern-Simons per la connessione di gauge e per la connessione 
di spin. L'invarianza dell'azione rispetto a 
trasformazioni di gauge e di Lorentz locali con parametri, rispettivamente,
$\L$ e $\W$, richiede inoltre che $B_{mn}$ trasformi secondo
\be
B \mapsto B + \frac{\k}{4} Tr(A \ww \L) - \frac{\k}{4} Tr(\w \ww \W)   \  .
\ee
Sottolineamo che il nuovo elemento introdotto dalla teoria di stringhe \`e la 
forma di Chern-Simons gravitazionale, in quanto l'inclusione della forma di 
Chern-Simons di gauge \`e richiesta gi\`a dall'accoppiamento della 
supergravit\`a con $N=1$ con la teoria di SYM.
Tra $\k$, $c$ e i parametri di stringa otteniamo le stesse relazioni 
del caso con $N=2$; inoltre $g_Y = g(2 \ap)^{3/2}$.

\section{Anomalie}
\markboth{} {}

Nella sezione precedente abbiamo visto come il limite di bassa energia delle 
varie teorie di stringhe sia dato da teorie di supergravit\`a; per la Tipo IIA 
abbiamo la supergravit\`a non chirale $2A$, per la Tipo IIB 
abbiamo la supergravit\`a chirale $2B$, per la Tipo I la supergravit\`a 
chirale $N=1$ accoppiata a SYM con gruppo di gauge $O(32)$; per le due 
teorie eterotiche, che verranno introdotte nel prossimo capitolo, abbiamo 
la supergravit\`a chirale con $N=1$ accoppiata a SYM con gruppo di 
gauge $O(32)$ o $E_8 \times E_8$.

La consistenza della fisica di bassa energia richiede che queste teorie non 
abbiano anomalie di gauge e gravitazionali. Questo \`e ovvio per la IIA che non 
\`e chirale; inoltre, in \cite{An1} \`e stato mostrato come il contenuto di campi della 
IIB sia tale da cancellare tutte le anomalie. Abbiamo infatti una quattro forma 
autoduale, due gravitini di Majorana-Weyl della stessa chiralit\`a
e due fermioni di Majorana-Weyl di chiralit\`a opposta a quella dei gravitini.
Ricordiamo che l'anomalia gravitazionale di una particella di spin $1/2$ 
in $4k+2$ dimensioni si pu\`o ottenere da:
\be
I_{1/2}(R) = \prod_{i=1}^{2k+1}(\frac{x_i}{\sinh{x_i}})   \  ,
\ee
estraendo il termine omogeneo di grado $k+1$; le $y_i$ sono definite in modo 
tale che $\sum_i^{2k+1}y_i^{2m} = \frac{1}{2}(-\frac{1}{4})^m trR^{2m}$.
Se il fermione \`e carico rispetto ad un campo di gauge, le anomalie 
gravitazionali, di gauge e miste sono date da
\be 
I_{1/2}(R,F) = tr(e^{iF})I_{1/2}(R)   \  ,
\ee
dove la traccia \`e nella rappresentazione dei fermioni.
Infine, le anomalie gravitazionali di un gravitino e di una forma antisimmetrica 
autoduale si ottengono da
\beq
I_{3/2} = I_{1/2}(-1 + 2 \sum_{i=1}^{2k+1} \cosh{2x_i})  \   ,  \\
I_{A} = - \frac{1}{8} \prod_{i=1}^{2k+1} \frac{2x_i}{\tanh{2x_i}}  \   .
\eeq
\`E facile verificare che in dieci dimensioni:
\be
I_{3/2} - I_{1/2} + I_{A} = 0  \  ,
\label{2b}
\ee
un risultato notevole, dato che nessun coefficiente nella (\ref{2b}) pu\`o essere 
aggiustato in modo arbitrario.

Allo stesso modo si vede che la pura supergravit\`a $N=1$ \`e una teoria 
anomala in dieci dimensioni.
Le anomalie possono comunque essere cancellate accoppiando la teoria a dei 
supermultipletti di Yang-Mills ed utilizzando il meccanismo 
di Green-Schwarz \cite{an2}. L'anomalia \`e infatti data da una 12-forma
\beq
I_{12} &=& -\frac{1}{720}TrF^6 + \frac{1}{24 \cdot 48}TrF^4trR^2
- \frac{1}{256}TrF^2[\frac{1}{45}trR^4 + \frac{1}{36}(trR^2)^2] \nonumber \\
&+& \frac{n-496}{64}[\frac{1}{2 \cdot 2835}trR^6 +
\frac{1}{4 \cdot 1080}trR^2trR^4 + 
\frac{1}{8 \cdot 1296}(trR^2)^3] \nonumber \\
&+& \frac{1}{384}trR^2trR^4 + \frac{1}{1536}(trR^2)^3   \  .
\eeq
dove $Tr$ indica la traccia nell'aggiunta e $n$ la dimensione del gruppo di 
gauge; se $I_{12}$ fattorizza secondo 
$I_{12} = (trR^2 + kTrF^2)X_8$, l'anomalia pu\`o essere cancellata da un 
controtermine locale $B \ww X_8$ se $\d B = Tr(\L F) - tr(\W R)$. Questa 
trasformazione di $B$ \`e proprio quella indotta dalla ridefinizione della sua 
curvatura in (\ref{hb}). Una condizione necessaria per questo, dovuta alla 
presenza di un Casimir del sesto ordine per $SO(10)$, \`e che sia nullo il 
coefficiente della parte irriducibile dell'anomalia gravitazionale $trR^6$; 
questo implica $n=496$. \`E poi necessario che
\be
TrF^6 = \frac{1}{48}TrF^2TrF^4 - \frac{1}{14400}(TrF^2)^3  \  ,
\ee
relazione soddisfatta dai gruppi $SO(32)$,  $E_8 \times E_8$  \ ,
$E_8 \times U(1)^{248}$ e $U(1)^{497}$. Come vedremo, nella Tipo I si ottiene 
il gruppo di gauge $O(32)$, mentre nell'eterotica si pu\`o avere sia $O(32)$ che 
$E_8 \times E_8$; le altre due teorie di 
supergravit\`a non anomale, non hanno ancora un'interpretazione nell'ambito 
della teoria di stringhe.
La presenza del controtermine necessario per la cancellazione della anomalia 
\`e stata verificata con un calcolo ad un loop in \cite{hod1}, 
dove si \`e anche 
mostrata la relazione tra invarianza modulare e assenza di anomalie. 

\`E importante osservare che, integrando $dH = trR^2 - \frac{1}{30}TrF^2$ 
su di una sottovariet\`a chiusa quadridimensionale, si ottiene un vincolo sui 
campi di background:
\be
\int trR^2 = \frac{1}{30}\int TrF^2   \  .
\ee

\`E importante considerare anche correzioni all'azione effettiva di bassa 
energia dovute ai loops di stringa, che possono essere derivate 
calcolando funzioni di correlazione su superfici di genere $g>0$. Metodi 
generali per estrarre in modo consistente questi contributi dalle ampiezze e 
in particolare per regolarizzare le divergenze infrarosse senza violare 
l'invarianza modulare, sono stati sviluppati in \cite{tc1,tc2}.

\chapter{Compattificazioni}
\markboth{} {}

Abbiamo considerato finora la dinamica della teoria di stringhe in vuoti 
invarianti di Poincar\`e in dieci dimensioni. Molti degli 
aspetti pi\`u profondi della teoria si manifestano quando si utilizzano teorie 
conformi pi\`u generali del tipo $ \it{C}_{st} \times \it{C}_{int} $, dove 
$ \it{C}_{st} $ descrive $ d $ supercoordinate $ X^\m (x,\t) $ ed \`e pertanto 
ancora associabile ad uno spazio minkowskiano con $ d < 10 $ dimensioni.
$ \it{C}_{int} $ rappresenta una generica teoria conforme con 
carica centrale $ c = 15 - \frac{3}{2} d $, in modo da eliminare
l'anomalia nella traccia del tensore energia-impulso.
Se $ \it{C}_{\it{int}} $ \`e un modello $ \s $
non lineare con una particolare variet\`a come spazio bersaglio, abbiamo 
ancora una naturale interpretazione geometrica della teoria; questa 
interpretazione non \`e tuttavia necessaria, e proprio in questa fusione di 
concetti algebrici e geometrici risiede una delle caratteristiche pi\`u 
affascinanti della teoria. 
In effetti nozioni geometriche spaziotemporali sono rilevanti solo a scale 
molto pi\`u grandi della scala di Planck; la scelta  essenziale \`e 
quella della teoria conforme, che pu\`o anche corrispondere a geometrie diverse, 
ma equivalenti fisicamente.

Una volta deciso come trattare i gradi di libert\`a interni si possono studiare 
vari aspetti del modello risultante: l'invarianza modulare, il numero di 
supersimmetrie, le simmetrie di gauge, lo spettro a massa nulla, gli 
accoppiamenti di Yukawa tra gli stati. Lo studio delle compattificazioni su 
variet\`a non banali si \`e basato inizialmente sulle propriet\`a della 
supergravit\`a effettiva, ed in questo contesto molte delle propriet\`a 
appena elencate possono essere descritte in modo completo ed elegante in quanto 
sono legate a caratteristiche topologiche della variet\`a considerata (che 
influenzano pertanto solo gli zero modi, il che spiega perch\`e sia sufficiente 
considerare la sola azione di bassa energia).

L'approccio pi\`u soddisfacente resta comunque una trattazione analitica 
completa; perch\`e questo sia possibile la teoria conforme interna deve essere 
esattamente risolubile. 
Una classe di teorie ovviamente adatta allo scopo \`e 
quella che utilizza campi liberi; in questa classe rientrano molte costruzioni 
importanti che discuteremo in dettaglio in questo capitolo: le 
compattificazioni toroidali, gli orbifolds e le costruzioni con fermioni 
liberi. Lo studio di queste compattificazioni ci consentir\`a inoltre 
di costruire
una nuova teoria di stringa consistente in dieci dimensioni, la 
stringa $ \bf{eterotica} $   $\cite{het1} $,
e di introdurre un primo esempio di spazio di moduli. Questi spazi sono di 
grande importanza, perch\`e descrivono le possibili deformazioni di un modello;
studiandoli con attenzione si possono
stabilire relazioni tra 
diverse teorie di stringhe che mostrano come la perdita di unicit\`a della 
teoria che si riscontra dopo la compattificazione sia meno grave
di quando non sembri inizialmente.
Le relazioni che discuteremo in questo capitolo, chiamate
$ \bf{T-dualita^{`} } $, sono state le prime ad essere scoperte in quanto sono 
valide ordine per ordine nell'espansione perturbativa.
Altre relazioni, valide non perturbativamente, verranno discusse nel quarto 
capitolo.

Ad energie molto inferiori alla scala di Planck, come abbiamo visto, la natura 
estesa della stringa non si manifesta e la teoria pu\`o essere ben 
descritta con 
una lagrangiana effettiva contenente campi con spin diversi (ma $\le 2$) e diversi numeri 
quantici; questi campi non sono altro che i modi degeneri di vibrazione della 
stringa al livello di massa pi\`u basso. Spettro, simmetrie e accoppiamenti di 
questa lagrangiana effettiva sono determinati dalla teoria di stringa. Il passo 
successivo, necessario per raggiungere l'intervallo di 
energie attualmente accessibile sperimentalmente
e legare quindi  i parametri di stringa  a quelli del modello 
standard, consiste nel ricavare da $ L_{eff}(M_P) $ una lagrangiana effettiva 
alla scala elettrodebole, $ L_{eff}(M_W) $. Il problema principale 
\`e che il numero di vuoti classici \`e enorme e 
le caratteristiche di bassa energia della teoria, 
come $ L_{eff} $, sono particolarmente sensibili alla scelta del vuoto, in 
quanto descrivono le piccole fluttuazioni attorno ad esso.

Estrarre informazioni fenomenologiche da questi vuoti richiede qualche cautela, 
in quanto la fisica di un sistema non \`e necessariamente descritta in modo 
accurato da una approssimazione semiclassica. 
Inoltre il modello, nella sua trasformazione dalla scala di Planck a quella 
elettrodebole deve subire una serie di rotture di simmetrie, in particolare
la rottura della supersimmetria, che non sono ancora comprese a fondo.
Nonostante questo \`e affascinante osservare quanto vicini si possa arrivare 
con queste semplici costruzioni a descrivere la fenomenologia attualmente nota.
Nella costruzione fermionica, ad esempio, c' \`e un modello che realizza 
un gruppo $SU(5) \times U(1)^4 $ "flippato" con tre generazioni di fermioni e 
Higgs in grado di rompere la simmetria di gauge a quella del modello standard
\cite{ff4};
dopo la rottura, lo spettro \`e una variante di quello del modello standard 
supersimmetrico minimale; la vita media del protone \`e compresa tra $ 10^{33}$ 
e $ 10^{57} $ anni, gli ulteriori fattori $U(1)$ necessari per il modello 
flippato emergono in modo naturale e in modo naturale si ha anche un meccanismo 
di "see-saw" per le masse dei neutrini.
Un'altra classe di modelli con tre generazioni \`e stata costruita nel contesto 
delle compattificazioni su spazi di Calabi-Yau da Tian e Yau e 
numerosi altri esempi pi\`u o meno plausibili fenomenologicamente sono noti.
In teorie di stringhe aperte, infine, modelli con tre generazioni appaiono in modo
naturale \cite{4d}.

\section{Compattificazioni toroidali}
\markboth{} {}

\subsection{Compattificazione su $ S^1 $}
Esaminiamo il caso di un'unica coordinata 
compatta $ X^{\m} \simeq X^{\m}+2\p R$,
calcolando la funzione di partizione. L'azione \`e
\be
S = \frac{1}{4\p}\int{d^2 \s \sqrt{g}g^{ij}\partial_{i}{X}\partial_{j}{X}} \ .
\ee
dove $ g_{ij} $ \`e la metrica sul toro con modulo $ \t $. A differenza del 
caso non compatto sono ora presenti soluzioni classiche delle 
equazioni del moto interpretabili come configurazioni solitoniche; 
esse descrivono 
stringhe con $n$ unit\`a di momento lungo la direzione circolare che si 
avvolgono $m$ volte attorno ad essa.
Nel calcolo della funzione di partizione bisogna tenere conto di tutti questi 
settori e si ottiene
\be
Z(R) = \frac{R}{\sqrt{\t_2}|\h |^2} \sum_{nm}{e^{-\frac{\p R^2}{\t_2}
 |m-n\t|^2}}  \  .
\label{torp}
\ee
La proporzionalit\`a ad R deriva dall'integrazione sullo zero modo 
dell'operatore di Laplace su di uno spazio compatto (il modo costante), mentre 
il termine $ \sqrt{\t_2}|\h |^2 $ \`e quello usuale per una coordinata 
bosonica; la sommatoria mostra la presenza nello 
spettro dei nuovi stati $ |mn\rangle $. Utilizzando la formula di Poisson,
\be
\sum_{n \in Z} f(2 \p n) = \sum_{m \in Z} \tilde{f}(m)   \ ,
\ee
dove $ \tilde{f}(n) = \frac{1}{2 \p} \int_{- \infty}^{\infty} f(x)e^{ikx} dx$
\`e la trasformata di Fourier della funzione $f(x)$,
la (\ref{torp}) pu\`o essere scritta in modo pi\`u trasparente: 
\be
Z(R)= \sum_{nm}{\frac{q^{\frac{p^{2}_{L}}{2}} \bar{q}^{\frac{p^{2}_{R}}{2}}}{\h
 \bar{\h}}} \  , \label{part}
\ee
dove $ q = e^{2\p i \t } $ e
\be
p_{L} = \frac{1}{\sqrt{2}}(\frac{m}{R} + nR)  \ , \hspace{2cm} 
p_{R} = \frac{1}{\sqrt{2}}(\frac{m}{R} - nR)  \ .
\ee
Questa funzione di partizione mostra che la teoria contiene infinite 
rappresentazioni di un algebra affine $  U(1)\times U(1) $ caratterizate dalle 
cariche $ (p_L,p_R) $. Agli stati $ |mn\rangle $ corrispondono gli operatori di 
vertice $ V_{mn} = : e^{ip_L X(z)+ip_R \bar{X}(\bar{z})} : $ .
La (\ref{part})  \`e invariante modulare e inoltre 
\be
\lim_{R \rightarrow \infty} \frac{Z(R)}{R} = \frac{1}{\sqrt{\t_2}|\h|^2} \  .
\ee
\`E possibile fare due interessanti osservazioni su
questa teoria. La prima \`e legata alla presenza degli operatori $ V_{mn} $: la 
loro dimensione conforme \`e $ ( \frac{p^2_L}{2}, \frac{p^2_R}{2} ) $, e quando 
R = 1 gli stati con $ m = n = \pm 1 $ e quelli con $ m = -n = \pm 1 $ hanno 
massa nulla mentre gli operatori corrispondenti hanno rispettivamente dimensione (1,
0) e (0,1). Questi operatori sono
\be
j^{\pm}(z)= \frac{1}{\sqrt{2}} e^{\pm i \sqrt{2} X} \ ,
\ee
e insieme a
\be
j^3(z) = \frac{1}{\sqrt{2}} \partial{X} \ ,
\ee
realizzano un'algebra affine $ SU(2)_L $ al livello k=1; in modo del tutto
analogo nel settore destro si ha una simmetria $  SU(2)_R $ . Questo 
allargamento dell'algebra di correnti sul worldsheet si traduce in
un'estensione 
della simmetria di gauge nello spaziotempo. Se infatti consideriamo lo spettro, 
accanto agli stati associati a $ \part X^\m \bar{\part} X^\n $,
che descrivono 
gli usuali campi bosonici, abbiamo, per  $ R $  generico, 
$ \part X^\m \bar{\part} X^{25} $
e $ \part X^{25} \bar{\part} X^\n $ che sono due nuovi campi di gauge e lo 
scalare $ \part X^{25} \bar{\part} X^{25} $; per $R=1$ abbiamo altri quattro 
campi di gauge
$ \part X^\m e^{ \pm i \sqrt{2} \bar{X}^{25}}$,
$ \bar{\part} X^\m e^{ \pm i \sqrt{2} X^{25}}$
che determinano un aumento della simmetria di gauge da $U(1)_L \times U(1)_R$ a 
$SU(2)_L \times SU(2)_R $, come si pu\`o verificare calcolando la funzione a
tre punti, e otto scalari 
$ \part X^{25} e^{ \pm i \sqrt{2} \bar{X}^{25}}$,
$ \bar{\part} X^{25} e^{ \pm i \sqrt{2} X^{25}}$,
$ e^{ \pm i \sqrt{2} X^{25}} e^{ \pm i \sqrt{2} \bar{X}^{25}}$,
che formano un campo di Higgs nella rappresentazione 
$(3,3)$ del gruppo di simmetria.
L'altro aspetto notevole \`e l'invarianza della (\ref{part}) 
sotto la trasformazione
\beq
R \rightarrow \frac{1}{R} &  ,  &
m\longleftrightarrow n  \ .
\label{Tdual}
\eeq
Dal punto di vista della teoria conforme si tratta semplicemente di un 
cambiamento di segno nel settore destro della carica: $ p_L
 \rar p_L $ e $ p_R \rar -p_R $; dal punto di vista spaziotemporale 
questa \`e invece una 
trasformazione geometrica non banale: la prima teoria descrive 
infatti la propagazione 
su un cerchio di raggio R mentre la seconda la propagazione 
su un cerchio di raggio $ \frac{1}{R} $. 
In effetti, per verificare che le (\ref{Tdual}) rappresentano 
un'equivalenza nella teoria di stringa completa bisogna verificare l'invarianza 
della funzione di partizione a genere arbitrario. Per far questo indichiamo con
$ a_i $ e $ b_j$   $ ( i,j=1,...,g  )$ la base canonica per i cicli di omologia 
di una superficie di Riemann di genere g e con $ \w_i $ delle (1,0) forme 
caratterizzate da
\be
\int_{a_i}{\w_j}= \d_{ij} \ , \hspace{2cm}   
\int_{b_i}{\w_j}= \tau_{ij} \ ,
\label{perio}
\ee
dove $ \tau_{ij} $ \`e la matrice dei periodi della superficie e  dipende da
$3(g-1)$ moduli. Sulla coordinata $X$ imponiamo le condizioni
\be
\int_{a_i} dX = 2 \p m_i  \ , \hspace{2cm} 
\int_{b_i} dX = 2 \p n_i  \ .
\ee
La funzione di partizione a genere $g$ \`e data da
\beq
Z_g &=& \int_{\it{M}_g} dm D(m) R \sum_{m_i,n_i} e^{-\p R^2(m+n \t)^t \t_2^{-1}
(m+n \t)} \nonumber \\ &=& 
R^{-g} det \t_2 \sum_{p_L,p_R} e^{i \p ( p_R \t p_R - p_L \bar{\t} p_L)} \ ,
\eeq
dove $\t_2 = Im \t$ e $\t$ \`e la matrice definita in (\ref{perio}); inoltre
\be
p_R= \frac{1}{\sqrt{2}} ( \frac{k}{R} - Rm )  \ , \hspace{2cm} ,  
p_L= \frac{1}{\sqrt{2}} ( \frac{k}{R} + Rm ) \  .
\ee
Se effettuiamo la trasformazione $ R \mapsto \frac{1}{R} $ , $ k \lrar m $ 
abbiamo $ Z_g( \frac{1}{R} ) = R^{2g-2} Z_g(R) $. La funzione 
di partizione completa
\be
Z( \f , R)= \sum_{g=0}^{\infty} e^{(1-g) \f} Z_g (R)
\ee
di una teoria di stringa compattificata 
su di un cerchio di raggio $R$ coincide quindi con quella della teoria su di un 
cerchio di raggio $ \frac{1}{R} $, scambiando momenti e avvolgimenti e 
ridefinendo il dilatone secondo la 
\be
\f \mapsto \f + 2 log R  \ .
\ee
Abbiamo quindi due differenti formulazioni per descrivere identici effetti 
fisici : la precisa relazione tra operatori vertice delle due teorie, 
ovvero tra i loro spettri, e l'invarianza di $ Z_g $ 
assicurano l'uguaglianza delle funzioni di correlazione collegate dalla 
trasformazione appena descritta. Questa relazione, valida ad ogni ordine 
dell'espansione perturbativa, prende il nome di T-dualit\`a.
\`E possibile dare un'elegante interpretazione della T-dualit\`a in termini di 
simmetria di gauge \cite{tor6}.
Quando $R=1$ la trasformazione $ J \mapsto J $ e 
$ \bar{J} \mapsto  - \bar{J} $ \`e una trasformazione di Weyl del gruppo
$ SU(2)_L \times SU(2)_R $; una perturbazione infinitesima della teoria
$ \e \int{ \part X \bar{\part} X } = \e \int{ J^3 \bar{J}^3 } $
ci allontana dal 
raggio autoduale, ma le teorie caratterizzate da $ \pm \e $ sono
equivalenti come conseguenza della simmetria di Weyl discreta che permane dopo 
la rottura della simmetria di gauge con l'usuale meccanismo di Higgs; questa
\`e proprio la versione infinitesima della trasformazione 
$ R \mapsto \frac{1}{R} $.

\subsection{Compattificazioni su $T^d$}

La discussione del paragrafo precedente
pu\`o essere generalizzata a tori di dimensione arbitraria 
con valori costanti di $ G_{ij} $ e $ B_{ij} $
\be
S = \frac{1}{4 \p} \int d^2 \x \sqrt{g}G_{ij} 
g^{\a\b} \part_\a X^i \part_\b X^j + 
\frac{1}{4 \p} \int d^2 \x  \e^{\a\b} B_{ij} \part_\a X^i \part_\b X^j   \ .
\ee
con $ X^i \sim X^i + 2 \p ( m^i \t + n^i \s ) $. $G_{ij}$ e $B_{ij}$ 
rappresentano i moduli della compattificazione, e se introduciamo un vielbein  
$e^a_i$ tale che $e^a_i e^a_j = G_{ij} $ e definiamo $X^a = e^a_i X^i$,
la condizione 
di periodicit\`a diviene $ X^a \sim X^a + 2 \p e^a_i m^i $. I vettori di 
avvolgimento appartengono al reticolo $\L$ generato sugli interi dai vettori 
$e_i$ e si verifica senza difficolt\`a che i momenti appartengono al reticolo 
duale $ \L^*$ generato da $E^i$ dove $E^{a i} = G^{ij} e^a_j $. Valutando 
l'integrale funzionale o utilizzando il formalismo canonico, possiamo anche in 
questo caso calcolare la funzione di partizione
\be
Z_{T^d} = \frac{\sqrt{G}}{ (\sqrt{ \t_2} \h \bar{\h})^d} 
\sum_{m,n} e^{ -\frac{\p}{ \t_2}( G_{ij}+B_{ij})(m_i+n_i \t)(m_j+n_j\bar{ \t})}
= \frac{1}{( \h \bar{\h})^d} \G_{d,d} (G,B)  \  ,
\ee
dove
\be
\G_{d,d}(G,B) = \sum_{m,n} q^{\frac{p_L^2}{2}} \bar{q}^{\frac{p_R^2}{2}} \  ,
\ee
si ottiene mediante una risommazione di Poisson e 
\be
p_L^i = \frac{G^{ij}}{\sqrt{2}}[m_j+(B_{jk}-G_{jk})n_k]  \ , \hspace{1cm}
p_R^i = \frac{G^{ij}}{\sqrt{2}}[m_j+(B_{jk}+G_{jk})n_k]  \  . 
\label{mom}
\ee
Possiamo individuare lo spettro dalle due condizioni $ L_0 = \bar{L}_0 =0$, che 
riscriviamo utilizzando gli operatori numero nella forma:
\beq
& & H = L_0 + \bar{L}_0 = \frac{1}{2} ( p_L^2 + p_R^2) +N 
+ \bar{N} -2 \  , \nonumber   \\
& & L_0 - \bar{L}_0 = p_L^2 - p_R^2  \ .  
\eeq
Gli zero modi sono caratterizzati dalla coppia $(p_L,p_R)$, che descrive il 
momento e l'av\-vol\-gi\-men\-to dello stato nelle dimensioni compatte. Questi vettori 
giocano il ruolo di cariche elettriche e magnetiche, ed \`e interessante 
osservare come da (\ref{mom}) segua che uno stato con avvolgimento non nullo ma
senza 
dipendenza temporale abbia momento diverso da zero; \`e questo un analogo 
dell'effetto di Witten \cite{wit} che verr\`a discusso in maggior dettaglio in seguito.
Variando il valore di aspettazione di $G$ e di $B$ modifichiamo lo spettro; non 
abbiamo dunque un'unica teoria ma una famiglia di teorie parametrizzata da
una variet\`a con $d^2$ dimensioni, il nostro primo esempio di uno 
spazio di moduli. L'esistenza di questi scalari con valore d'aspettazione 
lasciato indeterminato dalla teoria segue dall'esistenza di $d^2$ deformazioni 
marginali della teoria conforme di una stringa su 
di un toro.
Per descrivere lo spazio dei moduli osserviamo che le informazioni
sullo spettro sono interamente racchiuse nella struttura del reticolo
al quale appartiene la coppia di vettori $(p_L,p_R)$; per trattarli in modo 
equivalente definiamo un nuovo reticolo $W$ generato dai vettori 
\be
\l = \pmatrix{ E^jB_{kj}+e\cr E^jB_{kj}-e }  \ ,  \hspace{1cm}
\m = \pmatrix{ E^j\cr E^j }  \  . 
\ee
In questa base abbiamo
\be
p = \pmatrix{ p_L\cr p_R } = \m m + \l n  \  .
\ee
Se attribuiamo al reticolo una metrica lorentziana con segnatura (d,d)
e calcoliamo il prodotto scalare dei vettori di base, si verifica che il reticolo
$W$ \`e intero e pari; inoltre esso \`e chiaramente autoduale. Se viceversa
consideriamo la funzione di partizione associata ad un generico reticolo 
\be
Z = \frac{1}{ (\sqrt{ \t_2} \h \bar{\h})^d}  \sum_{m,n} q^{\frac{p_L^2}{2}}
\bar{q}^{\frac{p_R^2}{2}}  \ ,
\ee
e definiamo $ |(p_L,p_R)|^2 = p_L^2 - p_R^2 $ vediamo che l'invarianza rispetto 
alla trasformazione $ T: ( \t \mapsto \t +1  ) $ impone che il reticolo sia 
pari 
mentre l'invarianza rispetto alla trasformazione 
$ S:  ( \t \mapsto \frac{1}{ \t}  ) $
impone  che il reticolo sia 
autoduale. In definitiva, la richiesta che il reticolo della compattificazione 
sia un reticolo lorentziano pari e autoduale segue dall'invarianza modulare 
della teoria.
Se indichiamo con $(p,q)$ la segnatura di questi reticoli, \`e noto che 
essi possono 
esistere solo per $ |p-q| = 8n $ e che sono unici se $p,q \neq 0 $ a meno di 
trasformazioni in $O(p,q)$. Nel nostro caso il gruppo che collega tra loro i 
vari reticoli \`e $O(d,d)$, 
ed effettuare trasformazioni di questo tipo equivale 
a modificare i valori dei campi di background $G$ e $B$; tuttavia 
trasformazioni appartenenti al sottogruppo compatto massimale 
$O(d) \times O(d)$ lasciano lo spettro inalterato. Possiamo quindi identificare 
localmente lo spazio dei moduli con lo spazio simmetrico 
$ \frac{O(d,d)}{O(d) \times O(d)} $, che ha proprio dimensione $d^2$; bisogna 
tuttavia tenere conto anche di una serie di identificazioni discrete che 
generalizzano le trasformazioni di T-dualit\`a discusse precedentemente, e che 
corrispondono al gruppo degli automorfismi del reticolo $O(d,d;Z)$ 
\cite{tor5}.
Indichiamo con $g = \pmatrix{a & b\cr c & d} \in O(d,d;Z)$  
una generica matrice 
del gruppo; possiamo specificare un punto dello spazio dei moduli con la 
matrice $ E = G + B $ alla quale associamo una nuova matrice
\be
M(E) = \pmatrix{ G-BG^{-1}B & BG^{-1}\cr -G^{-1}B & G^{-1}} \  .
\ee
Possiamo scrivere $ H= \frac{1}{2} Z^t M Z $ dove $Z=(m,n)$. L'azione di $g$ su 
$M$   ,   $ M \mapsto gMg^t$     , induce un'azione su $E$ che corrisponde a
\be
E \mapsto g(E) = (aE + b)(cE + d)^{-1}  \  .
\ee
L'azione di $g$ sugli oscillatori \`e fissata 
dalle relazioni di commutazione canoniche, ed \`e data da
\be
\a_n(E) \mapsto (d-cE^t)^{-1} \a_n(E^{'})  \  ,  \hspace{1cm}
\bar{\a}_n(E) \mapsto (d+cE)^{-1} \bar{\a}_n(E^{'}) \  . 
\ee
Una prima simmetria del reticolo \`e associata alle trasformazioni del tipo
\be
g_{\Theta} = \pmatrix{ 1 & \Theta\cr 0 & 1} \  ,
\ee
dove $\Theta$ \`e una matrice antisimmetrica a coefficienti interi. Abbiamo poi 
cambiamenti di base, descritti da
\be
g_A = \pmatrix{A & 0\cr 0 & (A^t)^{-1}}  \ ,
\ee
con $A \in GL(d,Z) $, ed infine le trasformazioni
\be
g_{D_i} = \pmatrix{ 1-e_i & e_i\cr e_i & 1-e_i} \  ,
\ee
con $e_i$ vettore unitario nella direzione i-esima; in particolare se $T^d$ \`e 
il prodotto di $D$ cerchi di raggio $R_i$ abbiamo 
$ g_{D_i} : R_i \mapsto \frac{1}{R_i} $. Insieme queste trasformazioni generano 
il gruppo $O(d,d;Z)$; come generatori si possono scegliere una riflessione
$ R_1 $, una dualit\`a $D_1$, una traslazione $T_{12}$, una matrice 
antisimmetrica $ \Theta_{12} $ e le permutazioni $P_{ij}$.
Lo spazio dei moduli delle compattificazioni toroidali \`e dato quindi da
$ \frac{O(d,d)}{O(d) \times O(d)}/O(d,d;Z) $. In generale, spazi dei moduli 
della forma 
\be
M_{pq} =  \frac{O(p,q)}{O(p) \times O(q)}/O(p,q;Z)    \ , 
\ee
sono noti come spazi di Narain \cite{tor1}.

\section{Stringa eterotica}
\markboth{}{}

Le teorie di stringhe chiuse costruite finora si basano su modelli 
bidimensionali che hanno lo stesso contenuto di campi destri e sinistri; \`e 
possibile tuttavia costruire una teoria consistente utilizzando 
i modi sinistri di una superstringa ed i modi destri di una stringa 
bosonica \cite{het1} o, equivalentemente, realizzando nel settore sinistro 
un'algebra superconforme e nel settore destro solo l'algebra di Virasoro. 
L'assenza di anomalie conformi richiede allora $c=10$ e $\bar{c}=26$.
Indichiamo con $X^{\m}(z,\bar{z})$ le coordinate bosoniche e con 
$\psi^{\m}(z)$ i fermioni sinistri, $\m = 0,...,9$; siano poi 
$\phi^I(\bar{z})$ , $I=1,...16$ sedici ulteriori bosoni destri. Per invarianza 
modulare i $\phi^I$ devono appartenere ad un redicolo 16-dimensionale
$\L_{16}$. 
Le dimensioni non compatte dello spazio tempo sono dieci, in quanto i 
campi $\phi^I$ vanno interpretati come coordinate interne.
Calcoliamo la funzione di partizione per questo modello, effettuando 
nel settore sinistro la proiezione GSO su stati con $(-1)^F = -1$; il risultato 
\`e:
\be
Z_{Het} = \frac{1}{(\sqrt{\t_2}\h\bar{\h})^8} 
\frac{\bar{\G}_{16}}{\bar{\h}^{16}}
\frac{1}{2}\sum_{a,b=0}^1(-1)^{a+b+ab} \frac{\th^4\pmatrix{a\cr b}}{\h^4} \   .
\ee
Il contributo delle coordinate compatte:
\be
\frac{\bar{\G}_{16}}{\bar{\h}^{16}} = \sum_{\L_{16}} 
\frac{\bar{q}^{\frac{p_R^2}{2}}}{\bar{\h}^{16}}  \  ,
\ee
\`e invariante modulare solo se il 
reticolo corrispondente \`e pari e autoduale; in sedici dimensioni esistono due 
soli reticoli di questo tipo, $\G_8 \times \G_8$ e $\G_{16}$, e ad essi
corrispondono due distinti modelli eterotici: la stringa 
$E_8 \times E_8$ e la stringa $Spin(32)/Z_2$ (HE ed HO, per brevit\`a).

Il reticolo $\G_8 \times \G_8$ \`e il reticolo delle radici del gruppo
$E_8 \times E_8$ ed \`e formato dalle radici di $O(16)$, che sono 
112 vettori 
8-dimensionali $\e_{ij}$, $i \neq j$,  con $\pm1$ in posizione $i$, 
$\pm1$ in posizione $j$, e $0$ nelle altre posizioni, e dai pesi di una delle 
due spinoriali di $O(16)$, ad esempio i $128$ vettori 
$\e^{\a}_s = \frac{1}{2}(\z_1,...,\z_8)$ con $\z_i = \pm 1$ e 
$\sum \z_i = 0 \ mod(4)$, ovvero con un numero pari di segni meno. La funzione di 
partizione pu\`o essere scritta in termini di funzioni $\th$:
\be
\bar{\G}_{E_8 \times E_8} = \bar{\G}_8^2 = [\frac{1}{2}
\sum_{a,b=0}^1(-1)^{a+b+ab} 
\frac{\bar{\th}^8\pmatrix{a\cr b}}{\bar{\h}^8}]^2  \ .   
\ee
Tenendo conto del contributo degli oscillatori, abbiamo $2 \cdot 248$ stati 
con $ \bar{L}_0 =1 $ che formano l'aggiunta di $E_8 \times E_8$; infatti i modi 
sinistri realizzano l'algebra di correnti $E_8 \times E_8$ a livello 1. 

Il reticolo $\G_{16}$ \`e 
il reticolo delle radici di $O(32)$ con l'aggiunta dei 
pesi di una delle due rappresentazioni spinoriali; questi vettori hanno sedici 
componenti e sono definiti in modo del tutto analogo a quelli di $\G_8$ :
$\e_{i,j}$ con $i,j = 1,...,16$ e $\e^{\a}_s$, $\a = 1,...,2^{15}$. La funzione 
di partizione si scrive:
\be
\bar{\G}_{Spin(32)/Z_2} = \bar{\G}_{16} = \frac{1}{2}
\sum_{a,b=0}^1(-1)^{a+b+ab} 
\frac{\bar{\th}^{16}\pmatrix{a\cr b}}{\bar{\h}^{16}}  \ .      
\ee

Combinando questi stati con quelli del settore destro possiamo determinare lo 
spettro della teoria. Nel settore NS i vincoli sono $L_0 = \frac{1}{2}$ e 
$\bar{L}_0 = 1$; il tachione non compare, in quanto 
non soddisfa la condizione di level matching; abbiamo gli stati
$ \psi^{\m}_{-1/2} \bar{a}^{\n}_{-1} |p \ran $
che corrispondono a dilatone, gravitone  e tensore antisimmetrico, e gli stati
$\psi^{\m}_{-1/2}\bar{J}^a_{-1}|0 \ran$ che sono 
vettori per il gruppo di gauge 
$E_8 \times E_8$ per HE e $Spin(32)/Z_2$ per HO.
Dal settore di Ramond, con i vincoli $G_0 =0$ e $\bar{L}_0 = 1$ otteniamo gli 
stati fermionici necessari per completare lo spettro
della supergravit\`a con $N=1$ accoppiata con SYM, che \`e 
l'azione effettiva di bassa energia per questi modelli.
La teoria \`e chirale ma non anomala; nel capitolo precedente, infatti, abbiamo 
visto che le anomalie si cancellano proprio per i due gruppi di gauge 
fissati dall'invarianza modulare. 
\`E questo un ulteriore esempio di come la consistenza della teoria di 
bassa energia segua dalla consistenza della teoria di stringa.

\section{Relazioni in d=9}
\markboth{}{}

La distinzione tra i due modelli eterotici in dieci dimensioni deriva 
dall'esistenza di due distinti reticoli euclidei, pari ed autoduali; quando 
compattifichiamo toroidalmente $d$ coordinate interne, la teoria \`e 
specificata da un reticolo $\G_{d,16+d}$ che come sappiamo \`e unico a meno di 
trasformazioni in $SO(d,16+d)/SO(d) \times SO(16+d)$. Una conseguenza immediata 
di questo fatto \`e che le due teorie HE ed HO sono connesse in modo continuo 
dopo compattificazione toroidale \cite{tor7}. 
A genere uno, l'azione per le $p$ coordinate bosoniche compattificate e per i 
sedici fermioni interni \`e :
\beq
S &=& 
\frac{1}{4\p} \int d^2 \xi \sqrt{g}g^{ab}G_{\a\b}\part_a X^{\a} \part_b X^\b
+ \frac{1}{4\p} \int d^2 \xi 
\e^{ab}B_{\a\b}\part_a X^{\a} \part_b X^\b \nonumber \\
&+& \frac{1}{4\p} \int d^2 \xi \sqrt{g}\sum_{i=1}^{16}
\psi^i [ \bar{\btd} + Y^i_\a(\bar{\btd}X^{\a})]\bar{\psi}^i \ ,
\eeq
e d\`a origine alla funzione di partizione
\beq
Z_{p,p+16} &=& \frac{\sqrt{detG}}{\t_2^{p/2}\h^p\bar{\h}^{p+16}}
\sum_{m,n \in Z^n} exp[-\frac{\p}{\t_2}(m+\t n)(G+B)(m+\bar{\t} n)] \nonumber \\
& & e^{-i \p \sum_{i=1}^{16}n^{\a} Y^i_{\a} Y^i_{\b} (m^\b+ \bar{\t} n^\b)}
\frac{1}{2}\sum_{a,b=0}^1\prod_{i=1}^{16} \bar{\th}\pmatrix{a\cr b}
(Y^i_\g(m^\g+\bar{\t}n^\g)|\bar{\t})  \  .
\label{hettor}
\eeq
Questa espressione \`e invariante sotto $S$ mentre sotto $T$ trasforma secondo 
$Z_{p,p+16} \mapsto e^{4 \p i /3}Z_{p,p+16}$  .
La si pu\`o riscrivere come
\be
\G_{p,p+16}(G,B,Y) = \sum_{m,n,Q} q^{\frac{p_L^2}{2}} 
\bar{q}^{\frac{p_R^2}{2}}  \  ,
\ee
dove $Q$ appartiene al reticolo $\G_{16}$. Se si utilizza la matrice
\be
M = \pmatrix{G^{-1} & G^{-1}C & G^{-1}Y^T\cr
             C^TG^{-1} & G + C^TG^{-1}C+Y^TY & C^TG^{-1}Y^T+Y^T\cr
             YG^{-1} & YG^{-1}C+Y & !_{16}+YG^{-1}Y^T}   \   ,
\ee
dove $C_{\a \b} = B_{\a \b} - \frac{1}{2} Y^i_\a Y^i_\b$. 
$M \in O(p,p+16)$ e se 
$L = \pmatrix{0 & 1_p & 0\cr 1_p & 0 & 0\cr 0 & 0 & 1_{16}}$, si ha
\beq
p^2_L &=& \frac{1}{2}(m,n,Q)(M-L)(m,n,Q)  \  , \nonumber \\
p^2_R &=& \frac{1}{2}(m,n,Q)(M+L)(m,n,Q)   \   .
\eeq 
Osserviamo inoltre che $p^2_L - p^2_R = 2mn - Q^2$ e che per $Y=0$ si ottiene
\be
\G_{p,p+16}(G,B,Y=0) = \G_{p,p}(G,B)\bar{\G}_{16} \ .
\ee

Possiamo adesso verificare la T dualit\`a tra HE ed HO in modo esplicito. 
Consideriamo il caso pi\`u semplice, $d=9$; compattifichiamo HO su di un 
cerchio di raggio R e introduciamo anche delle linee di Wilson $Y^{I}$. 
la funzione di partizione (\ref{hettor}) \`e:
\be
Z_{HO} = \frac{1}{(\sqrt{\t_2}\h\bar{\h})^7} 
\frac{\bar{\G}_{1,17}(R,Y^{I})}{\h \bar{\h}^{17}}
\frac{1}{2}\sum_{a,b=0}^1(-1)^{a+b+ab} \frac{\th\pmatrix{a\cr b}^4}{\h^4}  \  .
\ee
Scegliamo otto $Y^i=0$ ed altri otto $Y^j=\frac{1}{2}$; nel nostro caso la
(\ref{hettor}) diviene
\be
\G_{1,17}(R) = \frac{1}{2} \sum_{h,g=0}^{1} \G_{1,1}(2R)\pmatrix{h\cr g}
\frac{1}{2}\sum_{a,b=0}^{1} \bar{\th}^8 \pmatrix{a\cr b}
\bar{\th}^8 \pmatrix{a+h\cr b+g}   \   ,
\ee
dove
\beq
\G_{1,1}(R)\pmatrix{h\cr g} &=& R \sum_{m,n \in Z}e^{-\frac{\p R^2}{\t_2}
|(m+\frac{g}{2})+ \t(n+\frac{h}{2})|^2} \nonumber \\
&=& \frac{1}{R} \sum_{m,n \in Z}(-1)^{mh+ng} e^{- \frac{\p}
{ \t_2 R^2} |m+ \t n|^2}  \  .
\eeq
Nel limite $R \rar \infty$ contribuisce solo $(g,h)=(0,0)$ e si ottiene 
HO in dieci dimensioni; nel limite $R \rar 0$ contribuiscono tutti i settori
$(h,g)$, la somma su $(a,b)$e $(g,h)$ fattorizza e si ottiene HE in dieci 
dimensioni.

Abbiamo una relazione simile anche per la IIA e la IIB; quando compattifichiamo 
queste teorie su di un cerchio, sotto una trasformazione di T-dualit\`a 
$R \mapsto \frac{1}{R}$:
\beq
& & \part X^9 \mapsto \part X^9  \ ,  \hspace{2cm} \psi^9 \mapsto \psi^9 \ , \nonumber \\
& & \bar{\part} X^9 \mapsto - \bar{\part} X^9  \ , \hspace{1.7cm} \bar{\psi}^9 \mapsto 
- \bar{\psi}^9   \  .
\eeq
Il cambiamento di segno in $\bar{\psi}^9$ inverte la proiezione nel settore 
$\bar{R}$ e di conseguenza la T-dualit\`a scambia IIA e IIB, ovvero il limite 
per $R \rar \infty$ della IIA compattificata su di un cerchio di raggio $R$ \`e 
la IIA in dieci dimensioni, mentre il limite per $R \rar 0$ \`e la IIB in dieci 
dimensioni e viceversa.

\section{Compattificazioni su variet\`a complesse}
\markboth{}{}

Come gi\`a accennato, \`e possibile studiare  le compattificazioni della teoria 
di stringhe anche dal punto di vista della teoria di campo effettiva 
\cite{cy1,cy2,cy3,cy5}. 
Perch\`e 
l'analisi abbia successo \`e importante richiedere che la teoria effettiva 
abbia almeno una supersimmetria non rotta. La condizione che una supercarica Q 
annulli lo stato di vuoto $|0 \ran$ equivale all'annullarsi di 
$ \lan  0| \{Q,U\} |0 \ran $ per ogni operatore fermionico U. Ora 
$\{Q,U\}$ coincide, per gli operatori fermionici, con la variazione di 
supersimmetria $\d U$ e pertanto trovare una supersimmetria non rotta ad albero 
equivale a trovare una trasformazione di supersimmetria tale che $\d U =0$  per 
tutti i campi fermionici elementari della teoria. 
Effettuiamo l'analisi per la supergravit\`a N=1 accoppiata a SYM; i campi 
fermionici sono il gravitino $\psi_M$, il dilatino $\l$ e i gaugini
$\chi^a$. Le variazioni di questi campi determinate da uno spinore $\h$ sono
\beq
\d \psi_M &=& \frac{1}{\kappa}D_M \h + \frac{\kappa}{32g^2 \phi}
(\G_M{}^{NPQ} - 9 \d^N_M \G^{PQ})\h H_{NPQ} + (Fermi)^2  \  , \nonumber \\
\d \chi^a &=& - \frac{1}{4g \sqrt{\phi}}\G^{MN}F^a_{MN}\h + (Fermi)^2  \  ,
\nonumber \\
\d \l &=& - \frac{1}{\sqrt{2}\phi}(\G \part \phi)\h + 
\frac{\kappa}{8\sqrt{2}g^2\phi}\G^{MNP}\h H_{MNP} + (Fermi)^2  \  ;
\eeq
ricordiamo che $ dH = tr R \ww R - tr F \ww F$. Queste equazioni si 
semplificano se si considerano vuoti con $H=0$ e $\phi$ costante; le condizioni 
non banali seguono allora dalla variazione del gravitino e dalla variazione del 
gaugino:
\be
D_M \h = 0 \label{tan}  \ ,
\ee    
\be
\G^{ij} F^a_{ij} \h =0 \ .
\label{vec}
\ee
Cominciamo con lo studiare la prima equazione, che \`e molto restrittiva; da 
essa segue che $R_{MNPQ}\G^{PQ}\h = 0$ e se cerchiamo una soluzione del tipo 
$M \times K$, con $K$ spazio compatto seidimensionale e $M$ spazio massimamente 
simmetrico, ovvero con tensore di Riemann 
$R_{\m\n\a\b} = \frac{r}{12}( g_{\m\a}g_{\n\b} -g_{\m\b}g_{\n\a})$, otteniamo 
subito $r=0$. Lo spazio non compatto \`e quindi uno spazio di Minkowski, e dalla
(\ref{tan}) segue che $\h$ \`e indipendente dalle coordinate spaziotemporali;
la (\ref{tan}) diviene quindi la condizione che lo spazio interno ammetta uno 
spinore covariantemente costante.

Perch\'e questo sia possibile, $K$ deve avere olonomia $SU(3)$.
Osserviamo che la 16 
di $SO(1,9)$ si decompone rispetto a $SO(1,3) \times SU(4)$ come 
$(2,4) \oplus (2^{'},\bar{4})$, e quindi se esiste un solo spinore 
covariantemente costante, esso trasforma rispetto ad $SO(1,3)$ nella 
$2 \oplus 2^{'}$, ovvero come uno spinore di Majorana, proprio come deve una 
carica di supersimmetria.

Utilizzando lo spinore $\h$ possiamo formare i tensori
\be
k_{ij} = \bar{\h} \G_{ij} \h  \ , \hspace{1cm}   J^i_j = g^{ik}k_{kj}   \  , \hspace{1cm} 
\w_{ijk} = \h^T \G_{ijk} \h  \  .
\ee  
$J^{i}_{j}$ \`e un campo tensoriale (1,1) con $J^2 = -1$, una struttura 
quasi complessa. Fissato un punto $p$ esistono coordinate complesse
$z^a$ con $a=1,2,3$ tali che $J$ in $p$ 
assume la forma canonica $J^a_b = i \d^a_b$,  
$J^{\bar{a}}_{\bar{b}} = -i \d^{\bar{a}}_{\bar{b}}$; la struttura quasi 
complessa si dice integrabile se \`e possibile introdurre un sistema di 
coordinate tale che $J$ assuma forma canonica nell'intorno di ogni punto. 
Un problema analogo \`e quello che si incontra in relativit\`a 
generale quando si cerca un sistema di coordinate in cui la metrica assuma 
la forma canonica $\h_{\m\n}$ nell'intorno di un punto $P$; in ogni punto \`e
possibile scegliere coordinate in cui $g_{\m\n}(P) = \h_{\m\n}$ e il tensore di
Riemann $R_{\m\n\r\s}$ misura l'ostruzione ad estendere questo sistema di
coordinate ad un intero intorno del punto $P$.
In questo caso il teorema di Newlander-Nirenberg afferma che la struttura 
quasi complesa $J$ \`e integrabile se e solo se il tensore di Nijenhuis
\be
N^k_{ij} = J^l_i ( \part_l J^k_j - \part_j J^k_l)  -
J^l_j ( \part_l J^k_i - \part_i J^k_l) \ ,
\ee
\`e nullo. 
La variet\`a ammette allora una struttura 
complessa e pu\`o essere ricoperta con carte locali complesse con funzioni di 
transizione olomorfe che conservano la forma canonica di $J$. Se $K$ ha 
olonomia $U(n)$, in ogni punto esiste $J$ in forma canonica; poich\`e $J$
commuta manifestamente con gli elementi del gruppo di olonomia, essa \`e anche 
covariantemente costante. Le sue derivate covarianti, e quindi N, sono 
nulle e la variet\`a \`e complessa.

Abbiamo tuttavia una struttura pi\`u fine. Se$k_{ij}=g_{ik}J^k_j$, k \`e 
covariantemente costante, in coordinate complesse 
$g_{ab}=g_{\bar{a}\bar{b}}=k_{ab}=k_{\bar{a}\bar{b}}=0$ e 
$k_{a \bar{b}} = -ig_{a \bar{b}} = - k_{\bar{a} b}$; k \`e allora una (1,1) 
forma ed essendo covariantemente costante, $\part k = \bar{\part} k =0$, si ha 
localmente
$k = -i \part \bar{\part} \phi$, con $\phi$ funzione scalare. La metrica della 
variet\`a \`e quindi k\"ahleriana e
$g_{a \bar{b}}= \frac{\part^2 \phi}{\part z^a \part z^{\bar{b}}}$. Viceversa 
una variet\`a k\"ahleriana ha olonomia $U(n)$. Infatti dalla forma della 
metrica segue che solo $\G^a_{bc}$ e $\G^{\bar{a}}_{\bar{b}\bar{c}}$ sono non 
nulli, quindi $J$ \`e covariantemente costante, e poich\`e il sottogruppo di 
$SO(2n)$ che lascia $J$ invariante \`e $U(n)$, K ha al pi\`u olonomia $U(n)$.
A questo punto la variet\`a
K \`e caratterizzata dalla sua struttura complessa e dalla scelta 
della classe di K\"ahler, che \`e la classe di $k_{ij}$ in $H^2(K,R)$. Dal fatto 
che $k \ww...\ww k = n! \e$, dove $\e$ \`e la forma di volume, segue che nessun gruppo 
di coomologia di grado pari di una variet\`a k\"ahleriana pu\`o essere 
banale. Infine, perch\'e l'olonomia sia $SU(n)$ e non $U(n)$, deve annullarsi 
la prima classe di Chern : $c_1(K) = 0$. 
L'anullarsi della classe di Chern equivale 
all'esistenza di una n-forma olomorfa covariantemente costante.

Siamo quindi portati in modo naturale a considerare compattificazioni su 
variet\`a di Calabi-Yau; 
una variet\`a di Calabi-Yau di dimensione complessa $d$ (detta un d-fold) \`e
infatti
una variet\`a k\"ahleriana compatta con olonomia $SU(d)$, o equivalentemente 
dotata di una metrica Ricci piatta. In uno spazio di Calabi-Yau \`e 
automaticamente 
nulla la prima classe di Chern del fibrato tangente; un importante teorema 
di Yau afferma che se una variet\`a k\"ahleriana compatta con classe 
di K\"ahler $J$ ha prima classe di Chern nulla, esiste ed \`e unica una 
metrica Ricci piatta la cui forma di K\"ahler \`e nella stessa classe di 
coomologia di $J$. In questo modo, per stabilire se una data variet\`a 
k\"ahleriana \`e un CY, basta calcolarne la prima classe di Chern senza dover 
esibire esplicitamente una metrica Ricci piatta (a tutt'oggi nessuna \`e nota 
in casi non banali).
Le informazioni topologiche sulla variet\`a vengono espresse tramite 
le dimensioni dei gruppi di coomologia di Dolbeault 
\be
h^{r,s} = dim H^{r,s}_{\bar{\part}} \ , 
\ee
riunite nel cosiddetto diamante di Hodge. 
In particolare la caratteristica di Eulero 
\`e data da $ \chi = \sum (-1)^{r+s} h^{r,s} $. Dato che la variet\`a \`e 
connessa si ha $h^{0,0}=1$, mentre per l'hermiticit\`a 
$h^{r,s} = h^{s,r}$ e per la dualit\`a di Hodge $h^{r,s} = h^{d-r,d-s}$. Come 
conseguenza dell'olonomia $SU(d)$ si pu\`o inoltre mostrare che 
$h^{s,0}=h^{0,s}=0$ per $1<s<d$ e $h^{0,d}=h^{d,0}=1$; 
la corrispondente  d-forma olomorfa 
viene usualmente indicata con $\W$ . Per $d=1$ e $d=2$ il diamante di Hodge \`e 
completamente fissato
\be
\pmatrix{ 1 & 1\cr 1 & 1} \ ,   \hspace{1cm} 
\pmatrix{ 1 & 0 & 1\cr 0 & 20 & 0\cr 1 & 0 & 1} \ .
\ee
I numeri riportati sono gli $ h^{r,s} $, con $r$ numero di riga e $s$ numero di 
colonna.
La superficie complessa con $h^{1,1}=20$ \`e nota come $K3$, ed \`e l'unica 
variet\`a di CY in due dimensioni complesse; per d=3 la variet\`a \`e 
caratterizzata topologicamente 
da due numeri, $h^{1,1}$ e $h^{2,1}$. 
Gli elementi di questi gruppi di coomologia 
possono essere identificati con le perturbazioni della  
variet\`a; se consideriamo infatti una variazione della metrica
$ \d g = \d g_{ij}dz^i dz^j + \d g_{i \bar{j}} dz^i d \bar{z}^j $,
la condizione che la metrica perturbata sia ancora Ricci-piatta 
impone che $ \d g_{i \bar{j}} dz^i \wedge d \bar{z}^j $ e
$ \W_{ijk}g^{k \bar{k}}g_{\bar{k} \bar{l}}dz^i \wedge dz^j \wedge dz^{\bar{l}}$ 
siano forme armoniche. Esse quindi appartengono rispettivamente ad
$H^{1,1}_{\bar{\part}}$ e ad $H^{2,1}_{\bar{\part}}$; perturbazioni 
appartenenti alla stessa classe sono equivalenti a meno di trasformazioni di 
coordinate e pertanto i gruppi di coomologia racchiudono l'informazione sulle 
perturbazioni non banali. In particolare,
gli elementi di $H^{1,1}_{\bar{\part}}$
corrispondono a variazioni della classe di K\"ahler e quelli di 
$H^{2,1}_{\bar{\part}}$ a variazioni della struttura complessa. Gli  spazi dei 
moduli delle strutture complesse e k\"ahleriane sono a loro volta variet\`a di 
K\"ahler ed hanno rispettivamente come potenziale 
$-ln(i \int \W \wedge \bar{ \W})$ e $ \int J \wedge J \wedge J $. 

Definiamo infine la forma di tripla intersezione 
\be
I^{1,1} : H^{1,1} \times H^{1,1} \times H^{1,1} \rar R \ , \hspace{1cm}  
I^{1,1} (A,B,C) = \int_M A \wedge B \wedge C  \ ,
\label{trint}
\ee
che conta il numero di intersezione dei 4-cicli duali alle 2-forme ed \`e un 
invariante topologico; analogamente la forma
\be
I^{2,1} : H^{2,1} \times H^{2,1} \times H^{2,1} \rar C \ , \hspace{1cm}  
I^{2,1} (A,B,C) = \int_M \W_{ljk} 
\label{pseud}
\tilde{A}^l \wedge \tilde{B}^j \wedge \tilde{C}^k \wedge \W \ ,
\ee
con 
\be
\tilde{A}^l \frac{\part}{\part z^l} = 
\W^{lij} A_{ij \bar{k}} dz^{ \bar{k}} \times \frac{\part}{\part z^l}  \ , 
\ee
\`e un invariante pseudotopologico, 
in quanto dipende dalla struttura complessa. 
Utilizzando $\W$ si possono stabilire due utili isomorfismi
\be
H^{r,s}(X) \lrar H^{0,s}( \L^r T^{*}_X) \  ,  \hspace{1cm}  
H^{d-r,s}(X) \lrar H^{0,s}( \L^r T_X) \  .
\ee
Si conoscono vari metodi per costruire spazi di Calabi-Yau, elegantemente 
unificati dalla geometria torica; un modo particolarmente semplice di procedere 
consiste nel considerare intersezioni di ipersuperfici in spazi proiettivi.
Ad esempio consideriamo in $CP^4$ la variet\`a $X$
definita come luogo degli zeri dell'equazione:
\be
\sum_{i=1}^5 z_i^n = 0 \   .
\ee
Il carattere di Chern di $X$ pu\`o essere calcolato utilizzando tecniche 
della geometria algebrica ed \`e dato da:
\be
ch(X) = \frac{(1+J)^5}{(1+nJ)} = 1 + (5-n)J + ...   \ ,
\ee
dove $J$ \`e la forma di K\"ahler; si vede che $c_1(X) = 0 $ quando $n=5$, 
ovvero se il polinomio considerato \`e di quinto grado. L'ipersuperficie appena 
descritta appartiene ad una famiglia di spazi di CY con 
$h_{11} = 101$ e $h_{12}=1$, indicata generalmente con $Y_{4;5}$
(il primo indice \`e la dimensione dello spazio proiettivo in cui si immerge la 
variet\`a ed il secondo il grado del polinomio usato per definirla). $h_{11}$, 
come sappiamo d\`a il numero delle deformazioni della struttura complessa e 
pu\`o essere calcolato semplicementeperch\'e in questo caso coincide con le 
possibili deformazioni del polinomio; $h_{12}$ pu\`o essere determinato 
una volta nota la caratteristica di Eulero $\chi = -200$. Con la stessa tecnica 
\`e facile ottenere una rappresentazione di $K3$ come luogo degli zeri di 
$\sum_{i=1}^4z_i^4$ in $CP^3$.

Finora abbiamo discusso le conseguenze della (\ref{tan}); per l'eterotica 
bisogna considerare anche la (\ref{vec}), che pone restrizioni sul fibrato di 
gauge. Una restrizione importante segue dalla condizione di cancellazione delle 
anomalie menzionata alla fine del capitolo precedente: 
$TrF^2 = trR^2$; questa equazione implica che la seconda classe di Chern del 
fibrato tangente deve coincidere con quella del fibrato di gauge:
$c_2(T_X) = c_2(V_X)$. Il modo pi\`u semplice di soddisfare questa condizione 
consiste nell'immergere la connessione di spin nella connessione di gauge; 
in termini concreti, se consideriamo la formulazione fermionica della HE, per 
evitare un'anomalia imponiamo che sei dei sedici fermioni destri 
trasformino come i sei fermioni sinistri, partners delle  
coordinate della variet\`a interna, ovvero rispetto ad un sottogruppo 
$SU(3)$ di $E_8$ che viene identificato con il gruppo di olonomia del CY. In 
questo modo il gruppo di gauge si rompe  a $E_6 \otimes SU(3) \otimes E_8$ 
($E_6$ \`e il massimo sottogruppo di $E_8$ che commuta con $SU(3)$); l'aggiunta
di $E_8$ si decompone quindi rispetto a $E_6 \otimes SU(3)$ come
\be
\bf{248} = (\bf{78},\bf{1}) \oplus(\bf{27},\bf{3}) 
\oplus(\bf{\bar{27}},\bf{\bar{3}}) \oplus (\bf{1},\bf{8})  \ .
\ee
Il primo termine d\`a i vettori di gauge per $E_6$ in quattro dimensioni,
l'ultimo d\`a dei singoletti di $E_6$. 
I termini centrali danno origine, come \`e 
ben noto dalla costruzione di modelli di grande unificazione con gruppo $E_6$,
a generazioni di materia; la $\bf{27}$ e la $\bf{\bar{27}}$ di $E_6$ possono 
infatti contenere le usuali $15$ particelle di una generazione del modello 
standard e altre $12$ particelle ( questo si vede in modo pi\`u chiaro 
scomponendo la $\bf{27}$ rispetto al sottogruppo massimale $SU(3)^3$ secondo
$(\bf{3},\bf{\bar{3}},1)\oplus (\bf{\bar{3}},1,\bf{3})
\oplus (1,\bf{3},\bf{\bar{3}})$  ed identificando uno 
degli $SU(3)$ con il gruppo di colore). Si pu\`o dimostrare che i
campi nella $\bf{27}$ corrispondono ad elementi di $H^{2,1}$ e che i campi 
nella $\bf{\bar{27}}$ ad elementi di $H^{1,1}$; da questo segue che il numero 
di generazioni corrisponde a met\`a della caratteristica di Eulero della 
variet\`a.

Un fatto notevole \`e che gli accoppiamenti di Yukawa tra i campi nella 
$\bf{27}$ e nella $\bf{\bar{27}}$ sono determinati essenzialmente dalla 
topologia del CY \cite{cy5}; in particolare gli accoppiamenti 
$(\bf{\bar{27}})^3$ sono determinati dalla forma di tripla intersezione 
introdotta in (\ref{trint}), mentre gli accoppiamenti 
$(\bf{27})^3$ sono determinati da (\ref{pseud}).

Tutti i nostri ragionamenti sono stati fatti prendendo in considerazione 
soltanto l'ordine pi\`u basso in $\ap$; \`e naturale chiedersi cosa accade
quando si includono gli ordini successivi della serie perturbativa. \`E 
stato dimostrato \cite{cy2}, utilizzando un teorema di non rinormalizzazione 
per il  superpotenziale, che partendo da spazi di CY \`e possibile costruire 
soluzioni classiche della teoria delle stringhe ordine per ordine in $\ap$; 
dato che il superpotenziale non viene modificato, i valori di aspettazione dei 
campi non massivi non vengono alterati e il vuoto \`e quindi stabile. D'altra 
parte questo risultato permette variazioni dei valori di aspettazione dei campi 
massivi, e in particolare correzioni alla metrica. Questo significa che, 
costruendo perturbativamente la soluzione, lo spazio risultante pu\`o anche non 
essere Ricci piatto. Un modo per arguire la presenza di queste modifiche 
consiste nell'osservare che se la funzione $\b$ di un modello $\s$ con 
bersaglio Ricci piatto fosse nulla ad ogni ordine in teoria delle 
perturbazioni, una variet\`a Ricci piatta sarebbe una soluzione della teoria di 
stringhe; consideriamo ora lo scattering di gravitoni: classicamente 
\`e governato da $R_{\m\n} =0$, e lo stesso sarebbe vero in teoria di stringhe 
se la funzione $\b$ si annullasse ad ogni ordine, in contrasto con il 
comportamento profondamente diverso delle ampiezze in teoria di stringhe ed in 
relativit\`a generale. In realt\`a, \`e stato verificato esplicitamente \cite{cy6}
che modelli 
$\s$ con supersimmetria $(2,0)$ e $(2,2)$ sviluppano una funzione $\b$ non 
nulla a partire da due e quattro loops rispettivamente, in accordo con i 
risultati ottenuti con il calcolo delle modifiche di stringa all'azione 
effettiva con il metodo della matrice S \cite{g1}.

Il passo successivo consiste nel chiedersi cosa accada non perturbativamente; 
\`e noto \cite{cy4} che i modelli di tipo $(2,2)$ sono soluzioni anche non 
perturbativamente mentre i modelli $(2,0)$ vengono destabilizzati dall'effetto 
degli istantoni del modello $\s$ ("worldsheet instantons"). 
Gli istantoni del modello $\s$ sono mappe non banali dal worldsheet nel CY, 
e descrivono 
superfici d'universo della stringa avvolte attorno a cicli della variet\`a;
in particolare, gli 
istantoni rilevanti per le correzioni al superpotenziale sono istantoni 
olomorfi ed antiolomorfi, ovvero mappe olomorfe che hanno per immagine 
$CP^1 \subset X$. Per quanto riguarda gli accoppiamenti di Yukawa,
i termini $(\bf{\bar{27}})^3$  non vengono rinormalizzati n\`e 
perturbativamente n\`e non perturbativamente mentre i termini
$(\bf{27})^3$ ricevono correzioni non perturbative, che torneremo a considerare
quando discuteremo la simmetria speculare.

\section{Orbifolds}
\markboth{}{}

Gli orbifolds, che hanno fatto la loro prima comparsa 
nella letteratura matematica \cite{orb0} con il nome di V-manifolds, sono
spazi che localmente hanno la struttura $M/G$, dove $M$ \`e 
una variet\`a e $G$ un sottogruppo discreto del gruppo delle isometrie di $M$. 
Se $M=R^d$,  $G$ consiste di rotazioni e traslazioni, ma nel caso
della stringa eterotica bisogna specificare anche l'azione di $G$ sui gradi di 
libert\`a di gauge. I punti fissi rispetto all'azione di G rappresentano delle 
singolarit\`a, ed \`e per questo motivo che un orbifold non \`e una variet\`a
liscia.
Le singolarit\`a possono essere riparate con il cosiddetto "blow-up": si 
rimuove un intorno del punto singolare e lo si sostituisce con una variet\`a 
non compatta che presenta il corretto andamento asintotico. In questo modo si 
ottiene una variet\`a liscia ed i moduli di blow-up non descrivono 
altro che i parametri caratterizzanti la 
variet\`a utilizzata per riparare la singolarit\`a.
Nel limite in cui questi moduli tendono a zero si riottiene l'orbifold.
L'importanza di questi modelli sta nel fatto che essi 
ammettono una descrizione in 
termini di teorie conformi, e sono quindi soluzioni esatte delle equazioni del 
moto classiche della teoria. Dal punto di vista della teoria conforme i moduli 
di blow-up sono deformazioni marginali che nella teoria effettiva di bassa 
energia corrispondono a campi scalari non massivi con un potenziale piatto.

Denotiamo un elemento di $G$ con una coppia $(\t ,v)$, dove $\t \in P$ 
rappresenta una rotazione ($P$ \`e il gruppo puntuale) e $v \in \L$ 
rappresenta una traslazione; dalla relazione
$(\t ,u)(1,v)(\t ,u)^{-1} = (1,\t v)$ segue che l'azione di $P$ sul reticolo 
$\L$ deve essere cristallografica. $P$ rappresenta anche il gruppo di olonomia 
dell'orbifold che \`e discreto perch\`e la curvatura \`e 
concentrata nelle singolarit\`a: quando queste vengono riparate $P$ viene 
immerso in un gruppo di Lie. Inoltre l'azione di G sull'algebra di correnti 
della teoria eterotica \`e ristretta 
dalla necessit\`a di rispettare la struttura del
gruppo spaziale e dall'invarianza modulare. La costruzione di un 
orbifold richiede due modifiche della teoria originaria; anzitutto bisogna 
ampliare lo spazio di Hilbert degli stati introducendo diversi settori $H_g$,
 uno per ogni classe di coniugio in $G$, 
che si ottengono quantizzando i campi bidimensionali $X$ con condizioni 
di (quasi)periodicit\`a
$X(\s + 2\p) = gX(\s)$. Questa modifica \`e una semplice conseguenza del 
fatto che ora i punti $X$ e $gX$ sono identificati, e bisogna quindi considerare 
anche stringhe chiuse a meno di trasformazioni in $G$. Questi settori sono 
detti settori $twisted$. Gli stati fisici devono poi essere invarianti rispetto 
a $G$ e questo porta a proiettare ciascun settore $H_g$ nel sottospazio 
$G$-invariante. Infine, quando si combinano modi destri e sinistri bisogna 
richiedere che lo stato destro abbia autovalori inversi rispetto a $G$ dello 
stato sinistro; questo equivale a restringere i campi conformi ad un 
sottoinsieme mutuamente locale.
Due semplici esempi di orbifold sono il segmento $I = S^1/Z_2$, dove l'azione 
sulla coordinate del cerchio \`e $X \mapsto -X$, con due punti fissi che sono 
gli estremi, e il 
tetraedro $T = T^2/Z_2$, dove $Z_2$ inverte 
le coordinate ed ha 4 punti fissi, i vertici.

Costruiamo la funzione di partizione per un bosone $X$ su un cerchio 
quozientato con l'inversione di coordinate. Dobbiamo anzitutto calcolare il 
contributo del settore untwisted, e questo si ottiene semplicemente introducendo 
nella traccia sugli stati il proiettore $\frac{1+P}{2}$, dove $P$ \`e 
l'operatore che realizza l'involuzione $PXP^{-1} = -X$; l'azione di $P$ sugli  
stati \`e:
\be 
P \prod_{i=1}^N a_{n_i} \prod_{j=1}^{\bar{N}} \bar{a}_{\bar{n}_j}|m,n \ran =
(-)^{N+\bar{N}} \prod_{i=1}^N a_{n_i} \prod_{j=1}^{\bar{N}} 
\bar{a}_{\bar{n}_j}|-m,-n \ran  \  .
\ee
Il contributo del settore untwisted \`e dato da:
\be
Z_{unt} = \frac{1}{2}tr(1+P)q^{L_0 - 1/24}\bar{q}^{\bar{L}_0 - 1/24} =
\frac{1}{2} Z(R) + | \frac{\h}{\th_2} |   \  ,
\label{untw} 
\ee
dove $Z(R)$ \`e come in (\ref{part}). 
Nel calcolo di 
$trPq^{L_0 - 1/24}\bar{q}^{\bar{L}_0 - 1/24}$ si \`e tenuto conto che 
contribuiscono solo gli stati con $m = n = 0$. Per determinare il contributo 
del settore twistato dobbiamo quantizzare il campo $X$ con condizioni al bordo 
antiperiodiche; l'espansioni in modi corrispondente \`e:
\be
X(\s,\t) = x_0 + \frac{i}{\sqrt{4 \p T}} \sum_{n \in Z}
\left ( \frac{a_{n+1/2}}{n+1/2}e^{i(n+1/2)(\s + \t)} +
\frac{\bar{a}_{n+1/2}}{n+1/2}e^{-i(n+1/2)(\s - \t)} \right )  \  .
\ee
Lo zero modo $x_0$ deve essere necessariamente uno dei due punti fissi: 
$x_0 = 0 , \p R$; calcolando la traccia del proiettore in questo settore si 
ottiene:
\be
Z_{tw} = \frac{1}{2}tr(1+P)q^{L_0 - 1/24}\bar{q}^{\bar{L}_0 - 1/24} =
| \frac{\h}{\th_4} |  + | \frac{\h}{\th_3} |   \  .
\ee
Complessivamente:
\be
Z_{orb} = \frac{1}{2}Z(R) + | \frac{\h}{\th_2} |  + | \frac{\h}{\th_4} | 
+ | \frac{\h}{\th_3} |   \  .
\label{orb}
\ee
Si verifica facilmente che la (\ref{orb}) \`e invariante modulare.

Un interessante modello eterotico \cite{tor4} si pu\`o ottenere come 
orbifold di HE. Indichiamo con $S_1$ la simmetria che agisce sul primo
$E_8$ fissando il vettore del sottogruppo $O(16)$ e cambiando segno allo 
spinore e con $S_2$ l'analogo operatore per il secondo $E_8$; il gruppo $Z_2$ 
\`e generato da $(-)^{F+1}S_1S_2$. \`E semplice calcolare la funzione di 
partizione:
\be
Z_{O(16) \times O(16)} = \frac{1}{2} \sum_{h,g=0}^1 
\frac{\bar{Z}_{E_8}^2\pmatrix{h\cr g}}{(\sqrt{\t_2}\h \bar{\h})^8}
\frac{1}{2} \sum_{a,b=0}^1(-1)^{a+b+ab+ag+bh+gh}\frac{\th^4\pmatrix{a\cr b}}{\h^4}  \  ,
\label{o16}
\ee
dove
\be
\bar{Z}_{E_8}^2\pmatrix{h\cr g} = \frac{1}{2} \sum_{\g,\d=0}^1(-1)^{\g g+\d h}
\frac{\bar{\th}^8\pmatrix{\g\cr \d}}{\bar{\h}^8}   \  .
\ee
La (\ref{o16}) descrive una teoria in dieci dimensioni con gruppo di gauge 
$O(16) \times O(16)$; la teoria non \`e supersimmetrica e non contiene 
tachioni. Un diretto analogo di questo modello pu\`o essere realizzato con 
stringhe aperte, come vedremo nel prossimo capitolo, costruendo il discendente 
di $OB$, una teoria non supersimmetrica di Tipo II.

Una classe generale di orbifolds si ottiene quozientando $T^{2n}$ con un gruppo 
ciclico $Z_N$ generato da una rotazione $\t \in SO(2n)$ di ordine $N$. Se
scegliamo n=3  ( che corrisponde a $ T^6 $ ) 
e consideriamo la stringa eterotica, le 
condizioni al bordo per i campi bosonici e fermionici sono
\be
X^i( \s+2\p )= \t^{ij}X^{j}( \s)   \  ,  \hspace{1cm}
\psi^i(\t_-+2\p)= \pm \t^{ij} \psi^{j}(\t_-) \ .
\ee
\`E comodo scegliere le coordinate $X^i$ in modo che 
$\t = e^{2\p i ( \frac{k_1}{N}J_{12}+\frac{k_2}{N}J_{34}+\frac{k_3}{N}J_{56})}$
e porre $X^i= \frac{1}{\sqrt{2}}(X_{2i-1}+iX_{2i})$ con $i=1,2,3$ in modo da 
avere $\t X = e^{2 \pi \frac{k_i}{N}}X^i$.
Quando si quantizzano questi campi l'unica variazione rispetto al risultato 
usuale \`e che gli oscillatori sinistri e destri di $X^i$ subiscono uno shift 
nella frequenza : $\a^i_l \mapsto \a^i_{l-\frac{k^i}{N}}$ e
$\tilde{\a}^i_l \mapsto \tilde{\a}^i_{l+\frac{k^i}{N}}$; per 
$\bar{X}^i$ si ha $\bar{\a}^i_l \mapsto \bar{\a}^i_{l-\frac{k^i}{N}}$ e
$\tilde{\bar{\a}}^i_l \mapsto \tilde{\bar{\a}}^i_{l-\frac{k^i}{N}}$. La 
stessa modifica si ha per la frequenza degli oscillatori fermionici; le 
relazioni di commutazione divengono
\be
[\bar{\a}^i_{m+\frac{k^i}{N}},\a^i_{n-\frac{k^i}{N}}]=(m+\frac{k^i}{N})
\d_{m+n,0}\d^{ij} \  .
\ee
\`E importante notare che se $\frac{k^i}{N} \ne 0$ nell'espansione di $X^i$ non 
c'\`e lo zero modo $\a_0^i = p^i$, mentre le condizioni al bordo impongono che $x^i$ 
corrisponda ad un punto fisso di $\t$; questo significa che gli stati dei 
settori twisted sono vincolati ai punti fissi.
Quando si calcolano i vari operatori di Virasoro, bisogna fare attenzione alla 
costante di ordinamento normale. Questa
energia di punto zero, che per un bosone twistato con fase $e^{2 \p ia}$ \`e 
data da $ -\frac{1}{24} + \frac{a}{4}(1-a)$ e per un fermione dalla stessa 
espressione con il segno opposto, 
pu\`o essere calcolata o regolarizzando la somma che compare nell'ordinamento 
normale utilizzando la funzione $\z$ o calcolando la dimensione conforme dei 
campi di twist che generano il vuoto dei vari settori da quello 
untwisted, 
in modo simile a quanto fatto per determinare la dimensione dei campi di spin.

Tecniche per il calcolo delle interazioni sugli orbifolds, ed in particolare per 
la determinazione delle funzioni di correlazione dei campi di twist, sono state 
sviluppate in \cite{orb2,orb3}.

\subsection{Invarianza modulare e torsione discreta}

L'invarianza modulare della funzione di partizione ad un loop si riduce ad una 
condizione di $\it{level~matching}$ per i fermioni; questa condizione \`e inoltre 
necessaria e sufficiente per l'invarianza modulare anche a loops superiori nel 
caso di orbifolds abeliani \cite{orb4}. 
Indichiamo con $(g,h)$ un settore twistato con $g$ 
lungo $ \t$ e con $h$ lungo $ \s$; una trasformazione modulare 
$ \t \mapsto \frac{ a \t + b}{ c \t +d }$, trasforma tale settore in 
$ ( g^d h^{-b} , g^{-c} h^{a} ) $. 
Quando un settore viene lasciato fisso bisogna 
imporre che $(g,h)$ e $ ( g^d h^{-b} , g^{-c} h^{a} )$ coincidano senza 
differire neanche per una fase, altrimenti si avrebbe una anomalia globale e 
non ci sarebbe modo di definire la fase di $(g,h)$ \cite{spin3}. 
Dobbiamo quindi verificare, 
per un rappresentante $(g,h)$ di ciascuna orbita,
l'invarianza modulare rispetto al sottogruppo $ \G_{(g,h)} $ che lascia fisse 
le condizioni al bordo. 
Consideriamo l'eterotica $E_8 \times E_8$ nella formulazione fermionica e un 
twist di ordine n, $ g = ( \th , w , \th_1, \th_2 )$, che agisce sui campi 
$ (X^i , \chi^i , \psi^a , \phi^a) $ con
\be
g(X^i , \chi^i , \psi^a , \phi^a) = 
(\th X + w, \th \chi, \th_1 \psi, \th_2 \phi)   \  .
\ee
Utilizziamo una base diagonale in cui
$\th = (e^{2 \p i v^i},e^{-2 \p i v^i})$ ,
$\th_1 = (e^{2 \p i v_1^i},e^{-2 \p i v_1^i})$ ,
$\th_2 = (e^{2 \p i v_2^i},e^{-2 \p i v_2^i})$ con $v^i = \frac{r^i}{n}$,
$v^i_1 = \frac{r^i_1}{n}$,  $v^i_2 = \frac{r^i_2}{n}$; si pu\`o mostrare che
la condizione $g^n =1$ impone, per $n$ dispari:
\be
\sum r_i^2 - \sum r_{1i}^2 - \sum r_{2i}^2 = 0 \ mod(n)  \   ,
\label{rcon1}
\ee
e per $n$ pari
\beq
& & \sum r_i^2 - \sum r_{1i}^2 - \sum r_{2i}^2 = 0  \ mod(2n)  \  , \nonumber \\
& & \sum r_i - \sum r_{1i} - \sum r_{2i} = 0 \  mod(2)   \ .
\label{rcon}
\eeq
Al fine di stabilire l'invarianza modulare possiamo trascurare i determinanti 
bosonici dei modi destri e sinistri, in quanto sono l'uno il complesso 
coniugato dell'altro. Consideriamo allora il contributo dei fermioni nel 
settore (-1,h), fissato da $ \t \mapsto \t + n $; il contributo di uno stato 
alla funzione di partizione \`e $ e^{2 \p i ( \t E_L - \bar{\t}E_R )} $ ed \`e 
invariante se e solo se $ n(E_L-E_R) = k$ , $k \in Z$. 
L'energia dei modi destri deve quindi differire da quella dei modi 
sinistri per multipli interi di $ \frac{1}{n} $; basta imporlo per lo stato 
fondamentale, con il risultato
\be
\frac{n}{2} \left [ \sum v_i^2 - v_{1i}^2 - v_{2i}^2 - (v_i - v_{1i} - v_{2i}) \right ]
= 0 \  mod(1) \ .
\ee
Si verifica facilmente che questa condizione equivale alla condizione 
quadratica in (\ref{rcon1}-\ref{rcon}); 
la condizione lineare per $n$ pari si ottiene 
considerando il settore twistato con $ (-1)^F \cdot g$.

Per dimostrare che le (\ref{rcon1}-\ref{rcon}) sono necessarie e sufficienti, 
calcoliamo 
il determinante dell'operatore di Dirac per un fermione complesso olomorfo,
$d(u,v; \t) = Tr_h (-1)^F \cdot g q^H $, 
con $g = e^{2 \p iu} $ e $h = e^{2 \p iv } $ :
\be
d(u,v; \t) = e^{i \p u (v-1)} q^{\frac{1}{2}(v^2 - v + \frac{1}{4})}
\prod_{n=1}^{\infty}(1-q^{n-v}e^{2 \p i u})(1-q^{n+v-1}e^{-2 \p i u})  \ .
\label{dir}
\ee
Se $\a \in SL_2(Z)$, il determinante (\ref{dir}) trasforma come:
\be
d(u,v;\a(\t)) = \e(\a)d(\a^{-1}(u,v); \t)   \  ,
\label{modtra}
\ee
dove $\e(\a)$ \`e una radice 12-esima dell'unit\`a e $(u,v)$ va considerato 
come un vettore colonna rispetto all'azione di $\a$; dato che complessivamente 
abbiamo quattro modi complessi destri e sedici modi complessi sinistri, 
possiamo trascurare la fase $\e$.
Abbiamo inoltre:
\be
d(u+1,v; \t)= -e^{-i \p v}d(u,v;\t) \ , \hspace{1cm}
d(u,v+1; \t)= -e^{-i \p u}d(u,v;\t)   \  .
\label{tra}
\ee
Per verificare l'invarianza modulare, dobbiamo assicurarci che 
il determinante non aquisti fasi per 
$\a^{-1} \in \G_{(g,h)}$. Per caratterizzare $\G_{(g,h)}$, osserviamo che, dato 
che $g$ ed $h$ commutano, essi generano un gruppo abeliano; un gruppo abeliano 
generato da $k$ elementi \`e della forma 
$Z_{m_1} \times...\times Z_{m_k}$, con $m_i < m_j$ per $i < j$ ( $m_i=1$ 
corrisponde al gruppo con un solo elemento e $m_i = \infty$ corrisponde a 
$Z$).
Nel nostro caso abbiamo tre possibilit\`a: 
\be
Z \times Z \ , \hspace{1cm} Z_m \times Z \  , \hspace{1cm} Z_m \times Z_n  \  ;
\ee
l'elemento generico per ciascuno dei tre casi \`e dato da:
\be
\pmatrix{1 & 0\cr 0 & 1}  \ , \hspace{1cm}
\pmatrix{1 & 0\cr 0 \ mod(m) & 1}  \ , \hspace{1cm}
\pmatrix{1 \ mod(n) & 0 \ mod(n)\cr 0 \ mod(m) & 1 \ mod(n)}  \  .
\label{stab}
\ee
Perch\`e $\a^{-1}=\pmatrix{a & b\cr c &d} \in \G_{(g,h)}$, si deve avere 
$\a^{-1}(u,v) = (u+b_1,v+b_2)$ con $b_1 = (a-1)u + bv$ e  
$b_2 = cu + (d-1)v$ interi. In base alle (\ref{modtra}-\ref{tra}), il determinante 
trasforma secondo:
\be
d(u,v;\a(\t)) = (-1)^{b_1+b_2+b_1b_2}e^{-i \p(b_1v-b_2u)}\e(\a)d((u,v); \t) \   .
\ee
Utilizzando le matrici (\ref{stab}) ed imponendo l'assenza di fasi, si 
ottengono per i twist $u$ e $v$ le stesse condizioni ricavate precedentemente
in (\ref{rcon1}-\ref{rcon}).

Possiamo generalizzare ulteriormente questo tipo di compattificazioni,
intoducendo quella che \`e chiamata "torsione discreta" \cite{orb4}. Questa 
generalizzazione  
consiste nell' assegnare fasi $\e(g,h)$ alle varie classi di mappe con 
condizioni al bordo 
$(g,h)$ che compaiono nella costruzione dell'orbifold in modo consistente con 
l'invarianza modulare e la fattorizzazione delle ampiezze.
Vediamo quali
restrizioni vengono imposte sulle fasi da queste richieste.
Se $(a_1,b_1;...;a_n,b_n)$ rappresenta una base 
canonica di cicli e se indichiamo con $(g_1,h_1;...;g_n,h_n)$ un generico 
settore twistato, dall'invarianza modulare ad uno e a due loops e dalla
fattorizzazione abbiamo:
\beq
\e(g,h) &=& \e (g^a  h^b  ,  g^c  h^d)  \  , \hspace{1cm} 
\e( g_1,h_1  ; g_2,h_2) = \e(g_1,h_1) \e(g_2,h_2) \ , \nonumber \\
\e( g_1,h_1  ; g_2,h_2) &=&  \e( g_1h_2h_1^{-1},h_1   ; g_2h_1h_2^{-1},h_2)  \ ,
\hspace{1cm} \e(1,1) = 1 \   ,
\label{cond}
\eeq
dove l'ultima condizione fissa la normalizzazione. Le (\ref{cond}) non 
vengono alterate a loops superiori; basta definire 
$ \e(g_1,h_1;g_2,h_2;...;g_n,h_n) = \e(g_1,h_1)\e(g_2,h_2)...\e(g_n,h_n)$.
La terza delle (\ref{cond}) pu\`o essere riscritta, dopo una trasformazione 
modulare
\be
\e(g_1,h_1)\e(g_2,h_2) = \e(g_1h_2,h_1) \e(g_2h_1,h_2)  \  .  
\label{dlop}
\ee
Consideriamo alcuni casi particolari: per $g_1=g_2=h_2=1$, si ha
$\e(h,1)=1$ e anche $\e(h,h)=\e(1,h)=1$; posto $g_1=g_2=1$ in (\ref{dlop})
si ha $\e(h_1,h_2) \e(h_2,h_1) =1$ e con $g_1=1$
si ha $\e(g_2,h_2) = \e(h_2,h_1) \e(g_2h_1,h_2)$. Utilizzando queste relazioni 
si vede che le (\ref{cond}) equivalgono alle:
\beq
\e(g_1g_2,g_3) &=& \e(g_1,g_3) \e(g_2,g_3)  \  ,  \nonumber \\
\e(g,h) &=& \e(h,g)^{-1}  \  ,  \nonumber \\
\e(g,g) &=& 1  \  .
\label{grup}
\eeq

La possibilit\`a di introdurre torsione discreta in un orbifold \`e analoga 
alla possibilit\`a di introdurre un campo $B_{\m\n}$ di background in una 
compattificazione toroidale (da qui il nome).
Consideriamo infatti
un campo $B_{\m\n} $ di background; quando $dB = 0$ il termine 
$e^{\frac{i}{2} \int B_{\m\n} \part_\a x^\m \part_\b x^\n \e^{\a\b}} $ \`e 
topologico e rappresenta una fase per le diverse classi di funzioni dal 
worldsheet nello spaziotempo. Per un toro $T^d$, le classi di omotopia sono 
classificate da $2d$ interi $(n_i,m_i)$ con $i=1,...,d$ che danno 
i numeri di avvolgimento del worldsheet 
sui $d$ cicli $W^\m_i$ del toro spaziotemporale
e si possono rappresentare con la famiglia di funzioni 
$X^\m(\t,\s) = \frac{1}{2 \p} ( n_i W^\m_i \t + m_j W^\m_j \s)$, 
alle quali corrisponde una fase
$ \e(n_i,m_j) = e^{i B_{\m\n}n_iW^\m_i m_jW^\n_j}$.
Dal punto di vista della torsione discreta, $G = Z^d$ e, detti
$W_i$ i generatori del reticolo del toro, possiamo porre
\be
\e(W_i,W_j) = \e_{ij}  \  ,   
\ee
dove $\e_{ij}$ sono fasi arbitrarie per $i<j$ con $\e_{ji} = \e_{ij}^{-1}$ e 
$\e_{ii}=1$. Si verifica facilmente che le (\ref{grup}) sono soddisfatte,  e se 
si sceglie $B_{\m\n}$ tale che $\e_{ij} = e^{i B_{\m\n}W^\m_i W^\n_j}$, si 
ottiene:
\be
\e(n_iW_i,m_jW_j) = e^{i B_{\m\n}n_iW^\m_i m_jW^\n_j} \  .
\ee
Le fasi della torsione discreta equivalgono quindi alla possibilit\`a 
di introdurre un tensore antisimmetrico sul toro.

Per un gruppo abeliano finito 
$G = Z_{n_1} \times Z_{n_2} \times ...\times Z_{n_k}$ possiamo dare una 
soluzione generale delle (\ref{grup}).
Se $\a_i$ \`e il generatore di $Z_{n_i}$, detta 
$\e_{ij}$ una radice $n_i$-esima 
dell'unit\`a, $i < j$ e posto $\e_{ji} = \e^{-1}_{ij}$ , $\e_{ii}=1$, 
definiamo:
\be
\e(\a_i,\a_j) = \e_{ij}  \  .
\ee
La fase deve essere una radice $n_i$-esima dell'unit\`a in quanto
$ \e(\a_i,\a_j)^{n_i} = \e(\a_i^{n_i},\a_j) = \e(1,\a_j) = 1$.
Nel caso non abeliano \`e possibile scrivere delle analoghe condizioni di 
consistenza per la torsione discreta e presentare delle soluzioni, ma non \`e 
possibile mostrare, come nel caso abeliano, che non ve ne sono altre.

Osserviamo che, nel caso di una variet\`a liscia $M$, 
la torsione discreta \`e un 
omomorfismo tra $H_2(M,Z)$ e $U(1)$, in quanto stiamo associando fasi a classi 
di immersioni di superfici in $M$; dato che la topologia di un orbifold \`e 
completamente specificata dal gruppo $G$, questa idea si pu\`o estendere 
considerando la coomologia del gruppo, definita in termini dello spazio di 
classificazione $BG$: la torsione discreta pu\`o infatti essere interpretata 
come un omomorfismo tra $H_2(G,Z)$ e $U(1)$.

Modelli con torsione discreta non sono connessi in modo continuo a 
compattificazioni su spazi di Calabi-Yau lisci; per risoluzione o deformazione 
le singolarit\`a vengono rimosse solo parzialmente.
Come esempio consideriamo un orbifold $Z_k \times Z_k$ \cite{orb4b}; 
se $\w$ \`e il generatore di $Z_k$, il generico settore sar\`a caratterizzato 
da un twist $T_{\s} = (\w^n,\w^m)$ lungo $\s$ e da un twist
$T_{\t} = (\w^{n^{'}},\w^{m^{'}})$ lungo $\t$;
in accordo con (\ref{grup}), possiamo associare a questo settore una fase
\be
\e(T_{\s},T_{\t}) = \z^{nm^{'}-mn^{'}}  \ ,  \hspace{1cm} \z = e^{2 \p i \frac{m}{k}}  \  , \hspace{1cm}
m = 0,...,k-1  \  .
\ee
Abbiamo quindi k teorie, in corrispondenza delle radici k-esime dell'unit\`a; 
per $k=0$ si ottiene l'orbifold usuale. Dal punto di vista della formulazione 
hamiltoniana, la fase aggiuntiva dovuta alla torsione discreta pu\`o essere 
attribuita alla rappresentazione degli operatori $T$ sugli stati. Se la 
rappresentazione standard \`e $ T_{\t} \mapsto \hat{T}_{\t}$, la condizione di 
fattorizzazione assicura che anche 
$ T_{\t} \mapsto \hat{T}_{\t} \e(T_{\s},T_{\t})$ \`e una rappresentazione;
valutando la traccia sui vari settori dello spazio di Hilbert 
si ottengono esattamente i contributi dell'orbifold con torsione discreta.
L'esempio pi\`u semplice \`e l'orbifold $\G = Z_2 \times Z_2$ di $T^6$; se 
scegliamo $T^6 = E_1 \times E_2 \times E_3$, dove gli $E_i$ sono tori 
bidimensionali con coordinate complesse $z_i$, l'azione dei generatori $h$ e 
$g$ del  gruppo \`e:
\be
h(z_1,z_2,z_3) = (-z_1,-z_2,z_3) \  , \hspace{1cm}  g(z_1,z_2,z_3) = (z_1,-z_2,-z_3)\  .  
\ee
Ciascuna delle tre trasformazioni contenute in $Z_2 \times Z_2$ lascia fissi 
$16$ tori; esistono inoltre $64$ punti fissi rispetto all'azione dell'intero 
gruppo, localizzati all'intersezione dei 48 tori fissi.
Per studiare la geometria dell'orbifold \`e opportuno determinare la struttura 
dello stato fondamentale nel settore di Ramond, che rappresenta l'analogo della 
coomologia. Cominciamo dal settore untwisted; il suo contributo alla coomologia 
\`e dato dalla parte $\G$-invariante di
$H^{*}(E_1 \times E_2 \times E_3) = H^{*}(E_1) \times H^{*}(E_2)
\times H^{*}(E_3)$, ovvero
\be
\pmatrix{1 & 0 & 0 & 1\cr 0 & 3 & 3 & 0\cr 0 & 3 & 3 & 0\cr 1 & 0 & 0 & 1\cr } 
\ .
\ee
Abbiamo poi tre settori twistati; dato che essi hanno la stessa struttura, 
concentriamoci sul settore twistato da $h$. Lo spazio degli zero modi 
nel settore di Ramond 
twistato \`e dato dalla quantizzazione delle configurazioni di stringa fisse
che, per l'elemento considerato, sono 16 tori; 
ci aspettiamo quindi un contributo pari 
a sedici volte la coomologia di $T^2$. Dobbiamo tuttavia 
osservare che, dal punto di vista della 
teoria conforme, gli indici $(p,q)$ che 
caratterizzano le forme su una variet\`a complessa,  sono delle cariche $U(1)$ che nei settori 
twistati subiscono uno shift \cite{orb8}. 
Nel nostro caso $H^{p,q} \mapsto H^{p+1,q+1}$ e il contributo del settore 
h-twisted \`e :
\be
\pmatrix{0 & 0 & 0 & 0\cr 0 & 16 & 0 & 0\cr 0 & 0 & 16 & 0\cr 
0 & 0 & 0 & 0\cr } \   . \label{htw}
\ee
Nello scrivere la (\ref{htw}), abbiamo mantenuto solo le forme 
in $H^{00}$ e $H^{11}$ che sono $\G$-invarianti e contribuiscono 
rispettivamente a $H^{11}$ e $H^{22}$ dell'orbifold. 
Complessivamente, il diamante di Hodge \`e dato da:
\be
\pmatrix{1 & 0 & 0 & 1\cr 0 & 51 & 3 & 0\cr 0 & 3 & 51 & 0\cr 
1 & 0 & 0 & 1\cr }   \  .
\ee
In presenza di torsione discreta, abbiamo una fase $\e( \a, \b) = -1$ in tutti 
quei settori con $ \b \neq 1, \a$; di conseguenza, nel settore h-twisted
l'azione di $g$ ed $hg$ su $z_3$ viene invertita e la proiezione sugli 
stati $\G$-invarianti mantiene ora i contributi di $H^{01}$ e $H^{10}$. il 
diamante di Hodge risultante \`e:
\be
\pmatrix{1 & 0 & 0 & 1\cr 0 & 3 & 51 & 0\cr 0 & 51 & 3 & 0\cr 
1 & 0 & 0 & 1\cr }   \  .
\ee
I moduli twisted nel settore senza torsione discreta sono in $H^{11}$, ovvero 
sono deformazioni della struttura k\"ahleriana; questo significa che le 
singolarit\`a vengono rimosse tramite blow-up. Nel modello con torsione 
discreta i moduli sono in $H^{12}$, 
ovvero sono deformazioni della struttura complessa.
In entrambi i casi \`e possibile rimuovere le singolarit\`a di codimensione 
due; nel caso senza torsione discreta vengono 
automaticamente rimosse anche 
le singolarit\`a di codimensione tre, i 64 punti fissi. Questo non \`e 
possibile nel modello con torsione discreta. Per convincersi di questo \`e 
sufficiente determinare il numero di deformazioni della struttura complessa 
dell'orbifold $Z_2 \times Z_2$. Rappresentiamo ogni toro $E_i$ mediante
un'equazione in $CP \times CP$
\be
y_i^2 = F_i(u_i,v_i)  \  ,
\ee
dove $F_i$ \`e un polinomio omogeneo di quarto grado e $y_i$ \`e una 
coordinata omogenea di 
grado due; la simmetria $Z_2$ agisce sulle coordinate secondo 
$(u_i,v_i,y_i) \mapsto (u_i,v_i,-y_i)$. 
Se $y = y_1y_2y_3$, l'orbifold \`e dato da:
\be
y^2 = \prod_{i=1}^3 F_i(u_i,v_i)  \  ,
\ee
Le deformazioni della struttura complessa, in questo caso, concidono con le 
deformazioni del polinomio $P = \prod_{i=1}^3 F_i(u_i,v_i)$ in un generico 
polinomio $P^{'} = F(u_1,u_2,u_3,v_1,v_2,v_3)$ e sono $125-9-1 = 115$, tenendo 
conto della simmetria $SL_2(C)$ dei tre tori e della normalizzazione del 
polinomio. Se tutte queste deformazioni fossero possibili per il nostro 
modello, dovremmo avere $H^{12}=115$ e non $H^{12}=51$. 
Ora il numero di deformazioni pu\`o essere ridotto se la variet\`a finale $M$
conserva ancora delle singolarit\`a; infatti se $M$ ha $n$ singolarit\`a con 
$k$ operatori rilevanti, il numero di moduli \`e $115-nk$. 
Rappresentiamo la singolarit\`a con un' equazione algebrica in $CP^4$, 
localizzandola nell'origine:
\be
F(x_1,x_2,x_3,x_4) = 0  \ , \hspace{1cm} \part_i F(0) = 0  \  ,
\ee   
le deformazioni 
rilevanti sono rappresentate dai polinomi nelle $x_i$ modulo l'ideale generato 
dalle derivate $\part_i F$; l'identit\`a \`e sempre un operatore rilevante e 
l'unica singolarit\`a che ha solo un operatore rilevante \`e il conifold:
\be
x_1^2 + x_2^2 + x_3^2 + x_4^2 = 0 \ .
\ee
Nel nostro caso, le 64 
deformazioni mancanti possono essere associate ai 64 punti fissi,
assumendo che l'orbifold appartenga ad una famiglia di spazi di 
Calabi-Yau con 64 singolarit\`a di conifold, che ha precisamente 51 moduli.
In modo pi\`u esplicito, possiamo deformare 
$y^2 = \prod_{i=1}^3 F_i(u_i,v_i)$ in
$y^2 = \prod_{i=1}^3 F_i(u_i,v_i) + \e \d F$; tenendo conto che ciascun $F_i$ 
appartiene ad uno spazio con cinque dimensioni e trascurando deformazioni in 
cui $\d F$ \`e un multiplo di $F_1F_2F_3$, abbiamo 12 deformazioni, 
ridotte a tre dalla simmetria $SL_2(C)$ che agisce su ciascuna 
coppia di coordinate $(u_i,v_i)$, che alterano 
un solo $F_i$ : $ \d F = \d F_1F_2F_3 + F_1 \d F_2F_3 + F_1F_2 \d F_3 $; 
queste deformazioni possono essere identificate con i tre modi del settore 
untwisted che preservano la struttura di orbifold osservando che,
al primo ordine in $\e$, il polinomio deformato \`e 
$y^2 = \prod_{i=1}^3 (F_i(u_i,v_i)+ \e \d F_i)$.
Abbiamo poi 48 deformazioni del tipo 
$ \d F = \d F_1 \d F_2F_3 + F_1 \d F_2 \d F_3 + \d F_1F_2 \d F_3 $, che sono i 
modi che risolvono le singolarit\`a di codimensione due, in quanto si annullano 
sui 64 punti fissi. Infine ci sono 64 deformazioni del tipo 
$ \d F = \d F_1 \d F_2 \d F_3$; queste deformazioni eliminano i punti fissi ma 
non sono presenti nel modello con torsione discreta, come si pu\`o arguire 
osservando che i modi del settore $h$ twisted deformano le 
singolarit\`a dei tori 
fissati da $h$ e devono quindi annullarsi 
per $F_1=F_2=0$ e per $F_1=F_3=0$ ma non 
per $F_1=F_2=0$ e corrispondono quindi a $\d F_1 \d F_2 F_3$; in modo analogo 
dai settori twistati con $g$ ed $hg$ si ottengono le altre deformazioni della 
seconda classe, ma nessun settore fornisce i modi della terza. 

Si pu\`o anche mostrare che la torsione discreta \`e supportata proprio in 
queste singolarit\`a; essa produce, infatti, un effetto non nullo in un intorno 
arbitrariamente piccolo del punto $P$ e nullo se questo punto viene rimosso.
Tutto questo fa pensare che la torsione discreta, che come abbiamo visto 
reintroduce nell'orbifold la possibilit\`a di accendere un campo $B_{\m\n}$ di 
background, sia una sorta di campo H concentrato nella singolarit\`a, in 
perfetta analogia con
le usuali singolarit\`a di curvatura presenti negli orbifold; proprio la 
presenza di questo campo impedirebbe la risoluzione della singolarit\`a.

Un ultimo aspetto interessante di questi orbifolds \`e che il modello con 
torsione discreta e quello senza torsione discreta costituiscono una coppia 
speculare.
Una simmetria speculare, come vedremo tra breve, 
\`e un isomorfismo tra due teorie 
superconformi $(2,2)$ che consiste nell'inversione della carica $U(1)$
dei soli modi sinistri,
interpretabile anche come un'inversione della struttura complessa, 
proprio come la T-dualit\`a consiste in una 
trasformazione di parit\`a sui soli modi sinistri.
Verificare l'isomorfismo 
richiede un'identificazione tra gli operatori delle due teorie
sotto la quale coincidano le funzioni di correlazione e la funzione di 
partizione. Generalmente si considera sufficiente, per questo scopo, 
mostrare l'uguaglianza dello 
spettro e delle funzioni a tre punti sulla sfera o della funzione di partizione 
a genere arbitrario.

Consideriamo un toro con coordinate $(X,Y)$ e con raggi $(R,R^{'})$; se 
$Z = X + iY$, l'effetto di una T-dualit\`a lungo $Y$ \`e quello di una 
trasformazione speculare, infatti
\beq
\part Z \mapsto \part Z , &  & \bar{\part} Z \mapsto \bar{\part} \bar{Z} ;
\eeq
il volume $RR^{'}$ viene inoltre scambiato con $\frac{R}{R^{'}}$. Per 
supersimmetria cambia solo il segno del partner fermionico 
sinistro di $Y$, $\h$. Questa costruzione pu\`o essere immediatamente 
generalizzata ad un prodotto di tori $T_i$ con coordinate $(X_i,Y_i)$. 
La trasformazione della misura 
nell'integrale funzionale dipende dal numero degli zero modi degli $\h_i$; 
se a fermioni con struttura di spin $a$ a genere $g$ viene cambiato segno, la 
misura $\m_{g,a}$ trasforma in $(-1)^{\s_a} \m_{g,a}$ dove 
$\s_a = 0,1$ a seconda che la struttura di spin sia pari o dispari. Limitandoci 
ad un loop si pu\`o mostrare, verificandolo esplicitamente per i settori 
$(1,1)$ , $(g,1)$ e $(g,gh)$ ed utilizzando quindi l'invarianza modulare, che 
per il modello con torsione discreta si ha:
\be
\m_a(x,y) \mapsto (-1)^{\s_a} \e(x,y) \m_a(x,y) \ ,
\ee
dove $(x,y) = (g^rh^s,g^th^u)$ ed $ \e(x,y) = (-1)^{ru-st}$. Dal primo fattore 
si vede che la trasformazione \`e una simmetria speculare; il secondo fattore
rimuove la torsione discreta mostrando esplicitamente che i due modelli sono 
mirror l'uno dell'altro.

\section{Costruzioni con fermioni liberi}
\markboth{}{}

Un'altra classe di modelli si ottiene considerando fermioni 
liberi di Majorana-Weyl \cite{ff1}; l'azione 
$S = \frac{1}{2} \int dz d \bar{z} \psi^A \bar{ \part} \psi^A$ \`e invariante 
rispetto alla trasformazione 
\be
\d \psi^A = \tilde{f}^{ABC} \psi^B \psi^C \e  \ ,
\label{var} 
\ee
con $\tilde{f}$ tensore totalmente 
antisimmetrico. Questa \`e una supersimmetria 
se e solo se $ \tilde{f}$ \`e proporzionale alle 
costanti di struttura di un gruppo di Lie 
semisemplice \cite{ff2}. 
Il teorema di Wick mostra infatti che la corrente 
associata a (\ref{var}) $T_F = \frac{1}{3} \tilde{f}^{ABC} \psi^A  
\psi^B  \psi^C$
soddisfa 
\be
\{ T_F(z),T_F(w) \} = 2T_B(z) \d(z-w) + cost \  , 
\ee
dove
$T_B(z) = \frac{1}{2} \psi^A \part \psi^A$ se
\be
\tilde{f}^{AE[B} \tilde{f}^{CD]E} = 0   \ ,  \hspace{1cm} \tilde{f}^{ACD}\tilde{f}^{BCD} 
= \frac{1}{2}\d^{AB} \   .
\ee
La prima condizione implica che le quantit\`a $\tilde{f}^{ABC}$ sono 
le costanti di struttura di un gruppo 
di Lie $G$, mentre la seconda implica che $G$ non ha sottogruppi normali abeliani. La 
relazione tra $ \tilde{f}^{ABC} $ e le costanti $f^{ABC}$ con la 
normalizzazione 
convenzionale \`e $ \tilde{f} = \frac{f}{\sqrt{2c(G)}}$ 
con $c(G)$ Casimir quadratico.
Il modello \`e quindi caratterizzato da una carica 
centrale $c = \frac{N}{2} = \frac{1}{2}dimG$ e da un'algebra di Kac-Moody 
associata a G con supercorrenti
\be
J^A(z, \th) = \sqrt{\frac{c}{2}} \psi^A + \th J^A(z)  \ ,
\ee
dove $J^A(z) = \frac{1}{2} f^{ABC} \psi^B \psi^C$ con livello 
$k= \frac{c(G)}{2}$.

Possiamo ora associare un nuovo modello supersimmetrico ad ogni sottogruppo $H 
\subset G$ tale che $G/H$ sia uno spazio simmetrico, imponendo condizioni 
periodiche su tutti i fermioni che non trasformano nell'aggiunta di $H$. In 
questo modo si rompe la simmetria del modello ad $H$  : le correnti 
corrispondenti a $G/H$ hanno $(-1)^{F_2}=-1$,  e quindi non compaiono nella 
teoria.
Dividiamo infatti
i fermioni in due gruppi $\psi^a$   $ ( a=1,...,n  ) $ e 
$\psi^i$    $  (  i=n+1,...,N=dimG   ) $ 
e assumiamo che i fermioni del primo gruppo compaiano sempre in 
coppia nei generatori di supersimmetria. Questa condizione richiede che
\be
f^{ijk} = f^{iab}=0  \ .
\label{rot}
\ee
Se $F_1$ e $F_2$ indicano i numeri fermionici per il primo ed il secondo 
gruppo, $(-1)^{F_2}$ \`e conservato dalle trasformazioni di supersimmetria 
e la teoria pu\`o essere consistentemente ristretta al sottospazio 
$(-1)^{F_2}=1$; dalle (\ref{rot}) segue che i fermioni nel primo gruppo 
trasformano nell'aggiunta di un sottogruppo $H \subset G$ tale che $G/H$ 
\`e uno spazio simmetrico.

Limitandoci a fermioni periodici ed antiperiodici e imponendo l'invarianza 
modulare sulle ampiezze di ogni genere per assicurare l'esistenza di una serie 
perturbativa consistente, si possono determinare le regole che generano tutte le 
soluzioni. In particolare, fattorizzazione ed invarianza modulare assicurano la 
presenza di un gravitone a massa nulla e la corretta relazione tra spin 
e statistica, 
mentre la presenza di un gravitino a massa nulla assicura l'assenza del 
tachione e l'esistenza 
ad ogni livello di massa di un numero uguale di bosoni  e fermioni. 

La descrizione di un modello di Tipo II con $D$ dimensioni compatte 
in termini di fermioni liberi richiede 
l'introduzione, accanto ai campi 
spaziotemporali, di $3D$ fermioni sinistri e $3D$ fermioni destri; un modello 
eterotico richiede invece l'introduzione di
$3D$ fermioni sinistri e $2(16 + D)$ fermioni destri.

Deriviamo le condizioni di consistenza per 
un modello eterotico in quattro dimensioni; la discussione per le superstringhe
di Tipo II 
o per un numero diverso di dimensioni \`e del tutto analoga.
Nel nostro caso il settore sinistro 
supersimmetrico contiene, oltre alle $4$ supercoordinate, 
$18$ fermioni $\chi^i$ , 
$y^i$ , $\w^i$ con $i=1,...,6$ che realizzano la supersimmetria in modo non 
lineare; il settore destro,
non supersimmetrico, contiene, invece, $44$ 
fermioni $\bar{\f}^A$ oltre alle $4$ coordinate bosoniche. 
La supercorrente deve avere la forma:
\be
G(z) = \sum \psi^{\m} \part X_\m + \sum \chi^i y^i \w^i \ ,
\label{supc}
\ee
perch\`e si abbiano fermioni non massivi e la supersimmetria non sia rotta.

\`E immediato scrivere la funzione di partizione del modello:
\be
Z = \int \frac{d^2 \t}{\t_2} \frac{1}{( \sqrt{\t_2} \h(\t) \h(\bar{\t}))^2}
\sum_{spin}C \pmatrix{ \bf{a} \cr \bf{b}} Z\pmatrix{a_{\psi}\cr b_{\psi}} 
\prod_{i=1}^{64} Z_F\pmatrix{a_f\cr b_f}   \  ,
\label{zff}
\ee
dove la somma \`e su tutte le strutture di spin per i fermioni, interni e 
spaziotemporali, e dove il termine $Z\pmatrix{a_{\psi}\cr b_{\psi}}$ rappresenta il
contributo dei due fermioni longitudinali $\psi^0$ e $\psi^1$ e dei ghosts del 
gravitino. Fermioni spaziotemporali, supercorrente e gravitino, devono avere 
tutti la stessa struttura di spin. 
$Z_{F} \pmatrix{a\cr b} $ rappresenta
il contributo alla funzione di partizione di un fermione con condizioni al bordo
$\psi (z+1) = e^{2i \p a} \psi(z)$ e 
$\psi (z+\t) = e^{2i \p b} \psi(z)$, e si esprime in termini di funzioni $\th$:
\be
Z_{F} \pmatrix{a\cr b} =  \left ( \frac{\th \pmatrix{a\cr b}}{\h(\t)}  \right )^{\frac{1}{2}}  \  .
\ee
Infine, i coefficienti $C\pmatrix{\bf{a}\cr \bf{b}}$ pesano il contributo 
delle varie strutture di spin.
Utilizzando le propriet\`a di trasformazione delle funzioni $\th$ rispetto a 
$SL_2(Z)$, si verifica facilmente che la (\ref{zff}) 
\`e invariante modulare quando:
\beq
& & C\pmatrix{\bf{a}\cr \bf{b}} = - e^{\frac{i \p}{8}\sum a_f}
C\pmatrix{\bf{a}\cr \bf{a} + \bf{b} +\bf{1}}  \  , \nonumber \\
& & C\pmatrix{\bf{a}\cr \bf{b}} = e^{\frac{i \p}{4} \sum a_fb_f}
C\pmatrix{\bf{b}\cr \bf{a}}  \  ,
\eeq
dove $\sum a_f = \sum_{modi \ sinistri} a_f - \sum_{modi  \ destri} a_f$.
Un' ampiezza di genere $g \geq 2$ \`e essenzialmente una catena di tori, e nel 
limite in cui questi tori vengono separati infinitamente l'uno dall'altro, 
fattorizza in un prodotto di ampiezze sul toro; questo implica che i 
coefiicienti a genere $g$ devono essere il prodotto di coefficienti a genere 
uno:
\be
C\pmatrix{\bf{a}^1 & \bf{a}^2 & ... & \bf{a}^g\cr
\bf{b}^1 & \bf{b}^2 & ... & \bf{b}^g} =
C\pmatrix{\bf{a}^1 & \bf{b}^1}C\pmatrix{\bf{a}^2 & \bf{b}^2} ...
C\pmatrix{\bf{a}^g & \bf{b}^g}   \   .
\ee
Infine, dato che trasformazioni modulari su superfici con pi\`u loops vengono 
generate da twist di Dehn attorno ai cicli dei vari tori e da un ulteriore 
twist di Dehn per ogni coppia di tori confinanti, si ottiene come nuova 
condizione per assicurare l'invarianza modulare a genere arbitrario:
\be
(-1)^{a_{\psi}+a^{'}_{\psi}} e^{\frac{i \p}{4} \sum a_fa^{'}_f}
C\pmatrix{\bf{a}\cr \bf{b}}C\pmatrix{\bf{a^{'}}\cr \bf{b^{'}}} =
C\pmatrix{\bf{a}\cr \bf{a^{'}} + \bf{b}}
C\pmatrix{\bf{a^{'}}\cr \bf{a} + \bf{b}}  \  .
\label{tloops}
\ee
La seconda fase deriva dalle propriet\`a note delle funzioni $\th$ su superfici 
di genere superiore; la prima \`e legata alle propriet\`a di trasformazione
del determinante del 
gravitino su queste superfici, note solo a meno di una fase. La fase 
pu\`o essere fissata richiedendo che i modelli in dieci dimensioni soddisfino
la (\ref{tloops}) ed \`e stata derivata in \cite{bs}.

Le condizioni di consistenza appena derivate possono essere risolte 
considerando l'insieme $F$ di tutti i fermioni e l'insieme delle parti di $F$,
$2^F$, un gruppo commutativo rispetto al prodotto simmetrico:
\be
\a\b = \a \cup \b - \a \cap \b  \ ,   \hspace{1cm} \a   ,  \b \in 2^F  \ .
\ee
Una struttura di spin \`e completamente specificata assegnando una 
coppia $(\a | \b)$, dove $\a$ e $\b$ sono insiemi di fermioni periodici lungo 
$1$ e $\t$ rispettivamente. I fermioni con condizioni al bordo 
$\pmatrix{1\cr1}$  , $\pmatrix{1\cr0}$  , $\pmatrix{0\cr0}$  e
$\pmatrix{0\cr1}$  corrispondono agli insiemi 
$[A_1;A_2;A_3;A_4] = [ \a \cap \b ;  \a - \a \cap \b ; F - \a \cup \b ;
\b - \a\cap\b]$.
\`E conveniente introdurre le quantit\`a
\be
\e_X = e^{\frac{i\p}{8}n(X)} \ ,  \hspace{1cm}  \d_X = \pm 1    \  ,
\ee
dove $n(X) = n_L(X) - n_R(X)$ \`e il numero netto di fermioni in $X$.
Inoltre, $\d_X$ per la stringa eterotica 
vale $-1$ se $\psi^{\m} \in X$ e $1$ in caso 
contrario, mentre per la stringa di Tipo II vale $-1$ se $\psi^{\m} \in X$ o
$\bar{\psi}^{\m} \in X$ ma non entrambi e $1$ in caso contrario. Introduciamo 
infine l'operatore di parit\`a $(-)^X$:
\be
(-)^X f = - f(-)^X  \ \ \ {\rm se~ } \ f \in X  \ , \hspace{1cm} 
(-)^X f = f(-)^X  \  \ \ {\rm altrimenti} \  .
\ee
Assegnata la struttura di spin $(\a | \b)$, bisogna assicurarsi che la 
supercorrente (\ref{supc}) sia periodica o antiperiodica per trasporto 
parallelo lungo i cicli del toro. Questo significa che gli insiemi 
$F - \a$ e $F - \b$ devono contenere o un numero dispari o un numero pari
di fermioni da ciascun addendo della supercorrente; equivalentemente si deve 
avere:
\be
[G(z),(-)^\a]_{-\d_\a} = [G(z),(-)^\b]_{-\d_\b} = 0   \  .
\label{comm}
\ee
Il sottoinsieme $\S \subset 2^F$ degli insiemi di fermioni che soddisfano
$(-)^X G(z) = \d_X G(z) (-)^X$, costituisce un sottogruppo di $2^F$, 
dato che $(-)^{XY} = (-)^X(-)^Y$ e $\d_{XY} = \d_X \d_Y$, e quindi le scelte di 
strutture di spin compatibili con la supercorrente sono elementi di 
$\S \times \S$. Se indichiamo con $(\a | \b)$ anche il contributo dei fermioni 
alla funzione di partizione:
\be
(\a | \b) = \left ( \frac{\th\pmatrix{1\cr 1}}{\h} \right )^{\a \cap \b}
 \left (\frac{\th\pmatrix{1\cr 0}}{\h} \right )^{\a - \a \cap \b}
 \left (\frac{\th\pmatrix{0\cr 0}}{\h} \right )^{F - \a \cup \b}
 \left (\frac{\th\pmatrix{0\cr 1}}{\h} \right )^{\b - \a \cap \b}   \  ,
\ee
con $\th^X = \th^{n_L(X)}(\bar{\th})^{n_R(X)}$. La funzione di partizione si 
scrive:
\be
Z = \sum C_{(\a | \b)} ( \a | \b)   \   ,
\ee
e le condizioni sui coefficienti divengono:
\beq
C_{(\a | \b)}  &=& - \e_\a C_{(\a | F \a \b)}   \  , \nonumber \\
C_{(\a | \b)}  &=& \e_{\a \cap \b}^2 C_{(\b | \a)}    \ , \nonumber \\
C_{(\a_1 | \b_1)} C_{(\a_2 | \b_2)}  &=& 
\d_{\a_1} \d_{\a_2} \e^2_{\a_1 \cap \a_2}
C_{(\a_1 | \a_2 \b_1)} C_{(\a_2 | \a_1 \b_2)}    \   .
\label{cons}
\eeq
Alcuni casi particolari delle (\ref{cons}) sono:
\beq
C_{(0|0)}C_{(\a|0)} &=& \d_\a C^2_{(\a|0)}   \  , \nonumber \\
C_{(\a|0)}C_{(\b|0)} &=& \d_\a C_{(\a|0)}C_{(\a\b|0)}    \ , \nonumber \\
C_{(\a|\b)}^2 &=& \d_\a \d_\b C_{(\a|0)}C_{(\b|0)}  \   , \nonumber \\
C_{(\a|\b)}C_{(\g|\d)} &=& \d_\a \d_\g C_{(\a|\b \g)}C_{(\g|0)}   \   .
\label{conp}
\eeq
Da queste equazioni
segue che, posto $C_{(0|0)} =1$, per ogni insieme fermionico si 
pu\`o avere o $C_{(\a|0)}= 0$ o 
$C_{(\a|0)}= \d_\a$. 
La collezione $\Xi = \{ \a_1, \a_2, ..\}$ 
di insiemi fermionici per i quali $C_{(\a_i|0)} = \d_{\a_i}$,
\`e un sottogruppo di $\S$, generato da una base 
$\{ b_0 =F, b_1, ..., b_N \}$ e quindi con cardinalit\`a $2^{N+1}$; da 
(\ref{conp}) segue che $(\a | \b)$ contribuisce all'ampiezza solo se 
$\a , \b \in \Xi$ e che in questo caso $C_{(\a|\b)}^2 = 1$. Risolvendo i 
vincoli (\ref{cons}), si dimostra che tutti i modelli consistenti 
che utilizzano come gradi di libert\`a interni fermioni periodici e 
antiperiodici si ottengono fissando un sottogruppo 
$\Xi$ di $2^F$ generato da una base
$\{ b_0 =F, b_1, ..., b_N \}$ che soddisfa:
\beq
n(b_i) &=& 2n(b_i \cap b_j) = 4n(b_i \cap b_j \cap b_k \cap b_l) = 
0 \  mod(8)   \   ,   \nonumber \\
(-)^{b_i}G(z) &=& \d_{b_i}G(z)(-)^{b_i}    \  .
\eeq
Per ogni $\Xi$ ci sono $2^{\frac{N(N+1)}{2}+1}$ teorie consistenti, in 
corrispondenza delle possibili scelte di segno $\pm 1$ per
$C_{(F|F)}$ e $C_{(b_i|b_j)}$   , $i > j =0,...,N$, scelte che possono essere 
estese in modo unico ad una mappa $C_{(\a|\b)} : \Xi \times \Xi \rar Z_2$
che soddisfa alle (\ref{cons}).

Questa costruzione pu\`o essere interpretata come una generalizzazione della 
proiezione GSO \cite{GSO}; il contributo della struttura $(\a|\b)$ equivale a
$(\a|\b) = (-)^\b R^\a N^{f \a}$, ovvero alla somma dei contributi degli stati 
nel settore $R^\a N^{F \a}$ con segno fissato da $(-)^\b$. La funzione di 
partizione si scrive:
\be
Z = \frac{1}{2^{N+1}} \sum_{\a \in \Xi} \left [ \sum_{\b \in \Xi} C_{(\a|\b)} \right ]
R^\a N^{F \a}   \    .
\ee
Utilizzando le (\ref{cons}) e scrivendo il generico elemento di $\Xi$ in 
termini degli insiemi di base, si ottiene:
\be
Z = \sum_{\a \in \Xi} \d_{\a}  \left [ \prod_{i=0}^N \frac{1}{2}(1 + \d_\a
C_{(\a|b_i)}(-)^{b_i}) \right ]
R^\a N^{F \a}    \   ,
\ee
ovvero, dato che le ulteriori proiezioni sono ridondanti,
\be 
Z = \sum_{\a \in \Xi} \d_{\a}  \left [ \prod_{\b \in \Xi}^N \frac{1}{2}(1 + \d_\a
C_{(\a|\b)}(-)^{\b}) \right ]
R^\a N^{F \a}    \   .
\label{ggso}
\ee
Quindi ad ogni elemento $ \b \in \Xi$ corrisponde un settore $R^\b N^{F \b}$
dello spazio di Hilbert e una proiezione GSO generalizzata che mantiene solo 
gli stati in $R^\a N^{F \a}$
con $(-)^{\b} = C_{(\a|\b)}\d_\a$. 
Se $\a \cap \b = 0$, la proiezione $(-)^\b = \r$ mantiene gli stati 
costruiti da $|0 \ran_\a$ con un numero pari di $\b$-oscillatori, per $\r = 1$,
e con un numero dispari, per $\r = -1$; se $\a \cap \b$ \`e non vuoto, si 
definisce un operatore di chiralit\`a 
$\G_{\a \cap \b} = \prod_{f \in \a \cap \b} f_0$ e si mantengono gli stati 
costruiti con un numero pari di $\b$-oscillatori sulle componenti del vuoto 
con chiralit\`a $\r$ e gli stati costruiti con un numero dispari di oscillatori 
sulle componenti del vuoto con chiralit\`a opposta.

Osserviamo che se $\psi^\m \in \a$ 
lo stato fondamentale $|0 \ran_\a$ del settore $R^\a N^{F \a}$
\`e uno spinore spaziotemporale 
e tutti gli altri stati hanno spin semiintero, mentre  se $\psi^\m \in F\a$ 
lo stato fondamentale \`e uno scalare e tutti gli altri 
stati hanno spin intero. Da 
questa osservazione segue che, per la presenza del 
fattore $\d_\a$ in (\ref{ggso}), il contributo all'ampiezza di vuoto \`e 
negativo nel primo caso e positivo nel secondo, in accordo con il fatto che 
eccitazioni con spin intero e semi-intero devono essere quantizzate 
rispettivamente secondo la statistica di Bose e secondo la statistica di Fermi.

La proiezione nei settori puramente Neveu-Schwarz $N^F$ \`e completamente 
fissata: $(-)^\b = C_{(0|\b)} = \d_\b$; questo significa 
che gli stati $ \psi^{\m}_{-1/2} \bar{X}^{\n}_{-1} | 0 \ran {}_{0} $, 
che contengono dilatone, gravitone e tensore antisimmetrico, non vengono mai 
eliminati dallo spettro di un modello consistente.

La presenza nello spettro di almeno 
una particella di spin $3/2$, infine, richiede 
che ci sia un 
insieme $S$ di fermioni contenente precisamente  $8$ modi sinistri, inclusi 
i fermioni $\psi^\m$ e tale che, per tutti i $\b \in \Xi$ disgiunti da $S$, 
si abbia $C_{(S|\b)} = -1$;
ad esempio $S= \{ \psi^\m, \chi^1,...,\chi^6 \}$.
Se \`e presente almeno un gravitino, si pu\`o provare l'esistenza di un uguale 
numero di bosoni e fermioni ad ogni livello di massa, l'annullarsi della 
costante cosmologica ad un loop e l'assenza di tachioni.

Effettuiamo questa costruzione direttamente in dieci dimensioni per una 
teoria di Tipo II; se scegliamo come insiemi di base $0$ , $F$ ed $S$, con $S$ 
insieme dei fermioni sinistri, otteniamo la superstringa di Tipo IIA e la 
superstringa di Tipo IIB. Se scegliamo come insiemi di base $0$ ed $F$, si 
ottengono due nuovi modelli, discussi per la prima volta in \cite{spin1}, e
chiamati comunemente $OA$ e $OB$. La funzione di partizione per questi modelli 
si scrive in termini di caratteri di $SO(8)$:
\beq
Z_{OA} &=& \int \frac{d^2 \t}{\t_2} 
\frac{1}{(\sqrt{\t_2} \h \bar{\h})^8}
 \left [|O_8|^2 + |V_8|^2 + S_8\bar{C}_8 + C_8\bar{S}_8 \right ]  \  ,  \nonumber \\
Z_{OB} &=& \int \frac{d^2 \t}{\t_2} 
\frac{1}{(\sqrt{\t_2} \h \bar{\h})^8}
\left [|O_8|^2 + |V_8|^2 + |S_8|^2 + |C_8|^2 \right ]    \  .
\eeq
Questi due modelli non sono supersimmetrici e contengono un tachione nel loro 
spettro, che indica l'instabilit\`a del vuoto; ci\`o nonostante,
hanno un alto grado di consistenza, in quanto sono modelli di stringhe 
interagenti, invarianti di Lorentz ed invarianti modulari a genere arbitrario.
Una discussione del tutto analoga pu\`o essere condotta anche per modelli 
contenenti fermioni complessi \cite{ff3}.

\section{Calabi-Yau e mirror symmetry}
\markboth{}{}

Dal punto di vista della teoria effettiva, come sappiamo, la ricerca di un 
vuoto con supersimmetria spaziotemporale porta a considerare 
la propagazione della stringa su una variet\`a di Calabi-Yau.
Posiamo affrontare lo stesso problema 
dal punto di vista della teoria conforme: scelta una compattificazione
$C_{st} \times C_{int}$, possiamo infatti chiederci quali 
condizioni deve soddisfare la teoria conforme interna perch\`e
esista un fissato numero di cariche di supersimmetria. In 
\cite{n2a,n2b,n2c} si \`e verificato che la supersimmetria spaziotemporale 
richiede l'estensione dell'algebra superconforme sul worldsheet;
in particolare una teoria spaziotemporale con $N=1$ richiede una 
teoria superconforme con $N=2$ e una teoria spaziotemporale con $N=2$ 
richiede una teoria superconforme con $N=4$. Sottolineiamo che queste 
estensioni dell'algebra superconforme sul worldsheet sono da interpretare come
simmetrie globali, in quanto estensioni 
locali altererebbero la dimensione critica della teoria. 
Le algebre superconformi con $N=2,3,4$ sono state studiate in dettaglio.
Il caso $N=4$, $c=6$ corrisponde alla compattificazione della 
teoria di stringhe su $K3$; l'analogo delle compattificazioni geometriche su CY 
6-dimensionali sono le teorie con $N=2$, $c=9$.
Queste classi di teorie sono state molto studiate, 
e se ne conoscono diverse realizzazioni: come modelli di 
Landau-Ginzburg, come modelli $\s$ con spazio bersaglio un CY,
come prodotto di modelli minimali (modelli di Gepner). Tra queste diverse 
realizzazioni esistono delle relazioni spesso sorprendenti. 
Descriviamo brevemente
l'algebra superconforme con $N=2$; si tratta in realt\`a di
una famiglia di algebre, isomorfe tra loro, caratterizzate dal parametro  $\h$ 
che specifica le condizioni al bordo per le supercorrenti $G^{pm}(z)$.
L'isomorfismo esplicito \`e detto flusso spettrale, e consiste nella 
moltiplicazione per l'operatore $e^{i \sqrt{\frac{c}{3}} \h \f}$, dove
$ j =  i \sqrt{\frac{c}{3}} \part \f$ . Quando $\h= \frac{1}{2}$ il flusso 
spettrale interpola tra settore NS e settore R, realizzando un operatore di 
supersimmetria spazio-temporale. I campi conformi sono caratterizzati, oltre che 
dalla loro dimensione $h$, anche dalla loro carica $Q$ rispetto a $J$, ed in 
effetti la supersimmetria spazio-temporale \`e assicurata se si proietta la 
teoria nel settore con carica $U(1)$ dispari. Di grande importanza sono i campi 
chirali ed antichirali, che hanno un prodotto regolare rispettivamente con $G^+$ 
e $G^-$, i due generatori fermionici dell'algebra superconforme con N=2; 
l'algebra superconforme implica che per gli operatori chirali 
$h= \frac{Q}{2}$ e $h \le \frac{c}{6}$. Utilizzando queste relazioni si 
verifica che i campi chirali sono in numero finito e sono chiusi rispetto al 
prodotto; tenendo conto sia del settore destro 
che di quello sinistro essi formano quindi quattro 
anelli $(c,c)$, $(c,a)$, $(a,c)$, $(a,a)$. Le deformazioni marginali sono 
descritte da operatori del tipo $\F_{1,1}$ e $\F_{-1,1}$; quando il modello ha 
un'interpretazione geometrica, si pu\`o vedere che le prime deformazioni 
corrispondono a variazioni della struttura complessa e le seconde a variazioni 
della struttura k\"ahleriana. 

Dal punto di vista della teoria conforme, gli 
operatori differiscono solo per un cambiamento di segno nel settore sinistro 
del modello, ma dal punto di vista geometrico essi controllano propriet\`a 
profondamente diverse dello spazio di CY; questa asimmetria ha portato a 
congetturare che esistano coppie di spazi di Calabi-Yau, $M$ e $\tilde{M}$, 
caratterizzate da 
\be
h^{1,1} = \tilde{h}^{2,1}   \ ,  \hspace{1cm}  h^{2,1} = \tilde{h}^{1,1}  \ ,
\ee
corrispondenti alla stessa teoria conforme. Queste coppie sono dette coppie 
speculari ("mirror pairs")
e la trasformazione che le lega simmetria speculare. Una serie di esempi di 
tali coppie \`e stata costruita in modo esplicito \cite{mm}.
Questa simmetria \`e di grande importanza perch\`e 
fornisce una dualit\`a tra forte e debole accoppiamento del modello $\s$; 
infatti il parametro di espansione $\frac{\a^{'}}{R^2}$ \`e determinato dalla 
classe di K\"ahler, e se in un modello $\frac{\a^{'}}{R^2}$ \`e tale da non 
consentire un'espansione perturbativa, modificando la struttura complessa si 
pu\`o fare in modo che la classe di k\"ahler del modello speculare, che \`e del 
tutto equivalente, assicuri un debole accoppiamento. Consideriamo come esempio 
un'identit\`a notevole dovuta alla simmetria speculare. Valutiamo nel modello 
$M$ la funzione a tre punti di operatori del tipo $\F_{1,1}$; per queste 
funzioni di correlazione \`e possibile provare un teorema di non 
rinormalizzazione, in base al quale esse sono date semplicemente da 
\be
\int_M \W^{abc} b_a^{(i)} \wedge b_b^{(j)} \wedge b_c^{(k)} \wedge \W   \ ,
\ee
dove $b_a^{(i)}$ sono gli elementi in $H^1(M,T)$ corrispondenti agli operatori
$\F_{1,1}^i$. Consideriamo ora nel modello speculare $\tilde{M}$ una funzione a 
tre punti degli operatori $\F_{-1,1}$ che corrispondono, tramite simmetria 
speculare, ai $\F_{1,1}$, e sono identificabili con elementi di 
$H^1(\tilde{M},T^*)$; in questo caso non si hanno teoremi di 
non rinormalizzazione 
e bisogna tener conto delle correzioni dovute agli istantoni del modello $\s$
che 
richiedono una conoscenza delle curve razionali in $\tilde{M}$. L'espressione 
che si ottiene \`e 
\be
\int_{\tilde{M}} b^{(i)} \wedge b^{(j)} \wedge b^{(k)} +
\sum_{m,\{u \}} e^{-\int_{CP^1} u^*_m(J)}
( \int u^*(b^{(i)}) \int u^*(b^{(j)}) \int u^*(b^{(k)}) )   \  ,
\label{f12}
\ee
dove $\{ u \}$ \`e l'insieme delle funzioni olomorfe $u : CP \rar \G$, 
con $\G$ una curva razionale in $\tilde{M}$,  $ \p_m$ \`e un 
ricoprimento m-esimo di $CP$
e $u_m = u \circ \p_m$. Il primo termine in (\ref{f12}), la forma 
d'intersezione su $\tilde{M}$,  \`e familiare dalla 
geometria classica; la serie di correzioni tiene conto della natura estesa 
della stringa.
L'esistenza della simmetria speculare assicura che queste due quantit\`a 
sono uguali. Una conseguenza immediata \`e che il calcolo degli accoppiamenti 
$(\bf{27})^3$ in una variet\`a $M$, che sono rinormalizzati dai worldsheet 
instantons, pu\`o essere effettuato in modo esatto utilizzando la variet\`a 
speculare $\tilde{M}$; questa uguaglianza \`e stata sfruttata anche in 
geometria enumerativa per il calcolo del numero di curve razionali di 
grado arbitrario su $\tilde{M}$.
Utilizzando la mirror symmetry si pu\`o anche concludere che 
singolarit\`a presenti nello spazio dei moduli k\"ahleriani quando una curva 
razionale degenera non danno origine a singolarit\`a nella teoria conforme, ma 
corrispondono a semplici deformazioni con operatori marginali. 
La forma di K\"ahler $J$ deve soddisfare una naturale condizione di 
positivit\`a:
\be
\int_{M_r} J^r > 0  \ ,
\label{cono}
\ee
dove $M_r$ \`e una sottovariet\`a $r$-dimensionale del CY; le forme che 
soddisfano la (\ref{cono}) definiscono un sottospazio di $H^2(X,R)$ noto come 
{\it cono~di~K\"ahler}. Questa restrizione non vale tuttavia
per il modulo della teoria conforme che corrisponde a $J$; a questo punto sorge 
il problema di interpretare geometricamente cosa accade quando il modulo viene 
variato fino a raggiungere la parete del cono di K\"ahler. In matematica 
esiste un'operazione detta "flop", in cui una curva razionale viene fatta 
degenerare e quindi viene riespansa 
in una direzione "trasversa" in modo da avere volume positivo rispetto alla 
metrica di un nuovo spazio ambiente. Si pu\`o verificare utilizzando la mirror 
symmetry che questa transizione viene realizzata in modo non singolare in 
teoria di stringhe proprio variando il modulo corrispondente a $J$
(\`e necessario ricorrere alla mirror symmetry perch\`e durante la 
transizione si attraversa una regione in cui 
il modello $\s$ \`e fortemente accoppiato).
Non c'\`e quindi
nessuno ostacolo dal punto di vista della teoria delle stringhe al'estensione
della 
descrizione geometrica da un cono di K\"ahler ad un altro, anche se questo
comporta 
un cambiamento di topologia della variet\`a (un cambiamento 
mite, dato che i numeri di Hodge restano inalterati).
Lo spazio dei moduli geometrici viene quindi esteso ed unificato nello spazio 
dei moduli della teoria conforme, che \`e lo spazio rilevante per capire le 
caratteristiche della teoria di stringhe; la geometria classica 
non riesce infatti a dare sempre una descrizione completa: molti limiti in cui 
la descrizione classica degenera, come ad esempio per la metrica singolare sugli 
orbifolds, sono perfettamente consistenti dal punto di vista della teoria di 
stringhe.

Per la costruzione di coppie speculari \`e essenziale la corrispondenza 
esistente tra modelli di Gepner e compattificazioni su spazi di Calabi-Yau; i 
modelli di Gepner sono semplicemente delle teorie conformi realizzate 
utilizzando i pi\`u semplici modelli interagenti, i modelli della serie 
minimale. 
In particolare, Gepner \cite{gep1} ha mostrato che si possono costruire 
modelli di stringhe in $d$ dimensioni formando il prodotto 
tensoriale di modelli minimali della serie $N=2$, che hanno carica centrale
$c = \frac{3k}{k+2}$, in modo da ottenere $c = \frac{3}{2}(10-d)$; proiettando 
la teoria sul settore con cariche $U(1)$ dispari, si ottiene un modello con 
supersimmetria spaziotemporale.
Questa costruzione di compattificazioni della teoria di stringhe utilizzando 
teorie conformi esattamente risolubili presenta evidenti vantaggi rispetto alla 
costruzione geometrica. In particolare, il calcolo delle
funzioni di correlazione tra operatori vertice per  dedurre gli 
accoppiamenti tra i campi della teoria effettiva pu\`o essere condotto in modo 
esatto, anche per quegli accoppiamenti che non sono protetti da un teorema di 
non rinormalizzazione; inoltre questi accoppiamenti sono opportunamente 
normalizzati, dato che per i campi della teoria conforme si scelgono funzioni a 
due punti canoniche.

Tra i modelli puramente algebrici di Gepner e le compattificazioni geometriche 
esiste tuttavia una relazione sorprendente, notata per la prima volta in 
\cite{gep1}. Il modello $(3,3,3,3,3)$ contiene $101$ ipermultipletti nella
$27$ di $E_6$ ed uno nella $\bar{27}$; ha inoltre un gruppo discreto di 
simmetria $Z_5^4 \times S_5$. Come abbiamo visto, lo stesso spettro e le stesse 
simmetrie discrete si ottengono compattificando la teoria su $Y_{4;5}$. Si 
pu\`o verificare l'accordo tra i due modelli a livello delle 
interazioni \cite{gep2} 
confrontando il calcolo geometrico all'ordine pi\`u basso 
degli accoppiamenti $( \bf{27})^3$, che \`e il risultato completo grazie al teorema di 
non rinormalizzazione, con il calcolo in teoria conforme. L'accordo \`e 
perfetto e questo, insieme ad altre verifiche, ha portato a congetturare che i 
due modelli siano identici.

La connessione tra modelli di Gepner e compattificazioni su CY pu\`o essere 
giustificata come segue. Consideriamo una teoria bidimensionale con 
supersimmetria $N=2$; l'azione ha la forma generale:
\be
\int d^2 z d^4 \th K(\bar{\F}_i,\F_i) +  ( \int d^2 z d^2 \th W( \F_i) + 
c.c.)  \  ,
\label{n2}
\ee
dove $K$ \`e detto potenziale k\"ahleriano e $W$, una funzione 
olomorfa dei supercampi chirali $\F_i$, \`e detto superpotenziale. Per ottenere una teoria invariante 
conforme, possiamo fare evolvere la (\ref{n2}) fino a raggiungere un punto 
fisso del gruppo di rinormalizzazione; dato che i termini $D$ sono delle 
perturbazioni irrilevanti ( hanno dimensione maggiore di due ), il flusso della 
teoria \`e determinato dal superpotenziale. In particolare per
$W(\F) = \F^{P+2}$ si ottiene al punto fisso il modello minimale della serie 
$N=2$ con $c = \frac{3P}{P+2}$ e con 
$W(\F_1,...,\F_r)= \F_1^{P_1+2}+...+\F_r^{P_r+2}$ si ottiene il prodotto 
tensoriale $(P_1,...,P_r)$. Questa relazione tra modelli minimali e teorie di 
Landau-Ginzburg \`e stata stabilita in \cite{mmlg} ed \`e alla base della 
relazione tra modelli minimali e compattificazioni su spazi di CY 
\cite{mmcy}.
Osserviamo infatti che  il potenziale per il modello
$(3,3,3,3,3)$ \`e molto simile all'equazione che definisce $Y_{4;5}$,
$z_1^5+z_2^5+z_3^5+z_4^5+z_5^5=0$. Per rendere la connesione pi\`u precisa, 
consideriamo il path integral:
\be
\int D\F_1...D\F_5 e^{i \int d^2z d^4 \th ( \F_1^5+...+ \F_1^5)}    \ ,
\ee
dove abbiamo trascurato il termine $D$; il cambiamento di variabili
$\xi_1 = \F_1^5$  , $\xi_i = \frac{\F_i}{\F_1}$ per $i=2,..,4$ permette di 
riscrivere l'integrale:
\be
\int D \xi_1...D \xi_5 e^{i \int d^2z d^4 \th 
\xi_1 ( 1+\xi_2^5+...+ \xi_1^5)}    \ ,
\ee
ed effettuando l'integrazione su $\xi_1$ si ottiene il vincolo
\be
\d ( 1+ \xi_2^5 + ... + \xi_5^5 )   \  ,
\ee
che \`e proprio l'equazione che definisce $Y_{4;5}$ in coordinate non omogenee; 
gli altri supercampi chirali descrivono quindi un modello $\s$ con proprio 
questa variet\`a come spazio bersaglio (la componente $\th$ della funzione $\d$ 
vincola i campi fermionici ad essere tangenti a questo spazio bersaglio). 
Naturalmente non abbiamo nessun 
controllo sui termini D nei quali \`e racchiusa l'informazione sulla metrica 
del CY che rende il modello $\s$ invariante conforme.

Prodotti di modelli minimali con livelli diversi sono legati a
CY descritti come sottovariet\`a di spazi proiettivi pesati. Ricordiamo che 
uno spazio proiettivo pesato $WCP^n_{w_1,...,w_{n+1}}$ \`e definito come il 
quoziente di $C^{n+1}$ tramite l'identificazione
$[z_1,...,z_{n+1}] \sim [\l^{w_1}z_1,...,\l^{w_n}z_n]$. 
Consideriamo ora un modello $(l_1,...,l_5)$ e riscriviamo il superpotenziale 
corrispondente come:
\be
W = \F_1^{l_1}(1 + \frac{\F_2^{l_2}}{\F_1^{l_1}} + ... +
\frac{\F_5^{l_5}}{\F_1^{l_1}})    \  ;
\ee
se effettuiamo il cambiamento di variabili
$\xi_1 = \F_1^{l_1}$ , $ \xi_i^{l_i} = \frac{\F_i^{l_i}}{\F_1^{l_1}} $ 
l'integrale funzionale diviene:
\be 
\int D \xi_1 ... D \xi_5 J e^{i \int d^2z d^2 \th \xi_1
(1 + \xi_2^{l_2} + .. + xi_5^{l_5})}  \   ,
\ee
dove lo Jacobiano $J \propto \F_1^{1-l_1 + l_1( \sum_{i=2}^5 \frac{1}{l_i})}$ 
si annulla solo quando
\be
\sum_{i=1}^5 \frac{1}{l_i} =1   \  .
\label{wei}
\ee
Se questa condizione \`e soddisfatta, l'integrazione su $\xi_1$ vincola gli 
altri campi sulla ipersuperficie definita da 
$1 + \xi_2^{l_2} + ... + \xi_5^{l_5} = 0$, che \`e proprio la variet\`a di CY
$\sum_{i=1}^{5} z_i^{l_i} =0$ in $WCP^4_{\frac{d}{l_1},...,\frac{d}{l_n}}$,
dove 
$d$ \`e il  minimo comune multiplo degli $l_i$. La condizione (\ref{wei}) 
equivale 
all'annullarsi della prima classe di Chern; inoltre la carica centrale del
modello considerato \`e $c = \sum_{i=1}^5 \frac{3(l_i-2)}{l_i}$, e utilizzando 
la (\ref{wei}) si ha $c = 9$.

Osserviamo che il cambiamento di variabili effettuato \`e ad un valore solo se 
si identifica:
\be
(\F_1,...,\F_5) \sim ( e^{2\p i/l_1} \F_1,...,e^{2\p i/l_5} \F_5)   \ .
\ee
Questa identificazione equivale alla proiezione $U(1)$ che viene effettuata nella costruzione 
minimale per assicurare la supersimmetria spaziotemporale.

\chapter{Stringhe aperte}
\markboth{}{}

\section{Discendenti aperti}
\markboth{}{}

Nella stringa eterotica le simmetrie di gauge spaziotemporali derivano dalla 
presenza sul worldsheet di un'algebra di correnti; un altro modo naturale, e 
storicamente antecedente, di 
introdurre gruppi di simmetria \`e fornito dalle
stringhe aperte e segue 
dall'osservazione, fatta originariamente da Chan e Paton \cite{CP}
nel contesto dei 
modelli duali, che le ampiezze ad albero,
ciclicamente simmetriche nelle gambe esterne, restano tali se 
moltiplicate per tracce di prodotti di matrici. Come notato da Schwarz
\cite{grap1}, 
queste ampiezze sono compatibili con l'unitariet\`a se soddisfano 
una serie infinita di condizioni, risolte poi da Marcus e Sagnotti \cite{grap2}
che hanno 
mostrato come le uniche algebre possibili siano quelle di Lie classiche
$U(n)$ , $O(n)$ e $Usp(n)$.
I gruppi di Chan-Paton possono anche essere rappresentati introducendo 
delle 
variabili fermioniche sul bordo della stringa \cite{grap3}
che, in d dimensioni, danno 
origine a gruppi di ordine $2^{d/2}$; gi\`a questo tipo di costruzione fa 
pensare che la
teoria bosonica con gruppo $SO(8192)$ e la superstringa aperta con $SO(32)$ 
abbiano propriet\`a particolari, cosa che mostreremo tra breve.
Sempre da condizioni di unitariet\`a segue che teorie di sole stringhe aperte 
non sono consistenti, ma devono contenere anche un settore chiuso.

La struttura dei modelli aperti \`e pi\`u complicata rispetto a quella dei 
modelli chiusi; la loro 
formulazione perturbativa coinvolge una somma su superfici di Riemann con bordi 
e non orientabili la cui caratteristica di Eulero \`e data da 
\be
\chi = 2 - 2h - b - c  \ , 
\ee
dove $h$ indica il numero di manici, $b$ il numero di bordi e $c$ il numero di 
crosscaps; 
ad esempio a genere $1/2$ abbiamo il disco e a genere $1$ oltre al toro abbiamo 
altre tre superfici : la bottiglia di Klein, l'anello e la striscia di M\"obius.

Un metodo efficace per trattare superfici non orientabili o con bordi consiste 
nel definirle come quozienti, tramite involuzioni antiolomorfe, partendo 
dai loro ricoprimenti doppi orientabili \cite{aa}. 
Possiamo descrivere ad esempio la bottiglia di Klein con modulo $ i \t_2$ 
partendo da un 
toro con modulo $2i \t_2$ soggetto all'involuzione $K(\s_1,\s_2)=
(1-\s_1, \s_2 + \t_2)$ che non ha punti fissi; in coordiante complesse 
$i_K(z) = 1 - \bar{z} +i \frac{\t_2}{2}$.
L'anello, sempre con modulo $ i \t_2$, si ottiene da un toro con modulo 
$ i \frac{\t_2}{2}$ con l'identificazione $A(\s_1,\s_2) = (2-\s_1,\s_2)$ che 
fissa i due bordi localizzati a $\s_1=0$ e $\s_1=1$; in coordinate complesse 
$i_A(z) = 1 - \bar{z}$.
Infine la striscia di M\"obius si ottiene da un toro con modulo 
$ \frac{1}{2} + i \t_2 $ e 
definito da
$M_1(\s_1,\s_2)=(\s_1+1,\s_2+\t_2)$ , $M_2(\s_1,\s_2)=(\s_1,\s_2+2\t_2)$, 
tramite l'involuzione $M_3(\s_1,\s_2)=(1 - \s_1,\s_2+\t_2)$; in coordinate 
complesse $i_M(z) =  \frac{1}{2} - \bar{z} +i \t_2$.
Tutte queste superfici sono quindi orbifold $Z_2$ del toro e orbifold non 
abeliani del piano complesso (il gruppo non abeliano \`e dato dall'unione 
delle traslazioni che definiscono il toro con l'involuzione).

Grazie a questa immersione, \`e possibile studiare i vincoli di consistenza
della teoria \cite{bs}
introducendo un sottogruppo del gruppo modulare, dato dal commutante 
dell'in\-vo\-lu\-zio\-ne, detto "gruppo modulare relativo".
Studiando la funzione di partizione a genere arbitrario, si stabiliscono quindi 
le condizioni
di cancellazione dei tadpoles, la relazione spin-statistica, la 
condizione che la funzione di partizione sia esprimibile in termini 
dell'algebra chirale massimale della teoria chiusa e che gli spettri 
consistenti siano sottoalgebre dell'algebra di fusione.
Nei vari modelli studiati le condizioni di cancellazione dei tadpoles 
assicurano anche la cancellazione delle anomalie nella teoria;
questa stretta relazione tra tadpoles ed 
anomalie \`e dovuta al fatto che queste derivano da tadpoles di particelle non 
fisiche \cite{od7}. 
Nella Tipo I, ad esempio, le condizioni di tadpole sono legate ad 
inconsistenze nelle equazioni del moto della 10-forma di Ramond.

Come sappiamo,
lo spettro perturbativo di una teoria di stringhe chiuse orientate \`e dato da 
una combinazione invariante modulare di teorie conformi che rispetta la 
relazione spin-statistica per i contributi bosonici e fermionici all'ampiezza 
di vuoto e satura l'anomalia conforme. L'identificazione di principi 
costruttivi per modelli consistenti di stringhe aperte \`e complicata 
dal fatto che non si ha per superfici con bordi 
e crosscaps una generalizzazione immediata del concetto 
di invarianza modulare. L'idea che ha consentito di sviluppare un ben definito 
algoritmo per la costruzione di modelli aperti \`e ispirata dalla descrizione 
delle nuove superfici in termini dei loro ricoprimenti orientabili e consiste
nel considerare tali
modelli come "discendenti" di modelli chiusi con uno spettro 
simmetrico nei modi destri e sinistri \cite{od1}.
Discendenti \`e inteso nel senso di "orbifold nello spazio dei parametri", dove 
la simmetria rispetto alla quale la teoria chiusa viene quozientata \`e la 
parit\`a sul worldsheet $\W$. La costruzione del modello parte 
dalla definizione del settore "untwisted", che consiste nel settore chiuso 
non orientato: la proiezione dello spettro chiuso viene effettuata dalla 
bottiglia di Klein. Bisogna quindi introdurre i settori "twisted",
che in questo 
caso vengono identificati con settori aperti determinati dall'ampiezza di 
anello ed ulteriormente proiettati dalla striscia di M\"obius.

Prima di descrivere la teoria di Tipo I come 
discendente aperto della IIB, vediamo come 
gi\`a nella teoria bosonica richiedere l'assenza di tadpoles fissi il gruppo di 
gauge. La difficolt\`a consiste nel fatto che lo spazio dei moduli per l'anello
, la Klein e la M\"obius \`e essenzialmente l'asse immaginario, 
ed include pertanto l'origine, 
che determina una divergenza ultravioletta per le ampiezze. 
Per i modelli chiusi, abbiamo gi\`a visto che l'origine non fa parte dello 
spazio dei moduli, perch\`e non \`e nella regione fondamentale del gruppo 
modulare. 
Sfruttiamo ora un'idea che si riveler\`a pi\`u volte utile nel seguito; le 
ampiezze ad un loop di stringa aperta possono essere reinterpretate, dopo una 
trasformazione S $( \t \mapsto - \frac{1}{\t})$, come descriventi la propagazione di stati di 
stringa chiusa tra bordi e crosscaps; in questo modo le divergenze 
ultraviolette divengono divergenze infrarosse associate agli stati non massivi 
dello spettro chiuso.
Scriviamo le varie ampiezze che contribuiscono alla funzione di partizione a 
genere uno
\beq
T &=& \frac{1}{2} \left (\frac{T}{2 \p} \right )^{13} \int_F \frac{d^2 \t}{(Im \t)^{14}}
\frac{1}{[\h(\t) \bar{\h}(\bar{\t})]^{24}}  \ ,  \nonumber \\
K &=& \frac{1}{2} \left (\frac{T}{2 \p} \right )^{13} 
\int_0^{\infty} \frac{d^2 \t_2}{\t_2^{14}}
\frac{1}{\h(2i \t_2)}   \ ,   \nonumber \\
A &=& \frac{N^2}{2} \left (\frac{T}{2 \p} \right )^{13} 
\int_0^{\infty} \frac{d^2 \t_2}{\t_2^{14}}
\frac{1}{\h(i \frac{\t_2}{2})}    \ ,   \nonumber \\
M &=& \frac{\e N}{2} \left (\frac{T}{2 \p} \right )^{13}
\int_0^{\infty} \frac{d^2 \t_2}{\t_2^{14}}
\frac{1}{\h(i \frac{\t_2}{2}+ \frac{1}{2})}   \ .   \nonumber \\
\eeq
Il segno di $M$ per ora \`e indeterminato mentre quello di $K$ \`e fissato 
dall'ampiezza di toro.
Riferiamo tutte le ampiezze al ricoprimento doppio : K in $D$ 
dimensioni prende un fattore $2^{D/2}$, A un fattore 
$2^{-D/2}$ ed M un fattore $2$. Con 
$q= e^{-2 \p \t}$
\be
Z= \frac{ \e N}{2} \left (\frac{T}{2 \p} \right )^{13} \int_0^1 \frac{dq}{2 \p q} 
\frac{1}{ \h(-q)^{24}} + 
\frac{N^2 + 2^{26}}{4 \cdot 2^{13}} \left (\frac{T}{2 \p} \right )^{13}
\int_0^1 \frac{dq}{2 \p q} \frac{1}{ \h(q)^{24}}  \ ,
\ee
e nel limite $q \rar 0$ vediamo che i diversi diagrammi si sommano dando una 
prescrizione di parte principale 
\be
Z= 24 \frac{(\e N + 2^{13})^2}{2^{13}} \frac{T^{14}}{(2 \p)^{13}}
\int_0^1 \frac{dq}{8 \p Tq}  \ .
\ee
Da questa espressione possiamo anzitutto leggere l'ampiezza per la scomparsa 
del dilatone nel vuoto
\be
\G = 4 \sqrt{3} \frac{T^7}{\sqrt{ \p^{13}}} \left | \frac{\e N}{2^{13}} + 1 \right | \ ,
\ee
e vediamo inoltre che la condizione di cancellazione, $\e = -1 $ e $ N = 
2^{13}$, seleziona come gruppo di gauge $SO(8192)$.

Per illustrare in modo pi\`u sistematico la costruzione
definiamo i caratteri per l'algebra affine $SO(2n)$
\beq
O_{2n} \!\!\! &=& \!\!\! \frac{1}{2 \h^n} \left [\th^n \pmatrix{0\cr 0} + \th^n \pmatrix{0\cr 1/2} \right ]
 \ , \hspace{0.4cm} 
S_{2n} = \frac{1}{2 \h^n}  \left [\th^n \pmatrix{1/2\cr 0} + i^n 
\th^n \pmatrix{1/2\cr 1/2} \right ] \ , \nonumber \\
V_{2n}\!\!\! &=&\!\!\! \frac{1}{2 \h^n}  \left [\th^n \pmatrix{0\cr 0} - \th^n \pmatrix{0\cr 1/2} \right ] \ ,  \hspace{0.4cm} 
C_{2n} = \frac{1}{2 \h^n}  \left [\th^n \pmatrix{1/2\cr 0} - i^n 
\th^n \pmatrix{1/2\cr 1/2} \right ]  \ .
\eeq
L'azione del gruppo modulare su questi caratteri \`e rappresentata dalle 
matrici
\be
S_{2n} = \frac{1}{2} \pmatrix{1 & 1 & 1 & 1\cr 1 & 1 & -1 & -1\cr 
1 & -1 & i^{-n} & -i^{-n}\cr 1 & -1 & -i^{-n} & i^{-n}}
\ , \hspace{1cm}
T_{2n} = e^{- \frac{2i \p}{3}} \pmatrix{1 & 0 & 0 & 0\cr 0 & -1 & 0 & 0\cr 
0 & 0 & e^{\frac{in \p}{4}} & 0\cr 0 & 0 & 0 & e^{\frac{in \p}{4}}} \ ,
\ee
che soddisfano le condizioni $S^2 = (ST)^3 = C$.

Costruiamo ora 
i discendenti aperti della $IIB$ e dei due modelli non supersimmetrici in dieci 
dimensioni \cite{spin1}. Le ampiezze di toro sono:
\beq
T_{IIB} &=& |V - S|^2  \ , \nonumber \\
T_{0A} &=& |O|^2 + |V|^2 + S \bar{C} + C \bar{S} \ , \nonumber \\
T_{0B} &=& |O|^2 + |V|^2  + |S|^2  + |C|^2   \ .
\eeq 
Il primo passo consiste nello scrivere la bottiglia di Klein, che 
contiene tutti 
i caratteri che compaiono in T con segni 
soggetti a restrizioni dovute all'algebra di fusione; 
il risultato generale \cite{ocon}
\`e che le possibili scelte di segno corrispondono agli automorfismi $Z_2$ 
dell'algebra di fusione compatibili con la proiezione GSO del toro. Scegliendo 
come base di caratteri $(O_8,V_8,-S_8,-C_8)$ si ottengono le ampiezze
\beq
K_{IIB} &=& \frac{1}{2}(V - S ) \ , \nonumber \\
K_{0A} &=& \frac{1}{2}( O + V ) \ , \nonumber \\
K_{0B} &=& \frac{1}{2}( O + V - S - C )  \ .
\eeq
\`E tuttavia possibile scegliere come base anche $(O_8,V_8,S_8,C_8)$ o
$(-O_8,V_8,-S_8,C_8)$ o $(-O_8,V_8,S_8,-C_8)$; queste scelte sono compatibili 
con la proiezione GSO solo per il modello $0B$ e danno origine alle nuove 
ampiezze:
\beq
K^{'}_{0B} &=& \frac{1}{2}(O + V + S + C)  \ ,  \nonumber \\
K^{''}_{0B} &=& \frac{1}{2}(-O + V + S - C)   \ .
\eeq
Alla prima ampiezza corrisponde lo spettro chiuso della Tipo I: gravitone, 
dilatone e 2-forma di Ramond.
Per quanto riguarda i tre discendenti della $0B$, lo spettro determinato da 
$K_{0B}$ non \`e supersimmetrico e contiene, nel settore $NS^2$, 
un tachione, un gravitone e un dilatone, nel settore $R^2$ 
due scalari e due 2-forme. Anche lo spettro corrispondente a $K_{0B}^{'}$ 
non \`e supersimmetrico e contiene, nel settore $NS^2$, 
un tachione, un gravitone e un dilatone, nel settore $R^2$ 
una 4-forma. Infine la proiezione $K_{0B}^{''}$ determina uno spettro 
contenente, nel settore $NS^2$, 
un gravitone e un dilatone, nel settore $R^2$, una 4-forma autoduale, una 
2-forma e uno scalare. Quest'ultimo spettro \`e il pi\`u interessante in quanto 
oltre ad essere chirale non contiene il tachione comune agli altri modelli. Siamo di fronte ad un 
analogo del modello eterotico non supersimmetrico $O(16) \times O(16)$ discusso 
nel secondo capitolo.
 
Nel canale trasverso le ampiezze diventano:
\beq
\tilde{K}_{IIB} &=& \frac{2^5}{2}( V - S )   \ ,  \nonumber \\
\tilde{K}_{0A} &=& \frac{2^5}{2}( O + V ) \ ,\nonumber \\
\tilde{K}_{0B} &=& \frac{2^6}{2} V  \ ,  \nonumber \\
\tilde{K}^{'}_{0B} &=& \frac{2^6}{2} O  \ ,  \nonumber \\
\tilde{K}^{''}_{0B} &=& - \frac{2^6}{2} C  \ .
\eeq

Per la costruzione del settore aperto,
bisogna individuare quale porzione dello spettro chiuso pu\`o fluire 
nell'anello trasverso. Le ampieze A ed M sono polinomi di secondo e primo grado 
nelle molteplicit\`a dei vari settori di carica, vincolati dalle condizioni di 
fattorizzazione dell'ampiezza sul disco. In particolare, i coefficienti 
dell'anello nel canale 
trasverso possono essere 
interpretati come moduli quadri della funzione ad un punto 
sul disco. Data l'ampiezza quasidiagonale 
$T = \sum_{ij} N_{ij} \chi_i \tilde{\chi}_j$, i caratteri che fluiscono 
nell'anello sono quelli accoppiati con i loro coniugati dalla proiezione GSO 
del modello chiuso; se il bordo deve rispettare una data simmetria, la 
riflessione \`e possibile solo se il coniugato GSO \`e anche coniugato rispetto 
a questa simmetria.
Per $ N_{ij} = C_{ij} $, Cardy \cite{od5}
ha stabilito un isomorfismo tra condizioni ai bordi e settori dello spettro. Questo isomorfismo
consente di
esprimere l'ampiezza di anello tramite i coefficienti dell'algebra di 
fusione $N^k_{ij}$ che determinano il contenuto del bulk (k) corrispondente 
alle condizioni al bordo (i) , (j), e quindi 
\be
A = \frac{1}{2} \sum_{ijk} N^k_{ij} n^i n^j \chi_k  \ .
\label{cardy}
\ee
Passando al canale trasverso 
\be
\tilde{A} = \frac{2^{-d/2}}{2} \sum_{ijkl} N^k_{ij} n^i n^j S^l_k \chi_l  \ ,
\ee
e infine, utilizzando la formula di Verlinde \cite{od6}
\be
N^k_{ij} = \sum_l \frac{S_{il}S_{jl}S^{\dag}_{kl}}{S_{1l}}  \ ,
\ee
si ottiene
\be
\tilde{A} = \frac{2^{-d/2}}{2} \sum_{jm}( \frac{S^m_jn^j}{\sqrt{S^m_0}})^2 
\chi_m  \ .
\ee

Osserviamo che l'ansatz di Cardy per l'anello non pu\`o essere
utilizzato per il modello $OA$, perch\'e nel suo canale di vuoto 
fluiscono solo due caratteri. Questo modello va quindi risolto
direttamente. Inserendo quindi coefficienti di riflessione per i 
vari settori dello spettro otteniamo, nel canale diretto e trasverso:
\beq
\tilde{A}_{IIB} &=& 2^{-5}\frac{n^2}{2}(V - S)  \ ,  \nonumber \\ 
A_{IIB} &=& \frac{n^2}{2}(V - S)   \ ,
\eeq
\beq
\tilde{A}_{0A} &=& \frac{2^{-5}}{2}[(n_b+n_f)^2V + (n_b-n_f)^2O]  \ , \\
A_{0A} &=& \frac{n_b^2+n_f^2}{2}( O + V) - n_b n_f ( S + C )   \ ,
\eeq
\beq
\tilde{A}_{0B} &=& \frac{2^{-6}}{2}[(n_o+n_v+n_s+n_c)^2V 
+ (n_o+n_v-n_s-n_c)^2O  \nonumber \\
&+& (n_o-n_v+n_s-n_c)^2S + (n_o-n_v-n_s+n_c)^2C]  \ , \\
A_{0B} &=& \frac{n_o^2+n_v^2+n_s^2+n_s^2}{2}V + (n_on_v+n_sn_c)O \nonumber \\
&-&(n_on_c+n_vn_s)S-(n_on_s+n_vn_c)C  \ .
\eeq
\beq
\tilde{A}^{''}_{0B} &=& \frac{2^{-6}}{2}[(n_1+\bar{n}_1+n_2+\bar{n}_2)^2V 
- (n_1-\bar{n}_1+n_2-\bar{n}_2)^2O 
\nonumber \\
&+& (n_1-\bar{n}_1-n_2+\bar{n}_2)^2S - (n_1+\bar{n}_1-n_2-\bar{n}_2)^2C]  \ , \\
A_{0B}^{''} &=& (n_1\bar{n}_1+n_2\bar{n}_2)V + 
(n_1\bar{n}_2+n_2\bar{n}_1)O \nonumber \\
&-&
(n_1n_2+\bar{n}_1\bar{n}_2)S-\frac{n_1^2+n_2^2+\bar{n}_1^2+\bar{n}^2_2}{2}C  \ .
\eeq

Resta da costruire l'ampiezza sulla striscia di M\"obius; nel canale trasverso, 
essa \`e data in ogni settore dalla media geometrica dei corrispondenti contributi di $\tilde{A}$ e
$\tilde{K}$, come segue da uno studio attento della misura \cite{bs}:
\beq
\tilde{M}_{IIB} &=& -2 \frac{n}{2}(V - S)  \ , \nonumber \\
\tilde{M}_{0A} &=& - [(n_b + n_f) \hat{V} + (n_b - n_f) \hat{O}] \ , \nonumber \\
\tilde{M}_{0B} &=& (n_o+n_v+n_s+n_c)\hat{V} \ , \nonumber \\
\tilde{M}_{0B}^{''} &=& (n_1+\bar{n}_1-n_2-\bar{n}_2)\hat{C}  \ .
\eeq
Si noti che i caratteri che compaiono nell'ampiezza di M\"obius 
sono caratteri reali, 
definiti da $ \hat{\chi} = e^{-i \p ( h - \frac{c}{24})} \chi$,  e che
il passaggio dal 
canale trasverso a quello diretto \`e ora regolato dalla matrice 
$P= T^{1/2}ST^2ST^{1/2}$, che soddisfa $P^2 =C$ :
\be
P_{2n} = \pmatrix{c & s & 0 & 0\cr s & -c & 0 & 0\cr 
0 & 0 & \z c & i \z s\cr 0 & 0 & i \z s  & \z c}  \ ,
\ee
dove $s = \sin{ \frac{ \p n}{4}}$ , $c = \cos{ \frac{ \p n}{4}}$ e
$\z = e^{-in \frac{\p}{4}}$.
La trasformazione $P$ rappresenta sui caratteri della teoria conforme lo 
scambio tra la proiezione di M\"obius dello spettro aperto e 
l'ampiezza bordo-crosscap della stringa chiusa.
Si ottiene quindi
\beq
M_{IIB} &=& - \frac{n}{2}(V - S)  \ ,  \nonumber \\
M_{0A} &=& - \frac{1}{2} (n_b + n_f ) \hat{V} + ( n_b - n_f ) \hat{O} \ , \\
M_{0B} &=& - \frac{1}{2} (n_o+n_v+n_s+n_c)\hat{V}  \ ,  \nonumber \\
M_{0B}^{''} &=& - \frac{1}{2} (n_1+\bar{n}_1-n_2-\bar{n}_2)\hat{C}  \ .
\eeq
Scriviamo ora le condizioni di cancellazione dei tadpoles. Nel caso della Tipo 
I ne abbiamo una sola che fissa come gruppo di gauge $SO(32)$; analogamente
per il 
discendente di $0A$ si ha: $n_b + n_f = 32$. Otteniamo cos\`\i \, una 
classe di modelli con gruppo di gauge $SO(n_b) \times SO(32 - n_b)$; 
l'interpretazione \`e che abbiamo introdotto due tipi di numeri quantici e 
nuovi settori di stringhe aperte i cui estremi portano l'uno o l'altro tipo di 
carica. Per $0B$ abbiamo tre condizioni
\be
n_o+n_v+n_s+n_c = 64 \hspace{1cm} n_o = n_v \hspace{1cm} n_v = n_c  \ .
\ee
Le due ultime condizioni equivalgono alla cancellazione delle anomalie del 
modello; il gruppo di gauge in questo caso \`e 
$SO(p) \times SO(p) \times SO(32-p) \times SO(32-p)$.
Per la proiezione $K^{''}$ otteniamo la condizione 
$n_1 + \bar{n}_1 - n_2 - \bar{n}_2 = 64$, che assicura la cancellazione delle 
anomalie di gauge e gravitazionali; non abbiamo tuttavia la possibilit\`a di 
annullare il tadpole del dilatone e pertanto la dimensione del gruppo di gauge 
resta indeterminata, ma se $n_2 = 0$ non c'\`e tachione ed abbiamo $U(32)$.

Possiamo dare regole equivalenti per orbifolds geometrici.
Quando 
consideriamo un orbifold nello spaziotempo, \`e naturale infatti 
aspettarsi che la 
funzione di partizione debba contenere la somma dei contributi provenienti da 
tutte le sezioni del fibrato che ha per base la superficie data e per fibra il 
gruppo di orbifold, in modo analogo a quanto abbiamo fatto per orbifold di 
stringhe chiuse o nel caso della proiezione GSO.
In generale, contribuiscono solo le sezioni compatibili con
l'involuzione; nel caso $Z_2$ esse sono quattro
per K ed M, come segue dalle relazioni $K^2_2=K_3$ e $M^2_2 = M_3$, e sei 
per A. Quando la superficie ha bordi, il carattere della sezione pu\`o 
essere specificato in termini di condizioni al bordo per le coordinate.
In questo modo si verifica facilmente che compattificazioni di teorie di 
stringhe aperte su orbifold $T^4/Z_2$ contengono in genere settori con 
condizioni al bordo $DD$ (Dirichlet-Dirichlet) e $ND$ (Neumann-Dirichlet).
Questo metodo per costruire discendenti aperti, introdotto in 
\cite{od2} e riconsiderato in \cite{or2}, \`e particolarmente utile per 
chiarire gli aspetti geometrici di questi modelli.

Osserviamo infine che la tecnica illustrata per la costruzione di discendenti 
aperti di teorie 
conformi \`e del tutto generale e pu\`o essere applicata, ad esempio, 
alla serie minimale e ai modelli $SU(2)$ WZW \cite{ocon}. 

\subsection{Modelli in sei e in quattro dimensioni}

Utilizziamo il metodo appena descritto per costruire 
classi di modelli in quattro e sei dimensioni come 
discendenti di orbifold della Tipo IIB \cite{od3}. 
In sei dimensioni si tratta essenzialmente di compattificazioni della Tipo I
su $K3$ che
danno origine a teorie effettive di supergravit\`a con 
supersimmetria $N=(1,0)$.
Una caratteristica peculiare di 
questi modelli \`e la presenza nello spettro di un numero variabile di 
multipletti tensoriali; 
anche compattificazioni dell'eterotica danno infatti origine a 
vuoti con lo stesso numero di supersimmetrie, ma, almeno perturbativamente, 
questi vuoti possono contenere un solo multipletto tensoriale.

Per discutere i modelli in sei 
dimensioni \`e opportuno introdurre una base di otto caratteri ``untwisted"
$\chi_i$ e di otto caratteri ``twisted'' $\tilde{\chi}_i$:
\beq
\chi_1 = Q_oO_4O_4 + Q_vV_4V_4 \ , &  & \tilde{\chi}_1 = Q_sS_4O_4 + Q_cC_4V_4  \  ,
\nonumber \\
\chi_2 = Q_oO_4V_4 + Q_vV_4O_4 \ , &  & \tilde{\chi}_2 = Q_sS_4V_4 + Q_cC_4O_4  \  ,
\nonumber \\
\chi_3 = Q_oC_4C_4 + Q_vS_4S_4 \ , &  & \tilde{\chi}_3 = Q_sV_4C_4 + Q_cO_4S_4 \   ,
\nonumber \\
\chi_4 = Q_oC_4S_4 + Q_vS_4C_4 \ , &  & \tilde{\chi}_4 = Q_sV_4S_4 + Q_cO_4C_4 \   ,
\nonumber \\
\chi_5 = Q_oV_4V_4 + Q_vO_4O_4 \ , &  & \tilde{\chi}_5 = Q_sC_4V_4 + Q_cS_4O_4 \   ,
\nonumber \\
\chi_6 = Q_oV_4O_4 + Q_vO_4V_4 \ , &  & \tilde{\chi}_6 = Q_sC_4O_4 + Q_cS_4V_4 \   ,
\nonumber \\
\chi_7 = Q_oS_4S_4 + Q_vC_4C_4 \ , &  & \tilde{\chi}_7 = Q_sO_4S_4 + Q_cV_4C_4 \   ,
\nonumber \\
\chi_8 = Q_oS_4C_4 + Q_vC_4S_4 \ , &  & \tilde{\chi}_8 = Q_sO_4C_4 + Q_cV_4S_4 \   ,
\nonumber \\
\label{chi}
\eeq
dove 
\beq
Q_o &=& V_4O_4 - C_4C_4 \ , \hspace{1cm}   Q_v = O_4V_4 - S_4S_4  \ , \nonumber \\
Q_s &=& O_4C_4 - S_4O_4 \ ,  \hspace{1cm}  Q_c = V_4S_4 - C_4V_4  \ .
\eeq
I caratteri massless sono $\chi_1$ , $\chi_5$ ,
$\tilde{\chi}_1$  , $\tilde{\chi}_6$  , $\tilde{\chi}_7$  , $\tilde{\chi}_8$   
e contengono fermioni chirali.

Consideriamo la IIB compattificata sul toro di $SO(8)$:
\be
T_B = |V_8 - S_8|^2(|O_8|^2+|V_8|^2+|S_8|^2+|C_8|^2) \  ,
\label{tb}
\ee
e utilizziamo l'involuzione che agisce cambiando 
segno a $V_4$ ed $S_4$ nel secondo e 
quarto fattore di ciascun termine che si ottiene dalla (\ref{tb}) 
quando si esprimono i caratteri di $SO(8)$ in termini dei 
caratteri di $SO(4)$.  
Proiettando il settore untwisted e costruendo il 
settore twisted, si ottiene come funzione di partizione 
dell'orbifold un invariante modulare diagonale
nei caratteri introdotti in (\ref{chi}): 
\be
T^{'}_B = \sum_{i=1}^{8}( |\chi_i|^2 +  |\bar{\chi}_i|^2 )   \ .
\label{toro}
\ee
L'ampiezza di Klein nel canale diretto \`e:
\be
K = \sum_{i=1}^{8}( \chi_i + \bar{\chi}_i)  \  .
\label{klein}
\ee
Da (\ref{toro}) e (\ref{klein}) si pu\`o leggere immediatamente lo spettro 
chiuso del modello: dal settore untwisted si ottiene il multipletto del 
gravitone (G), un multipletto tensoriale (T) e quattro ipermultipletti (H); dal 
settrore twisted si ottengono quattro multipletti tensoriali  e dodici 
ipermultipletti. Complessivamente lo spettro chiuso contiene
$(G,5T,16H)$. Dobbiamo ora costruire il settore aperto del modello; dalla Klein 
trasversa:
\be
\tilde{K} = \frac{2^5}{2} \chi_1   \  ,
\ee
segue che $\tilde{M}$ pu\`o contenere solo $\chi_1$, l'identit\`a 
dell'algebra di fusione:
\be
\tilde{M} = - \frac{2 \a}{2} (\hat{Q_o}\hat{O}\hat{O}-
\hat{Q_v}\hat{V}\hat{V})   \  .
\ee
Utilizzando $P_4$ otteniamo $M$ :
\be
M = \frac{\a}{2} (\hat{Q_o}\hat{O}\hat{O}-\hat{Q_v}\hat{V}\hat{V})  \   .
\ee
Parametrizzando l'anello secondo la (\ref{cardy}), nel canale trasverso 
abbiamo:
\be
\tilde{A} = \frac{2^{-3}}{2} \sum_{i,j=1}^{16}
(\frac{n_iS^j_i}{\sqrt{S_0^j}})^2\chi_j  \ ,  
\ee
\be
\tilde{M} = \frac{1}{2} \sum_{i=1}^8 (n_i + \tilde{n}_i)  
(\hat{Q_o}\hat{O}\hat{O}-\hat{Q_v}\hat{V}\hat{V})   \  ,
\label{m6}
\ee
e possiamo determinare le condizioni di cancellazione dei tadpoles. Dai due 
tadpoles del settore untwisted, relativi a $\chi_1$ e $\chi_5$, si ottiene:
\be
\sum_{i=1}^8 (n_i + \tilde{n}_i ) = 32   \  ,   \hspace{1cm}
\sum_{i=1}^8 n_i = \sum_{i=1}^8 \tilde{n}_i  \  .
\label{t1}
\ee
Le condizioni che derivano dal settore twisted sono:
\beq
& & n_5-n_1+\tilde{n}_1+\tilde{n}_6+\tilde{n}_7+\tilde{n}_8 = 8  \  ,
\hspace{0.6cm}
n_6-n_2+\tilde{n}_3+\tilde{n}_5+\tilde{n}_6+\tilde{n}_8 = 8  \ ,  \nonumber \\
& & n_7-n_3+\tilde{n}_2+\tilde{n}_5+\tilde{n}_7+\tilde{n}_8 = 8  \  ,
\hspace{0.6cm}
n_8-n_4+\tilde{n}_4+\tilde{n}_5+\tilde{n}_6+\tilde{n}_7 = 8  \  .
\label{t2}
\eeq
Una semplice soluzione delle (\ref{t1}-\ref{t2}) si ottiene prendendo
$n_1 = n_2 = \tilde{n}_7 = \tilde{n}_8 = 8$. Lo spettro aperto di questo 
modello si legge da:
\beq
A &=& \frac{n_1^2+n_2^2+\tilde{n}_7^2+\tilde{n}_8^2}{2} \chi_1   +
(n_1n_2+\tilde{n}_7 \tilde{n}_8) \chi_2  \nonumber \\
&+& (n_1\tilde{n}_7 + n_2\tilde{n}_8)\tilde{\chi}_7 
+ (n_1\tilde{n}_8+n_2\tilde{n}_8)\tilde{\chi}_8 \ , \nonumber \\
M &=& \frac{1}{2}(n_1+n_2+\tilde{n}_7+\tilde{n}_8)
(\hat{Q_o}\hat{O}\hat{O}-\hat{Q_v}\hat{V}\hat{V})    \ .
\eeq
Gli indici di Chan-Paton dei vettori di gauge, che sono contenuti nel carattere
$\chi_1$, vengono simmetrizzati e il gruppo di simmetria risulta pertanto
$Usp(8)^4$; lo spettro contiene inoltre gaugini destrorsi nell'aggiunta
e ipermultipletti, con fermioni sinistrorsi, nelle rappresentazioni
$(8,1,8,1)$  ,  $(1,8,1,8)$   ,   $(8,1,1,8)$    ,   $(1,8,8,1)$.
La cancellazione delle anomalie per questo spettro verr\`a discussa nel 
prossimo paragrafo.

Si possono ottenere gruppi di gauge unitari introducendo linee di Wilson 
discrete \cite{od3}. 
L'idea pu\`o essere illustrata in modo semplice considerando la Tipo 
I compattificata a $d=8$ sul toro di $SO(4)$. I caratteri della M\"obius si 
decompongono rispetto a $SO(2) \times SO(2)$ secondo:
\be
\hat{O}_4 = \hat{O}_2\hat{O}_2 - \hat{O}_2\hat{O}_2  \ , \hspace{1cm}
\hat{V}_4 = \hat{O}_2\hat{V}_2 + \hat{V}_2\hat{O}_2    \  .
\label{sm}
\ee
Le (\ref{sm}) sono il risultato del calcolo del determinante funzionale sulla 
striscia di M\"obius, ovvero sul ricoprimento doppio con la condizione 
che i fermioni sinistri sono legati ai 
fermioni destri nel punto immagine da:
\be
\psi[i(P)] = i \tilde{\psi}(P)   \  .
\label{inv}
\ee
\`E tuttavia consistente con l'esistenza di una supercorrente cubica sul 
ricoprimento doppio anche porre
\be
\hat{O}^{'}_4 = \hat{O}_2\hat{O}_2 + \hat{O}_2\hat{O}_2  \ , \hspace{1cm}
\hat{V}^{'}_4 = \hat{O}_2\hat{V}_2 - \hat{V}_2\hat{O}_2    \  ;
\label{sm}
\ee
che corrisponde ad avere due fermioni legati secondo la (\ref{inv}) e due 
secondo:
\be
\psi[i(P)] = -i \tilde{\psi}(P)  \  .
\ee
Ora ad ogni livello di massa gli stati in una data rappresentazione del 
gruppo di Chan-Paton non formano pi\`u multipletti di $SO(4)$ ma solo di 
$SO(2) \times SO(2)$; inoltre la matrice $P$ viene trasformata: sui caratteri 
$\hat{O}_4$ , $\hat{V}_4$ agisce infatti come 
$\pmatrix{0 & 1\cr 1 & 0}$, mentre sui caratteri 
$\hat{O}^{'}_4$ , $\hat{V}^{'}_4$  agisce come
$\pmatrix{1 & 0\cr 0 & -1}$. A questo punto \`e facile convincersi che, 
per la compattificazione toroidale considerata, se si effettua la 
scelta usuale di caratteri, i vettori di gauge non fluiscono nella M\"obius e 
si ottengono gruppi di gauge unitari, mentre se si scelgono i caratteri 
primati, i 
vettori di gauge fluiscono nella M\"obius e i gruppi di gauge 
sono ortogonali.

Torniamo ora al nostro modello in sei dimensioni; se modifichiamo la
(\ref{m6}) introducendo caratteri primati, passando al canale diretto si 
ottiene:
\be
M = - \frac{1}{2} \sum_{i=1}^8 (n_i + \tilde{n}_i)  
(\hat{Q_v}\hat{O}^{'}\hat{O}^{'}-\hat{Q_v}\hat{V}^{'}\hat{V}^{'})  \   .
\ee
Ora le stringhe non orientate non corrispondono pi\`u
all'identit\`a dell'algebra di fusione e si ottengono quindi gruppi di 
Chan-Paton unitari. Si pu\`o costruire un semplice modello con gruppo di gauge 
$U(8) \times U(8)$ e ipermultipletti nelle rappresentazioni
$(28 \oplus \bar{28},1)$ , $(1,28 \oplus \bar{28})$ , 
$(8,8)$ e $(\bar{8},\bar{8})$.

Una seconda classe di modelli discende da orbifold di:
\be
T_A = |V_8 - S_8|^2( |O_8|^2 + |V_8|^2 + S_8\bar{C_8}+C_8\bar{S_8})  \  .
\ee
La funzione di partizione dell'orbifold \`e adesso:
\beq
T  &=& |\chi_1|^2+|\chi_2|^2+|\chi_5|^2+|\chi_6|^2
+|\tilde{\chi}_1|^2+|\tilde{\chi}_2|^2+|\tilde{\chi}_5|^2+|\tilde{\chi}_6|^2
\nonumber \\
&+& \chi_7 \bar{\chi}_8 + \chi_3 \bar{\chi}_4 + \chi_8 \bar{\chi}_7 + 
\chi_4 \bar{\chi}_3 
+\tilde{\chi}_7\bar{\tilde{\chi}}_8+\tilde{\chi}_3\bar{\tilde{\chi}}_4
+\tilde{\chi}_8\bar{\tilde{\chi}}_7+\tilde{\chi}_4\bar{\tilde{\chi}}_3  \  .
\eeq 
Dalla Klein, che ora si scrive:
\be
K = \frac{1}{2}(\chi_1+\chi_2+\chi_5+\chi_6+\tilde{\chi}_1+
\tilde{\chi}_2+\tilde{\chi}_5+\tilde{\chi}_6)  \  ,
\ee
determiniamo lo spettro chiuso: il settore untwisted contribuisce con
$(G,T,4H)$ e quello twisted con $(6T,10H)$; complessivamente
$(G,7T,14H)$. Come si pu\`o apprezzare da questo esempio, cambiando la funzione 
di partizione del modello di origine, si ottengono in genere discendenti con 
un numero differente di multipletti tensoriali. Per completare la costruzione 
del modello osserviamo che ora
\be
\tilde{K} = \frac{2^4}{2}(\chi_1 + \chi_2)  \ ,
\ee
e quindi
\be
\tilde{M} = -\frac{2\a}{2}(\hat{Q}_0 \hat{O} \hat{O} - 
\hat{Q}_v \hat{V} \hat{V})
+\frac{2\a}{2}(\hat{Q}_0 \hat{O} \hat{V} - Q_v \hat{V} \hat{O})   \   .
\ee
Anello e M\"obius nel canale diretto e trasverso sono:
\beq
A &=& \frac{1}{2}(\sum_{i=1}^8n_i^2)(\chi_1 + \chi_2) 
+(n_1n_2+n_3n_4+n_5n_6+n_7n_8)(\chi_3+\chi_4)  \nonumber \\
&+& (n_1n_3+n_2n_4+n_5n_7+n_6n_8)(\chi_5+\chi_6)
+(n_1n_4+n_2n_3+n_5n_8+n_6n_7)(\chi_7+\chi_8)   \nonumber \\
&+& (n_1n_5+n_2n_6+n_3n_7+n_4n_8)(\tilde{\chi}_1+\tilde{\chi}_2) 
+(n_1n_6+n_2n_5+n_3n_8+n_4n_7)(\tilde{\chi}_3+\tilde{\chi}_4)   \nonumber \\
&+& (n_1n_7+n_2n_8+n_3n_5+n_4n_6)(\tilde{\chi}_5+\tilde{\chi}_6) \nonumber \\
&+& (n_1n_8+n_2n_7+n_3n_6+n_4n_5)(\tilde{\chi}_7+\tilde{\chi}_8)  \  ,
\eeq
\be
M = \frac{1}{2}(\sum_{i=1}^8n_i)(\hat{Q}_0 \hat{O} \hat{O} - 
\hat{Q}_v \hat{V} \hat{V})
-frac{1}{2}(n_1-n_2+n_3-n_4+n_5-n_6+n_7-n_8)
(\hat{Q}_0 \hat{O} \hat{V} + Q_v \hat{V} \hat{O})   \   ,
\ee
\beq
\tilde{A} &=& 2^{-5}[(n_1+n_2+n_3+n_4+n_5+n_6+n_7+n_8)^2\chi_1  \nonumber \\
&+& (n_1-n_2+n_3-n_4+n_5-n_6+n_7-n_8)^2\chi_2   \nonumber \\
&+& (n_1+n_2+n_3+n_4-n_5-n_6-n_7-n_8)^2\chi_5 \nonumber \\
&+& (n_1-n_2+n_3-n_4-n_5+n_6-n_7+n_8)^2\chi_6   \nonumber \\
&+& (n_1+n_2-n_3-n_4+n_5+n_6-n_7-n_8)^2\tilde{\chi}_1 \nonumber \\
&+& (n_1-n_2-n_3+n_4+n_5-n_6-n_7+n_8)^2\tilde{\chi}_2   \nonumber \\
&+& (n_1+n_2-n_3-n_4-n_5-n_6+n_7+n_8)^2\tilde{\chi}_5   \nonumber \\
&+& (n_1-n_2-n_3+n_4-n_5+n_6+n_7-n_8)^2\tilde{\chi}_6  \  ,
\eeq
\be
\tilde{M} = -(\sum_{i=1}^8n_i)(\hat{Q}_0 \hat{O} \hat{O} - 
\hat{Q}_v \hat{V} \hat{V})
+(n_1-n_2+n_3-n_4+n_5-n_6+n_7-n_8)
(\hat{Q}_0 \hat{O} \hat{V} + Q_v \hat{V} \hat{O})   \   .
\ee
Possiamo quindi determinare le condizioni di cancellazione dei tadpoles:
\beq
& & n_1+n_2+n_3+n_4 = 8   \ , \hspace{1cm} n_5+n_6+n_7+n_8 = 8 \ ,\nonumber \\
& & n_3+n_4+n_7+n_8 = 8  \ , \hspace{1cm} n_2+n_3+n_5+n_8 = 8  \ .
\eeq
Scegliendo ad esempio 
$n_1=n_3=n_7=0$ , $n_2=n_6=n_8=2$ , $n_5=4$ e $n_4=6$ si ottiene
un modello con gruppo di gauge $Usp(6) \otimes Usp(2)^3 \otimes Usp(4)$
ed ipermultipletti nelle rappresentazioni
$(6,2,1,1,1)\oplus(1,1,2,2,1)$ , $(6,1,2,1,1)\oplus(1,2,1,2,1)$ , 
$(6,1,1,2,1)\oplus(1,2,2,1,1)$  e $(6,1,1,1,4)$.
L'introduzione di linee di Wilson discrete permette di ottenere, ad esempio, un 
gruppo di gauge $U(4) \otimes U(4)$ ed ipermultipletti nelle rappresentazioni
$(6 \oplus \bar{6},1)\oplus(1,6\oplus\bar{6})$  ,  
$(4,\bar{4})\oplus(\bar{4},4)$ e $(4,4)\oplus(\bar{4},\bar{4})$.

In \cite{od3} sono state studiate altre due classi di modelli che discendono da 
orbifold $Z_2$ di
\beq
T_{AA} &=& |V_8 - S_8|^2(|O_4|^2 + |V_4|^2 + |S_4|^2 + |C_4|^2)^2  \  ,
\nonumber \\
T_{BB} &=& |V_8 - S_8|^2(|O_4|^2 + |V_4|^2 + S_4 \bar{C}_4 
+ C_4\bar{S}_4)^2  \  .
\eeq
La costruzione delle varie ampiezze per questi modelli richiede l'introduzione 
di $64$ caratteri. La prima classe contiene nello spettro chiuso sei 
multipletti tensoriali e nello spettro aperto presenta
caratteri che non sono autoconiugati.
La seconda classe non contiene altri multipletti tensoriali oltre a quello 
proveniente dal settore non twistato ed \`e pertanto la pi\`u simile ai modelli 
perturbativi eterotici. Si possono costruire modelli con gruppo di gauge 
$Usp(16) \otimes Usp(16)$ che contengono, oltre ai gaugini, fermioni sinistri
nelle rappresentazioni $(\bf{120},\bf{1})$ e $(\bf{1},\bf{120})$ dai settori 
non twistati e nella rappresentazione $(\bf{16},\bf{16})$ dai settori twistati.
Introducendo linee di Wilson discrete si ha un aumento di simmetria; il gruppo 
di gauge \`e ora $U(16) \otimes U(16)$ e i fermioni sono nelle rappresentazioni
$(\bf{120 \oplus 120^{*}},\bf{1})$  , $(\bf{1},\bf{120 \oplus 120^{*}})$ e  
$(\bf{16},\bf{16})$. Questo stesso modello \`e stato ritrovato in \cite{or2}.

\subsection{Anomalie}

Abbiamo visto nel primo capitolo che la cancellazione delle anomalie in dieci
dimensioni richiede la scelta di un gruppo di gauge appropriato, per eliminare 
i termini contenenti tracce irriducibili dal polinomio di anomalia, e 
l'esistenza di un tensore antisimmetrico che trasformi in modo opportuno 
rispetto a simmetrie di Lorentz e di gauge, per eliminare il polinomio residuo.
In teorie di stringhe chiuse, la cancellazione delle anomalie segue 
dall'invarianza modulare della ampiezza di toro, come \`e stato verificato in 
\cite{war}. In teoria di stringhe aperte la cancellazione deriva da una delicata combinazione 
dei contributi delle ampiezze di toro, bottiglia di Klein, anello e striscia di 
M\"obius. 

Nei modelli considerati nel paragrafo precedente, l'imposizione delle 
condizioni di tadpole elimina i termini irriducibili dal polinomio di 
anomalia, tuttavia il polinomio residuo non fattorizza nel prodotto
di due 4-forme e l'usuale meccanismo di Green-Schwarz non pu\`o essere 
applicato. L'anomalia pu\`o essere cancellata grazie ad un meccanismo di
Green-Schwarz generalizzato \cite{6d3} che coinvolge le forme 
antisimmetriche messe a disposizione dal modello. Illustriamo questo meccanismo 
per la classe di modelli con cinque multipletti tensoriali. Riportiamo 
anzitutto delle semplici identit\`a che permettono di esprimere 
$TrF^n$, la traccia nella rappresentazione aggiunta di un 
gruppo $Usp(n)$ o $SO(n)$ di una forma $F^n$ con 
$F = \frac{1}{2}F_{\m\n}dx^\m dx^\n$, in termini di tracce 
nella rappresentazione fondamentale:
\be
TrF^2 = (n \pm 2)trF^2  \  ,  \hspace{1cm} TrF^4 = (n \pm 8)trF^4 + 3(trF^2)^2  \ , 
\ee
dove il primo segno si riferisce a $Usp(n)$ e il secondo a $SO(n)$.
Le stesse formule collegano tracce nella simmetrica o antisimmetrica di $SU(n)$ 
a tracce nella fondamentale; per l'aggiunta di $SU(n)$ abbiamo:
\be
TrF^2 = 2ntrF^2  \ , \hspace{1cm} TrF^4 = 2ntrF^4 + 6(trF^2)^2   \  .
\ee
Abbiamo, infine, per tracce nella rappresentazione 
$(m,n)$ di $G_1 \otimes G_2$:
\beq
Tr_{(m,n)}F^2 &=& mtr_nF^2 + ntr_mF^2  \ ,   \nonumber \\
Tr_{(m,n)}F^4 &=& mtr_nF^4 + ntr_mF^4 + 6 tr_mF^2tr_nF^2   \  .
\eeq
Utilizzando queste identit\`a possiamo calcolare il polinomio di anomalia per 
il modello con $n_1=n_2=\tilde{n}_7=\tilde{n}_8=8$:
\beq
I &=& \frac{1}{8}[(trF_1^2)^2+(trF_2^2)^2+(trF_{\bar{7}}^2)^2+
(trF_{\bar{8}}^2)^2]
+\frac{1}{16}[trF_1^2+trF_2^2+trF_{\bar{7}}^2+trF_{\bar{8}}^2]trR^2
\nonumber \\
&-& \frac{1}{4}[trF_1^2trF_{\bar{7}}^2+trF_1^2trF_{\bar{8}}^2+
trF_2^2trF_{\bar{7}}^2+trF_2^2trF_{\bar{8}}^2]
-\frac{1}{32}(trR^2)^2  \   ,
\label{pan}
\eeq
dove $F_i$ si riferisce all' i-esimo fattore del gruppo di gauge $Usp(8)^4$.
Il polinomio (\ref{pan}) pu\`o essere considerato una forma quadratica nelle 
tracce delle field strenghts; scritto in forma diagonale:
\beq
I &=& -\frac{1}{32}[trF_1^2+trF_2^2+trF_{\bar{7}}^2+trF_{\bar{8}}^2
-trR^2]^2
+\frac{3}{32}[trF_1^2+trF_2^2-trF_{\bar{7}}^2-trF_{\bar{8}}^2]^2 \nonumber \\
&+& \frac{1}{32}[trF_1^2-trF_2^2+trF_{\bar{7}}^2-trF_{\bar{8}}^2]^2
+\frac{1}{32}[trF_1^2-trF_2^2-trF_{\bar{7}}^2+trF_{\bar{8}}^2]^2  \   .
\label{paq}
\eeq
Osserviamo che il tensore antiautoduale che appartiene al multipletto del 
gravitone \`e l'unico che si accoppia alla forma di Chern-Simons 
gravitazionale.
Questa espressione mostra chiaramente che la cancellazione delle anomalie 
richiede l'azione congiunta di varie forme antisimmetriche e dato che la forma 
quadratica in (\ref{paq}) ha esattamente sei autovalori non nulli, in questo 
caso sono necessari tutte le forme presenti nello spettro.
Se consideriamo il caso generale con sedici settori di carica, 
si vede che la struttura del polinomio di anomalia \`e fissata dalla matrice
$S$:
\be
I = - \frac{1}{2}[\sum_m S_{1m}trF^2_m - 4trR^2]^2 
+\frac{1}{2}\sum_k[\sum_m S_{km}trF^2_m]^2  \  ,
\label{pag}
\ee
dove $m$ varia tra $(1)-(8)$ e $(\bar{1})-(\bar{8})$ mentre $k$ solo sui 
settori che contribuiscono ai tadpoles: 
$(5), (\bar{1}) , (\bar{5}) , (\bar{7}) , (\bar{8})$. 
Introducendo una metrica $\h_{ij}$ con segnatura $(1-n)$, dove $n$ 
\`e il numero di tensori autoduali, e definendo i polinomi $F^{2(i)}$ che 
corrispondono alle varie righe di (\ref{pag}), possiamo scrivere:
\be
I = - \frac{1}{2} \sum_{ij} \h_{ij}F^{2(i)}F^{2(j)}  \  .
\ee
Se modifichiamo la field strenght della  i-esima 2-forma
$B_{(i)}$  secondo $H_{(i)} = dB_{(i)}  + \w_{(i)}$, dove $\w_{(i)}$ \`e la 
combinazione di forme di Chern-Simons di gauge e gravitazionali corrispondenti 
alle tracce in (\ref{pag}),  
possiamo finalmente costruire un controtermine di Green-Schwarz 
generalizzato:
\be
\D L = \frac{1}{2} \sum_{ij} \h_{ij}F^{(i)}B_{(j)}  \  ,
\ee
che mostra come tutte le forma antisimmetriche che possono fluire nel canale 
trasverso contribuiscono alla cancellazione delle anomalie.

Come per i modelli in dieci dimensioni, l'accoppiamento alla forma di Chern-
Simons gravitazionale non compare nella lagrangiana effettiva di bassa energia; 
gli accoppiamenti alla forma di Chern-Simons di gauge in sei dimensioni sono 
invece termini con due derivate e possono essere studiati costruendo le 
equazioni del moto della teoria di supergravit\`a corrispondente \cite{6d3}.
L'accoppiamento della supergravit\`a $(1,0)$ in sei dimensioni ad un numero 
generico di multipletti tensoriali \`e stato considerato in \cite{6d1}, mentre 
l'accoppiamento ad un multipletto tensoriale e a materia arbitraria \`e stato 
studiato in \cite{6d2}; le teorie effettive dei nostri modelli di stringhe 
aperte richiedono l'accoppiamento ad un numero generico sia di multipletti 
tensoriali che di multipletti vettoriali. In \cite{6d3} questo studio \`e stato 
condotto all'ordine pi\`u basso nei campi fermionici. 
Per scrivere le equazioni di queste teorie \`e opportuno
considerare il coset $SO(1,n)/SO(n)$ parametrizzato dagli scalari nei 
multipletti tensoriali ed introdurre la matrice pseudoortogonale
\be
V = \pmatrix{v_0 & v_M\cr x^m_0 & x^m_M}  \  .
\ee
Gli elementi $v^r$ e $x^r_s$,  $(r = 0,...,n)$ di questa matrice soddisfano:
\beq
\tilde{v}^rv_s + \tilde{x}^r_mx^m_s &=& \d^r_s \ ,
\hspace{1cm} v_r\tilde{v}^r =1 \ , \nonumber \\
x^m_r\tilde{x}^r_n &=& \d^m_n  \ ,
\eeq
dove $\tilde{v}^r = \h^{rs}v_s$ e $\tilde{x}^r_m = - \h^{rs}x^m_s$.
Un risultato notevole 
riguarda la relazione tra gli accoppiamenti di Chern-Simons e i termini 
cinetici di gauge. Le field strenghts degli $n$ tensori 
$A^r_{\m\n}$ vanno infatti modificate 
ponendo $F^r = dA^r - c^{rz}\w_z$, dove $z$ \`e un indice per i vari gruppi di 
gauge e $c^{rz}$ \`e una matrice di costanti legata agli elementi della matrice 
$S$ della teoria conforme; si ha, inoltre, sotto trasformazioni di gauge 
$\d A^r = c^{rz}tr_z(\L dA)$. 
Quando si costruiscono le equazioni del moto per i 
vettori di gauge e per i gaugini del gruppo $G_z$, 
si vede che i termini cinetici 
$\g^{\m}D_{\m} \l$ e $D^\m F_{\m\n}$ sono moltiplicati proprio da 
$v_r c^{rz}$. Questo significa che i campi scalari vanno ristretti alla regione 
dello spazio dei moduli in cui $v_r c^{rz} > 0$, poich\`e ai bordi di questa 
regione le costanti di accoppiamento di gauge divergono.
Questo comportamento delle costanti di accoppiamento di gauge \`e stato 
collegato in \cite{6d5} alla presenza di stringhe solitoniche la cui tensione 
si annulla nel punto corrispondente dello spazio dei moduli e che sono state 
chiamate pertanto "tensionless strings".

Le variazioni dei campi e le equazioni del moto complete per queste teorie sono 
state calcolate in \cite{6df} lungo le linee suggerite in \cite{6d4}; l'intera 
costruzione si \`e rivelata un interessante laboratorio di algebra di correnti.
 
Possiamo studiare in modo del tutto analogo le anomalie della seconda classe di 
modelli considerata nel paragrafo precedente; in particolare,  
questo ci permetter\`a di apprezzare quantitativamente che
in generale alla cancellazione delle anomalie partecipano solo le forme 
antisimmetriche appartenenti a settori che contribuiscono ai tadpoles.
Consideriamo un modello con gruppo di gauge $U(4)^4$ e ipermultipletti nelle 
rappresentazioni $(4,1,4,1)$ , $(1,4,1,4)$ , $(4,1,1,4)$ e $(1,4,4,1)$.
Il polinomio di anomalia in forma diagonale \`e:
\beq
I &=& - \frac{1}{16}[\sum_itrF_i^2 + \sum_{\bar{i}}trF_{\bar{i}}^2
- \frac{1}{2}trR^2]^2
+ \frac{1}{16}[\sum_itrF_i^2 - \sum_{\bar{i}}trF_{\bar{i}}^2]^2 \nonumber \\
&+& \frac{1}{16}[trF_1^2+trF_2^2-trF_3^2-trF_4^2
+trF_{\bar{1}}^2+trF_{\bar{2}}^2-trF_{\bar{3}}^2-trF_{\bar{4}}^2]^2 
\nonumber \\
&+& \frac{1}{16}[trF_1^2-trF_2^2-trF_3^2+trF_4^2
+trF_{\bar{1}}^2-trF_{\bar{2}}^2-trF_{\bar{3}}^2+trF_{\bar{4}}^2]^2   \  .
\eeq
Da questa espressione si vede che il meccanismo di Green-Schwarz utilizza solo 
quattro degli otto tensori antisimmetrici disponibili e questo consistentemente 
con il fatto che gli unici caratteri massless che in questo modello possono 
fluire nell'anello trasverso sono $\chi_1$ , $\chi_5$ , 
$\bar{\chi}_1$  e $\bar{\chi}_6$.

\subsection{Compattificazioni toroidali e discendenti dei modelli di Gepner}

Compattificazioni toroidali della Tipo I sono state descritte in
\cite{od4} e, pi\`u recentemente in \cite{vettor}. La funzione di partizione 
della teoria di Tipo IIB su $T^d$ si scrive, limitandoci alla parte interna
\be
T = \frac{1}{\h(\t)\h(\bar{\t})} \sum_{\bf{m}\bf{n}} 
e^{i \p \t p^2_L} e^{- i \p \bar{\t} p^2_R}  \   ,
\ee 
dove
\be
p^i_{L,R} = m_a \tilde{e}^{ai} \pm \frac{1}{2}n^a e_a^i 
- \frac{1}{2}B_{ab}\tilde{e}^{ai}n^b  \  ,
\ee
con $e^i_a e^i_b = g_{ab}$. Per un $B_{ab}$ generico non si ha simmetria tra 
modi destri e sinistri; perch\`e questa sussista \`e necessario che, per ogni 
coppia di vettori di interi $\bf{m}$, $\bf{n}$, esistano due vettori
$\bf{m^{'}}$, $\bf{n^{'}}$, tali che
\be
m_a + \frac{1}{2}( - B_{ac} + g_{ac})n^c = m^{'}_a 
- \frac{1}{2} ( B_{ac} + g_{ac})n^{'c}  \ .
\label{bquan}
\ee
La (\ref{bquan}) \`e in effetti una condizione di quantizzazione per $B_{ab}$ e 
quindi gli unici moduli chiusi continui compatibili con la costruzione del 
discendente aperto sono gli elementi di $g_{ab}$ consistentemente con il fatto 
che la bottiglia di Klein rimuove $B_{ab}$ dallo spettro. Lo stesso risultato 
si ottiene se, partendo dal reticolo con $g=1$ e $B=0$ si determinano le 
trasformazioni in $SO(d,d)/SO(d) \otimes SO(d)$ che preservano l'identit\`a 
dei reticoli destro e sinistro; si pu\`o verificare che si tratta di un 
sottogruppo con $\frac{d(d+1)}{2}$ parametri. La possibilit\`a di introdurre un 
campo $B_{ab}$ quantizzato \`e essenziale per compattificare su reticoli di 
algebre di Lie, quando sia $g$ che $B$ sono legati alla matrice di 
Cartan; inoltre se il rango di $B$ \`e $r$, la dimensione del gruppo di 
Chan-Paton viene ridotta da un fattore $2^{r/2}$. Infine valori di aspettazione 
continui per i campi di gauge interni, permettono di legare modelli che 
differiscono per la presenza di linee di Wilson discrete, descritti nel 
paragrafo precedente. \`E anche possibile interpolare in modo continuo tra 
gruppi di gauge ortogonali e simplettici.

Modelli chirali in quattro dimensioni sono stati costruiti come discendenti 
dello $Z$ orbifold \cite{4d}. Un'altra classe di modelli naturale da studiare 
\`e rappresentata dai discendenti dei modelli di Gepner ed \`e particolarmente 
interessante in quanto coinvolge teorie interne interagenti
\cite{og}; le serie $(k=2)^4$ con simmetria $Z_2$ e $Z_4$, costruite 
ancora con teorie libere, corrispondono alle quattro classi di modelli 
discusse precedentemente.
Numerosi altri modelli sono stati costruiti \cite{orienti} e ne sono state 
messe in luce sia le 
relazioni con altre compattificazioni perturbative di teorie di stringhe, sia
le relazioni con compattificazioni della M teoria ( ad esempio su
$K3 \times S^1/Z_2$ ) e della F teoria.

\section{Stringhe aperte in un campo elettromagnetico uniforme}
\markboth{}{}

Come le stringhe chiuse sono l'analogo della gravit\`a, cos\`\i \, le stringhe 
aperte sono l'analogo dei campi di Yang-Mills. \`E pertanto
interessante studiarne la dinamica in campi elettrici e magnetici 
di background \cite{ob1,ob2}.
L'azione che descrive una stringa aperta in un campo di background, con
una metrica euclidea sia per il worldsheet che per lo spaziotempo, \`e 
\be
S = \frac{1}{4 \p \a^{'}} \int d \s d \t \part_a X_\m \part^a X^\m
+ i \int d \t A_\m \part_\t X^\m  \ .
\ee
Utilizzando il metodo del campo di background con $ X^\m = \bar{X}^\m + \x^\m $
otteniamo
\beq
S( \bar{X}+ \x ) &=& S( \bar{X}) + \frac{1}{4 \p \a^{'}} \int d \s d \t 
\part^a \x^\m \part_a \x_\m  
+ i \int d \t ( \frac{1}{2} \btd_\n F_{\m\l} \x^\n \x^\l \part_\t \bar{X}^\m \nonumber \\
&+& \frac{1}{2} F_{\m\n} \x^\n \part_t \x^\m + \frac{1}{3} \btd_v F_{\m\l}
\x^\n \x^\l \part_\t \x^\m+...)  \  .
\eeq
Calcoliamo il controtermine $\Delta S = i \int d \t \G_\m \part_\t \bar{X}^\m $
e ricaviamo le equazioni del moto da 
$ \b_\m = \L \frac{\part \G_\m}{\part \L}$. Se utilizziamo il propagatore 
definito nel semipiano complesso superiore che soddisfa
\be
\frac{1}{2 \p \a^{'}} \part \part  G(z,z^{'}) = - \d (z-z^{'}) \ , \hspace{1cm}
\part_\s G(z,z^{'})|_{ \s = 0} = 0 \ ,
\ee
ed \`e dato da 
\be 
G(z,z^{'}) = - \a^{'}(ln|z-z^{'}|+ln|z-\bar{z}^{'}|) \  ,
\ee
dobbiamo sommare tutti i grafici ad un loop con un $ \part_\t \bar{X}$ 
esterno e 
tutte le possibili inserzioni di $ F_{\m\n} \x^\n \part_\t \x^\m$. 
Alternativamente possiamo in questo caso calcolare il propagatore esatto che 
soddisfa le condizioni al bordo
\be
\part_\s G_{\m\n}(z,z^{'})+iF_\m{}^\l \part_\t G_{\l\n}(z,z^{'})|_{ \s = 0} = 0 \ ,
\ee
ed \`e dato da
\beq
G_{\m\n}(z,z^{'}) &=& - \a^{'}[ \d_{\m\n}ln|z-z^{'}| + 
\frac{1}{2}(\frac{1-F}{1+F})_{\m\n} ln(z-\bar{z}^{'}) \nonumber \\
&+& \frac{1}{2}(\frac{1+F}{1-F})_{\m\n} ln(\bar{z}-z^{'})]  \  .
\eeq
\`E quindi necessario calcolare solo un grafico, con il risultato
\be
\Delta S = - \frac{i}{2 \p \a^{'}} \int d \t \frac{1}{2} \btd_\n F_{\m\l} 
\part_\t \bar{X}^\m G^{\n\l}( \t , \t^{'})|_{ \t \rar \t^{'} }   \ .
\ee
Otteniamo quindi le equazioni del moto
\be
\btd^\n F^\l{}_\m (1 - F^2)^{-1}_{\l\n} = 0  \  .
\ee
Queste equazioni non sono ottenibili dalla variazione di un funzionale d'azione 
ma \`e possibile scrivere un funzionale che ha come equazioni del moto 
$ \h_{\m\n}(F) \b^\n $, equivalenti a $\b =0 $ in quanto $\h(F)_{\m\n}$ \`e 
un tensore invertibile. Il risultato \`e l'azione di Born-Infeld
\be
L_{BI} = \sqrt{det(1+F)} \ ,
\ee
con equazioni del moto $ \sqrt{det(1+F)}(F^2)^{-1}_{\m\n}\b^\n=0$. Vale la 
pena osservare che la funzione $\b$ per i 
campi di stringa chiusa non viene alterata all'ordine pi\`u basso dalla presenza 
di campi di gauge sul 
bordo. Inoltre, l'anomalia conforme \`e un effetto locale e quindi la funzione 
$ \b$ non dovrebbe dipendere dalla topologia del worldsheet. Questo pu\`o 
essere verificato in modo esplicito ad un loop calcolando il propagatore esatto 
sull'anello. 

In un campo di gauge di background \`e facile calcolare lo spettro 
della teoria; usando ora una metrica minkowskiana
\be
S = - \frac{1}{4 \p \a^{'}} \int d \s d \t \part^a X^\m \part_a X_\m
- \int d \t [ q_1 A_\m \dot{X}^\m(0) + q_2 A_\m \dot{X}^\m( \p)]  \  .
\ee
L'equazione del moto \`e l'usuale equazione d'onda, con condizioni al bordo
\beq
X^{' \m} &=& 2 \p \a^{'} q_1 F^\m{}_\n \dot{X}^\n \ \hspace{1.3cm}  {\rm per } \
\s =0  \  ,  \nonumber \\
X^{' \m} &=& -2 \p \a^{'} q_2 F^\m{}_\n \dot{X}^\n 
\ \hspace{1cm}  {\rm per } \ \s = \p  \ .
\eeq
Consideriamo un campo con componenti non nulle solo nelle direzioni $1$ e $2$ :
$F_{\m\n} = \pmatrix{0 & f\cr -f & 0}$ , $ \m , \n =1 , 2$;
definendo $X^{\pm} = \frac{1}{\sqrt{2}}(X_1 \pm iX_2)$  riscriviamo le 
condizioni al bordo
\beq
X^{'}_+ &=& -i \a \dot{X}_+ \ , \hspace{1cm} X^{'}_- = i \a \dot{X}_-  \ , \hspace{1cm}
\s =0   \ ,  \nonumber \\
X^{'}_+ &=& i \b \dot{X}_+  \ , \hspace{1cm} X^{'}_- = -i \b \dot{X}_-  \ , \hspace{1cm}
\s = \p  \ , 
\eeq
dove $ \a = 2 \p \a^{'} q_1f$ e $ \b = 2 \p \a^{'} q_2f$. Quando 
$ \a + \b \ne 0$ abbiamo le  autofunzioni 
\be
\psi_n( \s , \t) = \frac{1}{\sqrt{|n- \e|}}cos[(n- \e) \s + \g]
e^{-i(n- \e) \t }  \ ,
\ee
dove $ \e = \frac{ \g + \g^{'}}{ \p} $ e $ \g = arctg( \a) $ , 
$ \g^{'} = arctg( \b) $. Possiamo allora scrivere
\be
X_+ = x_+ +i \left ( \sum_{n \ge 1} a_n \psi_n - \sum_{n \ge 0} b_n^{\dag} \psi_{-n} \right )   \ ,
\ee
\be
X_- = x_- +i \left ( \sum_{n \ge 0} b_n \bar{\psi}_{-n} - 
\sum_{n \ge 1} a_n^{\dag} \bar{ \psi}_{n} \right ) \  ,
\ee
dove le espansioni non contengono termini lineari in $ \s , \t$. 
Possiamo quantizzare in modo canonico; il momento coniugato a $X_+$ \`e
$ P_+ = \frac{\dot{X}_-}{2 \p \a^{'}}-A_-(q_1 \d(\s)+q_2\d(\p-\s))$ che nel 
gauge $A_\m =  - \frac{1}{2}F_{\m\n}X^{\n}$   diviene 
$2 \p \a^{'} P_+ = \dot{X}_- + \frac{i}{2}X_-(\a \d(\s)+\b \d(\p-\s))$. Per gli 
oscillatori si ottengono quindi le relazioni di commutazione
\be
[a_n,a_m]=[a_n^{\dag},a_m^{\dag}]=[b_n,b_m]=
[b_n^{\dag},b_m^{\dag}]=0  \ ,
\ee
\be
[a_n,a_m^{\dag}] = 2 \a^{'} \d_{n,m} \ , \hspace{1cm}  n,m \ge 0 \ , 
\ee
\be
[b_n,b_m^{\dag}] = 2 \a^{'} \d_{n,m} \ , \hspace{1cm} n,m \ge 1 \ , 
\ee
\be
[x_+,x_-] = \frac{ \p \a^{'}}{\a+\b} \  .
\ee
Si possono costruire nel modo solito i generatori di Virasoro, che realizzano 
l'algebra
\be
[L_n,L_m] = (n-m)L_{n+m} + \d_{n+m,0}[\frac{1}{12}n(n^2-1)+n \e(1-\e)] \  .
\ee
Il termine centrale pu\`o essere ricondotto a quello usuale ridefinendo 
$L_0 \mapsto L_0 + \frac{\e}{2}(1-\e)$, che equivale ad alterare la costante 
di ordinamento normale in modo tale che 
$a=1 \mapsto a = 1 - \frac{ \e}{2}(1- \e)$. 
Il campo di background altera quindi la frequenza degli 
oscillatori $a$ di $- \e$, degli oscillatori $b$ di $ \e$ e modifica
l'energia di vuoto e 
la struttura degli zero modi della teoria. Non abbiamo pi\`u un operatore di 
momento totale, ma possiamo scegliere gli stati come autostati di $x_+$; dato 
che l'hamiltoniana non dipende da $x_\pm$, 
abbiamo una degenerazione infinita se 
le coordiante $X^1,X^2$ non sono compatte e finita se sono 
compatte. La situazione 
\`e analoga a quella di una particella in un campo magnetico: esistono livelli 
di Landau ugualmente spaziati che  gli operatori $b_0^{\dag} , b_0 $ collegano tra loro 
creando e distruggendo quanti di frequenza $ \e$. Infatti
\be
L_0 = \sum_{m=1} \left (|m- \e| a^{\dag}_ma_m+|m+ \e|b^{\dag}_mb_m \right ) 
+ \e b^{\dag}_0b_0  \  .
\ee
dove i primi due termini controllano il livello delle eccitazioni di stringa 
mentre l'ultimo controlla i livelli di Landau. 
Nel caso compatto, 
$X^{1,2} \sim X^{1,2} + L$ la degenerazione \`e
$n = \frac{ \a^{'}}{ \p}(q_1+q_2)fL^2$. 

Se $ \a + \b = 0$ cambia la struttura 
degli zero modi e
\be
X_+ = \frac{x_+ + p_-( \t - i \a( \s - \p /2))}{\sqrt{1+\a^2}}+
i\sum_{n=1}^{\infty}[a_n\psi_n-b_n^{\dag}\psi_{-n}] \ .
\ee
Si pu\`o verificare che lo spettro non viene alterato in questo caso dalla 
presenza del campo magnetico, mentre la funzione di partizione aquista una 
dipendenza da $F$ : $Z_F = det(1+F)Z$, con $Z$ l'usuale funzione di partizione 
per la stringa bosonica. \`E facile modificare questi risultati per studiare il 
caso di un campo elettrico \cite{ob2}; 
con $F_{01} = E$ ed utilizzando le coordinate
$X^{\pm} = \frac{1}{\sqrt{2}}(X^0 \pm X^1)$ abbiamo
\be
\part_\s X^{\pm} = \mp \b_1 \part_\t X^{\pm} \s=0   \  , \hspace{1cm}
\part_\s X^{\pm} = \pm \b_2 \part_\t X^{\pm} \s= \p  \ ,
\ee
dove $\b_{1,2} = 2 \p \ap q_{1,2}E$; di conseguenza
$X^{\pm}= x^{\pm} + i a_0^{\pm} \f_0^{\pm} + 
i\sum_{n=1}^{\infty}(a_n^{\pm} \f_n^{\pm} - h.c.)$ con
\be
\f_n^{\pm} = \frac{1}{n\pm i \e}e^{-i(n-i \e) \t}
cos[(n-i \e) \s \mp i arcth \b_1]  \  , 
\ee
dove 
\be 
\e = \frac{arcth \b_1 + arcth \b_2}{ \p} \ ,
\ee
e
\be
[a^+_n,a^{-*}_n]=[a_n^-,a_n^{+*}]=-1 \ , \hspace{1cm}
[x^+,x^-]= - \frac{i \p}{\b_1 + \b_2} \ , \hspace{1cm}
a_0^{\pm *}= \pm i a_0^{\pm}  \  .
\ee
\`E interessante calcolare la funzione di partizione, la cui parte immaginaria 
d\`a l'analogo in teoria di stringhe dell'effetto Schwinger in teoria di campi.
Per la superstringa di Tipo I l'azione
\beq
S &=& - \frac{1}{4 \p \a^{'}} \int ( \d^aX^\m \part_a X_\m - 
i \bar{\psi}^\m \r^a \part_a X_\m) + 
\frac{q_1}{2} \int d \t F_{\m\n}[X^\n \part_\t X^\m - \frac{i}{2} 
\bar{\psi}^\n \r^0 \psi^\m]|_{\s=0}  \nonumber \\
&+& \frac{q_2}{2} \int d \t F_{\m\n}
[X^\n \part_\t X^\m - \frac{i}{2} \bar{\psi}^\n \r^0 \psi^\m]|_{\s= \p} \  ,
\eeq 
contiene anche fermioni con condizioni al bordo
\be
\psi_R^\m - \psi_L^\m = \p q_1 F^\m {}_\n( \psi_R^\n + 
\psi_L^\n)|_{\s=0}  \  ,
\ee
\be
\psi_R^\m + (-1)^a \psi_L^\m 
= -\p q_2 F^\m {}_\n( \psi_R^\n - (-1)^a \psi_L^\n)|_{\s= \p}  \ ,
\ee
con $a = 0 , 1 $ a seconda che il settore sia NS o R. Per un campo nella 
direzione $X^1$, dividendo i fermioni in $ \psi^{\pm} $ e $ \psi $
\beq
(1 \mp \b_1) \psi_R^{\pm} &=& (1 \pm \b_1) \psi_L^{\pm} \ , \hspace{1cm}
  \psi_R = \psi_L \ , \hspace{0.5cm}  \s = 0   \ ,\nonumber \\
(1 \pm \b_2) \psi_R^{\pm} &=& - (-1)^a(1 \mp \b_2) \psi_L^{\pm}  \ , \hspace{1cm}
  \psi_R = -(-1)^a \psi_L  \hspace{0.5cm}     \s = \p \  ,
\eeq
e cos\`i $ \psi_{R,L}^{\pm} = \sum_{n=-\infty}^{n= \infty} d_n^{\pm}
\chi^{\pm}_{n R,L}(\s,\t)$ dove
\be
\chi_{n,R}^{\pm} = \frac{1}{\sqrt{2}} e^{-i(n\pm i\e)(\t - \s) \pm arcth \b_1}
 \  , \hspace{1cm}  
\chi_{n,L}^{\pm}= \frac{1}{\sqrt{2}} e^{-i(n\pm i\e)(\t + \s) \mp arcth \b_1} .
\ee
Le relazioni di commutazione sono quelle usuali,
$ \{d^\m_n,d^\n_m \} = \h^{\m\n} \d_{m+n} $ , $d^{\m*}_n = d^{\m}_{-n}$, ed $n$ 
\`e intero o semintero a seconda che il settore sia R o NS. L'ampiezza di 
vuoto \`e data dalla somma del contributo di superfici orientabili e non 
orientabili :
\be
-iFV = T + K + A + M \  .
\ee
$T$ e $K$ danno il contributo chiuso e non sono modificate dalla presenza del 
campo elettrico. Inoltre, fissate le cariche agli estremi della stringa aperta,
\beq
& & A(q_1,q_2) = - \frac{1}{2} \int_0^\infty \frac{dt}{t}
\sum_{a,b=0,1} C \pmatrix{a\cr b}tr(-1)^{bF}e^{- \p t(L_0-1)}   \\
& & =-i\frac{V}{2}\int_0^\infty\frac{dt}{t}(2 \p^2 t)^{-5} 
\h (\frac{it}{2})^{-8} f_A(t,\b_1,\b_2) 
\sum_{a,b=0,1}C \pmatrix{a\cr b}
[\frac{ \th \pmatrix{a\cr b}(0| \frac{it}{2})}{\h (\frac{it}{2})}]^4
g \pmatrix{a\cr b}(t,\e)  \   ,   \nonumber
\eeq
con $C \pmatrix{0\cr 0} =-C \pmatrix{0\cr 1} =
-C \pmatrix{1\cr 0} = \pm C \pmatrix{1\cr 1} = \frac{1}{2}$
e dove 
\be
\frac{\th \pmatrix{a\cr b} (0|\frac{it}{2})}{\h (\frac{it}{2})}=
q^{\frac{a^2}{8}-\frac{1}{24}} \prod_{n=1}(1+q^{n+\frac{a-1}{2}}e^{i \p b})
\prod_{n=1}(1+q^{n-\frac{a+1}{2}}e^{i \p b})  \   ,
\ee 
e $q = e^{- \p t}$. L'effetto del campo elettrico di background \`e racchiuso
nella funzione
\be
f_A= \frac{(\b_1+\b_2)t e^{- \frac{ \p t \e^2}{2}}}{2 sen( \frac{ \p t \e}{2})}
\prod_{n=1} \frac{(1-e^{-\p tn})^2}
{(1-e^{- \p t(n+i \e)})(1-e^{- \p t(n-i \e)})}  \   ,
\ee
che d\`a conto del contributo degli oscillatori bosonici $a_n^\pm$, dei 
ghosts, della variazione dell'energia di vuoto e della struttura degli zero 
modi e nelle funzioni 
\be
g \pmatrix{a\cr b}(t, \e) = \frac{\th \pmatrix{a-2i \e\cr b} (0|\frac{it}{2})}
{\th \pmatrix{a\cr b} (0|\frac{it}{2})}    \ ,
\ee
che deformano la proiezione GSO.
Dato che $g \pmatrix{a\cr b} $ non ha poli sull'asse $t$ positivo, il tasso di 
produzione di coppie si ottiene dal teorema di Cauchy sommando sui poli di $f_A$ in 
$t= \frac{2k}{|\e |}$, e il risultato\`e
\be
w= \frac{1}{2(2\p)^{d-1}}\sum_{Stati} \frac{\b_1 + \b_2}{\p}
\sum_{k=1}(-1)^{(k+1)(a+1)}(\frac{| \e |}{k})^{\frac{d}{2}} 
e^{- \frac{ \p k m_S^2}{| \e |}}  \   .
\ee

\section{D-brane}
\markboth{}{}

Abbiamo osservato precedentemente che gli stati perturbativi di stringa non 
sono carichi rispetto ai campi del settore R-R. Queste 
p-forme antisimmetriche possono tuttavia 
accoppiarsi in modo naturale ad eccitazioni
estese in $p$ dimensioni spaziali (p-brane) a mezzo di termini di Wess-Zumino
\be
I = \r_p \int d^{p+1} \z \hat{C}^{p+1}  \   ,
\ee
dove $Y^{\m}(\z^{\a})$ sono campi che descrivono l'immersione della brana nello 
spaziotempo e  $\hat{C}^{p+1} = C_{\m_1 ... \m_{p+1}}(Y) \part_0 Y^{\m_1}...
\part_p Y^{\m_{p+1}}$ \`e il pullback della p-forma sul volume della p-brana.

Vedremo in maggior dettaglio nel prossimo capitolo che le teorie effettive di 
supergravit\`a ammettono soluzioni descriventi p-brane; quando queste 
sono cariche rispetto a campi del settore NS-NS se ne conosce anche una 
descrizione in termini di teoria conforme \cite{br1,br2}. 
Per brane cariche rispetto a 
campi di R-R, comunque presenti nel limite di bassa energia, non si \`e avuta 
una descrizione in termini di teoria di stringhe fino a quando Polchinski
\cite{db1} ha intuito che esse possono essere descritte in modo semplice e 
trasparente utilizzando 
teorie conformi su superfici con bordi che includono campi con 
condizioni di Dirichlet. In altri termini, brane cariche rispetto a campi $R-R$ (D-brane) sono 
regioni spaziotemporali
sulle quali possono terminare stringhe aperte.

Mentre condizioni al bordo di Neumann rispettano l'invarianza di Poincar\'e, 
condizioni al bordo di Dirichlet descrivono quindi difetti spaziotemporali \cite{db2};
come abbiamo visto esse emergono in modo naturale studiando compattificazioni 
della Tipo I \cite{od2} e sono state considerate anche per generare un 
comportamento partonico in ampiezze di stringa \cite{db2a}. L'importanza di 
\cite{db1} sta nell'avere sottolineato il ruolo che queste condizioni al bordo 
giocano nel descrivere eccitazioni non perturbative.

Una Dp-brana ha $p+1$ coordinate con condizioni al bordo di Neumann, 
$X^{\a}$  , $\a = 0,...,p$
( le coordinate tangenti al worldvolume ) e $d-p-1$ coordinate con 
condizioni al bordo di Dirichlet
$X^{m}$  , $m = p+1,...,d$ (le coordinate normali al worldvolume ).
In coordinate complesse $z = e^{\t+i \s}$, queste condizioni al bordo 
si scrivono:
\beq
\part X^{\a} = \bar{\part} X^{\a} |_{Imz=0}  \   , & &   
\part X^{m} = - \bar{\part} X^{\m} |_{Imz=0}  \   ,  \nonumber \\
\psi^{\a} = \e \bar{\psi}^{\a}|_{Imz=0}   \   ,  &  &
\psi^{m} = -\e \bar{\psi}^{m}|_{Imz=0}    \    ,
\eeq 
dove $\e =1$ nel settore R e $\e = sign(Imz)$ nel settore NS. Da queste 
condizioni seguono quelle per i campi di spin:
\be
S = \P_p \bar{S}|_{Imz=0}  \   ,
\label{spin}
\ee
con $\P_p = (i\G_{11}\G^{p+1})(i\G_{11}\G^{p+2})...(i\G_{11}\G^{9})$; nella 
Tipo I inoltre la (\ref{spin}) deve essere invariante rispetto allo scambio di 
$S$ e $\bar{S}$, dal che segue l'ulteriore condizione $\P_p^2 =1$. 
Da (\ref{spin}) si ricava, inoltre, che la IIA contiene brane con 
$p=0,2,4,6,8$, la IIB brane con $p=-1,1,3,5,7,9$ e la Tipo I brane con $p=1,5,9$, 
in perfetto accordo con i campi R-R presenti nello spettro di ciascuna di 
queste teorie.
La presenza di una D9 corrisponde a condizioni di Neumann per tutte le 
coordinate ed \`e identificata con il settore aperto della Tipo I, 
in quanto la consistenza della teoria 
richiede l'inclusione di 16 brane e la proiezione con $\W$, la parit\`a sul 
worldsheet.
La D8 \`e un muro di dominio che separa regioni con valori distinti di $H^0$.

Vedremo nel prossimo capitolo il ruolo fondamentale che queste D-brane hanno avuto nello 
studio delle dualit\`a; in particolare, le D0 brane suggeriscono l'esistenza di 
una undicesima dimensione non perturbativa e della
M-teoria \cite{tw1,du2}, le D1 brane sono cruciali per la dualit\`a 
tra HO e Tipo I \cite{du2,du4} e la autodualit\`a della IIB \cite{pq2}, 
le D-1 brane per lo studio delle correzioni non perturbative alla IIB \cite{dist}, 
le D7 brane per le compattificazioni della F-teoria \cite{Fth,For}, le D2 brane per la 
dualit\`a $IIA/K3$ ed $Het/T^4$ \cite{du1} e le D3 brane per la transizione di 
conifold \cite{con,con2}.

Per verificare che le D-brane sono cariche rispetto ai campi di Ramond e 
sono stati BPS \footnotemark \footnotetext{Descriveremo in maggior dettaglio 
gli stati BPS nel prossimo capitolo}, 
determiniamo la loro tensione e la loro carica.
Questo pu\`o essere fatto calcolando funzioni ad un punto sul disco, prestando
attenzione all'appropriata normalizzazione degli operatori di vertice, oppure
calcolando l'energia di interazione tra due brane statiche. Questo secondo 
metodo \`e quello seguito in \cite{db1} ed esteso a D-brane dinamiche in
\cite{db3}.
Dobbiamo confrontare il risultato che si ottiene dalla teoria di 
campo effettiva con quello che si ottiene dal calcolo di 
stringa, in modo analogo a quanto fatto nel primo capitolo per determinare la 
relazione tra $g$ ed $\ap$ e i parametri della teoria di campo. 
Cominciamo con lo srivere l'azione spaziotemporale (supergravit\`a) 
e l'azione che descrive la brana come un difetto nello spaziotempo, 
caratterizzata in prima approssimazione dalla sua tensione $T_p$ e dal 
suo accoppiamento $ \r_p$ al corrispondente campo di Ramond 
\be
S_{st} = - \frac{1}{2 \k^2} \int d^{10}x \sqrt{-g} \left [R + \frac{1}{2}( \btd \f )^2
+ \frac{1}{12}e^{-\f}(dB)^2 + \sum_p \frac{1}{2(p+2)!}
e^{\frac{3-p}{2} \f}(dC)^2_{p+1} \right ]  \   ,
\ee
\be
S_{vol} = T_p \int d^{p+1} \z e^{\frac{p-3}{4} \f} \sqrt{det \hat{g}}
+ \r_p \int d^{p+1} \z \hat{C}_{p+1}  \   .
\ee
Siamo nel frame di Einstein $( g_{\m\n} = e^{-\frac{\f}{2}}G^{stringa}_{\m\n})$
e in $S_{st}$ $p$ \`e pari per la IIA e dispari per la IIB. 
Localizziamo la prima brana in un punto $x$ e la seconda in un punto 
$\tilde{x}$ con $|x-\tilde{x}| = r$.
L'energia di interazione ad albero \`e
\be
E(r)T = -2 \h^2 \int d^{10}x \int d^{10} \tilde{x}
[j_{\f} \D \tilde{j}_{\f} - j_C \D \tilde{j}_C + 
T_{\m\n} \D^{\m\n\r\s}T_{\r\s}]  \  .
\ee
Il propagatore dello scalare $\D$ e quello del gravitone 
$\D^{\m\n\r\s}$ si estraggono
dalla prima azione e le sorgenti dalla seconda; il risultato \`e:
\be
E(r) = 2V_p \k^2 [ \r_p^2 - T^2_p] \Delta_{9-p}(r)  \   ,
\label{energ}
\ee
dove $\Delta_{9-p}(r)$ \`e il propagatore scalare euclideo in $9-p$ dimensioni.
La (\ref{energ}) 
mostra che la forza complessiva \`e il risultato dell'attrazione causata da dilatone 
e gravitone e della repulsione dovuta al campo di Ramond (le due brane hanno la 
stessa carica). 

Dal punto di vista della teoria di stringhe, l'interazione tra le due D-brane 
\`e descritta dallo scambio di stringhe chiuse; il diagramma ad albero di 
stringhe chiuse,
come sappiamo, pu\`o essere interpretato come un diagramma ad un loop di 
stringhe aperte; questo equivale ad interpretare
l'interazione tra D-brane in modo 
analogo all'effetto Casimir: come due lastre superconduttrici,
le brane 
interagiscono perch\`e alterano le fluttuazioni quantistiche del vuoto.
Il contributo all'energia di vuoto di stringhe che hanno entrambi gli estremi
sulla stessa brana \`e indipendente da $r$ e pu\`o essere omesso; per stringhe 
tese tra le due brane si ha ( includendo un fattore $2$ che tiene conto delle 
due possibili orientazioni della stringa ) :
\beq
E(r) &=& -2 \frac{V_p}{2} \int \frac{d^{p+1}k}{(2 \p)^{p+1}}
\int_0^{\infty} \frac{dt}{t} Str e^{- \p t (k^2 + M^2)/2}  \\
&=& -2 \frac{V_p}{2} \int_0^{ \infty} \frac{dt}{t} (2 \p^2 t)^{(p+1)/2}
e^{- \frac{r^2 t}{2 \p}} \frac{1}{2 \h^{12}(\frac{it}{2})}
\left [ \th_3^4(0|\frac{it}{2}) - \th_2^4(0|\frac{it}{2}) 
- \th_4^4(0|\frac{it}{2}) \right ]  \   , \nonumber 
\label{enstr}
\eeq
dove $\ap = 1/2$ e si \`e tenuto conto che $M^2 = (\frac{r}{\p})^2 + 2N$. Come 
conseguenza della supersimmetria, questa energia si annulla e quindi da 
(\ref{energ}) ricaviamo che $ \r_p = T_p $. Per ottenere il valore preciso di 
$T_p$, separiamo in (\ref{enstr}) il contributo del settore di Ramond e quello 
del settore NS ed espandiamo l'integrando attorno a $t = 0$; il risultato \`e:
\be
E(r) = V_p (1-1) 2 \p ( 4 \p^2 \ap)^{3-p} \Delta_{9-p}(r) + 
o \left (e^{-r/ \sqrt{\ap}} \right )  \    ,
\ee
e quindi
\be
T^2_p = \r^2_p = \frac{\p}{\k^2}  ( 4 \p^2 \ap)^{3-p}   \  .
\label{ten}
\ee
Nella Tipo I le stringhe non sono orientate e quindi la tensione di una D-brana 
include un ulteriore fattore $\frac{1}{\sqrt{2}}$.

Possiamo mostrare la consistenza di questo risultato con l'analogo della 
condizione di quantizzazione di Dirac per oggetti estesi \cite{br4} e con la 
T-dualit\`a.
Se abbiamo una p-brana nell'origine, integrando il duale della sua 
field strenght sullo spazio trasverso si ottiene :
\be
\int_{S^{8-p}} *H^{p+2} = 2 \k^2 \r_p  \   .
\ee
Ora $*H^{p+2} = H^{8-p} \sim d C^{7-p}$ con $C^{7-p}$ non globalmente definita 
dato che la p-brana \`e una sorgente nell'identit\`a di Bianchi per 
$H^{8-p} $; si pu\`o definire un potenziale regolare su $S^{8-p}$ tranne che 
per una singolarit\`a su una ipersfera $S^{7-p} \subset S^{8-p}$. 
Una $6-p$ brana trasportata attorno a $S^{7-p}$ acquista una fase
$ \f = \r_{6-p} \int_{S^{7-p}} C^{7-p} = \r_{6-p} \int_{S^{8-p}} H^{8-p}$ e 
perch\`e la singolarit\`a risulti non osservabile deve valere la condizione di 
quantizzazione:
\be
\r_p \r_{6-p} = \frac{\p n}{\k^2}  \ , \hspace{1cm}    n \in Z   \  .
\label{quant}
\ee
Le D-brane soddisfano la (\ref{quant}) con $n=1$, e portano quindi la minima 
carica permessa dalla condizione di quantizzazione. Nella Tipo I la presenza 
del fattore $1/ \sqrt{2}$ non viola la (\ref{quant}), in quanto le eccitazioni 
dinamiche corrispondono non a singole 5-brane ma a coppie di 5-brane. La proiezione
$\W$ elimina infatti le coordinate collettive di una 
delle due brane,
come \`e stato arguito per consistenza con la dualit\`a Tipo I-HO in
\cite{hod3} e verificato poi in \cite{or2}.

Consideriamo ora la T dualit\`a. Abbiamo visto nel capitolo precedente che la T 
dualit\`a \`e un'operazione di parit\`a sul solo settore antiolomorfo;  
compattificando la teoria su di un cerchio di raggio $R_9$, si ha:
\be
\bar{\part} X^{9'} = - \bar{\part} X^9  \  , \hspace{1cm} H^{'} = iH \G^9 \G_{11}  \   ,
\ee
dove $H$ \`e un bispinore. Da queste equazioni segue, per le forme 
antisimmetriche :
\be
H^{'}_{\m_1...\m_n} = H_{9 \m_1...\m_n}  \ ,  \hspace{1cm}
H^{'}_{9 \m_1...\m_n} = - H_{\m_1...\m_n}   \  ,
\ee
per $\m_i \neq 9$. Forme di grado pari e dispari vengono quindi scambiate dalla 
T dualit\`a e lo stesso accade alle D-brane. 
Consideriamo infatti una 
$(p+1)$-brana avvolta attorno alla nona dimensione;  
le stringhe aperte che vivono sulla brana possono muoversi lungo la nona 
dimensione ma non possono avvolgersi attorno ad essa. Le condizioni di Neumann
$\part X^9 = \bar{\partial} X^9$ per $z$ reale impongono 
$\a_k = \tilde{\a}_k$ e $m_9=0$. 
Una T dualit\`a scambia condizioni di 
Neumann con condizioni
di Dirichlet e la (p+1)-brana avvolta diviene, nella teoria duale, 
una 
p-brana localizzata lungo il cerchio; infatti ora abbiamo
$\a_k = - \tilde{\a}_k$ e $n_9=0$,
e le stringhe aperte non possono pi\`u 
muoversi liberamente ma possono avvolgersi attorno al cerchio. 
Perch\`e la tensione della brana
misurata in nove dimensioni resti invariante rispetto a questa trasformazione 
si deve avere:
\be
2 \p R_9 T_{p+1} = T^{'}_p \ .
\ee
Questa relazione \`e soddisfatta dalla (\ref{ten}).

Possiamo studiare la dinamica delle D-brane utilizzando l'equivalenza tramite T 
dualit\`a tra campi di background dal punto di vista delle stringhe aperte e 
coordinate dal punto di vista della brana.
Consideriamo una D0-brana; l'azione sul disco \`e
\be
S = \frac{1}{4 \p \a^{'}} \int d^2z \part^a X^\m \part_a X_\m +
\frac{1}{2 \p \a^{'}} \int d \t Y^j(X^0)\part_\s X_j \ ,
\ee
e nelle coordinate duali
\be
S = \frac{1}{4 \p \a^{'}} \int d^2z \part^a X^\m \part_a X_\m +
ie \int d \t A^j(X^0)\part_\t X_j  \ ,
\ee
perch\`e $2 \p \a^{'} e A^j = Y^j $ e quindi $2 \p \a^{'} e E^j = v^j $. 
Otteniamo un primo risultato notevole: l'azione di Born-Infeld che per la 
stringa aperta rappresenta il risultato della somma di tutte le correzioni in 
$ \a^{'}$, $L_{BI} \sim \sqrt{1-(2 \p \a^{'}eE)^2} $, da questo punto di vista 
\`e semplicemente conseguenza del fatto che i solitoni della teoria di tipo II 
devono comportarsi come particelle relativistiche ed avere quindi una 
lagrangiana $ L \sim \sqrt{1-v^2}$. L'azione di Born-Infeld 
riceve tuttavia ulteriori correzioni quando il campo elettromagnetico 
varia spazialmente ( anche se, nel caso supersimmetrico, i termini dominanti a 
due derivate $ (\part F)^2F^n$ sono assenti ), e quindi l'azione della 
particella su scale molto pi\`u piccole di $\sqrt{\a^{'}}$ contiene termini 
con derivate della posizione superiori alla prima; questo non rappresenta una 
difficolt\`a in quanto \`e proprio in questa regione che le variabili 
$Y^j(X^0)$ perdono la loro interpretazione di coordinate spaziotemporali.
In modo del tutto analogo abbiamo per una Dp-brana una corrispondenza tra 
azione di Born-Infeld 
e azione di Nambu-Goto
\be
L_{BI} \sim \sqrt{-det(\h_{\m\n}+2 \p \a^{'}eF_{\m\n})} \lrar
L_{NG} \sim \sqrt{-det(\h_{\a\b}+\part_\a Y^\m \part_\b Y_\m)} \ ,
\ee
nel gauge fisico.

Sappiamo gi\`a che due brane parallele non si attraggono in quanto stati BPS; 
calcoliamo ora la metrica sullo spazio dei moduli e mostriamo che \'e piatta a 
tutti gli ordini in $\a^{'}$. 
Consideriamo per questo una configurazione di due 
Dp-brane parallele in moto relativo lungo la nona dimensione,
descrivendole attraverso le condizioni al 
bordo per le stringhe aperte tese tra di esse; se indichiamo con $b$ il 
parametro di impatto
\`e facile verificare che $X^0$ e $X^9$ in $\s = 0$ soddisfano rispettivamente
a condizioni di Neumann e di Dirichlet mentre 
$X^{\pm}(\t, \p) = e^{\pm \p \e} X^{\pm}(\t,0)$; in altri termini un estremo 
della stringa \`e fisso mentre l'altro \`e trasformato con un boost con 
velocit\`a $v$, consistentemente con il fatto che \`e fissato alla brana in 
movimento. 
L'espansione per le coordinate di Neumann 
o Dirichlet \`e nota, e sappiamo anche che
\be
X^0 \pm X^d = i \sqrt{ \a^{'}} \sqrt{ \frac{1 \pm v_1}{1 \mp v_1}}
\sum_{n=-\infty}^{\infty} \left [ \frac{a_n}{n+i \e}e^{-i(n+i \e)( \t \pm \s)}+
\frac{\bar{a}_n}{n-i \e}e^{i(n-i \e)( \t \mp \s)} \right ]   ,
\ee
dove ora il parametro $ \e = \frac{arcthv_2 -arcthv_1}{\p} $ emerge 
naturalmente come la composizione relativistica delle velocit\`a delle brane. 

L'ampiezza di anello, dalla quale abbiamo ricavato l'effetto Schwinger per 
teorie di stringhe aperte, corrisponde ora allo spostamento di fase per lo 
scattering in avanti delle due D-brane
\beq
\d(b,v) &=& -2 \frac{V_p}{2} \int_0^{\infty} \frac{dt}{t}(2 \p^2 t)^{-p/2}
e^{-b^2t/2 \p} \frac{1}{2} \sum_{s=2}^4(-1)^{s+1}
\frac{\th_s(\frac{\e t}{2}|\frac{it}{2})}{\th_1(\frac{\e t}{2}|\frac{it}{2})}
\frac{\th^3_s(0|\frac{it}{2})}{\h^9(\frac{it}{2})}    \nonumber \\
&=& -2 \frac{V_p}{2} \int_0^{\infty} \frac{dt}{t}(2 \p^2 t)^{-p/2}
e^{-b^2t/2 \p} 
\frac{\th_1^4(\frac{\e t}{4}|\frac{it}{2})}{\th_1(\frac{\e t}{2}|\frac{it}{2})
\h^9(\frac{it}{2})}  \  .
\eeq
Nel limite $v \rar 0$ si ottiene 
\be
\d (b,v) \sim - \e^3 \left ( \frac{2 \p^2 \ap}{b^2} \right )^3 + o(\e^7)  \  .
\ee
Poich\'e il tempo di interazione \`e $\frac{1}{|v|} \sim \epsilon$, da
$\d (v,b) \sim \frac{\d_0(b)}{v} + \d_1(b)v + \d_2(b)v^3+...$ si vede che si 
annulla
non solo il termine costante, come conseguenza del bilanciamento tra 
attrazione gravitazionale e repulsione $R-R$, ma anche 
il termine quadratico; dato che lo scattering ad ordine $o(v^2)$ tra 
solitoni pesanti pu\`o essere descritto come il moto geodetico sullo spazio dei 
moduli degli zero modi, otteniamo che, almeno a quest'ordine dell'espansione in 
loop e a tutti gli ordini in $\ap$, la metrica su tale spazio dei moduli \`e 
piatta. Un altro fatto notevole, conseguenza della supersimmetria, \`e che la 
dipendenza dei termini $o(v^4)$ dal parametro di impatto $b$ \`e la stessa nei 
due regimi $b >> \sqrt{\ap}$, in cui domina la supergravit\`a, e 
$ b << \sqrt{\ap}$, ovvero a scale inferiori a quella di stringa; questo 
comportamento \`e alla base della congettura formulata in \cite{mm1}
che la M teoria possa essere 
descritta da un modello di matrici.
Nel limite $b \rar \infty$, estraibile dal comportamento 
asintotico per $t \sim 0$ della funzione di partizione, 
lo spostamento di fase coincide, come deve, 
con quello che si pu\`o calcolare, in modo 
simile a quanto fatto in precedenza, nella teoria effettiva di supergravit\`a.

La presenza di una parte immaginaria nell' ampiezza di anello, legata alla 
creazione di
coppie di stringhe aperte, d\`a allo spostamento di fase una parte 
d'assorbimento che nel caso di D0 brane \`e
\be
Im ( \d ) = \sum_{s} \frac{dim(s)}{2} \sum_{k} e^{-\frac{2 \p \ap k}{\e}
\left ( \frac{b^2}{(2 \p \ap)^2} + M(s)^2 \right )}   \    ,
\ee
dove la somma \`e su tutti i multipletti dello spettro di dimensione $dim(s)$ e 
massa dovuta agli oscillatori $M(s)$; il fenomeno T duale alla creazione di 
coppie \`e quindi la creazione di stringhe aperte tese tra due D-brane 
in movimento.
Ci si aspetta che il fenomeno di creazione venga soppresso esponenzialmente a 
bassa velocit\`a; infatti per $\p \e \sim v << 1$, solo stati non massivi 
contribuiscono ad $Im \d$ e si ha
\be
Im ( \d ) \sim \frac{8 V^p}{(2 \p )^p} \sum_{k=1}^{\infty} 
\frac{1}{k} \left ( \frac{v}{\p k} \right )^{p/2} e^{-b^2k/v}  \ ,
\ee
una quantit\`a esponenzialmente piccola per parametri d'impatto 
$b \geq \sqrt{v \ap}$, sensibilmente inferiori alla scala di stringa. Nel 
limite ultrarelativistico si ha invece
$\e \sim - \frac{1}{2 \p} ln \frac{1-v}{2} \sim 
\frac{1}{m_p}ln{s}{m_p^2} >> 1$ ($m_p$ massa della p-brana). La parte 
d'assorbimento \`e dominata dal comportamento asintotico della funzione di 
partizione delle stringhe aperte in un intorno dell'origine e d\`a
\be
Im ( \d ) \sim \frac{V_p}{2(2 \p)^p \sqrt{\p}^{p-8}} \frac{s}{m_p^2}
\left ( ln \frac{s}{m_p^2} \right )^{\frac{p}{2}-4} e^{-b^2/ln \left ( \frac{s}{m^2_p} \right )}   \  ,
\ee
un comportamento analogo a quello delle ampiezze di stringa ad alta energia.
In questo limite le D-brane si comportano quindi come dischi neri assorbenti di 
area $b^2 \sim \ap ln \frac{s}{m^2_p}$; brane ultrarelativistiche non possono 
esere utilizzate per analizzare scale inferiori a $\sqrt{\ap}$.
Questa relazione tra dinamica delle D-brane e
scale in teoria di stringhe \`e stata analizzata in modo sistematico
in \cite{db5}. In particolare, studiando il comportamento delle D-particelle, 
si pu\`o mostrare che nel limite non relativistico queste sono sonde pi\`u 
precise delle stringhe per analizzare la struttura dello spazio tempo.
La dinamica dei solitoni non pu\`o essere separata dalla teoria alla quale 
appartengono, ma nel limite di bassa energia essa \`e ben approssimata dalla 
meccanica quantistica nello spazio dei moduli degli zero modi, cos\`\i \,
in questo limite la dinamica delle D0-brane \`e ben descritta da un modello 
di matrici ottenuto riducendo dimensionalmente la teoria di SYM da 
$9+1$ a $0+1$ dimensioni. Pi\`u in generale, per Dp-brane
gli zero modi danno origine
a campi non massivi confinati sul volume e la 
loro dinamica di bassa energia \`e descritta da una 
teoria di gauge in $p+1$ dimensioni. Questo sembrerebbe indicare che la 
struttura su piccola scala dello spazio tempo \`e intimamente connessa con le 
teorie di gauge supersimmetriche.
Infatti le eccitazioni di bassa energia di una Dp-brana formano un multipletto 
vettoriale ridotto dimensionalmente da dieci a (p+1) dimensioni:
\be
A^M( \z^b) \mapsto A^{\m}(\z^b) , Y^m(\z^b) \    ,
\label{mult}
\ee
con $\m =0,...,p$ e $m = p+1,...,9$. Gli scalari $Y^m$ rappresentano la 
posizione della brana nello spazio trasverso e sono i bosoni di Glodstone 
dell'invarianza per traslazioni; le altre eccitazioni fisiche di bassa energia corrispondono ad 
un campo di gauge. Supponiamo ora di avere $n$ Dp-brane parallele; gli stati di 
stringa aperta sono ora accompagnati da una coppia di indici di Chan-Paton che 
indicano su quale brana si trovano gli estremi della stringa. I campi in 
(\ref{mult}) vengono quindi generalizzati in modo naturale a matrici:
$A^{\m} \mapsto A^{\m}_{ij}$ e $Y^m \mapsto Y^m_{ij}$. Questo risultato
\`e naturale dal 
punto di vista della teoria di bassa energia che \`e SYM in $p$ dimensioni 
con gruppo di gauge
\be
U(n) \simeq U(1)_{cm} \times SU(n)  \  ;
\ee
gli $A^{\m}_{ij}$ sono infatti vettori di gauge non 
abeliani e gli $Y^m_{ij}$ campi di Higgs nell'aggiunta; \`e invece
notevole data 
l'interpretazione degli $Y^m_{ij}$ come coordinate trasverse della brana. 
L'usuale nozione di coordinate si recupera considerando lo spazio dei moduli 
della teoria. Il potenziale per gli scalari \`e infatti
\be
V \sim tr[Y^m,Y^r]^2   \  ,
\ee
e si annulla quando le matrici commutano l'una con l'altra e sono quindi 
diagonalizzabili simultaneamente; lo spazio dei moduli della teoria di gauge riflette allora 
la possibilit\`a per le brane di muoversi arbitrariamente nello spazio 
trasverso (ricordiamo che per questa configurazione non c'\`e forza statica
tra le brane), ma l'interpretazione dei campi $Y^m$ come semplici coordinate 
viene meno a piccole distanze quando bisogna tenere conto delle fluttuazioni 
attorno alle configurazioni di vuoto della teoria. \`E anche interessante 
notare che il gruppo di Weyl di $SU(n)$ contiene $S_n$, il gruppo delle
permutazioni di $n$ oggetti, e quindi l'indistinguibilit\`a delle brane 
diviene, da questo punto di vista, parte della simmetria di gauge. 

Nella Tipo I il gruppo di gauge $U(N)$ viene ridotto dalla proiezione $\W$; per 
D1 e D9-brane si hanno naturalmente gruppi ortogonali, in modo consistente 
con l'interpretazione 
della Tipo I come una teoria di nove brane, mentre per D5-brane si hanno naturalmente gruppi 
simplettici \cite{hod3,or2}, in modo consistente con la dualit\`a TipoI-HO e 
l'interpretazione delle D5 come piccoli istantoni eterotici. Osserviamo inoltre 
che la D1 non ha campi di gauge, come deve per essere identificata con la 
stringa eterotica, e che una singola D5 non ha coordinate trasverse: la minima 
eccitazione dinamica \`e costituita da una coppia di 5-brane con gruppo di 
gauge $Usp(2) \sim SU(2)$.

La natura non commutativa 
delle coordinate delle D-brane, evidenziata per la prima volta in \cite{br7}, 
ha portato a considerare 
attentamente il ruolo che la geometria non commutativa pu\`o rivestire 
in teoria di stringhe e, in generale, per la descrizione della struttura su piccola 
scala dello spazio tempo \cite{ncg}.
In particolare, teorie di gauge su tori non commutativi compaiono in modo 
naturale quando si considera la teoria di volume per D-brane su tori con un 
campo $B_{\m\n}$ di background.

La capacit\`a delle D-brane di sondare scale inferiori a quella di stringa \`e 
stata sfruttata in vari modi; si pu\`o ad esempio vedere in modo diretto come i 
settori twistati di un orbifold riparino le singolarit\`a \cite{subs}.
Consideriamo la IIB compattificata su $T^4/Z_2$, dove come al solito 
l'involuzione $R$ inverte tutte le coordinate,  
concentriamoci 
sull'intorno di uno dei sedici punti fissi e introduciamo una D-stringa nel 
piano $X^{2,...,9} = 0$. Perch\`e la D-stringa possa allontanarsi dal punto 
fisso, si devono avere due indici di Chan-Paton, uno per la stringa ed uno per 
la sua immagine tramite $R$; $R$ agisce allora nel modo usuale sugli 
oscillatori di una stringa aperta ma scambia i fattori di Chan-Paton tra la 
stringa e l'immagine. Si ha quindi
\be
R| \psi ,ij \ran = \s^1_{ii^{'}}|R \psi, i^{'}j^{'} \ran \s^1_{j^{'}j}  \quad ,
\ee
dove le $\s^i$ sono le matrici di Pauli.
Nel settore NS, gli stati non massivi R-invarianti sono
\beq
& & \part_t X^{\m} \s^{0,1}  \ , \hspace{1cm}   \m = 0,1  \  , \hspace{3cm} 
\part_n X^{i} \s^{0,1}  \ , \hspace{1cm}   i = 2,3,4,5  \  ,  \nonumber \\
& & \part_n X^{m} \s^{2,3}  \ ,  \hspace{1cm}   m = 6,7,8,9  \  ,
\eeq
e descrivono 
un campo di gauge, la posizione della stringa nello spazio non compatto e le 
coordinate trasverse. Indichiamo con $A^\m$, $x^i$, $x^m$ i campi 
corrispondenti, che sono matrici $2 \times 2$; l'azione di bassa energia
\`e la riduzione di $SYM$ con gruppo di gauge $U(2)$ da dieci a due dimensioni, 
proiettata con $R$, il che rompe il gruppo a $U(1) \times U(1)$. Il potenziale 
\`e
\be
U = 2 \sum_{i,m} Tr([x^i,x^m]^2) + \sum_{m,n} Tr([x^m,x^n]^2) \   .
\label{unpot}
\ee
Lo spazio dei moduli ha due branche. La branca con $x^m =0 $ e 
$x^i = u^i \s^0 + v^i \s^1$, corrisponde al moto delle due D-stringhe nelle 
sei dimensioni non compatte, con posizione $u^i \pm v^i$; la simmetria di gauge 
non \`e rotta e si hanno due fattori $U(1)$ indipendenti per ciascuna stringa.
Nella branca $x^m \neq 0$ e $x^i = u^i \s^0$ la simmetria di gauge associata a
$\s^1$ \`e rotta; posto $x^m = w^m \s^3$, si vede che si sta descrivendo il 
moto della D-stringa al di fuori del punto fisso, con la D-stringa e la sua 
immagine localizzate in $(u^i,\pm w^i)$. Per studiare i moduli dei settori 
twistati, definiamo le variabili complesse $q^m$ ponendo
$x^m = \s^3 Re(q^m) + \s^2 Im(q^m)$ ed introduciamo i due doppietti
\be
\F_0 = \pmatrix{q^6 + iq^7\cr q^8 + iq^9} \ , \hspace{1cm}
\F_1 = \pmatrix{\bar{q}^6 + i\bar{q}^7\cr \bar{q}^8 + i\bar{q}^9}  \ . 
\ee
Rispetto al gruppo $U(1)$ generato da $\s^1$, $\F_{0,1}$ hanno carica $\pm1$; i 
moduli $(NS)^2$ possono essere scritti come un vettore $\bf{D}$ e il potenziale 
\`e proporzionale a 
\be
(\F_0^{\dagger} \t \F_0 - \F_1^{\dagger} \t \F_1 + D)^2  \  ,
\ee
dove le $\t^i$ sono sempre le matrici di Pauli. Quando $\bf{D}=0$, il 
potenziale si riduce a (\ref{unpot}) mentre per $\bf{D} \neq 0$ i punti 
singolari vengono riparati. Lo spazio dei moduli della D-stringa, ottenuto 
imponendo l'annullarsi del potenziale, non \`e altro 
che l'insieme delle possibili posizioni, vale a dire lo spazio ALE che si 
ottiene rimuovendo le singolarit\`a.
La metrica sullo spazio dei moduli, contenuta nei termini cinetici dei campi 
della D-stringa, coincide con la metrica sull'orbifold riparato, la metrica di 
Eguchi-Hanson. Questi risultati possono essere estesi ad orbifold pi\`u 
generali \cite{subs}, fornendo una realizzazione fisica della costruzione degli 
spazi ALE come quozienti hyperk\"ahleriani di Kronheimer e Nakajima
\cite{KN}.

Tecniche analoghe permettono di interpretare istantoni nella teoria di gauge 
sul volume di una p-brana come stati legati di una $p$ e di una $(p-4)$ brana, 
riproducendo la descrizione lagrangiana della costruzione ADHM, e permettono 
anche di 
studiare cambiamenti di topologia come la transizione di flop \cite{subs}.

In tutte queste applicazioni, sono cruciali gli stati di stringa aperta tesi
tra le brane, perch\`e sono questi a dominare la dinamica di piccola scala.
La relazione tra teorie di gauge e struttura dello spazio tempo riflette
in ultima analisi i due modi che possono essere utilizzati per calcolare
l'interazione tra brane, come uno scambio di stringhe chiuse o come un loop di 
stringhe aperte. Accanto alla descrizione usuale in termini di 
supergravit\`a  abbiamo una 
nuova descrizione basata sulla teoria effettiva che descrive le stringhe aperte 
pi\`u leggere, nella quale lo spaziotempo emerge come un concetto derivato, 
come spazio dei moduli a bassa energia di una teoria di gauge 
supersimmetrica. Le due descrizioni hanno due distinti domini di 
applicabilit\`a : l'una per distanze molto pi\`u grandi della scala di 
stringa, l'altra per 
distanze molto inferiori, ed \`e possibile connettere in modo continuo l'idea 
classica di spaziotempo con la nuova idea di brane in moto sullo spazio dei 
moduli della loro teoria di volume.

Osserviamo, per concludere, che configurazioni di brane sono state utilizzate 
per realizzare teorie di gauge supersimmetriche come teorie di volume
\cite{bft1,elit,bqcd}, consentendo una migliore comprensione delle loro 
propriet\`a. Notevole \`e anche il loro utilizzo per lo studio delle 
propriet\`a termodinamiche dei buchi neri \cite{bh1}.


\chapter{Dualit\`a}
\markboth{}{}

L'espansione perturbativa della teoria di stringhe presenta un'elegante 
unificazione di molti concetti noti dalla teoria dei campi; fissata la teoria 
conforme possiamo calcolare lo spettro, abbiamo un modo unico di introdurre 
interazioni e sappiamo costruire una serie perturbativa con una chiara 
interpretazione geometrica. Gran parte delle informazioni riguardanti le 
interazioni degli stati non massivi pu\`o essere racchiusa in una azione 
effettiva definita in modo tale che un'ampiezza calcolata
ad albero utilizzando questa azione riproduca il risultato di stringa; 
in generale si mantengono solo i termini contenenti due derivate, costruendo 
quella che viene chiamata azione effettiva di bassa energia, valida su scale 
molto maggiori di $ \sqrt{\a^{'}} $. 
Le congetture di dualit\`a  sono ispirate dalle simmetrie di queste 
lagrangiane effettive che vengono estese, tipicamente in una forma discreta 
a causa della quantizzazione della carica, all'intera teoria di stringhe.
Iniziamo pertanto ad analizzare queste lagrangiane, delineando quelle che 
dovrebbero essere le relazioni tra i vari modelli, 
e successivamente illustriamo 
diverse verifiche che sono state fatte per metterle alla prova; discuteremo 
quindi l'attuale punto di vista sulla struttura complessiva 
della teoria. Per ora ci limitiamo ad osservare che
l'importanza delle dualit\`a sta nel fatto che esse 
eliminano la non unicit\`a della 
teoria e permettono di studiare fenomeni a forte accoppiamento in un modello 
utilizzando informazioni perturbative di un altro.
Ogni teoria \`e caratterizzata da una serie di campi di background; le 
trasformazioni di dualit\`a collegano tra loro regioni dello spazio dei moduli 
di due teorie differenti, come nel caso della Tipo I e dell'eterotica SO(32), o 
della stessa teoria, come nel caso della Tipo IIB.
La T-dualit\`a funziona allo stesso modo senza tuttavia scambiare i contributi 
ai diversi ordini nella serie perturbativa delle due teorie.
La supersimmetria \`e stata sino ad ora uno strumento essenziale per la 
verifica di queste congetture, 
dato che in teorie supersimmetriche \`e possibile 
provare teoremi di non rinormalizzazione che permettono di estendere  a forte
accoppiamento risultati ottenuti a debole accoppiamento, e quindi di 
confrontali con quelli della teoria supposta duale. In particolare, con un 
numero di supersimmetrie $N \ge 16 $ l'azione effettiva di bassa energia, dato 
lo spettro, \`e completamente fissata e non riceve pertanto correzioni dai 
loops di stringa. Quindi ogni valida simmetria della teoria deve essere una 
simmetria dell'azione effettiva.
In effetti, \`e sufficiente che siano invarianti le equazioni 
del moto, dato che l'azione 
effettiva viene utilizzata solo per calcoli ad albero.
Inoltre la relazione tra la massa e la carica per una classe di stati, detti stati 
BPS, \`e determinata 
dall'algebra e non riceve correzioni. Si pu\`o anche arguire che la 
molteplicit\`a degli stati BPS non varia dal regime perturbativo 
a quello non perturbativo, 
e di conseguenza ogni simmetria della teoria deve essere una simmetria dello 
spettro degli stati BPS. 

Riassumiamo qui le relazioni di  dualit\`a in dieci dimensioni;
relazioni tra teorie compattificate seguono in modo pi\`u o meno rigoroso da 
queste. La teoria di Tipo I equivale, a forte accoppiamento, alla teoria 
eterotica con gruppo di gauge $SO(32)$; a debole accoppiamento la stringa 
eterotica fondamentale compare nella Tipo I come una D-stringa. 
La teoria di Tipo IIB \`e autoduale a 
forte accoppiamento; pi\`u precisamente \`e caratterizzata in dieci dimensioni 
da un gruppo $SL(2,Z)$ di dualit\`a e contiene nello spettro stringhe 
caratterizzate da una coppia di cariche $(p,q)$ che dominano la dinamica in 
regioni diverse dello spazio dei moduli. Il regime di forte accoppiamento 
della IIA e della eterotica con gruppo di gauge $E_8 \times E_8$, non \`e 
descritto, come nei casi precedenti, da un'altra teoria di stringhe debolmente 
accoppiata, ma da una nuova teoria 11-dimensionale, 
la cui struttura deve essere ancora chiarita 
e che \`e stata chiamata provvisoriamente M-teoria. La IIA corrisponde alla 
M-teoria compattificata su un cerchio e l'eterotica $E_8 \times E_8$ alla M-teoria 
compattificata sull'intervallo $S/Z_2$; i due fattori del gruppo di gauge sono 
localizzati nei due estremi dell'intervallo. \`E noto che il 
limite di bassa energia della M teoria \`e la supergravit\`a in 
undici dimensioni e che tra le sue eccitazioni compaiono membrane e 5-brane. 
L'idea \`e che tutte le teorie di stringhe corrispondono a compattificazioni 
dela M-teoria, i cui gradi di libert\`a devono essere quindi abbastanza versatili da 
riprodurre la grande variet\`a delle eccitazioni presenti in questi modelli.

Nella discussione che segue
indichiamo con $ g_{\m\n} = e^{ - \frac{2}{d-1} \f }G_{\m\n}$ la metrica di 
Einstein e con $ g_s^2 = e^\f $ la costante d'accoppiamento, dove 
$ \f = \f^{10} - lnV $ \`e il dilatone della teoria compattificata e 
$ (2 \p)^{9-d} V $ \`e il volume dello spazio interno misurato con $G_{\m\n}$.
Ricordiamo inoltre che il valore della costante di accoppiamento $g$  pu\`o  
essere variato
con la trasformazione  $ \f \mapsto \f - 2c $ ,  $g_{s} \mapsto e^c g_{s}$,
mentre il valore di $\a^{'}$ tramite  $\a^{'} \mapsto \l \a^{'}$ , 
$ G_{\m\n} \mapsto \l G_{\m\n} $.
Variando $ <\f> $ otteniamo tutti i possibili valori per $g_s$, e riscalando 
$ G_{\m\n} $ tutti i possibili valori per $ \a^{'} $. 


\section{Dualit\`a Tipo$I-SO(32)$}
\markboth{}{}

Iniziamo considerando la dualit\`a $ \bf{TipoI-SO(32)} $  \cite{du4}.
Per entrambe le teorie l'azione effettiva pu\`o essere ricavata con un 
calcolo ad albero, e comunque \`e fissata dalla supersimmetria.
L'eterotica contiene i campi $ G_{\m\n} $, $ B_{\m\n} $, $ \f $ e $ A_\m^a $;
la Tipo I  $ G_{\m\n} $ e  $ \f $ dal settore chiuso $ (NS)^2 $, $ B_{\m\n} $ 
dal settore chiuso $(R)^2$ e $ A_\m^a $ dal settore aperto.
Confrontiamo le due azioni, scrivendole entrambe nel frame di Einstein:
\beq 
S^{H} &=& \frac{1}{(2 \p)^7} \int d^{10}x \sqrt{-g} [ R - \frac{1}{8} g^{\m\n}
\part_\m \f \part_\n \f - \frac{1}{4} g^{\m\n}g^{\r\s} e^{-\frac{ \f }{4}}
trF_{\m\n}F_{\r\s} \nonumber \\
&-&  \frac{1}{12} g^{\m\n}g^{\r\s}g^{\e\z} e^{- \frac{\f}{2}}
H_{\m\n\r}H_{\s\e\z}] \  ,
\eeq
\beq 
S^{I} &=& \frac{1}{(2 \p)^7} \int d^{10}x \sqrt{-g} [ R - \frac{1}{8} g^{\m\n}
\part_\m \f \part_\n \f - \frac{1}{4} g^{\m\n}g^{\r\s} e^{ \frac{ \f }{4}}
trF_{\m\n}F_{\r\s} \nonumber \\
&-&  \frac{1}{12} g^{\m\n}g^{\r\s}g^{\e\z} e^{ \frac{\f}{2}}
H_{\m\n\r}H_{\s\e\z}] \  ,
\eeq
dove
\be 
F_{\m\n} = \part_\m A_\n - \part_\n A_\m + \sqrt{2}[A_\m , A_\n] \  ,
\ee
\be
H_{\m\n\r} = \part_\m B_{\n\r} - \frac{1}{2}Tr \left ( A_\m F_{\n\r} - 
 \frac{ \sqrt{2}}{3} A_\m [A_\n , A_\r] \right ) + cicl  \  .
\ee
Queste due azioni si ottengono chiaramente 
l'una dall'altra identificando tra loro i campi corrispondenti
delle due diverse teorie e ponendo $ \f^{H} = - \f^{I} $; il cambiamento di 
segno nel dilatone lega il regime 
perturbativo della Tipo I a quello non perturbativo dell'eterotica e viceversa.
La lagrangiana effettiva \`e unica -supersimmetria e cancellazione delle 
anomalie fissano completamente i termini con due derivate e gli accoppiamenti 
di Green-Schwarz con quattro derivate; l'esistenza di questa dualit\`a 
sembrerebbe indicare che anche l'estensione microscopica della teoria sia
 unica.

Una prima evidenza per questa congettura \cite{du4} 
viene fornita dall'esistenza di una 
soluzione dell'azione effettiva della Tipo I che descrive un oggetto 
unidimensionale identificabile con la stringa eterotica, 
conformemente con l'idea che quando si hanno queste relazioni tra regime 
perturbativo e non perturbativo di due teorie i gradi di libert\`a fondamentali
dell'una appaiano come solitoni nell'altra.
\`E semplice verificare che la D-stringa nella teoria di Tipo I ha le stesse 
eccitazioni fisiche
della stringa eterotica; costruendo con le tecniche usuali lo 
spettro, si vede infatti che nel settore $DD$ si hanno $8$ bosoni e $8$ 
fermioni con chiralit\`a negativa 
mentre nel settore $ND$ si hanno $32$ fermioni con chiralit\`a 
positiva.

Un test non banale \cite{hod2,ohdu,ft1}
consiste nel calcolo di termini successivi nell' azione 
effettiva non fissati in modo univoco dalle simmetrie e dalla 
consistenza della teoria di campo, ma tuttavia protetti da qualche 
teorema di non rinormalizzazione. Adatti a questo scopo sono i termini del tipo 
$ \bf{F^4}$ e $ \bf{R^4} $ in compattificazioni toroidali; si crede che essi
ricevano correzioni solo da multipletti corti ad un loop e da punti di sella 
massimamente supersimmetrici.
Nell'eterotica gli unici punti di sella noti sono quelli associati alla 5-brana 
NS, e se ci si limita a $ d>4 $ dimensioni non compatte non esistono 6-cicli 
finiti che possano dare contributi; pertanto il risultato ad un loop dovrebbe 
essere esatto.
Nella Tipo I si hanno due nuovi effetti; la supersimmetria spaziotemporale non 
commuta con l'espansione in superfici, e quindi termini diversi di un 
superinvariante possono essere generati ad ordini diversi; inoltre esistono 
correzioni non perturbative dovute alla presenza di D1-istantoni - 
configurazioni euclidee di D-brane -  avvolti 
attorno a cicli non banali che corrispondono attraverso la dualit\`a a stringhe 
eterotiche.
In altri termini, il risultato ad un loop nella stringa 
eterotica corrisponde alla somma 
di effetti non perturbativi e perturbativi nella Tipo I.
Questa dualit\`a \`e di particolare interesse, in quanto lega due espansioni 
perturbative del tutto differenti : da una parte abbiamo ad ogni ordine un solo 
diagramma, dall'altra la somma di diverse superfici con bordi e crosscaps. La 
consistenza da una parte \`e data dall'invarianza modulare, dall'altra dalla 
cancellazione di tadpoles.
Il calcolo in dieci dimensioni pi\`u che una verifica della dualit\`a \`e 
conseguenza della consistenza delle due teorie; il confonto diviene non banale 
quando si compattifica qualche dimensione, e come vedremo sar\`a cruciale il 
ruolo dei D-istantoni.


\section{Dualit\`a $SL_2(Z)$ della $IIB$}
\markboth{}{}

Consideriamo la situazione nella IIB. Questa teoria contiene una 4-forma autoduale,
e la descrizione lagrangiana di questo tipo di campi presenta delle
difficolt\`a, risolte solo recentemente grazie ad un metodo
sviluppato da Pasti, Sorokin e Tonin \cite{Tonin}.  In quanto segue
\`e comunque sufficiente limitare la nostra attenzione alle 
equazioni del moto, la cui forma covariante \`e stata ottenuta in 
\cite{sugy1}.
Queste equazioni sono invarianti 
rispetto a trasformazioni $ M = \pmatrix{ p & q\cr r & s} \in SL_2(R) $ che 
agiscono sullo scalare complesso $ \l = a + ie^{- \frac{ \f}{2}} $, formato 
dallo scalare di Ramond e dal dilatone, e su $ B = \pmatrix{ B_{NS}\cr B_R}$ 
nel modo seguente:
\be
\l \mapsto \frac{p \l+q}{r \l+s} \ , \hspace{2cm}   B \mapsto MB  \ .
\ee
Possiamo presumere che nella teoria di stringhe sopravviva un sottogruppo 
discreto $SL_2(Z)$ di questa simmetria \cite{du1}. 
La restrizione a trasformazioni con 
coefficienti interi pu\`o essere giustificata osservando che con una opportuna 
normalizzazione la stringa porta un'unit\`a di carica rispetto a $B_{NS}$ e 
dato che non esistono nello spettro stati di stringa frazionari, la carica di 
$B_{NS}$ \`e quantizzata in interi. Si pu\`o quindi verificare che, a meno di 
riscalare $B_{R}$, il sottogruppo massimale di $SL_2(R)$ per cui $p$ \`e sempre 
un intero \`e $SL_2(Z)$.
Utilizzando restrizioni dovute alla supersimmetria e richiedendo l'invarianza 
modulare degli accoppiamenti, \`e possibile determinare correzioni all'azione 
effettiva della IIB che includono contributi di ogni ordine nell'espansione in 
loops ed anche effetti non perturbativi dovuti a D-istantoni; questi termini 
possono essere collegati ad accoppiamenti analoghi presenti in M teoria e possono
anche essere calcolati partendo dalla supergravita' in undici dimensioni
\cite{dist}.

Se combiniamo la dualit\`a $SL_2(Z)$ della IIB con la T-dualit\`a delle 
stringhe di tipo II compattificate su $T^n$ otteniamo che i gruppi completi di 
dualit\`a delle compattificazioni toroidali delle teorie 
di tipo IIA e IIB sono identici. 
Questi gruppi \cite{du1}, detti gruppi di U-dualit\`a, si ottengono 
combinando la 
simmetria non perturbativa $SL_2(Z)$ con la 
simmetria perturbativa $SO(d,d;Z)$ e 
coincidono con le versioni discrete dei gruppi di simmetrie "nascoste" non 
compatte scoperti studiando le teorie di supergravit\`a \cite{hid1,11d}, 
dei quali il pi\`u 
famoso \`e il gruppo $E_{7,7}$ di Cremmer e Julia per la supergravit\`a 
massimamente supersimmetrica $N=8$ in $d=4$. 
Ancora una volta, la restrizione a gruppi discreti \`e legata alla 
quantizzazione della carica. La lista completa dei gruppi di U-dualit\`a \`e la 
seguente
\beq
& & D = 9 \hspace{1cm} SL(2,Z) \nonumber \\
& & D = 8 \hspace{1cm} SL(2,Z) \times SL(3,Z) \hspace{1cm} 
SL(2,Z) \times SL(2,Z) 
\nonumber \\
& & D = 7 \hspace{1cm} SL(5,Z) \hspace{3.1cm} 
SL(2,Z) \times SO(3,3,Z) \nonumber \\
& & D = 6 \hspace{1cm} SO(5,5,Z) \hspace{2.6cm} 
SL(2,Z) \times SO(4,4,Z) \nonumber \\
& & D = 5 \hspace{1cm} E_{6,6}(Z) \hspace{3.3cm} 
SL(2,Z) \times SO(5,5,Z) \nonumber \\
& & D = 4 \hspace{1cm} E_{7,7}(Z) \hspace{3.3cm} 
SL(2,Z) \times SO(6,6,Z) \nonumber \\
& & D = 3 \hspace{1cm} E_{8,8}(Z) \hspace{3.3cm} 
SL(2,Z) \times SO(7,7,Z) \nonumber \\
& & D = 2 \hspace{1cm} \hat{E}_{8,8}(Z) \hspace{3.3cm} 
SL(2,Z) \times SO(8,8,Z)    
\eeq
Il primo gruppo \`e il gruppo di U dualit\`a mentre il secondo \`e il gruppo di 
T dualit\`a. Per $D=1$ ci si aspetta un'algebra di dualit\`a molto vasta, 
basata su un algebra di Lie iperbolica, non ancora compresa in dettaglio.

\section{Eterotica su $T^6$}
\markboth{}{}

Un ragionamento del tutto analogo pu\`o essere fatto per le due teorie 
eterotiche \cite{sen1}; 
una volta compattificate su $T^n$ esse coincidono, ed hanno 
di conseguenza lo stesso gruppo di dualit\`a. In particolare, per $d>4$ il 
gruppo completo coincide con quello di T-dualit\`a, mentre in $d=4$ si ha una 
simmetria non perturbativa $SL_2(Z)$ che si estende anche a $d<4$. Nelle varie 
dimensioni abbiamo quindi
\beq
& & D=9 \hspace{1cm} O(1,17,Z) \hspace{3.1cm} O(1,17,Z) \nonumber \\
& & D=8 \hspace{1cm} O(2,18,Z) \hspace{3.1cm} O(2,18,Z) \nonumber \\
& & D=7 \hspace{1cm} O(3,19,Z) \hspace{3.1cm} O(3,19,Z) \nonumber \\
& & D=6 \hspace{1cm} O(4,20,Z) \hspace{3.1cm} O(4,20,Z) \nonumber \\
& & D=5 \hspace{1cm} O(5,21,Z) \hspace{3.1cm} O(5,21,Z) \nonumber \\
& & D=4 \hspace{1cm} O(6,22,Z)\times SL(2,Z) \hspace{1cm} O(6,22,Z) \nonumber \\
& & D=3 \hspace{1cm} O(8,24,Z) \hspace{3.1cm} O(7,23,Z) \nonumber \\
& & D=2 \hspace{1cm} \hat{O}(8,124Z) \hspace{3.1cm} O(8,24,Z) \nonumber \\
\eeq
La dualit\`a $SL_2(Z)$ dell'eterotica su $T^6$ \cite{sen1}
\`e stato il primo esempio di 
dualit\`a non perturbativa, cos\`\i \, come l'invarianza $SL_2(R)$ della 
supergravit\`a corrispondente \cite{hid1} 
\`e stata la prima simmetria nascosta ad essere
notata. 
Compattificando l'eterotica su $T^6$ abbiamo $132$ scalari - 21 dalla metrica, 
15 dal tensore antisimmetrico e 96 dai campi di gauge - che possono essere 
descritti introducendo una matrice simmetrica $M\in O(6,22)$, dove
$ MLM^t = L $ con 
\be
L = \pmatrix{ 0 & 1_6 & 0\cr 1_6 & 0 & 0\cr 0 & 0 & -1_{16}} \ . 
\ee
Altri due 
scalari si ottengono dal dilatone e dualizzando la forma $H_{\m\n\r}$ :
\be
H^{\m\n\r} = - \frac{1}{\sqrt{-g}} e^{2 \f} \e^{\m\n\r\s} \part_{ \s}a \ ,
\ee
e possono essere utilizzati per costruire il campo complesso 
$ \l = a+i e^{- \f} $. Nel generico punto dello spazio dei moduli abbiamo anche 
$16$ campi di gauge abeliani che corrispondono ai generatori della sottoalgebra di 
Cartan. L'azione effettiva per questi campi,
\beq
S &=& \frac{1}{2 \p} \int d^4x \sqrt{-g} [ R 
- g^{\m\n} \frac{ \part_\m \l \part_\n \bar{\l}}{2(\l_2)^2}
+ \frac{g^{\m\n}}{8}Tr(\part_\m ML \part_\n ML) \nonumber \\
&-& \frac{1}{4} \l_2 g^{\m\n}g^{\r\s}F^a_{\m\r}(LML)_{ab}F^b_{\n\s}
+\frac{1}{4} \l_1 g^{\m\n}g^{\r\s}F^a_{\m\r}(L)_{ab}F^b_{\n\s}]  \  ,
\eeq
\`e invariante se $ M \mapsto \W M \W^{t} $ , $ A^a_\m \mapsto \W A^a_\m$  
con $ \W \in O(6,22;R) $. Ora sappiamo che $O(6,22;Z)$ \`e una simmetria 
della completa teoria di stringa, ma l'azione $(4.11)$ \`e invariante anche 
rispetto a trasformazioni di un gruppo $SL_2(R)$
\be
F^a_{\m\n} \mapsto (r \l_1+s)F^a_{\m\n} + r \l_2(ML)_{ab} \tilde{F}^b_{\m\n}  \  ,
\ee
\be 
\l \mapsto \frac{p \l+q}{r \l+s}  \  .
\ee
Si tratta di una dualit\`a elettromagnetica, in quanto lega $F$ e il suo duale 
$\tilde{F}$. 
La congettura \`e ancora una volta che $SL_2(Z)$ sia una dualit\`a della teoria 
di stringa; infatti, a causa della condizione di quantizzazione di Dirac-
Schwinger-Zwanziger, le cariche magnetiche appartengono ad un reticolo $\L_{28}$,
i vettori delle cariche elettriche e magnetiche trasformano come un doppietto 
e quindi il sottogruppo di $SL_2(R)$ che 
rispetta la quantizzazione \`e $SL_2(Z)$.


\section{Dualit\`a $Het/T^4$-$IIA/K3$ }
\markboth{}{}

Un'altra importante congettura di dualit\`a \`e la relazione tra la stringa 
eterotica su $T^4$ e la stringa IIA sulla variet\`a $K3$ \cite{du1}.
Cosideriamo 
anzitutto lo spettro delle due teorie. L'eterotica contiene, oltre a metrica, 
tensore antisimmetrico e dilatone, $ 10 + 6 + 64 = 80 $ scalari e
$8 + 16 = 24$ vettori; con $M \in O(4,20) $ , $ M = M^t$ possiamo scrivere
\beq
S &=& \frac{1}{(2 \p)^3} \int d^6x \sqrt{-g} [ R 
- \frac{1}{2}g^{\m\n}  \part_\m \f \part_\n \f
+ \frac{g^{\m\n}}{8}Tr(\part_\m ML \part_\n ML) \nonumber \\
&-& \frac{1}{4} e^{ -\frac{ \f}{2}} g^{\m\n}g^{\r\s}F^a_{\m\r}(LML)_{ab}F^b_{\n\s}
-\frac{1}{12} e^{- \f} g^{\m\n}g^{\r\s}g^{\e\z}H_{\m\r\e}H_{\n\s\z}]   \ ,
\eeq
dove 
\be
H_{\m\n\r} = \part_\m B_{\n\r} + \frac{1}{2}A^a_\m L_{ab}F^b_{\n\r}+cicl \ .
\ee 
Il gruppo di dualit\`a \`e in questo caso $O(4,20;Z)$.

Quando compattifichiamo la IIA su $K3$ abbiamo $ 58 $ scalari che descrivono le 
fluttuazioni nella struttura complessa e k\"ahleriana della variet\`a; $22$ 
scalari che si ottengono decomponendo $B_{mn}$ rispetto alle $22$ 2-forme 
armoniche $ \w^p_{mn}$ di $K3$ :  
$ B_{mn}(x,y) \sim \sum_{p=1}^{22} \f_p(x)\w^p_{mn}(y) $ con $x$ coordinate su
$R^6$ e $y$ coordinate su $K3$. Complessivamente abbiamo $80$ scalari che 
parametrizzano un coset $O(4,20)/O(4) \times O(20) $ proprio come per 
l'eterotica. Decomponendo $C_{mnp}$ nella base di 2-forme, 
$ C_{mnp}(x,y) \sim \sum_{p=1}^{22} A^p_\m (x)\w^p_{mn}(y) $, otteniamo $22$ 
campi di gauge; un altro proviene da $A_\m$ ed un altro si ottiene 
dualizzando $C_{\m\n\r}$; in totale abbiamo $24$ campi di gauge. L'azione 
effettiva \`e
\beq
S &=& \frac{1}{(2 \p)^3} \int d^6x \sqrt{-g} [ R 
- \frac{1}{2}g^{\m\n}  \part_\m \f \part_\n \f
+ \frac{g^{\m\n}}{8}Tr(\part_\m ML \part_\n ML) \nonumber \\
&-& \frac{1}{4} e^{ \frac{ \f}{2}} g^{\m\n}g^{\r\s}F^a_{\m\r}(LML)_{ab}F^b_{\n\s}
-\frac{1}{12} e^{ \f} g^{\m\n}g^{\r\s}g^{\e\z}H_{\m\r\e}H_{\n\s\z} \nonumber \\
&-& \frac{1}{16} \frac{ \e^{\m\n\r\s\e\z}}{ \sqrt{-g}} 
B_{\m\n}F^a_{\r\s}L_{ab}F^b_{\e\z}]  \   ,
\eeq
dove $H_{\m\n\r} = \part_\m B_{\n\r} + cicl$.
Le cariche degli stati rispetto ai $24$ campi di gauge appartengono a due 
reticoli $ \L_{a} $ e $ \L_h $ legati da una rotazione $ \W_0 \in O(4,20) $ :
$ \L_h = \W_0 L_{a} $. Le equazioni del moto e le identit\`a di Bianchi delle 
due teorie coincidono se
\beq
& & M_h = \tilde{ \W}M_a \tilde{ \W}^t  \ , \hspace{1cm}  \f^h = - \f_a   \ ,  
\hspace{1cm}
A_h = \tilde{ \W}A_a   \  ,   \nonumber \\
& & \sqrt{-g} e^{ - \f_h}H_h^{\m\n\r} = 
\frac{1}{6} \e^{\m\n\r\s\e\z}H^a_{\s\e\z}  \ .   
\eeq
La matrice $ \tilde{\W} \in O(4,20) $ non \`e arbitraria ma pu\`o essere 
fissata effettuando una trasformazione di T-dualit\`a $\W$; nell'eterotica 
l'effetto della trasformazione \`e $ M_h \mapsto \W M_h \W^t $ e quindi 
$ M_a \mapsto 
\tilde{ \W}^{-1} \W \tilde{ \W} M_a ( \tilde{ \W}^{-1} \W \tilde{ \W})^t $; ora 
perch\`e $ \tilde{ \W}^{-1} \W \tilde{ \W} $ preservi il reticolo $ \L_a$
deve essere $ \tilde{ \W} = \W_0^{-1}$.
Una conseguenza immediata di questa dualit\`a \`e che IIA e IIB su 
$K3 \times T^n$ hanno lo stesso gruppo di dualit\`a dell'eterotica su 
$T^{n+4}$.
Questa congettura sembra portare ad un paradosso per quanto riguarda l'aumento 
della simmetria di gauge; nello spazio dei moduli della stringa eterotica si ha 
aumento di simmetria in assenza di valori di aspettazione per le componenti 
interne dei campi di gauge e in corrispondenza di raggi autoduali, mentre nella 
IIA su $K3$ tutti i campi vettoriali provengono dal settore di Ramond e 
un aumento di simmetria sembrerebbe impossibile proprio per l'assenza di stati 
carichi rispetto ad essi. La soluzione viene ancora una volta dalle D-brane. 
Consideriamo un 
generico punto dello spazio dei moduli dell'eterotica dove il gruppo di gauge 
\`e completamente rotto; i bosoni massivi sono stati BPS e la loro massa va a 
zero nell'avvicinarsi ad un punto di maggior simmetria; dato che la formula 
di massa \`e una conseguenza della supersimmetria, anche gli stati duali a 
questi nella IIA - che sono legati a 
D-brane - hanno massa nulla nel 
punto immagine dello spazio dei moduli. La formula BPS \`e data da 
$ m^2 = e^{- \frac{ \f}{2}} \a^t (LML + L) \a$, dove $ \a$ \`e un 
vettore con $24$ 
componenti che specificano lo stato; posto $M = \W^t \W$, si ha 
$ m^2 = e^{- \frac{ \f}{2}} \a^tL \W^t(I+L) \W L \a$ e al variare di $M$ il 
vettore $\W L \a$ pu\`o coincidere con uno dei $20$ autovalori nulli di $I+L$, 
dando origine ad uno stato a massa nulla se il numero di stati caratterizzato 
da quelle particolari cariche \`e diverso da zero.
Da un punto di vista geometrico si pu\`o stabilire una corrispondenza  
tra punti nei quali si ha aumento della simmetria nell'eterotica e punti nei 
quali la superficie $K3$ sviluppa delle singolarit\`a, singolarit\`a che 
ammettono una classificazione ADE. Ad esempio, nel caso di $SU(2)$ i bosoni 
$W^{\pm}$ corrispondono a D2-brane avvolte attorno a 2-cicli che degenerano nel 
limite in cui la massa di $W^{\pm}$ va a zero, dando origine ad una superficie 
$K3$ con singolarit\`a $A_1$; la massa delle D2-brane, che dal punto di vista 
delle dimensioni non compatte sono particelle, \`e data dal prodotto della 
tensione della brana per il volume del ciclo, e di conseguenza si annulla nel 
limite degenere. Questa relazione tra singolarit\`a della variet\`a utilizzata 
per compattificare la teoria e gruppi di gauge \`e alla base della F teoria.

Se compattifichiamo la teoria su $S^1$ possiamo cercare di interpretare queste 
singolarit\`a dal punto di vista della IIB su $K3 \times S^1$; in questo caso 
non abbiamo D2-brane che possano avvolgersi attorno ai 2-cicli, ma D3-brane 
che, 
nel limite in cui il ciclo degenera, danno origine non a particelle massless ma 
a $\bf{stringhe}$ $ \bf{tensionless}$ \cite{6d5}, 
gi\`a citate nel capitolo precedente,
che, avvolgendosi a loro volta attorno 
ad $S^1$, generano i bosoni di gauge necessari per l'aumento di simmetria.
Dato che la supergravit\`a di tipo IIB ha un gruppo di simmetria 
$SO(5,21;R)$ su $K3$, si congettura che la IIB su $K3$ abbia come gruppo di 
dualit\`a $SO(5,21;Z)$.


\section{M teoria}
\markboth{}{}

Uno degli aspetti pi\`u interessanti degli sviluppi legati alle dualit\`a \`e 
la crescente evidenza in favore di una teoria in 11 dimensioni le cui compattificazioni
danno luogo 
ad un vasto spazio dei moduli ai bordi del quale si trovano espansioni 
perturbative che riproducono le usuali teorie di stringa \cite{du2,tw1}. 
Uno dei problemi 
pi\`u urgenti consiste nel trovare una formulazione completa di questa teoria 
che renda finalmente possibile lo studio non perturbativo delle 
diverse teorie di stringa, e le presenti in modo chiaro come aspetti diversi 
di un'unica teoria fondamentale. La prima osservazione \`e che 
la supergravit\`a di tipo IIA, che rappresenta il limite di bassa energia  
di una teoria quantistica consistente, pu\`o essere ottenuta come riduzione 
dimensionale della supergravit\`a in 11 dimensioni che invece non rappresenta il 
limite di nessuna teoria di stringhe. L'azione bosonica \`e \cite{11d}
\be
S = \frac{1}{(2 \p )^8} \int d^{11}x \left [ \sqrt{-g} ( R - \frac{G^2}{48}) 
- \frac{1}{(12)^4} \e^{ \m_0 ... \m_10 } C_{\m_0\m_1\m_2}
G_{\m_3\m_4\m_5\m_6}G_{\m_7\m_8\m_9\m_{10}} \right ] \  ,
\ee 
dove $G = dC$ e dove si \`e posta ad uno la lunghezza di Planck in 11 dimensioni, $l_p$.
Compattificando su un cerchio di 
raggio $R \sim \sqrt{ g_{10,10}} $ misurato in unit\`a 10-dimensionali e 
trascurando i modi di Kaluza-Klein, si ottiene la tipo IIA, con le identificazioni
\beq
& & \sqrt{g_{10,10}} = e^{ \frac{\f}{3}} \  , \hspace{1cm}
g^{11}_{\m\n} \sim e^{- \frac{\f}{12}}g_{\m\n}  \  ,  \hspace{1cm}
g^{11}_{\m10} \sim e^{ \frac{\f}{3}} A_\m  \   , \nonumber \\
& & C_{\m\n\r} \sim C_{\m\n\r}  \  ,  \hspace{1cm}  C_{10 \m\n} \sim B_{\m\n}  \  ,
\hspace{1cm} { \rm per } \ \ \ 0 \le \m , \n \le 9   \  .
\eeq
Il limite di forte accoppiamento della IIA $ < \f > \rar \infty$ corrisponde quindi al
limite di decompattificazione $ R \rar \infty $, e questa osservazione \`e stata 
la prima a motivare la congettura \cite{du2,tw1}
che a forte accoppiamento la tipo IIA divenga 
una teoria invariante di Lorentz in 11 dimensioni, che ha come limite di bassa 
energia la supergravit\`a di Cremmer, Julia e 
Scherk \cite{11d} e che \`e stata chiamata M teoria.


\section{Solitoni in teoria di stringhe}
\markboth{}{}

Per la consistenza di queste relazioni di dualit\`a \`e essenziale l'esistenza 
di solitoni che forniscano i gradi di libert\`a necessari perch\`e vi sia  
accordo tra lo spettro e le simmetrie delle due teorie supposte duali.
Questi solitoni si manifestano nel limite di bassa energia come soluzioni delle 
equazioni del moto, localizzate in $(D-p-1)$ dimensioni ed 
indipendenti dalle altre e descrivono pertanto delle p-brane piatte, e in particolare
per p=1 delle corde tese 
infinitamente lunghe. Questi solitoni risolvono le equazioni 
non lineari della teoria e 
sono oggetti non perturbativi, dotati di una carica 
topologica e di un numero finito di moduli. 
Possono essere carichi rispetto ai vari campi della teoria; in particolare, se essi
portano cariche NS si possono costruire
\cite{br1,br2} delle teorie superconformi che a 
grandi distanze coincidono con le
soluzioni delle equazioni del moto di bassa energia.

In teorie supersimmetriche che 
possiedono cariche topologiche, l'usuale algebra di supersimmetria viene 
modificata a causa del contributo di termini di superficie \cite{OW}.
L'algebra completa contiene cariche centrali che consentono 
in alcune teorie di ottenere formule esatte per le masse: l'approssimazione ad 
albero per le fluttuazioni elementari e la formula semiclassica per i dioni 
sono esatte nella teoria quantistica completa.

Consideriamo un esempio bidimensionale:
\be
L= \int d^2x \left [ \frac{1}{2} \part_\m \f \part^\m \f + \frac{i}{2}\bar{\psi} 
\g^\m \part_\m \psi - \frac{1}{2}V^2( \g ) -\frac{1}{2} V^{'}( \f) \bar{\psi} 
\psi \right ] \ . 
\ee
Dalla supercorrente
\be
S^\m = ( \part_\a \f) \g^\a \g^\m \psi + i V(\f) \g^\m \psi  \  ,
\ee
si possono calcolare le cariche
\be
Q_{\pm} = \int dx \left |[\dot{\f} \pm \f^{'}] \psi_{\pm} \mp V(\f)\psi_{\mp} 
\right |  \  ,
\ee
e utilizzando queste espressioni si pu\`o verificare
che $Q_+^2=P_+$ e $Q_-^2=P_-$, ma ora
\be
\{Q_+,Q_-\}=\int dx 2V(\f)\frac{\part\f}{\part x}= \int dx 
\frac{\part}{\part x} (2H(\f))  \  ,
\label{sec}
\ee
dove $H^{'}(\f)=V(\f)$. L'anticommutatore \`e quindi 
una divergenza totale e d\`a un 
risultato diverso da zero in uno stato solitonico. Se indichiamo con $T$ il 
secondo membro della (\ref{sec}), con $V = \l(\f^2-a^2)$ si ottiene
$T= - \int_{- \infty}^{+ \infty} dx \frac{\part}{\part x}( 2a^2 \l \f - 
\frac{2}{3} \l \f^3)$ e con $V= -sen \f$, 
$T=\int_{- \infty}^{+ \infty} dx \frac{\part}{\part x}(2 cos \f)$.
L'algebra modificata $Q_+^2=P_+$ , $Q_-^2=P_-$ , $\{ Q_+,Q_- \}=T$ implica 
quindi che 
$M \ge \frac{|T|}{2}$. Questo limite \`e saturato da stati che soddisfano 
$(Q_+ + Q_-)| \a >=0$ o $(Q_+ - Q_-)| \a >=0$, ed \`e facile vedere che le 
soluzioni solitoniche, per le quali $\dot{\f} =0$ e $\f^{'}= \pm V$, saturano 
il limite e sono quindi invarianti rispetto all'azione di una 
combinazione delle cariche di supersimmetria.

Passiamo ad una teoria di gauge in quattro dimensioni; il termine dell'algebra 
di supersimmetria contenente cariche centrali \`e, utilizzando spinori di 
Majorana,
\be
\{ Q_{\a i},\bar{Q}_{\b j} \} = \d_{ij} \g^\m_{\a \b}P_\m + \d_{\a\b}U_{ij}
+\g_{5 \a \b} V_{ij}  \  ,
\ee
con $U$ e $V$ antisimmetriche. Per N=2 SYM
\beq
L &=& \int d^4x[ -\frac{1}{4}F^2 + \frac{i}{2}\bar{\psi}^a_i\g^\m\D_\m \psi^a_i
+\frac{1}{2}(D_\m S^a)^2 + \frac{1}{2}(D_\m P^a)^2 \nonumber \\
&+& \frac{g^2}{2}tr[S,P]^2 +\frac{ig}{2}\e_{ij}tr([\bar{\psi}^i,\psi^j]S+
[\bar{\psi}^i,\g_5 \psi^j]P)]    \  ,
\eeq
dove $S$ e $P$ sono campi scalari e pseudoscalari nell'aggiunta. La 
supercorrente 
\beq
S_{\m i} &=& tr ( \s^{\a\b}F_{\a\b}\g_\m \psi_i + \e_{ij}D_\a S \g^\a \g_\m 
\psi_j  \nonumber \\
&+& \e_{ij}D_\a P \g^\a \g_5 \psi_j + g \g_\m \g_5 [S,P] \psi_i)  \  ,
\eeq
permette di calcolare le cariche centrali $ U_{ij} = \e_{ij}U$  e 
$V_{ij}= \e_{ij}V$,  dove
\beq
U &=& \int d^3x \part_i \left (S^aF^a_{0i}+P^a \frac{\e_{ijk}}{2}F^a_{jk} 
\right )  \ , 
\nonumber \\
V &=& \int d^3x \part_i \left (P^aF^a_{0i}+S^a \frac{\e_{ijk}}{2}F^a_{jk} 
\right )  \ ,
\eeq
e il limite sulle masse \`e ora $M^2 \ge U^2 + V^2$. Se definiamo le 
cariche elettriche e magnetiche come
\beq
e &=& \frac{1}{<S>} \int d^3x \part_i(S^aF^a_{0i}) \  , \nonumber \\
g &=& \frac{1}{<S>} \int d^3x \e_{ijk}\part_i(S^aF^a_{jk}) \ ,
\eeq
dove la direzione del campo di Higgs $S^a$  determina il sottogruppo di gauge 
corrispondente all'elettromagnetismo, 
per un modello con $G = O(3)$ abbiamo
\be
M \ge <s> \sqrt{e^2 + g^2} \  ,
\ee
e questa formula \`e valida per tutti gli stati noti della teoria, sia 
elementari (fotoni, bosoni massivi, Higgs ... ),  sia solitonici come i dioni.
Che il modello contenga solitoni lo si pu\`o capire osservando che
la parte bosonica (ponendo $P^a$ a zero con una rotazione chirale) non \`e 
altro che il modello di Georgi-Glashow nel quale vennero per la prima volta 
identificati monopoli 
magnetici \cite{thp} e dioni \cite{jz}. 
Per la massa di questi monopoli esiste un limite \cite{Bog}
(limite di Bogomolny, 
pu\`o essere generalizzato senza difficolt\`a ai dioni)
\beq
E^2 &=& \int d^3x \left \{ \frac{1}{2} \left [(B^i_a)^2+(E^i_a)^2+(D^0 S_a)^2+(D^i S_a)^2 \right ]
+V(\f) \right \}  \nonumber \\
&\ge& \int d^3x (B^i_a - D^i_a S)(B^i_a - D^i_a S) + sg \  , 
\label{Bo}
\eeq
ovvero $M \ge sg$.
Prasad e Sommerfield \cite{BPS}
hanno ottenuto una soluzione esplicita per il caso in cui 
la disuguaglianza (\ref{Bo}) \`e saturata;
in generale questo limite viene detto quindi
limite BPS. L'uguaglianza richiede che il potenziale venga posto a 
zero e che sia soddisfatta l'equazione di Bogomolny, un'equazione del primo 
ordine pi\`u semplice da risolvere delle equazioni del moto del 
secondo ordine:
\be
B^i_a = \pm D^i_a S \  .
\ee
Quando il modello di Georgi-Glashow viene immerso in un modello 
supersimmetrico, 
la richiesta che il potenziale sia nullo diviene naturale, in quanto il potenziale 
ammette delle direzioni piatte che non vengono  eliminate dalle correzioni 
quantistiche; inoltre le cariche dei dioni appaiono come cariche centrali 
nell'algebra di supersimmetria e il limite BPS viene reinterpretato 
come una conseguenza di quest'algebra.
Le rappresentazioni di supersimmetria per stati che saturano il limite sono 
pi\`u corte di quelle per particelle massive perch\`e, in modo 
del tutto analogo a quanto avviene per particelle a massa nulla, questi stati 
sono invarianti sotto un certo 
numero di supercariche. Uno stato che saturi il 
limite BPS resta tale, e quindi la formula di massa resta valida nella teoria 
quantistica, se non ci sono meccanismi per generare gli stati che completino un 
multipletto massivo.
Montonen ed Olive \cite{MO}, 
ispirati dalla relazione tra modello di Thirring e 
modello di sine-Gordon nella  quale cariche topologiche ed elementari vengono 
scambiate passando da una descrizione all'altra, hanno 
congetturato che anche in questo caso esista una formulazione
nella quale cariche elettriche e magnetiche vengono scambiate, i monopoli 
divengono bosoni di gauge e i bosoni di gauge emergono come solitoni. La 
congettura, come originalmente formulata, incontrava una serie di difficolt\`a 
: il problema delle correzioni quantistiche al potenziale, il fatto che i 
monopoli hanno spin zero, la variazione della costante d'accoppiamento con la 
scala. Il primo problema viene risolto considerando teorie supersimmetriche; il 
secondo ed il terzo 
considerando il modello con N=4. SYM con N=4 \`e una teoria 
molto particolare: ha una funzione $\b$ nulla e la struttura degli 
zero modi fermionici identifica il supermultipletto del 
monopolo con quello di gauge. Non esiste ancora una prova di questa dualit\`a, 
ma una grande evidenza si \` accumulata in suo favore.
Come vedremo essa pu\`o anche
essere "dedotta" dalla dualit\`a $SL_2(Z)$ della IIB.

Vediamo ora come questa discussione si estende alla teoria di stringhe.
Le teorie di supergravit\`a sono il limite di bassa energia tanto delle 
varie teorie di stringhe quanto della M-teoria e 
possiedono una grande variet\`a 
di soluzioni solitoniche. Alcune di queste ammettono spinori covariantemente 
costanti e preservano parte della supersimmetria originaria, 
rappresentando quindi 
la diretta generalizzazione degli stati BPS gi\`a discussi in 
teoria di campo; anche in questo caso nell'algebra di supersimmetria sono 
presenti cariche centrali che sono ora generalmente delle forme antisimmetriche 
\cite{towcen}. 
Per costruire un solitone partiamo dall'azione di bassa 
energia; consideriamo anzitutto la stringa eterotica  \cite{br1,br2}
\be
S= \frac{1}{\a^{'}} \int d^{10}x \sqrt{-g} e^{-2 \f} \left ( R + 4(\btd \f)^2-
\frac{1}{3}H^2 -\frac{\a^{'}}{30}TrF^2 \right ) \ ,
\ee
dove
$H= dB + \a^{'}( \w^L_3(\W_-)-\frac{1}{30}\w_3^{YM}(A))$ e $\W_{\pm M}^{AB} = 
\w_M^{AB} \pm H_M^{AB}$. Se vogliamo soluzioni che preservino alcune delle 
supersimmetrie, dobbiamo trovare una configurazione di campi bosonici per la 
quale le variazioni dei campi fermionici
\beq
\d \chi &=& F_{MN} \g^{MN} \e \ , \nonumber \\
\d \l &=& \left ( \g^M \part_M \f - \frac{1}{6}H_{MNP} \g^{MNP} \right )\e \ ,  \nonumber \\
\d \psi_M &=& \left ( \part_M + \frac{1}{4}\W_{-m}^{AB}\g_{AB} \right )\e \ ,
\label{sus}
\eeq
si annullino.
Se vogliamo che la soluzione descriva una 5-brana, l'invarianza $SO(9,1)$ 
diviene $SO(5,1) \times SO(4)$ e lo spinore $\e$ si decompone in 
$(4_+,2_+) \oplus (4_-,2_-)$; assumiamo infine che i tensori abbiano componenti 
non nulle solo se tutti gli indici sono trasversi (gli 
indici trasversi sono indicati 
con $n,m=6...9$ sullo spazio tangente e con $\n,\m =6...9$ sulla variet\`a, 
gli indici tangenti con $N,M=0...5$). Si noti che la 
variazione del gaugino si annulla per 
spinori con chiralit\`a positiva di $SO(4)$ se il campo di gauge \`e un 
istantone, ovvero se $F_{mn} = \tilde{F}_{mn}$. Se utilizziamo per il tensore 
metrico e per il tensore antisimmetrico l'ansatz:
\beq
g_{\m\n} &=& e^{2 \f} \d_{\m\n} \  ,  \nonumber \\
H_{\m\n\r} &=& - \e_{\m\n\r}{}{}{}^{\l} \part_{\l} \f \  ,
\eeq
si verifica facilmente che anche la seconda delle (\ref{sus}) si annulla per
$\e \in (4_+,2_+)$. Infine da
\be
\W_{\pm \m mn} = \d_{m \m} \part_n \f - \d_{n \m} \part_m \f \mp 
\e_{\m mn}{}{}{}^{\l}\part_{\l} \f  \ ,
\ee
si vede che $\W_{\pm}$ \`e una connessione $SU(2)$, e la terza equazione \`e 
risolta se $\e \in (4_+,2_+)$ \`e costante. Si pu\`o mostrare che tutte 
le configurazioni di 5-brana sono descritte dall'ansatz considerato. 

La forma funzionale del dilatone si determina imponendo l' identit\`a di 
Bianchi
\be
dH = - \frac{1}{2} \* \Box e^{-2 \f} = \a^{'}(trR \wedge R - 
\frac{1}{30}trF \wedge F) \ .
\ee
Questa equazione pu\`o essere risolta perturbativamente in $\a^{'}$; essendo 
$ \part \f \sim O( \a^{'})$ e quindi $ R \sim O(\a^{'})$, trascurando al primo 
ordine $R \wedge R$ si ottiene
\be
e^{2 \f} = e^{2 \f_0} + \* \a^{'} 
\frac{(x^2+2 \r^2)}{(x^2 + \r^2)^2}+O(\a^{'}) \  .
\ee
La configurazione del campo di gauge \`e
\be
A_\m = - \frac{\S_{\m\n}x^\n}{(x^2+\r^2)} \  ,
\ee
dove $\r$ \`e la taglia dell'istantone.
Questa soluzione \`e detta soluzione di "gauge".
\`E anche possibile costruire una soluzione di 5-brana esatta in $ \ap$, detta 
soluzione "simmetrica", imponendo $dH=0$. Per cancellare tra loro il termine di 
curvatura e l'istantone basta immergere la connessione di spin nella connesione 
di gauge, e quindi $\W_-$ deve essere autoduale. Si verifica 
facilmente che questo \`e vero se $ \Box e^{2 \f}=0$. Questa scelta determina 
una supersimmetria $(4,4)$ del corrispondente modello $\s$, e 
il modello risultante \`e 
una teoria conforme e quindi una soluzione esatta. La soluzione 
dell'equazione di consistenza per il dilatone \`e
\be
e^{2 \f} = e^{2 \f_0} + \sum_{i=1}^{N} \frac{Q_i}{(x-x_i)^2} \  .
\label{dil}
\ee
Queste soluzioni possono essere caratterizzate da due cariche, il numero 
istantonico
\be
\n = \frac{1}{480 \p^2} \int d^4x Tr F \ww F \  ,
\ee
e la carica assionica
\be
Q= - \frac{1}{2 \p} \int_{S^3} H  \ .
\ee
La soluzione di gauge ha $\n=1$ e $Q= 8 \ap$; in generale $Q$ \`e 
quantizzata in multipli interi di $\ap$ perch\`e quando $dH \ne 0$ la 2-forma 
$B$ non \`e globalmente definita sulla sfera asintotica $S^3$. La soluzione 
(\ref{dil}) ha $\n=1$ e $Q=N \ap$. Anche in questo caso la carica \`e 
quantizzata in multipli di $\ap$. La configurazione del campo di gauge 
corrispondente a questa 5-brana simmetrica \`e un istantone di dimensione 
$\r = \sqrt{Q \ap} e^{- \f_0}$. L'andamento del dilatone in (\ref{dil})
\`e molto interessante, perch\`e tende ad 
un valore costante all'infinito e diverge in 
corrispondenza dei poli che sono identificati con la posizione delle 5-brane. 
In un intorno della singolarit\`a si pu\`o effettuare il cambiamento di 
variabili
\be
t = \sqrt{Q} ln \left (\frac{r}{\sqrt{Q}} \right ) \ ,
\ee
ottenendo come elemento di linea approssimato
\be
ds^2 \sim \frac{Q}{r^2}(dr^2 + r^2 d \W^2_3) = dt^2 + Qd \W^2_3 \  ,
\ee
e per gli altri campi
\be
\f = - \frac{t}{\sqrt{Q}}   \  ,  \hspace{1cm}   H = - Q \e_3  \ .
\ee
La geometria globale \`e quindi una collezione di cilindri semiinfiniti 
( "wormholes" ), 
uno per ogni polo, di sezione $S^3$, che si uniscono in 
uno spazio asintoticamente piatto. Per una singola 5-brana la topologia \`e 
quindi $M^6 \times S^3 \times R$.

Sono state costruite molte altre soluzioni, in particolare 
$\it{black-brane}$ \cite{bb} e stringhe \cite{pq1}, cruciali per la dualit\`a 
SL(2.Z) della IIB.
In generale, tanto pi\`u lo spettro di una teoria \`e ricco di forme 
antisimmetriche, tanto pi\`u numerose sono le brane che possono comparire.
Nelle teorie di Tipo II si ha una grande variet\`a di forme 
antisimmetriche e corrispondentemente una grande variet\`a di p-brane; di 
queste p-brane solo recentemente si \`e ottenuta una descrizione in termini di 
teorie  conformi grazie alle D-brane, dato che tutte queste forme provengono 
dal settore $(R)^2$.

Per studiare queste soluzioni consideriamo la parte dell' azione che contiene 
lo scalare di curvatura, il dilatone e la forma antisimmetrica
\be
I = \int d^Dx \sqrt{-g}\left [ R- \frac{1}{2}\btd_M \f \btd^M \f - \frac{1}{2n!}
e^{a \f }F^2_n \right ]  \  ,
\ee
le cui equazioni del moto sono
\beq
R_{MN} &=& \frac{1}{2}\part_M \f \part_N \f + S_{MN} \ ,   \nonumber \\
\part \f &=& \frac{a}{2n!}e^{a \f}F^2  \ , \hspace{1cm}
\btd_{M_1}(e^{a \f}F^{M_1...M_n}) = 0   \ ,  
\eeq
dove
\be
S_{MN} = \frac{1}{2(n-1)!}e^{a \f} \left (F_{M N_2...N_n}F_N{}^{N_2...N_n}- 
\frac{n-1}{n(D-2)}F^2g_{MN} \right )  \   .
\ee
Cerchiamo ora delle soluzioni che possiedano una simmetria 
${ \rm Poincare^{'}}_d \times SO(D-d)$ e 
suddividiamo le coordinate in due gruppi :
$x^M = (x^\m,y^m)$, dove $x^\m$ varia sul volume della brana 
( $\m = 0,...,p=d-1$) e $y^m$ parametrizza lo spazio trasverso 
($m = d,...,D-d$). Utilizziamo per la metrica l'ansatz
\be
ds^2 = e^{2A(r)}dx^\m dx^\n \h_{\m\n} + e^{2B(r)}dy^m dy^n \d_{mn}  \  ,
\ee
con $r^2 = y^my^m$, e per il dilatone poniamo $\f = \f(r)$. Per il campo 
di gauge $F_n$ abbiamo due scelte naturali : la prima consiste nello scegliere 
un potenziale $A_{n-1}$ con componenti non nulle solo lungo il volume della 
brana
\be
A_{\m_1...\m_{n-1}} = \e_{\m_1...\m_{n-1}}e^{C(r)}   \  ,
\ee
che d\`a come campo di gauge
\be
F_{m \m_1...\m_{n-1}} = \e_{\m_1...\m_{n-1}}  \part_m e^{C(r)}  \   .
\ee
Perch\`e questo sia possibile deve essere $n= p+2$. L'altra possibilit\`a \`e
quella di porre
\be
F_{m_1...m_n}= \l \e_{m_1...m_n p} \frac{y^p}{r^{n+1}}  \  ,
\ee
che ha componenti non nulle solo lungo le direzioni trasverse.

Le brane presenti in una determinata teoria sono determinate
dalle forme antisimmetriche che compaiono nello spettro. La M-teoria 
contiene quindi 
una 2-brana elettrica (M2) e una 5-brana magnetica (M5); la Tipo I e 
l'eterotica una 1-brana, che nel primo caso 
\`e la stringa eterotica, ed una 5-brana (le 
stringhe di Tipo I non sono stati BPS); nella IIA abbiamo la stringa 
fondamentale e la 5-brana NS, D0 e D6, D2 e D4. Infine 
nella IIB la D3 che \`e 
autoduale, i D-istantoni (D-1) e D7, D1 e D5. Queste ultime brane possono formare 
stati legati dando origine a 
famiglie infinite di $(p,q)$-stringhe, parametrizzate da una coppia di 
cariche elettriche, e 
di $(p,q)$  5-brane parametrizzate da una coppia di cariche magnetiche.

Possiamo costruire in modo esplicito queste stringhe $(q_1,q_2)$ partendo dall' 
azione (nel frame di Einstein)
\be
S = \int d^{10}x \sqrt{-g} \left [ R - \frac{1}{12}H^t_{\m\n\r}MH^{\m\n\r}
+\frac{1}{4}tr(\part^\m M \part_\m M^{-1}) \right ]   \ ,
\ee
con $B = \pmatrix{B_{NS}\cr B_R}$ , $H = dB$ , 
$M= e^{\f} \pmatrix{|\r|^2 & \chi\cr \chi & 1}$ e $\r = \chi + ie^{-\f}$. 

$S$ \`e invariante rispetto alle seguente azione di $\L \in SL_2R$ 
\be
M \mapsto \L M \L^t \ , \hspace{1cm} B \mapsto \L^{-t}B  \ ,
\ee
e la soluzione \cite{pq1} pu\`o essere quindi generalizzata, facendo uso di una
trasformazione $SL_2(R)$, nella seguente configurazione per una stringa 
$(q_1,q_2)$
\beq
ds^2 &=& A^{-3/4}(-dt^2+(dx^1)^2)+A^{1/4}(dx)^2   \  , \hspace{1cm} 
B_{01}^i = q_i \D^{-1/2}_{q_1q_2}A^{-1}  \  ,   \nonumber \\
\r &=& \frac{i(q_2 \chi + q_1|\p|^2)A^{1/2}-q_2e^{-\f}}
{i(q_1 \chi + q_2)A^{1/2}-q_1e^{-\f}}  \ ,    
\eeq
dove $A= 1 +Q \frac{\D_{q_1q_2}^{1/2}}{r^6}$ e 
\be
\D_{q_1q_2} = e^{\f}|q_1-q_2\r|^2  \ .
\ee
Tutti i moduli si intendono valutati all'infinito; quindi, ad esempio,
$\r = \r( \infty)$. 
La metrica \`e minkowskiana e possiamo quindi calcolare la 
tensione ADM : $T_{q_1q_2} = \D_{q_1q_2}^{1/2}T_1$, dove $T_1$ \`e la tensione 
della stringa fondamentale. Ponendo per semplicit\`a $ \chi = 0 $
\be
T_{q_1q_2} = \left ( gq_1^2 + \frac{q_2^2}{g} \right )T_1  \ ,
\ee
e tornando nella metrica di stringa 
$ ( g^s_{\m\n} = e^{\frac{\f}{2}}g_{\m\n} ) $
si ha $ T_{q_1q_2} = ( q_1 + \frac{q_2^2}{g^2})T_1 $ ovvero $T_{10} = T_1$, e 
$T_{01} = \frac{T_1}{g}$, come ci si aspetta per la stringa fondamentale e una 
D1 brana. La disuguaglianza $T_{p_1+q_1,p_2+q_2} \le T_{p_1p_2}+T_{q_1q_2}$ con 
segno di uguale valido solo se $(p_1,p_2)$ e $(q_1,q_2)$ sono proporzionali, 
mostra che una stringa con $(q_1,q_2)$ relativamente primi \`e assolutamente 
stabile.

Se compattifichiamo la IIB su un cerchio, le eccitazioni delle 
varie stringhe hanno massa
\be
M^2 = \left ( \frac{m}{R} \right )^2 
+ (2 \p RnT_{q_1q_2})^2 + 4 \p T_{q_1q_2} (N_L+N_R)  \ .
\ee
Questa espressione non sembra di grande utilit\`a, in quanto fissato il valore 
di $\r$ solo una delle stringhe \`e debolmente accoppiata. Gli stati 
con $N_L=0$ o $N_R=0$ sono tuttavia stati BPS, e la loro formula di massa \`e 
valida per ogni valore dei moduli; otteniamo cos\`\i \, un'espressione 
che descrive 
gran parte dello spettro e pu\`o essere utilizzata per confrontare 
l'interpretazione di questi stati in descrizioni duali e per 
legare tra loro i vari parametri. In particolare possiamo collegare 
la M teoria su $T^2$ e la IIB su $S^1$ e interpretare tutti questi stati BPS 
dal punto di vista undicidimensionale. Osserviamo anzitutto che ogni stato
\`e descritto, oltre che dagli oscillatori, da tre numeri $m$ , $l_1$ e $l_2$ 
con $(l_1,l_2) = n (q_1,q_2)$; infatti il termine di winding pu\`o essere 
scritto $N^2 T^2_{q_1q_2} = e^{\f}|l_1-l_2 \r|^2 T_1$. 

Caratterizziamo ora il toro della M-teoria con il parametro $ \t$ e l'area 
$A = (2 \p R_{11} )^2 \t_2 $, misurata con la metrica in undici dimensioni; 
dato 
che non si conosce una teoria quantistica di membrane non possiamo confrontare 
le fluttuazioni della M2 con quelle della stringa, ma possiamo cercare di 
collegare i modi di Kaluza-Klein e gli avvolgimenti nelle due teorie; se 
$z = x + iy $ \`e la coordinata sul toro, una funzione a un valore pu\`o essere 
scritta 
\be
\f_{l_1l_2} \sim e^{\frac{i}{R_{11}} \left [ xl_2-\frac{y}{\t_2}(l_2\t_1-l_1) 
\right ] }  \ ,
\ee
e d\`a alla massa al quadrato il contributo 
\be
\frac{1}{R_{11}^2} \left [ l_2^2+\frac{1}{\t_2^2}(l_2\t_1-l_1)^2 \right ]=
\frac{|l_1-l_2\t|^2}{(\t_2 R_{11})^2} \  .
\ee
Se M2 si avvolge $m$ volte sul toro questo d\`a un contributo $(AT_2^Mm)^2$, 
dove $T_2^M$ \`e la tensione di M2. Le eccitazioni KK della brana possono 
essere messe in relazione con il termine di avvolgimento della stringa, 
identificando $\t$ con $\r$  e 
il termine di avvolgimento della brana con le eccitazioni KK della 
stringa. Per rendere l'identificazione precisa dobbiamo ricordare che le due 
quantit\`a sono misurate con metriche diverse; se poniamo 
$g^M_{\m\n} = \b^2 g_{\m\n}$, da
\be
AT_2^M = \frac{\b}{R} \  ,  \hspace{1cm}
\frac{1}{\t_2 R_{11}} = 2 \p R \frac{T_1 \b}{\sqrt{\t_2}}  \ ,
\ee
ricaviamo
\be
\b = A^{1/4} \sqrt{\frac{T_2^M}{T_1}}  \  ,  \hspace{1cm} 
(T_1 R^2)^{-1} = A^{3/2}T_2^M  \ ,
\ee
Se identifichiamo la M teoria su $T^2$ con la IIB su $S^1$ ponendo $\r = \t$ 
otteniamo un'in\-ter\-pre\-ta\-zio\-ne 
geometrica della dualit\`a $SL_2(Z)$ di quest'
ultima in termini del gruppo dei diffeomorfismi di un toro; 
inoltre $R \sim A^{-3/4}$. Si pu\`o dare un'interpretazione in M-teoria anche 
degli altri stati della IIB e stabilire altre relazioni tra i parametri delle 
due teorie. Una stringa $(q_1,q_2)$ dovrebbe, ad esempio, corrispondere 
ad una 2-
brana avvolta attorno ad un ciclo di omologia $(q_1,q_2)$ del toro; in M-teoria 
la tensione di questa stringa \`e
\be
T^M_{q_1q_2}= 2 \p R_{11}|q_1-q_2\t|T_2 = (A \D_{q_1q_2})^{1/2}T^M_2  \  ,
\ee
e quindi $T_{q_1q_2} = \b^{-2}T^{M}_{q_1q_2} + \D_{q_1q_2}^{1/2}T_1$, in accordo 
con quanto visto precedentemente. La M2 dovrebbe corrispondere a D3 avvolta 
lungo il cerchio, e procedendo in modo analogo a quanto fatto finora si ottiene
\be
T^M_2 = \b^{3} 2 \p R T_3 \rar T_3 = \frac{1}{2 \p} T^2_1   \ ,
\ee
una relazione tra i paramteri della sola IIB. Possiamo confrontare con D3 anche 
M5 avvolta sul toro:
\be
T^M_5 A = \b^4 T_3 \rar T^M_5 = \frac{1}{2 \p} (T_2^M)^2  \ .
\ee
Infine avvolgendo una M5 lungo un ciclo $(q_1,q_2)$ e confrontando con D5 
avvolta sul cerchio
\be
T^M_5 = 2 \p R_{11}|q_1-q_2\t| = \b^5 2 \p R T_{5(q_1q_2)} \rar
T_{5(q_1q_2)} = \frac{1}{(2\p)^2} \D_{q_1q_2}^{1/2} T_1^3   \ .
\ee
Osserviamo che dalla formula precedente una brana con carica puramente NS ha 
una 
tensione che scala come $\frac{1}{g^2}$, mentre quella di una brana con carica
Ramond come $\frac{1}{g}$, come ci si aspetta per la brana NS e la D5 brana.
\`E interessante notare che questa relazione tra M-teoria e Tipo IIB lega tra 
loro anche la supergravit\`a in 11 dimensioni e la supergravit\`a di Tipo IIB, 
ovvero una teoria non chirale ed una chirale: uno dei principali problemi della 
teoria di Cremmer e Julia consisteva proprio nell'assenza di un vuoto chirale, 
ma ora  
il passaggio dalla supergravit\`a alla M-teoria e
l'introduzione delle 
brane ha portato a riconsiderare questa situazione.

Un ragionamento del tutto analogo lega i parametri 
della M-teoria a quelli delle teorie eterotiche e della Tipo I; in particolare 
la teoria $E_8 \times E_8$ compattificata su di un cerchio con $L_H=2\p R_H$ 
viene identificata con la M-teoria compattificata su di un cilindro con
$L_1$ e $L_2= 2\p R_2$. Dal punto di vista della teoria di stringhe abbiamo 
brane con p=1,5 e, avvolgendo queste sul cerchio, anche con p=0,4; dal punto di 
vista della M-teoria queste brane devono derivare da M2 e M5 avvolte o meno sul 
cilindro. Ora M2 non pu\`o non avvolgersi in quanto sui bordi non \`e presente 
una 3-forma e quindi lo stato non \`e BPS, pertanto d\`a origine a p=0,1 a 
seconda che avvolga tutto il cilindro o che si estenda solo lungo $L_1$. 
Procedendo come per la IIB otteniamo
\be
\g_H = \frac{L_1}{L_2} \hspace{1cm} 
L_1 L_2^2 T^M_2 = \left ( \frac{T_1 L_H^2}{2 \p} \right )^{-1}  \  .
\ee
Fissato il modulo $\frac{L_1}{L_2}$ vediamo che $L_H \sim A^{-3/4}$ e inoltre
\be
T^H_5 = \frac{1}{(2\p )^2} \left ( \frac{L_2}{L_1} \right )^2(T^H_1)^3  \  .
\ee
Nell'eterotica 
$T^h_1$ \`e costante e quindi $T^H_5 \sim g_H^{-2}$, mentre nella Tipo I 
$T^H_1 \sim g_H^{-1}$ e quindi $T^H_5 \sim g_H^{-1}$, in accordo con il fatto 
che entrambe sono D-brane. 

Abbiamo definito le D-brane come iperpiani sui quali le stringhe possono 
terminare; questo \`e un primo esempio di configurazioni pi\`u generali  
di brane che terminano su altre brane \cite{br5}. 
L'esistenza di queste configurazioni \`e 
legata alla possibilit\`a o meno di conservare la carica associata con la 
brana; infatti il bordo della brana che si collega ad un altra appare come una 
sorgente nella teoria di volume di quest'ultima, e 
deve essere possibile dare ai 
campi una configurazione che permetta alla carica di fluire.
Consideriamo la D5 nella IIB. La teoria di volume contiene un campo di gauge 
che pu\`o essere dualizzato in una 3-forma; ora se una D3 brana termina sulla 
D5 il suo bordo appare come una 2-brana contenuta nel volume, e questa pu\`o 
essere consistentemente accoppiata alla 3-forma: questo porta a concludere che 
una D3 pu\`o terminare su una D5 e pi\`u in generale, utilizzanto la 
dualit\`a $SL_2(Z)$ rispetto alla quale D3 \`e fissa, su una 5-brana con carica 
$(q_1,q_2)$ arbitraria. Analogamente, su M5 abbiamo un multipletto tensoriale 
che 
contiene una 2-forma autoduale che permette ad M2 di terminare su M5.
Una configurazione di brane parallele \`e stabile; supponiamo che tra due p-
brane parallele sia tesa una q-brana : nel limite in cui la distanza 
$l \rar 0 $ si ottiene una $q-1$ brana in uno spazio p-dimensionale con 
tensione $T = l T_q$ che tende a sua volta a zero. Un'eccezione \`e il caso 
della stringa eterotica come limite di M2 tese lungo l'intervallo: in questo 
caso si pu\`o effettuare un riscalamento di Weyl che assicura tensione finita.
\`E possibile, utilizzando configurazioni di brane parallele, 
legare la dualit\`a
della IIB ad un'altra dualit\`a $SL_2(Z)$ pi\`u familiare, 
quella di SYM in $d=4$ con 
$N=4$. Consideriamo infatti due D3 parallele separate da una distanza $l$; 
le stringhe tese tra di esse appaiono nella teoria di volume come cariche 
elettriche, ma una trasformazione di dualit\`a le trasforma in stringhe 
$(q_1,q_2)$ che danno per la massa degli stati pi\`u leggeri
\be
M = lT_{q_1q_2} = l \left ( q_1^2 + \frac{q_2^2}{g^2} \right )^{1/2} T_1  \  ,
\ee
in accordo con la formula BPS per i dioni in N=4 SYM. Questa configurazione 
pu\`o essere ottenuta in M-teoria da due M5 parallele con una M2 tesa tra loro, 
compattificando su un toro ed avvolgendo M2 lungo un ciclo $(q_1,q_2)$.
Un'altra caratteristica interessante di queste famiglie di stringhe \`e la 
possibilit\`a di realizzare con esse delle $\bf{giunzioni}$ $\bf{di}$ 
$\bf{stringhe}$, 
stati BPS dati dall'unione di tre di queste stringhe
con cariche $(q_i,p_i)$. Per la conservazione della 
carica si deve avere $ \sum q_i = \sum p_i = 0 $ e per la stabilit\`a la somma 
vettoriale delle tensioni deve dare zero. Ancora una volta dal punto di vista 
della M-teoria questi stati provengono da una M2
che avvolge ciascun ramo su di un ciclo diverso di 
omologia, 
il che \`e possibile solo se le precedenti condizioni sono soddisfatte. 
Utilizzando queste giunzioni possono essere realizzate delle configurazioni 
pi\`u complicate come delle reti di stringhe; inoltre sembra che utilizzando 
questi oggetti non perturbativi sia possibile costruire in modo ragionevolmente
esplicito modelli di stringhe 
aperte con gruppi di gauge eccezionali \cite{Zwi}.

Lo studio delle brane e delle teorie di campo sulle brane ha messo in luce 
varie regole che caratterizzano la loro dinamica. Un fenomeno particolarmente 
interessante \`e la creazione di una terza brana tra due brane parallele 
che si attraversano; questo effetto \`e stato considerato per la prima volta in 
\cite{bft1}, per la consistenza della teoria di campo effettiva, e giustificato 
con argomenti basati su anomalie in \cite{db6}.


\section{Spettro degli stati BPS}
\markboth{}{}

Abbiamo visto che
gli stati BPS formano multipletti dell'algebra di supersimmetria pi\`u corti 
rispetto a quelli degli stati massivi in quanto essi sono annullati da un certo 
numero di supercariche; un'altra loro importante caratteristica \`e che la 
supersimmetria lega le loro masse alle cariche che essi hanno rispetto ai diversi 
campi della teoria. Per una generica algebra di supersimmetria estesa con $N$ 
supercariche possiamo scrivere
\be
\{Q_\a,Q_\b\} = f_{\a\b}(m,Q_i,\{y\}) \   , 
\ee
dove $f_{\a\b}$ \`e funzione della massa e delle cariche dello stato e dei 
moduli della teoria. Se $f_{\a\b}$ \`e non singolare, ruotando le supercariche 
si ottiene un'algebra di Clifford N-dimensionale che ha una rappresentazione 
$2^{\frac{N}{2}}$-dimensionale, e in questo caso non ci sono stati BPS. 
Se $f_{\a\b}$ ha rango $M$, si pu\`o porre $f_{\a\b} = \d_{\a\b} $ per 
$ 1 \le \a , \b \le M $ e $ f_{\a\b} = 0 $ se $ \a >M $ o $ \b >M$. In questo 
caso otteniamo una rappresentazione irriducilbile imponendo che gli stati siano 
annullati da $Q_\a$ con $\a>M$ ed abbiamo un'algebra di Clifford di dimensione 
$M$ con una rappresentazione di dimensione $2^{\frac{M}{2}}$; si hanno cos\`i 
stati BPS appartenenti a multipletti tanto pi\`u corti quanto pi\`u numerose 
sono le cariche di supersimmetria che li lasciano invariati.
Risolvendo il vincolo sugli zeri di $f_{\a\b}$ si ottiene la formula che lega 
$m$, le $Q_i$ e i moduli $ \{y\}$ per gli stati BPS. La degenerazione di questi 
stati non dipende dai moduli e poich\`e tra questi figura la costante di 
accoppiamento, resta la stessa sia a debole che a forte accoppiamento; per 
verificarlo bisogna assumere che lo spettro vari in modo continuo al variare dei 
moduli, perch\`e in questo caso 
un multipletto corto non pu\`o trasformarsi in un altro tipo di 
multipletto. \`E importante che la massa di questi stati non sia in 
prossimit\`a di uno spettro continuo perch\`e in questo caso non avremmo una 
procedura ben definita per contare gli stati. Quindi la massa di uno stato BPS 
deve essere strettamente minore di quella di ogni insieme di due o pi\`u stati 
caratterizzati dagli stessi numeri quantici.

Un primo modo di mettere alla prova una congettura di dualit\`a \`e quindi il 
seguente:
cominciamo con l'identificare gli stati BPS presenti nello spettro perturbativo 
e vediamo a quali stati BPS essi corrispondono tramite la trasformazione di 
dualit\`a; stati individuati in questo modo e non presenti nello spettro 
perturbativo sono predetti dalla dualit\`a e si ottiene una prima evidenza per 
la congettura verificando che siano effettivamente presenti nella teoria.
 
Applichiamo questo metodo allo spettro BPS della IIB su $S^1$; con $32$ 
supercariche i multipletti lunghi hanno $(256)^2$ stati, quelli corti 
$16 \cdot 256 $ e quelli ultracorti $ 256$. Gli stati dello spettro 
perturbativo sono caratterizzati da 
$ k_{L,R} = \frac{1}{\sqrt{2}}
( \frac{k \l_2^{1/4}}{R} \mp \frac{wR}{ \l_2^{1/4}} )$  con $k,w$ interi, ed 
hanno massa
\be
m^2 = \frac{2}{\sqrt{ \l_2}}( k_L^2 + 2N_L )= 
 \frac{2}{\sqrt{ \l_2}}( k_R^2 + 2N_R )  \   .
\label{mas}
\ee
Gli stati con $N_L=N_R=0$ sono ultracorti e da (\ref{mas})  si vede che hanno 
$k=0$ 
oppure $w=0$, ovvero sono stati o con puro momento o con puro avvolgimento. Gli 
stati con $N_L=0$ oppure $N_R=0$ sono corti; se scegliamo $N_R=0$ da 
(\ref{mas}) 
vediamo che $N_L=kw$. Calcoliamo la degenerazione; \`e facile vedere che 
c' \`e un solo multipletto ultracorto, e infatti nel settore sinistro abbiamo 
$16$ stati - $8$ Ramond e $8$ Neveu-Scwhartz - e analogamente nel settore 
destro, per un 
totale di $256$ stati. Per quanto riguarda gli stati ultracorti, la loro 
degenerazione $d(N_L)$ \`e data da 
\be
\sum_{N_L}d(N_L)q^{N_L} = \frac{1}{16} \prod_{n=1}^{\infty}
\left ( \frac{1+q^n}{1-q^n} \right )^8  \   .
\ee
Denotiamo lo stato ultracorto $k=0,w=1$, che porta un'unit\`a di carica 
ripetto a $B_{NS}$ con $ \pmatrix{1\cr 0}$; 
una trasformazione $ \pmatrix{ p & q\cr r & s} \in SL_2Z $ lo trasforma in uno
stato $ \pmatrix{p\cr r} $, con $p,r$ relativamente primi. La predizione della 
dualit\`a \`e allora che per ogni coppia di interi relativamente primi lo 
spettro deve contenere un solo multipletto ultracorto con carica $(p,q)$ e con 
massa, ottenuta dall'algebra di supersimmetria, 
$ m^2 = \frac{R^2}{\l_2}|r \l - p|^2 $, invariante rispetto a $SL_2(Z)$.
Pi\`u in generale, caratterizziamo uno stato con le cariche rispetto a 
$B^{NS}_{9 \m}$,$B^{R}_{9 \m}$ e $G_{9 \m}$ e rappresentiamolo con un vettore
$(n, m, k)$. Gli stati $(0, 0, 0)$ , $(n, 0, 0)$ e $(0, 0, k)$ sono nello 
spettro 
perturbativo e sono ultracorti; il primo e il terzo sono invarianti mentre il 
secondo diviene $(np, nq, 0)$, e quindi per $(p,q)$ relativamente primi si deve 
avere un solo multipletto ultracorto. Gli stati $(n, 0, k)$ con $n,k \neq 0$
sono corti e divengono $(np, nq, k)$ con $(p,q)$ relativamente primi; indicata 
con $d(k,p,q)$ la degenerazione di questo multipletto, possiamo legarlo a
mezzo di
trasformazioni in $SL_2(Z)$ al multipletto con $(1, 0, k)$ che ha degenerazione 
$d(k)$ ( $N_L=k$ ); in altri termini $d(k,p,q)$ non dipende da $(p,q)$ e quindi
$\sum_{N_L}d(k,p,q)q^{k} = \frac{1}{16} \prod_{n=1}^{\infty}
( \frac{1+q^n}{1-q^n} )^8$.

Verificare queste predizioni della dualit\`a richiede l'identificazione degli 
stati con i numeri quantici appropriati. La difficolt\`a consiste nel fatto che 
lo spettro perturbativo non contiene stati carichi rispetto ai campi RR; una 
prima soluzione consiste nel costruire soluzioni delle equazioni del 
moto della supergravit\`a con le cariche opportune e nel calcolare la 
degenerazione degli stati quantizzando le fluttuazioni intorno a tali 
configurazioni di campo; questa procedura si \`e rivelata utile in alcuni casi, 
ad esempio per $Het/T^6$, ma generalmente queste soluzioni sono singolari e 
diviene difficile studiarne le fluttuazioni. Uno strumento essenziale sono 
allora le D-brane, che forniscono una descrizione microscopica delle brane che 
si presentano come oggetti estesi nelle equazioni della supergravit\`a; 
identificando gli stati non perturbativi con D-brane \`e facile effettuare il 
calcolo delle degenerazioni utilizzando i soliti metodi di teoria delle 
stringhe. Nel nostro caso una D-stringa porta un'unit\`a di carica rispetto a 
$B_R$,  e di conseguenza una D-stringa avvolta attorno ad $S^1$ \`e un candidato 
per lo stato $ \pmatrix{ 0\cr 1}$. Lo si pu\`o verificare 
calcolando lo spettro delle eccitazioni della D-stringa.
Siamo quindi portati ad identificare gli stati della forma $\pmatrix{p\cr 1}$
con D-stringhe che portano cariche diverse rispetto a $B$, e possiamo verificare
la correttezza di questa identificazione calcolando i numeri quantici e la
degenerazione di questi stati.
Poich\`e $X^9$ \`e una coordinata compatta e vogliamo studiare multipletti 
ultracorti, possiamo ricondurci ad un problema di meccanica quantistica, 
restringendoci al settore degli zero-modi; gli stati sono caratterizzati dai 
momenti coniugati alle $8$ coordiante bosoniche,  
dall'intero $y$ che specifica la carica rispetto 
a $B$ e da $16$ coordinate fermioniche che determinano una degenerazione pari a 
$256$, proprio come previsto dalla dualit\`a.
\`E ora immediato identificare gli stati $\pmatrix{p\cr q}$ con stati legati di 
stringhe fondamentali e D-stringhe \cite{br7}. 
Nello studio dei multipletti corti con momento lungo $S^1$ non possiamo 
effettuare la riduzione dimensionale. Gli stati BPS corrispondono a 
configurazioni con solo modi destri o sinistri eccitati e la degenerazione 
$d(k,p,r)$ si determina calcolando in quanti modi $k$ pu\`o essere suddiviso 
tra le varie eccitazioni sinistre bosoniche e fermioniche.
Tutto \`e in perfetto accordo con quanto predetto dalla dualit\`a $SL_2(Z)$.

Passiamo ora alla relazione tra M teoria e Tipo IIA. 
Per dare evidenza a questa 
congettura cerchiamo di individuare  nella teoria di 
stringhe di tipo IIA gli stati di Kaluza Klein 
che derivano dalla compattificazione della 
teoria in undici dimensioni; dal punto di vista della supergravit\`a si tratta 
di stati BPS che formano multipletti ultracorti con degenerazione pari a $256$ 
e carica $ \frac{k}{R}$ che rappresenta il momento lungo $S^1$.  
Nella teoria di stringhe $g_{\m 10}$ corrisponde al vettore di Ramond e quindi 
gli stati cercati, per $k=1$, sono le D0-brane. La dinamica delle  
coordinate collettive di una sola D0-brana \`e data dalla riduzione 
dimensionale di SYM in $9+1$ dimensioni con gruppo di gauge $U(1)$ a $0+1$ 
dimensioni, un modello di meccanica quantistica supersimmetrica;
la degenerazione \`e determinata dai sedici zero-modi fermionici.
Il problema per $k>1$ \`e pi\`u sottile, perch\`e 
stati legati corrispondono a stati 
fondamentali normalizzabili dell'hamiltoniana che descrive il sistema di $k$ 
D0-brane, sempre riduzione di SYM ma questa volta con gruppo di gauge 
$SU(k)$. L'energia di uno stato legato con carica $k$ coincide con quella di 
$k$ stati con carica 1 a riposo : si tratta di stati legati alla soglia, posti 
all'inizio di uno spettro continuo.
Il problema si semplifica se si considrera la M teoria su $T^2$ che \`e il 
limite della IIA su $S^1$; la massa degli stati KK 
$ \sqrt{ (\frac{k_1}{R_1})^2+(\frac{k_2}{R_2})^2}$ \`e strettamente minore 
della somma delle masse determinate da $k_1$ e $k_2$, con 
$k_1 , k_2$ relativamente primi e possiamo quindi 
verificare l'esistenza di tutti i 
multipletti con la degenerazione opportuna.
Con una T-dualit\`a possiamo inoltre 
collegare $M/T^2$ e $IIB/S^1$, ottenendo 
un'interpretazione geometrica del gruppo di dualit\`a della $IIB$ come gruppo 
dei diffeomorfismi globali del toro della M teoria.
La IIA \`e invariante di Lorentz in dieci dimensioni, mentre 
la IIB possiede una 
simmetria che scambia tra loro la direzione perturbativa e quella non 
perturbativa $( R_{10}$  e $ R_{11} )$, e quindi nel limite in cui il volume di 
$T^2$ va all'infinito recuperiamo l'invarianza di 
Lorentz in undici dimensioni, come ci si aspetta per la M teoria.

Compattificando la M teoria su diverse variet\`a a D dimensioni 
si ottengono nuovi vuoti 
della teoria di stringhe, non costruibili perturbativamente e che tipicamente 
rappresentano limiti della tipo IIA compattificata a D-1 dimensioni; 
le propriet\`a 
di questi vuoti possono essere studiate legandoli tramite dualit\`a a vuoti 
perturbativi.

Una relazione interessante \`e l'identificazione 
$ \bf{M/I} \bf{ \sim} \bf{E_8 \times E_8}$ \cite{du3,du3b}
dove I \`e l'in\-ter\-val\-lo $ S^1/Z_2 $; 
dal punto di vista della M teoria i campi di gauge vivono sui bordi dello 
spazio tempo, mentre la gravit\`a si propaga nel volume mettendo in contatto i 
due settori.

In M teoria, come nella IIA e nella IIB, esistono meccanismi di aumento 
della simmetria legati a brane che divengono non massive a causa della 
degenerazione della variet\`a sulla quale la teoria \`e stata compattificata.


\section{Relazioni tra le dualit\`a}
\markboth{}{}

\`E possibile ridurre tutte queste dualit\`a ad un insieme minimale facendo una 
serie di semplici ipotesi riguardo al loro comportamento in seguito a 
compattificazioni \cite{sen2}. 
Abbiamo gi\`a visto come la combinazione di una simmetria 
non perturbativa ( S-dualit\`a della IIB ) con una simmetria 
perturbativa ( T-dualit\`a ) consenta di dedurre la U-dualit\`a delle 
teorie di tipo II. Consideriamo ora un esempio differente : sappiamo che $Het/
T^4$ e $IIA/K3$ sono duali, e 
se compattifichiamo ulteriormente le due teorie su 
$T^2$ otteniamo $Het/T^6$ e $IIA/K3 \times T^2$, entrambe con gruppo di 
autodualit\`a $O(6,22) \times SL_2(Z)$ e con gruppi perturbativi di T-dualit\`a 
rispettivamente $O(6,22)$ e $O(4,20) \times SL_2(Z) \times SL_2(Z)$.
Se assumiamo che le due teorie restano 
duali anche dopo questa nuova compattificazione, si pu\`o verificare che sotto 
la trasformazione di dualit\`a il gruppo $O(6,22)$ della $IIA$ corrisponde a 
quello perturbativo dell'eterotica e, viceversa, il gruppo $SL_2(Z)$ 
dell'eterotica corrisponde a quello di T-dualit\`a della tipo IIA.
In altri termini gruppi di autodualit\`a non perturbativi 
possono essere collegati, tramite un'altra trasformazione di dualit\`a, 
a gruppi perturbativi del modello duale; in questo modo \`e ad esempio 
possibile, come abbiamo accennato, giustificare l'autodualit\`a di 
$Het/T^6$ e $IIA/K3 \times T^2$ dalla dualit\`a 
$Het/T^4$  $IIA/K3$.
Una tecnica molto importante \`e la cosiddetta $\bf{approssimazione}$ 
$\bf{adiabatica}$ \cite{dualp}: 
indichiamo con $(A,K_A)$ una teoria di tipo $A$ 
compattificata sulla variet\`a $K_A$ e supponiamo che $(A,K_A)$ e $(B,K_B)$ 
siano duali; costruiamo due nuove variet\`a fibrando su una base $M$ 
$K_A$ e $K_B$. Se i moduli delle due teorie variano lentamente 
sulla base ci si aspetta, applicando la dualit\`a fibra per fibra, che la 
coppia $(A,K_A \times M)$ e $(B,K_B \times M)$ sia a sua volta duale. 
Supponiamo ora che $M$ abbia una involuzione $g$, indichiamo con $h_A$ e 
$h_B$ due involuzioni rispettivamente di $A$ e di $B$ e costruiamo i due 
orbifolds $K_A \times M / g \cdot h_A$ e $K_B \times M / g \cdot h_B$; nel 
punto generico la fibra \`e $K_A$ ($K_B$) mentre se il punto $Q$ \`e fisso 
($g(Q)=Q$) la fibra degenera in $K_A/h_A$ ( $K_B/h_B $), e per quanto osservato 
precedentemente ci aspettiamo che le teorie restino duali. Consideriamo come 
esempio $(IIB,R^{10})$ con $h_A=(-1)^{F_L}$ e l'immagine della $IIB$ tramite 
$S$, l'elemento di $SL(2,Z)$ che manda $\l$ in $- 1/ \l$, con $h_B= \W$; 
sia inoltre $M=T^4$ e $g=I$, l'involuzione che cambia segno 
a tutte le coordinate del toro. Si pu\`o verificare che a $(-1)^{F_L}$ 
corrisponde tramite $S$ l'operazione di parit\`a sul worldsheet $ \W$;
infatti
le azioni di $S(-1)^{F_L}S^{-1}$ e $ \W$ coincidono sugli stati non massivi 
e l'azione di $S$ sugli stati massivi, che non \`e 
nota, pu\`o essere definita proprio identificando $S(-1)^{F_L}S^{-1}$ e $ \W$. 
Effettuando una trasformazione di T-dualit\`a sulla prima teoria IIB 
si ottiene, osservando 
che sotto T-dualit\`a $(-1)^{F_L} \cdot I$ corrisponde ad $I$,
la $IIA$ su $T^4/I$ che \`e un caso limite di $K3$. Se effettuiamo invece sulla 
seconda teoria una trasformazione di T-dualit\`a su tutte le coordinate del 
toro ed osserviamo che $\W \cdot g \mapsto \W$, otteniamo la TipoI su $T^4$ che 
\`e S-duale di $Het/T^4$; in questo modo abbiamo stabilito, utilizzando 
l'autodualit\`a della $IIB$ e la dualit\`a TipoI-eterotica, la dualit\`a 
$Het/T^4 - IIA/K3$ in un punto particolare dello spazio dei moduli.
Un ultimo argomento utile a legare tra loro le varie dualit\`a \`e il seguente: 
se una teoria ha un gruppo di autodualit\`a $G$ quando compattificata su 
$M_1 \times M_2$ e $H$ \`e un sottogruppo che lascia fisso il volume di $M_2$, 
la teoria su $M_1$ che si ottiene nel limite in cui $Vol(M_2) \rar \infty$ 
dovrebbe avere come gruppo di autodualit\`a $H$. Consideriamo 
ad esempio l'eterotica su 
$T^2$, che ha come gruppo di T-dualit\`a $O(2,18;Z)$, e 
sia $SL_2(Z) \times SL_2(Z)$ 
un sottogruppo, con un fattore associato a diffeomorfismi globali di $T^2$  
ed un altro alle trasformazioni $R \mapsto 1/R $; il primo di questi \`e un gruppo 
di dualit\`a anche per la TipoI su $T^2$ e quindi per la IIB su $T^2/ \W$.
Una T-dualit\`a su entrambe le coordinate del toro trasforma $\W$ in 
$(-1)^{F_L} \cdot \W \cdot I$, e si giunge quindi alla IIB sul tetraedro $T^2/I$ con 
un'ulteriore involuzione $(-1)^{F_L} \cdot \W$ quando si gira attorno ai punti 
fissi. La IIB sul tetraedro ha come sottogruppo di T-dualit\`a 
$SL_2Z \times SL_2Z$ e nel limite in cui il volume diverge uno dei due fattori 
diviene parte dei diffeomorfismi della IIB e 
l'altro il gruppo di autodualit\`a.
Vediamo cos\`\i \, che la struttura perturbativa della teoria e ipotesi
ragionevoli sul comportamento delle trasformazioni di dualit\`a al variare 
della geometria quali l'approssimazione adiabatica, l'ereditariet\`a del gruppo 
di dualit\`a e il limite di decompattificazione, consentono di 
ricondurre tutte le congetture di 
dualit\`a ad una sola, la dualit\`a TipoI-Eterotica 
in dieci dimensioni.


\section{F teoria}
\markboth{}{}

Con il nome di F teoria si indica una classe notevole di compattificazioni non 
perturbative della Tipo IIB che possono avere una descrizione di grande 
eleganza in termini di una teoria essenzialmente geometrica formulata in dodici 
dimensioni \cite{Fth}.
La costruzione si basa sulla naturale interpretazione 
geometrica del campo complesso $ \l = a + ie^{- \frac{ \f}{2}} $
come modulo di un toro, in modo da giustificare l'azione proiettiva del gruppo 
modulare su di esso discussa in precedenza.
Questo porta a considerare una variet\`a ausiliaria
in dodici dimensioni con segnatura $(10,2)$, alla quale in generale non viene 
data un'interpretazione fisica come nel caso delle 11 dimensioni della M 
teoria. Definiamo in modo preciso queste compattificazioni; sia quindi $M$ una 
variet\`a che ammette una fibrazione ellittica; sia  $B$  
la base, con coordinata $z$, e indichiamo con $\t (z)$ una famiglia 
di superfici di Riemann fibrata su $B$. La F teoria su $M$ \`e definita come la 
tipo IIB su $B$ con $ \l (z) = \t (z)$; la novit\`a di queste compattificazioni 
sta proprio nel fatto che il dilatone e lo scalare di Ramond hanno valori di 
aspettazione variabili. Osserviamo che $\t$ determina solo la struttura 
complessa del toro, e quindi la teoria non dipende da come la dimensione della 
fibra vari sulla base. Inoltre, $\t$ \`e definito a meno di trasformazioni in 
$SL_2(Z)$ quando si percorre 
un cammino chiuso in $B$. Se compattifichiamo su 
un cerchio di raggio R, poich\`e 
$IIB/S^1$ con un fissato valore di $\l$ \`e duale a 
$M/T^2$ con $\l$ modulo del toro possiamo identificare, ragionando fibra per 
fibra, la F teoria su $M \times S^1$ con la M teoria su $M$. Possiamo quindi 
determinare propriet\`a dei vuoti 
della F teoria da quanto \`e noto sulla M teoria considerando il
limite $ R \rar \infty$ . Una famiglia di curve ellittiche pu\`o essere 
descritta dall'equazione
\be
y^2 = x^3 + f(z)x + g(z)  \  ,
\label{efam}
\ee
dove $f$ \`e un polinomio di grado $8$ in $z$ e $g$ \`e un polinomio 
di grado $12$. Il modulo \`e definito implicitamente dalla funzione $j$ :
\be
j(\t) = \frac{4(24f)^3}{27g^2+4f^3} \ .
\ee
Se 
scegliamo come base $CP^1$, la variet\`a $M$ che si ottiene tramite fibrazione 
ellittica \`e $K3$, e quindi $F/K3 \sim IIB/CP^1$ con $\l (z) = \t (z)$. Si 
pu\`o calcolare la metrica sulla base $CP^1$ nel limite in cui il suo volume 
\`e grande utilizzando l'azione effettiva di bassa energia, e il risultato \`e
\be
ds^2 = F( \t, \bar{ \t})dzd\bar{z} \prod_{i=1}^{24}(z-z_i)^{ - \frac{1}{2}}
(\bar{z}-\bar{z}_i)^{ - \frac{1}{2}}   \   ,
\ee
dove $ F( \t, \bar{ \t}) = \t_2 \h^2 \bar{\h}^2$ e dove gli $z_i$ sono gli zeri 
di $  27 g^2 + 4 f^3 $. Se come base scegliamo $CP^1 \times CP^1$ e in 
(\ref{efam})
utilizziamo un polinomio $f(z,w)$ di grado $(8,8)$, ed un polinomio $g(z,w)$
di grado $(12,12)$, 
otteniamo una variet\`a di Calabi-Yau.

Un'importante dualit\`a che coinvolge la F teoria \`e $F/K3 \sim Het/T^2$.
Consideriamo un punto nello spazio dei moduli dove $\l$ \`e costante 
\cite{For}; questo significa che 
$ \frac{f^3}{g^2}$ \`e costante, e quindi $ f = \a \f^2$ e $g = \f^3$ con
$\f$ polinomio omogeneo di quarto grado che possiamo scegliere della forma
$ \prod_{i=1}^4(z-z_i)$. In questo caso 
\be
j = \frac{4(24   \a)^3}{27+4 \a^3}  \   ,
\ee
\be 
ds^2 =Fdzd \bar{z} \prod_{i=1}^4 (z-z_i)^{ - \frac{1}{2}}
(\bar{z}-\bar{z}_i)^{ - \frac{1}{2}}  \   ,
\ee
con $F$ costante. Con un cambiamento di 
coordinate $ dw = \prod_{i=1}^4 (z-z_i)^{ - \frac{1}{2}} dz $, la metrica 
diviene piatta, ovvero 
$ ds^2 = C dwd \bar{w}$, tranne nei punti $z_i$, dove \`e presente un 
angolo di deficit di $ \p$: lo spazio base \`e quindi un tetraedro $T^2/I$.
Se seguiamo un cammino chiuso attorno ad uno dei punti $z_i$, 
la coordinata $w$ 
cambia segno, e 
se $u$ indica la coordinata sul toro si ha anche $ u \mapsto -u$.

La trasformazione 
\be
\pmatrix{ -1 & 0\cr 0 & -1}  \ ,
\ee
appartiene ad  $SL_2Z$ e pu\`o essere  identificata con $(-1)^F \W$, 
e quindi nel particolare punto scelto dello spazio 
dei moduli abbiamo $IIB/T^2/I/(-1)^F \W$ \cite{For}; 
una doppia T-dualit\`a ci d\`a
$IIB/T^2/ \W$, ovvero $I/T^2$. A questo punto la dualit\`a tra 
eterotica e tipo I permette di stabilire la relazione cercata e da questa ne 
segue immediatamente un'altra : $F/CY \sim Het/K3$, che si pu\`o verificare 
fibrando su una base $CP^1$ la variet\`a $K3$ e utilizzando fibra per fibra 
la dualit\`a appena illustrata.


\section{Conifolds e cambiamenti di topologia}
\markboth{}{}

Abbiamo incontrato, finora, S-dualit\`a in cui il regime perturbativo di una 
teoria \`e 
legato al regime non preturbativo della stessa teoria, 
dualit\`a che collegano teorie 
di stringhe apparentemente differenti e dualit\`a che vanno oltre le 
costruzioni perturbative note coinvolgendo strutture come la M-teoria e la 
F-teoria. L'idea  \`e che tutti questi vuoti siano connessi tra loro, anche 
se questo non  
\`e ancora del tutto chiaro, in particolare per alcune classi di modelli
in sei dimensioni. 
L'unit\`a di molti di questi vuoti pu\`o comunque 
essere mostrata esibendo transizioni 
continue da una compattificazione ad un altra. 
Riportiamo qui, come esempio, il caso di
teorie di Tipo II compattificate a quattro dimensioni su spazi 
di Calabi-Yau; muovendosi lungo lo spazio 
dei moduli si pu\`o infatti modificare la topologia 
realizzando in modo continuo la transizione di conifold, in cui un 
3-ciclo collassa e viene riparato con un 2-ciclo, 
alterando i numeri di Hodge; l'interpretazione fisica di questa 
transizione \`e stata data da Strominger in termini di ipermultipletti di buchi 
neri che divengono non massivi nella singolarit\`a.

Nel nostro studio dello spazio dei moduli delle compattificazioni su spazi di 
Calabi-Yau abbiamo visto come lo spazio geometrico delle strutture complesse e 
k\"ahleriane viene esteso ed unificato 
nello spazio delle teorie $N=2$ superconformi, che \`e quello davvero rilevante 
dal punto di vista della teoria delle stringhe. In particolare, la mirror 
symmetry ci ha permesso di concludere che le
singolarit\`a che si incontrano quando ci si avvicina alle pareti del cono di 
K\"ahler sono apparenti e vengono rimosse dall'effetto dei worldsheet 
instantons. Ci sono tuttavia singolarit\`a anche nello spazio dei moduli 
complessi che costituiscono un sottospazio di codimensione due detto luogo 
discriminante. Queste singolarit\`a sono molto pi\`u gravi in quanto sono 
singolarit\`a della teoria conforme; si hanno ad 
esempio divergenze nelle funzioni di correlazione quando ci si avvicina alla 
regione singolare. Strominger \cite{con} ha proposto un meccanismo fisico per 
la risoluzione di queste singolarit\`a nelle teorie di tipo II,
la cui idea essenziale \`e del tutto 
analoga a quella utilizzata in \cite{sw} per determinare l'azione effettiva di 
bassa energia di SYM con $N=2$ in $d=4$. Questa risoluzione rende inoltre 
possibili transizioni fisicamente non singolari tra CY con diversi numeri di 
Hodge e grazie a questo risultato \`e possibile mostrare che molti, e 
probabilmente tutti, gli spazi di CY sono nello stesso spazio dei moduli. 

Cominciamo con l'osservare che la teoria conforme cattura la dinamica dei gradi 
di libert\`a che sono leggeri nel limite $g_s \rar 0$ e questo elimina gli 
stati non perturbativi, come le D-brane. La singolarit\`a nella teoria conforme 
pu\`o essere quindi attribuita alla comparsa, in quella particolare regione 
dello spazio dei moduli, di stati non perturbativi con massa nulla. In 
$N=2$ SYM si ha un fenomeno del tutto analogo \cite{sw}; 
sullo spazio dei moduli quantistico sono presenti singolarit\`a 
dovute a monopoli o dioni che divengono di massa nulla:
l'inclusione di questi nuovi gradi di libert\`a nella lagrangiana effettiva 
rimuove la singolarit\`a, che viene invece riprodotta esattamente quando questi
sono integrati via.

Ricordiamo che lo spazio dei moduli \`e una variet\`a k\"ahleriana 
speciale e che coordinate proiettive locali si ottengono integrando la 
3-forma olomorfa $\W$ su una base simplettica in $H^3(M,Z)$:
\beq
A_i \cap B^j &=& \d_i^j  \ ,  \hspace{1cm}  
A_i \cap A_j = B^i \cap B^j = 0  \ , \nonumber \\
z^i &=& \int_{B^i} \W \  ,  \hspace{1cm}   G_i = \int_{A_i} \W \ .
\eeq
Le $G_i$ possono essere espresse in funzione delle $z^i$,
che sono le coordinate locali. Il potenziale \`e 
$K = - ln(i\bar{z}^iG_i - iz^i\bar{G}_i)$ e, essendo la variet\`a speciale,
$G_i=\frac{\part F}{\part z^i}$, con $F$ il prepotenziale. La singolarit\`a di 
conifold corrisponde a $z^j = 0$, con $j$ fissato; 
se si compie un percorso chiuso attorno a 
questo punto, gli integrali della tre forma trasformano secondo
$z^j \mapsto z^j$ e $G_j \mapsto G_j + z^j$. Questo significa che localmente, a meno di
una parte regolare,
$ G = \frac{1}{2 \p i} z^j ln z^j$ e di conseguenza la componente
$g_{j \bar{j}}$ della metrica \`e singolare, 
con $g_{j \bar{j}} \propto ln{z^j \bar{z}^j}$. Questa metrica determina
i termini cinetici per i moduli nell'azione effettiva, 
$ \int d^4x g_{i \bar{j}} \btd \f^i \btd \f^{\bar{j}}$,
e la sua degenerazione 
implica una singolarit\`a nella descrizione fisica della teoria.

La tipo IIB contiene, come sappiamo, delle D3-brane, alle quali corrispondono
nel limite di bassa energia le "black-branes" descritte in \cite{bb}. Questi 
stati sono carichi rispetto alla 4-forma autoduale $A$, e la loro carica si 
ottiene integrando $F=dA$ su una classe di omologia di superfici in cinque 
dimensioni: $Q(\S_5) = \int_{\S_5} F$. Quando compattifichiamo su un CY, le 
D3-brane possono avvolgersi attorno ai cicli $A_i$ e $B^j$; in quattro 
dimensioni esse appaiono allora come oggetti puntiformi, dei buchi neri, con cariche 
date da:
\be
\int_{A_i \times S^2} F = n_i g_5  \ ,  \hspace{1cm}
\int_{B^i \times S^2} F = m^i g_5   \   ,
\ee
dove $S^2$ \`e una sfera che circonda il difetto nelle dimensioni non compatte.
Le cariche si riferiscono ai campi di gauge $U(1)$ presenti 
nell'azione effettiva della IIB, in quanto i  moduli complessi compaiono 
in multipletti vettoriali ed abbiamo assunto che siano quantizzate in unit\`a 
di $g_5$.
Questi oggetti sono BPS saturati e soddisfano quindi la relazione:
\be
M = g_5 e^{K/2} |m^i G_i - n_i z^i|   \ ,
\ee
e danno origine ad un ipermultipletto della superalgebra con $N=2$. Sia 
$n_i = \d_{ij}$ e $m_i =0$  $ \forall i$  , $j$ fissato; quando $z^j \rar 0$, 
la massa della D3-brana si annulla. Possiamo verificare che la singolarit\`a 
nell'azione effettiva \`e dovuta proprio all'avere escluso 
l'ipermultipletto del buco nero, mostrando che la stessa singolarit\`a 
viene prodotta se lo si rimuove dall'azione effettiva. La parte rilevante della 
lagrangiana \`e 
\be
L = \frac{1}{8 \p} Im \left [ \int d^2 \th \t_{ij} W^{i \a}W^j_{\a} + ... 
\right ]   \ ,
\ee
con $\t_{ij}= \part_i G_j$  e $G_j = \part_j F$, $F$ prepotenziale. La costante 
di accoppiamento del j-esimo $U(1)$ \`e $\t_{jj} = \part_j G_j$. Con un calcolo 
ad un loop possiamo determinare il contributo dell'ipermultipletto al running 
di $\t_{jj}$:
\be
\t_{jj} = \frac{1}{2 \p i} ln z^j + (reg.)  \   ,
\ee
e quindi
\be
G_j = \frac{1}{2 \p i} z^j ln z^j + (reg.)   .
\ee

La comparsa di questi ipermultipletti non perturbativi in corrispondenza del 
conifold permette di realizzare fisicamente \cite{con2} una costruzione 
matematica che lega tra loro spazi di moduli di CY lungo il sottospazio 
corrispondente ai conifolds e che consiste nel far degenerare un 3-ciclo e nel 
ripararlo con un 2-ciclo (piccola risoluzione). 
Se si considerano 
delle singolarit\`a pi\`u generali di quella appena descritta,
il potenziale
scalare degli ipermultipletti presenta, oltre alle direzioni piatte  
corrispondenti alla deformazione della struttura complessa e lungo le quali si 
riespande il 3-ciclo degenere, ridando contemporaneamente massa ai buchi neri, 
delle nuove direzioni piatte.
Vedremo infatti che, se
degenerano $P$ cicli legati da $R$ relazioni, si hanno $R$ nuove direzioni
che corrispondono ad una fase di Higgs in cui $P-R$ multipletti 
vettoriali si accoppiano con altrettanti ipermultipletti acquistando massa; 
fisicamente $R$ multipletti di buchi neri condensano e, dopo la transizione 
di conifold, si presentano come stati dello spettro perturbativo della stringa,
un risultato notevole  che d\`a forza all'idea che vi sia un legame profondo 
tra buchi neri e particelle elementari. 

I numeri di Hodge della variet\`a e la sua caratteristica di Eulero cambiano 
quindi nel modo seguente:
\beq
& & (h_{21},h_{11}) \mapsto (h_{21} - (P-R), h_{11} + R)  \   ,  \nonumber \\
& & \chi \mapsto \chi - 2P  \  .
\eeq
Descriviamo il conifold in maggiore dettaglio. Sappiamo che localmente pu\`o 
essere rappresentato dall'equazione $\sum_{i=1}^4 w^2_i = 0$ in $C^4$, ovvero 
come un cono con punto singolare nel vertice; analizzando
l'intersezione con 
$S^7 \subset R^8$ si pu\`o identificare la base del cono con $S^2 \times S^3$.
Rimosso un intorno del punto singolare, la singolarit\`a pu\`o essere 
riparata in due modi: per deformazione, incollando $B^3 \times S^3$, ovvero
riespandendo il ciclo degenere, o per risoluzione, incollando $S^2 \times B^4$ e 
determinando un drastico cambiamento di topologia in quanto i numeri di Hodge 
vengono alterati. Nella teoria effettiva, la deformazione corrisponde al moto 
lungo la branca coulombiana, e la risoluzione al moto lungo la branca di 
Higgs. 

Per studiare questa transizione abbiamo bisogno di un teorema di Lefschetz; 
se indichiamo con 
$\g^a$   $k$ 3-cicli che degenerano nel punto di conifold nello spazio dei 
moduli, un 3-ciclo $\d$ trasportato attorno a tale punto trasforma secondo:
\be
\d \mapsto \d + \sum_{a=1}^k (\d \cap \g^a) \g^a   \   .
\ee
Da questo segue che l'integrale della 3-forma $\W$ trasforma nel modo seguente
\be
\int_{\d} \W \mapsto \int_{\d} \W + \sum_{a=1}^k (\d \cap \g^a) 
\int_{\g^a} \W \ ,
\ee
e quindi:
\be
\int_{\d} \W = \frac{1}{2 \p i} \sum_{a=1}^k (\d \cap \g^a) \left (\int_{\g^a} 
\W \right ) ln \left ( \int_{\g^a} \W \right ) + (reg.)   \   .
\label{mon}
\ee

Prendiamo come esempio specifico $Y_{4;5}$ e scegliamo una famiglia che abbia 
un sottospazio $CP^2$ fisso, dato ad esempio da $x_3=x_4=0$; questo significa 
che il polinomio non deve contenere monomi nelle sole variabili $x_0$  , 
$x_1$  e $x_2$  e pu\`o quindi essere scelto della forma
\be
P = x_3 g(x) + x_4 h(x)   \  ,
\label{pol}
\ee
dove $g$ ed $h$ sono omogenei di quarto grado. Questa famiglia ha $86$ 
parametri, in quanto contiene $21$ monomi in meno e i cambiamenti di 
variabile che rispettano la forma (\ref{pol}) hanno $19$ parametri invece degli 
usuali $25$ di $GL(5)$.

I punti singolari sono i sedici punti di intersezione di $g$ ed $h$ in $CP^2$ :
$\{ x_3 = x_4 = g(x) = h(x) = 0 \}$; si pu\`o verificare che si tratta di 
conifolds valutando l'hessiano del polinomio. Se ora da $CP^4$ 
rimuoviamo un sottospazio $B^8$ dall'intorno di ogni punto singolare, 
corrispondentemente rimuoviamo dal CY una regione che ha per bordo 
$S^2 \times S^3$. $CP^2$ passa attraverso ciascuno dei sedici punti e quindi 
sappiamo pi\`u precisamente che stiamo rimuovendo un $B^4$ da questo spazio; la 
frontiera del $B^4$ \`e proprio l'$S^3$ del ciclo che degenera. Quando 
ripariamo le singolarit\`a otteniamo una variet\`a di CY liscia, e quindi
$CP^2 \setminus 16 B^4$ \`e una sottovariet\`a del nostro CY con
frontiera $\sum_{a=1}^{16} \g^a$ che \`e pertanto omologicamente banale. In 
sintesi vale la relazione
\be
\sum_{a=1}^{16} \g^a = 0   \ .
\ee
Se effettuiamo una piccola risoluzione della singolarit\`a, lo spazio 
risultante pu\`o essere descritto da
\be
y_0x_4 - y_1 x_3 = 0  \  ,   \hspace{1cm} y_0 g(x) + y_1 h(x) = 0   \  ,
\label{x2}
\ee
con
$[x] \in CP^4$ e $[y] \in CP^1$. C'\`e una nuova classe $(1,1)$ che misura 
l'area comune degli $S^2$.

Scegliamo ora la nostra base simplettica in modo che
$ \g^a = B^a $ , $a = 1,...,15 $ e $ \g^{16} = - \sum_{a=1}^{15} B^a $. Da 
(\ref{mon}) segue, con $\d = A_j$  ,
\be
G_j = \frac{1}{2 \p i} z^j ln z^j + \frac{1}{2 \p i} \left ( \sum_{i=1}^{15} z^i
\right ) ln \left ( \sum_{i=1}^{15} z^i \right ) + (reg.)  
\ \ \  j = 1,...,15    \ .
\label{mon2}
\ee
Supponiamo ora che ciascuno dei sedici stati che si ottengono avvolgendo 
D3-brane su cicli degeneri dia campi quantistici indipendenti e che si abbiano 
quindi sedici multipletti di buchi neri. Indichiamo con $H^a$ l'ipermultipletto 
corrispondente al ciclo $\g^a$; la carica di questo stato rispetto all'i-esimo 
campo di gauge $U(1)$ \`e $Q^a_i = A_i \cap \g^a$ e quindi, nel nostro caso,
$Q^a_i = \d^a_i$ per $a \neq 16$ e 
$Q^{16}_i = -1$ ,  $i = 1,...,15$. Infine la 
massa $m^a$ \`e proporzionale a $\sum_{i=1}^{15} Q^a_i z^i$. Abbiamo ora tutti 
gli elementi necessari per calcolare il contributo al running della costante 
d'accoppiamento:
\be
\t_{ij} = \frac{1}{2 \p i} \sum_{a=1}^{16} Q^a_i Q^a_j ln \left ( m^a \right ) \  ,
\ee
e quindi
\be
\t_{ij} = \frac{1}{2 \p i} \d_{ij} ln z^j + \frac{1}{2 \p i} 
ln \left ( \sum_{k=1}^{15} z^k \right ) + (reg.)   \  ;
\ee
integrando riproduciamo (\ref{mon2}). 

Il potenziale per gli ipermultipletti \`e 
$ V = \sum_{i=1}^{15} D^i_{\a\b} D_i^{\a\b}$ dove
\be
D_i^{\a\b} = \sum_{a=1}^{16} Q^a_i \left [ \e^{\a\g} \bar{h}^{(a)}_{\g} h^{(a)\b} +
\e^{\b\g} \bar{h}^{(a)}_{\g} h^{(a)\a} \right ]   \ ,
\ee
e gli $h^{(a)}_{\a}$ sono i due scalari complessi dell'ipermultipletto $H^a$;
l'indice $\a$ si riferisce ad una simmetria globale $SU(2)_R$.
Questo potenziale presenta delle direzioni piatte quando 
$ \lan h^{(a)}_\b \ran =0$ 
e si d\`a un valore 
d'aspettazione agli scalari nei multipletti vettoriali; queste 
direzioni corrispondono alla fase coulombiana : i buchi neri acquistano massa, 
la singolarit\`a viene deformata e torniamo nella regione liscia dello spazio 
dei moduli di $Y_{4;5}$. Grazie alla relazione tra i cicli d'omologia, che
impone $\sum_{a=1}^{16} Q^a_i = 0$    $ \forall i$, si ha un'altra direzione 
piatta, unica a meno di trasformazioni di gauge, data da $h^{(a)\a} = v^{\a}$
    $\forall a$, con $v$ vettore complesso arbitrario a due componenti.
Abbiamo infatti $32$ campi complessi soggetti a  
$45$ condizioni reali $D^i_{\a\b} =0$; tenendo conto delle $16$ trasformazioni 
di gauge, otteniamo $4$ parametri reali per lo spazio dei moduli.
Lungo questa direzione $15$ campi di gauge divengono massivi e ci si sposta in 
una nuova branca dello spazio dei moduli con $h_{21} = 86 $ e $h_{11} = 2$,
proprio i numeri di Hodge della piccola risoluzione (\ref{x2}).

Siamo quindi riusciti a connettere in modo continuo due spazi di CY con numeri 
di Hodge distinti; dato che la costruzione matematica realizzata fisicamente 
dalla condensazione di buchi neri permette di legare tra loro molti, e forse 
tutti, gli spazi di moduli di queste variet\`a, con ogni probabilit\`a lo 
spazio delle compattificazioni con $N=2$ in quattro dimensioni non ha 
componenti sconnesse. Questa osservazione permette, con argomenti di 
deformazione analoghi a quelli discussi nel terzo capitolo, 
di estendere la mirror 
symmetry a  regioni molto pi\`u vaste dello spazio dei moduli.

Possiamo fare due importanti osservazioni.
Anzitutto, la singolarit\`a di conifold si presenta, come abbiamo visto, a 
livello della teoria conforme, ovvero nello spazio dei moduli classico; ci si 
aspetterebbe allora che effetti quantistici non influenzino questa 
singolarit\`a in quanto vengono soppressi nel limite $g_s = e^{\f} \rar 0$. Il 
fatto notevole \`e che esistono effetti non perturbativi che persistono 
indipendentemente dal valore di $g_s$ e di conseguenza il limite $g_s \rar 0$ 
della teoria quantistica completa differisce dal risultato ad albero. In modo 
pi\`u concreto, l'azione classica nel frame di Einstein si scrive:
\be
S \sim \int \left ( R + (Vet) + (Hyp) \right )  \  .
\label{act}
\ee
Il dilatone appartiene ad un 
ipermultipletto e non ha accoppiamenti con i multipletti vettoriali 
\cite{spg1}; i primi due termini in (\ref{act}) non sono quindi 
sensibili al valore di $g_s$. I buchi neri che divengono non massivi sono 
soluzioni solitoniche che coinvolgono solo questi primi due termini e si 
presentano quindi come ipermultipletti con azione indipendente da $g_s$, 
producendo effetti quantistici indipendenti da $g_s$.

Un altro punto cruciale riguarda il conteggio degli ipermultipletti; perch\`e 
il meccanismo funzioni, non devono esserci stati di singola particella con 
carica multipla (brane avvolte ripetutamente attorno ad un ciclo), 
diversamente dal caso della relazione tra IIA e supergravit\`a in undici 
dimensioni che richiede una torre infinita di stati con carica multipla.
Evidenza per la correttezza del conteggio nei vari casi
deriva dal calcolo dell'indice di 
Witten per la teoria sul volume della brana \cite{winx}.


\section{Coppie duali in 4 e in 6 dimensioni}
\markboth{}{}

Le dualit\`a considerate finora riguardano teorie con 32 o 16 cariche di 
supersimmetria, e questo pone vincoli molto restrittivi sia 
sulla forma dell'azione 
effettiva che sullo spettro degli stati BPS; la situazione per teorie con 8 
cariche \`e molto pi\`u complicata e di conseguenza molto pi\`u ricca. 

Un metodo per costruire in modo sistematico
coppie di teorie duali in varie dimensioni \`e il seguente \cite{dualp}: 
partiamo dalla 
$IIA$ su $K3$ e costruiamo uno spazio di Calabi-Yau facendo variare $K3$ come 
una fibra su una base $CP^1$; applicando fibra per fibra la dualit\`a con $Het/
T^4$ otteniamo come possibile modello duale l'eterotica su una variet\`a 
ottenuta fibrando $T^4$ su $CP^1$, in accordo con la trasformazione 
di dualit\`a, che tipicamente risulta essere $K3 \times T^2$; inoltre dato 
che la relazione tra 
$IIA/K3$ e $Het/T^4$ d\`a una precisa relazione tra i moduli delle due teorie e 
che per l'eterotica \`e possibile avere anche campi di gauge di background, 
noto come $K3$ varia su $CP^1$ si pu\`o determinare come il campo di background 
vari su $K3 \times T^2$. Come esempio consideriamo $K3 \times T^2$ e 
quozientiamo con l'involuzione $I_E \cdot I_2$, dove $I_2$ cambia segno alle 
coordinate del toro e $I_E$ \`e l'involuzione di Enriques di $K3$; questa 
trasformazione scambia tra loro dieci delle 2-forme armoniche e cambia segno 
alle altre due, agisce in modo  analogo su $A^p_\m$ e lascia invariati i campi 
di 
gauge provenienti dalla 1-forma e dal duale della 3-forma. Dal punto di vista 
dell'eterotica questo equivale a scambiare i campi di gauge nei due fattori 
$E_8$, a scambiare $(G_{9 \m},B_{9 \m})$ e $(G_{8 \m},B_{8 \m})$ e a cambiare 
segno a $(G_{7 \m},B_{7 \m})$; geometricamente, oltre allo scambio dei fattori 
$E_8$, si ha $x^8 \lra x^9$ , $ x^7 \ra - x^7 $ ed uno shift di mezzo vettore 
in $ \L_{24} $, non visibile perturbativamente nella IIA ma 
necessario per l'invarianza modulare dell'eterotica. 

\`E pi\`u difficile verificare dualit\`a tra modelli con un numero minore di 
supersimmetrie, in quanto non \`e possibile confrontare le loro azioni 
effettive, che ricevono correzioni dai loops di stringa; inoltre lo 
spettro BPS pu\`o variare in modo discontinuo. Esiste per\`o un importante 
risultato sulla struttura della lagrangiana di teorie supersimmetriche 
con $N=2$ : mantenendo termini del secondo ordine nelle derivate non si hanno 
accoppiamenti tra multipletti vettoriali ed ipermultipletti.
Il termine cinetico per gli scalari \`e quindi dato da
\be
G^V_{mn}(\f ) \f^m \f^n + G^H_{mn}( \psi) \psi^m \psi^n \  ,
\ee
dove $G^V$ descrive una variet\`a k\"ahleriana speciale e $G^H$ uno spazio 
quaternionico. L'utilit\`a di questo risultato \`e dovuta al fatto che nelle 
teorie di tipo II il dilatone fa parte di un ipermultipletto mentre 
nell'eterotica fa parte di un multipletto vettoriale. Poich\`e le correzioni 
quantistiche in teoria di stringa sono pesate dal dilatone e quindi contengono 
un accoppiamento al dilatone, come conseguenza del disaccoppiamento il termine 
$G^V$ calcolato ad albero nella tipo II e il termine $G^H$ calcolato ad albero 
nell'eterotica sono esatti. Un modo di verificare la dualit\`a consiste quindi 
nel calcolare $G^V$ nella IIA, riesprimerlo in termini di campi dell'eterotica, 
espanderlo in potenze di $g_{Het}$ identificando i termini ad albero, ad un 
loop e non perturbativi e confrontare il risultato con calcoli espliciti 
effettuati nell'eterotica.
In tutti i casi studiati l'accordo \`e perfetto grazie ad identit\`a 
matematiche non banali che consentono una riscrittura dei risultati  
nei due modi descritti \cite{du8}. 
Si possono studiare anche i termini non perturbativi 
grazie ai metodi sviluppati per le teorie di campo da Seiberg e 
Witten \cite{sw}; se 
infatti consideriamo il limite in cui la teoria di stringhe si riduce ad una 
teoria di campo questi termini non perturbativi possono essere confrontati con 
i risultati di Seiberg e Witten; l'accordo \`e perfetto e mostra come la 
costruzione della curva algebrica ausiliaria per la soluzione di queste teorie 
di gauge supersimmetriche abbia un'elegante interpretazione in termini della 
geometria degli spazi di Calabi-Yau.

Utilizzando i vincoli imposti dalla supersimmetria sull'azione effettiva \`e possibile
dedurre informazioni sullo spazio dei moduli della teoria 
superconforme \cite{seib}. 
Infatti gli operatori esattamente marginali hanno funzioni $\b$ nulle e la 
famiglia di teorie conformi alle quali un dato modello appartiene si ottiene 
considerando la perturbazione $\d S = \sum g_i \int d^2z \F^i(z) $; lo spazio 
dei parametri $g_i$ \`e lo spazio dei moduli della teoria e possiamo definire 
la metrica di Zamolodchikov tramite la funzione a due punti
\be
< \F^i(z) \F^j(w) e^{- \d S} > = \frac{g_{ij}(\{g_i \})}{|z-w|^4} \ .
\ee

Quando interpretiamo la teoria conforme nello spaziotempo come una soluzione 
classica della teoria di stringhe, questi operatori marginali compaiono come 
campi non massivi $\f^i$ con operatori di vertice 
$V^i(k) = \int d^2z \F^i(z) e^{ikX}$ e con un potenziale che ha direzioni 
piatte. Lo spazio dei moduli \`e ora il luogo degli zeri di questo potenziale, 
e la metrica \`e data dal coefficiente del termine cinetico per i campi della 
teoria effettiva: $L_{eff} \sim g_{ij}(\f) \part_\m \f^i \part^\m \f^j$.
Si hanno singolarit\`a nei punti in cui, variando i moduli, uno stato massivo 
diviene non massivo - un operatore irrilevante diviene marginale - perch\`e 
l'azione effettiva diviene non locale. Consideriamo teorie superconformi di 
tipo $(4,4)$, ad esempio dei modelli $\s$ con spazio bersaglio $K3$. La 
teoria effettiva \`e una teoria in sei dimensioni con $N=2$. I
multipletti delle varie algebre di supersimmetria in sei dimensioni sono
\beq
& & \bf{(2,0)}  \nonumber \\
& & {\rm multipletto~gravitazionale~~} (g_{\m\n}, 2 \psi_{\m L} , 5B^{+}_{\m\n})
\ , \nonumber \\
& & {\rm multipletto~tensoriale~~} (B^{-}_{\m\n}, 2 \chi_R, 5 \f)
\ , \nonumber \\
& & \bf{(1,1)}     \nonumber \\
& & {\rm multipletto~gravitazionale~~} (g_{\m\n}, \psi_{\m L} , \psi_{\m R} ,
B^{+}_{\m\n},  B^{-}_{\m\n},  \chi_L,  \chi_R, 4 A_\m, \f)
\ , \nonumber \\
& & {\rm multipletto~vettoriale~~} (A_\m, \l_L , \l_R , 4 \f)
\ , \nonumber \\
& & \bf{(1,0)}    \nonumber \\
& & {\rm multipletto~gravitazionale~~} (g_{\m\n}, \psi_{\m L} , B^{+}_{\m\n}) \ ,
\nonumber \\
& & {\rm multipletto~tensoriale~~} (B^{-}_{\m\n}, \chi_R,  \f) \  , \nonumber \\
& & {\rm multipletto~vettoriale~~} (A_\m, \l_L)\  , \nonumber \\
& & {\rm ipermultipletto~~} (\psi_R, 4 \f) \   .
\eeq
Supergravit\`a di tipo $(1,0)$ si ottengono compattificando su $K3$ la Tipo I o 
l'eterotica, quelle di tipo $(2,0)$ e $(1,1)$ compattificando su $K3$ 
rispettivamente la IIB e la IIA.
Osserviamo che la prima e la terza teoria sono chirali; ricordando 
l'espressione per le anomalie di particelle con spin $ \frac{1}{2}$ , 
$ \frac{3}{2}$ e forme autoduali \cite{An1}
\be
I_{1/2} = \frac{1}{180}Y_4 + \frac{1}{72}Y_2^2 \   ,    \hspace{1cm}
I_{3/2} = \frac{245}{180}Y_4 - \frac{43}{72}Y_2^2 \  ,  \hspace{1cm}
I_{A} = \frac{7}{45}Y_4 - \frac{1}{9}Y_2^2  \  ,
\ee
con $Y_{2m} = \frac{1}{2}(- \frac{1}{4})^m tr R^{2m}$, si vede che
la teoria $(2,0)$ deve contenere $21$ multipletti di materia, e quindi il suo 
contenuto \`e completamente fissato dalla cancellazione delle 
anomalie. L'esistenza di uno spazio dei moduli per la supergravit\`a di bassa energia
implica che le teorie superconformi di tipo $(4,4)$ non sono isolate, e il 
numero di moduli \`e fissato da condizioni di consistenza della teoria 
spaziotemporale (microscopicamente, queste corrispondono all'invarianza 
modulare del modello).
Lo spazio dei moduli della teoria \`e quindi $M_{5,21}$ e il gruppo di 
dualit\`a \`e $SO(5,21;Z)$; dato che non ci sono altri scalari, questo \`e lo spazio 
dei moduli completo, e il dilatone va identificato con uno di questi $105$ 
scalari. Se compattifichiamo ancora su $S^1$ otteniamo solo un nuovo modulo, 
corrispondente al raggio del cerchio, dato che il modello non contiene campi 
vettoriali; lo spazio dei moduli diviene allora $M_{5,21} \times R$ e coincide 
con quello di $Het/T^5$. Questa dualit\`a, che pu\`o essere vista come una 
conseguenza della T-dualit\`a tra $IIA$ e $IIB$ e S-dualit\`a tra $IIA/K3$ e 
$Het/T^4$, \`e interessante perch\`e il dilatone dell'eterotica corrisponde al 
raggio del cerchio della $IIB$: abbiamo quindi una situazione analoga a quella 
tra IIA ed M-teoria in dieci dimensioni, il limite a forte accoppiamento di 
$Het/T^5$ \`e $IIB/K3 \times S^1$ a grande raggio, vale a dire una teoria sei 
dimensionale. 

La $IIB/K3$ pu\`o anche essere studiata come la M-teoria compattificata su 
$T^5/Z_2$ dove $Z_2$ inverte il segno di tutte le coordinate. Questo introduce 
$32$ punti fissi ciascuno dei quali ha una carica $-1/2$ in unit\`a nelle 
quali 
la 5-brana ha carica unitaria; per cancellare la carica nello spazio interno 
compatto \`e necessario introdurre $16$ 5-brane. Riusciamo cos\`\i \, a 
giustificare il contenuto di campi della teoria: dalla compattificazione della 
supergravit\`a in 11d su $T^5/Z_2$ si ottiene il multipletto gravitazionale e 5 
multipletti tensoriali; ciascuna delle 5-brane d\`a poi un ulteriore 
multipletto. La teoria in 6d contiene anche stringhe autoduali la cui tensione 
pu\`o essere arbitrariamente piccola e che derivano da D3-brane autoduali 
avvolte attorno a 2-cicli di K3 (come per i cicli, anche 
per queste stringhe 
esiste una classificazione ADE) ; dal punto di vista della M-teoria su 
$T^5/Z_2$, esse 
derivano da M2 tese tra le 16 M5 nel limite in cui la distanza tra queste 
tende a zero.

Poich\`e in dieci dimensioni l'oggetto duale alla stringa \`e la 5-brana, ci si 
aspetta che la stringa eterotica appaia in $IIA/K3$ avvolgendo una 5-brana con 
topologia $K3 \times S^1$ attorno a $K3$ dando origine ad una stringa in sei 
dimensioni; \`e possibile costruire in modo esplicito il solitone 
corrispondente in supergravit\`a \cite{hs}. 
Generalizziamo questa costruzione per 
ottenere vuoti con $n \ne 20$, utilizzando 
la M teoria e considerando quindi
$(K3 \times S^1)/ Z_h$, dove scegliamo una superficie $K3$ dotata di 
una simmetria discreta di ordine $h$ che combiniamo con una rotazione di $2 \p 
/ h$ sul cerchio, in modo da assicurare l'assenza di punti fissi e quindi una 
variet\`a regolare. Le simmetrie discrete di $K3$ sono state classificate da 
Nikulin e si pu\`o concludere che vuoti consistenti si hanno per le seguenti 
coppie $(h,n)$ : $(2,12)$ , $(3,8)$ , $(4,6)$ , $(5,4)$ ,
$(6,4)$ , $(7,2)$ , $(8,2)$ : lo spazio dei moduli \`e dato nei vari casi da
$[SO(4,n)/SO(4) \times SO(n) ] \times R$.

Abbiamo diversi modi di guardare ai modelli con $(1,0)$ supersimmetrie: come
compattificazioni dell'eterotica $SO(32)$ su $K3$ con eventuale  introduzione 
di 5-brane; come compattificazioni 
dell'eterotica $E_8 \times E_8$ con la possibilit\`a sia di 
suddividere gli istantoni tra i due fattori $E_8$ sia di includere 5-brane; 
come compattificazioni della Tipo I su K3; come vuoti dell'M-teoria su $K3 
\times S^1/Z_2$.
Per le teorie $(1,0)$ l'assenza di anomalie gravitazionali richiede
\be
n_h - n_v + 29n_t = 273 \  ,
\ee
dove $n_h$ , $n_v$ e $n_t$ sono il numero di ipermultipletti, multipletti 
vettoriali e tensoriali; la forma di anomalia allora fattorizza 
$I_8 = I_4 \wedge \tilde{I}_4$ con \cite{6d3}
\be
I_4 = trR^2 - \sum_a v_a trF^2_a \   ,   \hspace{1cm}
\tilde{I}_4 = trR^2 - \sum_a \tilde{v}_a trF^2_a \  ,
\ee
con $v_a$ , $\tilde{v}_a$ dipendenti dal gruppo di gauge assunto semisemplice 
$G= \prod_a G_a$. In presenza di altri fattori $U(1)$ nella forma d'anomalia 
pu\`o comparire il termine $F \ww Y_6$ con $F$ la field strenght del campo 
abeliano e $Y_6$ una 6-forma; le anomalie possono ancora essere cancellate se 
\`e presente un campo scalare $\chi$ che trasforma sotto $U(1)$ ( $\chi \mapsto 
\chi + \L$ ) : il campo di gauge diviene massivo, non \`e pi\`u presente una 
simmetria $U(1)$ non rotta e la teoria diviene consistente.

Quando il gruppo di gauge ha un'interpretazione perturbativa come nel caso 
della stringa eterotica, le costanti $v_a$ possono essere legate al livello 
$n_a$ dell'algebra di Kac-Moody
\be
v_a tr F^2_a = \frac{n_a}{h_a} Tr F^2_a \   ,
\ee
dove abbiamo indicato la traccia nella fondamentale e nell'aggiunta 
rispettivamente con $tr$ e $Tr$ e dove $h_a$ \`e il numero di Coxeter duale. 
Quando $n = 1$ si ha $v = 2$ per $SU(n)$ o $Sp(n)$, $v = 1$ per $SO(n)$ ,
$v = 1/3$ per $E_6$ ,$v = 1/6$ per $E_7$ ,$v = 1/30$ per $E_8$ .

Integrando l'identit\`a di Bianchi su $K3$ per
$H = dB + \w^L - \sum_a v_a \w_a^{YM}$ 
si ottiene la condizione di 
consistenza
\be
\sum_a n_a = \sum_a \int trF_a^2 = \int_{K3} trR^2 =24  \  .
\ee
Per la teoria $SO(32)$ questa diviene
\be
n_1+n_5 = 24 \   ,
\ee
dove $n_1$ \`e il numero di istantoni immersi nel gruppo $SO(32)$ e $n_5$ \`e il 
numero di 5-brane che corrispondono a sorgenti per $H$
localizzate su $K3$ ed estese nelle sei dimensioni non compatte. Per la teoria 
$E_8 \times E_8$ si ha
\be
n_1 + n_2 + n_5 = 24 \   ,
\ee
dove i numeri $n_1$ e $n_2$ indicano il numero istantonico del fibrato di gauge 
$E_8 \times E_8$ che viene quindi rotto ad un sottogruppo 
$G_{n_1} \times G_{n_2}$.
Possiamo utilizzare quanto sappiamo sulle dualit\`a in sei dimensioni per 
studiare vuoti dell'eterotica e della Tipo I su $K3$ e vuoti della Tipo II su 
un 3-fold $X^3$; queste dualit\`a permettono di interpretare molti fenomeni non 
perturbativi nell'eterotica in termini della geometria dello spazio di Calabi-
Yau, in particolare in termini di avvolgimento di brane intorno a cicli non 
banali. I vuoti dell'eterotica sono determinati essenzialmente dalla scelta del 
fibrato di gauge, mentre quelli della Tipo II sono determinati dalla scelta di $X^3$; 
compattificazioni su CY corrispondono a vuoti eterotici perturbativi solo se 
$X^3$ \`e una fibrazione di $K3$ su $CP^1$. 
I vuoti eterotici perturbativi hanno un unico multipletto tensoriale, quello 
contenente il dilatone; in generale $n_t = 1 + n_5$.

Torniamo alla teoria $SO(32)$ e consideriamo la decomposizione 
$SO(32) \supset SO(28) \times SU(2) \times SU(2)$. Possiamo immergere gli 
istantoni in uno dei due $SU(2)$ preservando una simmetria di gauge 
$SO(28) \times SU(2)$ e calcolare il numero di ipermultipletti nelle varie 
rappresentazioni utilizzando teoremi d'indice. Quando si passa a vuoti non 
perturbativi includendo 5-brane, queste possono essere identificate, come 
mostrato da Witten \cite{hod3}, 
con $ \it{piccoli}$  $\it{istantoni}$, istantoni la cui 
taglia  \`e stata ridotta a zero. 

Il massimo gruppo di gauge, $SO(32) \times Sp(24)$,
si ottiene quando $n_1=0$ e $n_5=24$ e tutte le 5-
brane coincidono.
Complessivamente si hanno 1672 multipletti vettoriali e 1916 ipermultipletti 
che ancora soddisfano $n_h - n_v =273$.

Abbiamo interpretato le costanti $v_a$ come legate ai livelli dell'algebra di 
Kac-Moody; per quanto riguarda $\tilde{v_a}$ con considerazioni di 
supersimmetria si pu\`o mostrare \cite{6d3}
che il termine cinetico dei campi di gauge \`e
\be
\sum_a ( v_a e^{- \f} + \tilde{v}_a e^{\f})tr(F_a^2) \   .
\ee
A debole accoppiamento domina il primo termine ma quando 
\be
e^{2 \f_0} = - \frac{v_a}{\tilde{v}_a} \   ,
\ee
la costante d'accoppiamento di gauge diverge (naturalmente si deve avere 
$\tilde{v}_a < 0$); 
in corrispondenza di questa singolarit\`a la tensione di una stringa solitonica 
si annulla \cite{6d5}.

Vediamo cos\`\i \, 
che nello spazio dei moduli della teoria eterotica in $d=6$ si possono avere 
singolarit\`a tanto nel regime di debole accoppiamento che in quello di forte 
accoppiamento. Il primo tipo di singolarit\`a \`e associato ai piccoli 
istantoni; muovendosi in $M_H$ uno degli istantoni pu\`o avere dimensione nulla 
e dato che $M_H$ e $M_V$ sono disaccoppiati la singolarit\`a pu\`o presentarsi 
per ogni valore di $g_s$. Per la teoria $SO(32)$ in corrispondenza della 
singolarit\`a si hanno degli ulteriori campi di gauge che divengono non massivi
determinando un aumento di simmetria. Per la teoria $E_8 \times E_8$ non si 
hanno campi vettoriali aggiuntivi ma una stringa non critica ha una tensione 
nulla; la teoria sul wordlsheet di questa stringa ha supersimmetria $(0,4)$ e 
contiene una due forma autoduale; nella 
singolarit\`a si ha un ulteriore multipletto tensoriale, un fenomeno suggerito dalle
condizioni di assenza delle anomalie, che mostrano anche che per 
avere un nuovo multipletto tensoriale bisogna fissare $29$ ipermultipletti. 
Consideriamo per l'eterotica un fibrato con $(n_1,n_2) = (12+k,12-k)$ ed 
effettuiamo una transizione $k \mapsto k-1$ riducendo un istantone nel primo 
$E_8$ e quindi aggiungendone uno nell'altro $E_8$: i due vuoti perturbativi 
sono connessi da una branca coulombiana intermedia con multipletti tensoriali 
per la quale non si ha una descrizione geometrica. Possiamo ottenerla 
considerando la M-teoria su $K3 \times S^1 / Z_2$; la posizione di una 5-brana 
che occupa le sei dimensioni non compatte \`e definita da 4 scalari per $K3$, 
che corrispondono ad un ipermultipletto, e da uno scalare per $S^1/Z_2$, che 
corrisponde ad un multipletto tensoriale. La transizione descritta 
precedentemente pu\`o essere interpretata come un istantone che alla 
singolarit\`a diviene una 5-brana, viene emesso nel bulk e si immerge 
nell'altra 9-brana; 
in altri termini la branca coulombiana non perturbativa parametrizza il moto 
della 5-brana nel bulk. In $d=4$ la branca non perturbativa ha ulteriori 
multipletti vettoriali con particolari accoppiamenti di Chern-Simons; nel 
modello duale di Tipo II la transizione corrisponde ad un cambiamento di 
topologia.

Sono stati studiati numerosissimi esempi di coppie duali, che hanno superato 
tutti i test di consistenza che le tecniche note attualmente hanno reso 
possibili; in particolare sono state studiate in grande dettaglio coppie con
$N=2$ in $d=4$, attraverso un attento esame non solo dello 
spettro e delle varie simmetrie, ma anche delle possibili correzioni di soglia 
alle azioni di bassa energia. Naturalmente un progresso reale sar\`a possibile 
solo quando si avr\`a una formulazione dinamica completa della M teoria che 
consenta di derivare tutte queste relazioni di dualit\`a e di
descrivere in modo completo le interazioni dei suoi gradi di 
libert\`a, oltre le 
approssimazioni semiclassiche fino ad ora necessarie. Nel 
prossimo paragrafo illustreremo brevemente alcune proposte riguardanti 
possibili formulazioni non perturbative della M teoria.


\section{Modello di matrici e relazione AdS/CFT}
\markboth{}{}

In \cite{mm1} \`e stato proposta una formulazione non perturbativa della M 
teoria in temini di un modello di meccanica quantistica di matrici. Abbiamo 
osservato nel terzo capitolo che alle D-particelle sono accessibili scale 
inferiori a quella di stringa. Il limite ricavato in \cite{db3,db5} per il 
parametro di impatto nel regime di bassa energia 
\`e $b \sim \sqrt{\frac{v}{\p T_F}}$, con $T_F = \frac{1}{2 \p \ap}$ 
tensione della stringa fondamentale; abbiamo anche un limite quantomeccanico
$b T_0 v \sim 1$ e nel regime in cui entrambi i limiti sono saturati otteniamo
$b \sim \frac{1}{T_FT_0} \sim \frac{1}{T_2}$ dove $T_2$ \`e la tensione della 
membrana della M teoria. La dimensione dinamica delle D-particelle \`e quindi 
confrontabile con la scala di Planck della M teoria; la teoria di stringhe 
perturbativa non cattura allora tutti i gradi di libert\`a presenti a piccola 
scala. L'idea di BFSS \`e che le D-particelle siano i gradi di libert\`a 
dominanti della M teoria e che questa possa essere descritta in modo completo 
nel gauge del cono di luce se si considera
il limite di grandi $N$ del modello di matrici supersimmetrico che 
descrive la loro dinamica di bassa energia. Il limite di grandi $N$ \`e 
necessario per decompattificare l'undicesima dimensione della teoria, ma \`e 
stato anche proposto \cite{sus}
che per N finito il modello di matrici corrisponda alla 
compattificazione della M teoria nel cono di luce discreto (DLCQ).
Tutta la costruzione ricorda profondamente il modello a partoni. \`E stato 
mostrato che questo modello ha molte delle caratteristiche della M teoria, e in 
particolare riproduce lo spettro della 
supergravit\`a in 11 dimensioni e pu\`o descrivere membrane e 5-brane. 
Compattificazioni toroidali del modello di matrici sono definite come limiti 
per grandi $N$ di teorie SYM definite su tori duali; in questo modo si 
riproducono i gruppi di U dualit\`a della teoria di stringhe. I principali 
problemi di questo approccio sono la sua formulazione non covariante e le serie
dificolt\`a che si incontrano quando si cerca di 
estendere la prescrizione per la 
costruzione di compattificazioni a tori di dimensione superiore a 5 
e a variet\`a non banali. 

Alla base del modello di matrici c'e' un'altra idea fondamentale, il 
principio olografico di 't Hooft e Susskind \cite{olo}. 
In base a questo principio, in una 
teoria consistente della gravit\`a quantistica, una regione macroscopica dello 
spazio e tutto ci\`o che contiene pu\`o essere rappresentata da una teoria 
definita sul bordo di tale regione. Inoltre, la teoria di bordo non deve 
contenere pi\`u di un grado di libert\`a per area di Planck. In modo pi\`u 
preciso, il numero di stati quantici distinti non deve superare 
$e^{\frac{A}{4 G_D}}$, dove $G_D$ \`e la costante gravitazionale e $A$ l'area 
del bordo. \`E stato recentemente congetturato \cite{mal} che la Tipo IIB 
compattificata su $AdS_5 \times S_5$ con $N$ unit\`a di flusso della cinque 
forma su $S^5$ \`e duale alla teoria di $SYM$ con sedici supercariche e gruppo 
di gauge $U(N)$ definita sul bordo di $AdS_5$. Lungo queste linee \`e stata 
stabilita una corrispondenza tra teorie di campo conformi definite 
sul bordo di una data variet\`a e teorie di stringhe compattificate su quella 
stessa variet\`a; in particolare sono state sviluppate tecniche per esprimere 
funzioni di correlazione per la teoria di bordo in termini di calcoli fatti 
utilizzando la teoria definita su tutta la variet\`a.

Sia la natura non commutativa dei gradi di libert\`a fondamentali del modello
di matrici,
sia la peculiare relazione tra teorie sul bordo e teorie nel bulk 
richiesta dal principio olografico e dalla corrispondenza AdS/CFT saranno 
con ogni probabilit\`a 
due elementi essenziali per la formulazione completa della M 
teoria.


\chapter{Stringhe aperte, M teoria e rottura della supersimmetria}
\markboth{}{}

\section{Rottura della supersimmetria e costante cosmologica}
\markboth{}{}

Tutti i modelli discussi in questa tesi fino ad ora, 
tranne poche eccezioni, 
sono invarianti rispetto ad una o pi\`u supersimmetrie.
Uno dei problemi pi\`u importanti consiste nello studiare i possibili 
meccanismi di rottura della 
supersimmetria in teoria di stringhe e le loro conseguenze per il valore della 
costante cosmologica.
Attualmente, infatti, la sola ovvia spiegazione per l'annullarsi 
della costante cosmologica \`e l'esistenza di una supersimmetria non rotta. 
La supersimmetria, se realizzata in natura, \`e rotta, 
dal momento che non esistono particelle bosoniche degeneri in massa
con quelle fermioniche presenti nello spettro di bassa energia.
\`E quindi cruciale comprendere
perch\`e la costante cosmologica sia tanto inferiore alla scala di rottura 
della supersimmetria, ovvero perch\`e $\L << (1 Tev)^4$.

\subsection{Supersimmetria globale}

Il teorema di Coleman e Mandula afferma che le 
simmetrie della matrice S possibili in una teoria di campo relativistica 
sono descritte, quando ci si limita a considerare algebre di Lie, 
dalla somma diretta dell'algebra di
Poincar\`e e di un algebra interna. Il teorema \`e stato esteso 
alle superalgebre 
da Haag, Lopuszansky e Sohnius che hanno mostrato come,
in questo caso, le algebre pi\`u generali sono le algebre di 
supersimmetria. Oltre ai generatori del 
gruppo di Poincar\`e, traslazioni $P_\m$ e rotazioni $M_{\m\n}$, 
queste contengono dei generatori 
anticommutanti $Q^i_\a$ e $\bar{Q}^i_{\dot{\b}}$, con $i=1,...,N$,
che trasformano come spinori di Weyl
\be
[P_\m,Q^i_\a] = [P_\m,\bar{Q}^i_{\dot{\b}}] = 0  \ , \hspace{0.5cm}
\{ Q_\a^i, \bar{Q}^j_{\dot{\b}} \} = 2 \s^\m_{\a \dot{\b}} P_\m \d^i_j
\  , \hspace{0.5cm}
\{ Q_\a^i, Q^j_\b \} = \e_{\a\b} Z^{ij}  \ .
\label{susy}
\ee
Le $Z^{ij}$, dette cariche centrali, sono state gi\`a incontrate nel quarto 
capitolo e sono essenziali per lo studio delle dualit\`a.
Per scrivere in modo elegante e compatto delle lagrangiane supersimmetriche \`e 
conveniente 
introdurre il concetto di superspazio, un'esten\-sione dello spazio di 
Minkowski che include un certo numero di coordinate fermioniche anticommutanti. 
Un punto del superspazio \`e 
caratterizzato da $(x^\m, \th_i, \bar{\th}_i)$, dove $x^\m$ sono le coordinate 
usuali e $\th^i_\a$, $\bar{\th}^i_{\dot{\b}}$ sono spinori di Weyl, 
$i=1,...,N$, dove $N$ \`e il 
numero di cariche di supersimmetria. Se trascuriamo le 
rotazioni, un elemento del gruppo di supersimmetria si pu\`o scrivere
\be
G(x, \th , \bar{\th}) = e^{i(-xP+ \th Q + \bar{\th} \bar{Q})}  \ .
\ee
Queste trasformazioni soddisfano la legge di moltiplicazione
\be
G(\a, \xi, \bar{\xi}) G(x, \th, \bar{\th}) = 
G(x^{'}, \th^{'}, \bar{\th^{'}}) \ ,
\ee 
con
\be
x^{'} = x + \a + i( \xi \s \bar{\th} - \th \s \bar{\xi}) \ , \hspace{1cm}
\th^{'} = \th + \xi \ , \hspace{1cm} \bar{\th}^{'} = \bar{\th} +
\bar{\xi} \ .
\ee 

Un supercampo \`e una funzione $F(x, \th , \bar{\th})$ 
definita sul superspazio. Le variabili $\th$ anticommutano e quindi 
l'espansione di F in $\th$ \`e un polinomio che
unifica in un'unica struttura campi con spin diverso. Trasformazioni 
infinitesime di supersimmetria sono definite da
\be
\d_S F = F(x^{'}, \th^{'}, \bar{\th}^{'}) - F(x, \th, \bar{\th}) =
-i(\xi Q + \bar{\xi} \bar{Q} - \a P) F   \ ,
\ee
dove gli operatori differenziali $P, Q, \bar{Q}$ sono
\be
P_\m = i \partial_\m \ , \hspace{1cm}
Q_\a = i( \frac{\partial}{\partial \th^\a} 
- i (\s^\m \bar{\th})_\a \partial_\m ) \ , \hspace{1cm}
\bar{Q}^{\dot{\a}} = i( - \frac{\partial}{\partial \th_{\dot{\a}}} 
+ i (\th \s^\m )^{\dot{\a}} \partial_\m ) \ .
\ee
Gli operatori 
\be
D_\a = i( \frac{\partial}{\partial \th^\a} 
+ i (\s^\m \bar{\th})_\a \partial_\m ) \ , \hspace{1cm}
\bar{D}_{\dot{\a}} = i( - \frac{\partial}{\partial \th^{\dot{\a}}} 
- i (\th \s^\m )^{\dot{\a}} \partial_\m ) \ ,
\ee
soddisfano l'algebra 
$\{ D_\a , \bar{D}_{\dot{\a}} \} = - 2 \s^\m_{\a \dot{\a}} P_\m$ e 
anticommutano con $Q_\a$ e $\bar{Q}_{\dot{\a}}$; sono quindi delle derivate 
covarianti che anticommutano con la variazione di supersimmetria
\be
D_\a(\d_S F) = - \d_S ( D_\a F) \ ,
\ee
e possono essere utilizzate per imporre 
vincoli sui supercampi in modo da ottenere rappresentazioni irriducibili 
dell'algebra di supersimmetria. Cominciamo con il considerare, ad esempio, un 
supercampo scalare $\F$; per definizione questo supercampo trasforma secondo
\be
\d_S  \F = -i[ - \a P + \xi Q + \bar{\xi}\bar{Q}, \F ] =
-i( - \a P + \xi Q + \bar{\xi} \bar{Q}) \F  \ .
\label{variaz}
\ee
Lo sviluppo in potenze di $\th$ \`e,
\be
\F(x, \th , \bar{\th}) = \f + \th \psi + \bar{\th} \bar{\chi} + \th^2 M
+ \bar{\th}^2 N + \th \s^m \bar{\th} V_\m + \th^2 \bar{\th} \bar{\l}
+ \bar{\th}^2 \th j + \th^2 \bar{\th}^2 F    \  ,
\label{scal}
\ee
e contiene campi di spin $ \le 1$.
Se si impone il vincolo 
\be
\bar{D}_{\dot{\a}} \F = 0 \ , 
\label{chir}
\ee
il numero di componenti 
indipendenti in (\ref{scal}) viene ridotto; in particolare ridefinendo la 
variabile indipendente $x$ come 
$ y = x + i \th \s^\m \bar{\th} $, le derivate covarianti divengono
\be
D_\a = i( \frac{\partial}{\partial \th^\a} 
+ 2i (\s^\m \bar{\th})_\a \partial_\m ) \ , \hspace{1cm}
\bar{D}_{\dot{\a}} = - i \frac{\partial}{\partial \th^{\dot{\a}}} \ ,
\ee
e il vincolo (\ref{chir}) si riduce all'indipendenza di 
$ \F(y, \th , \bar{\th})$
da $\bar{\th}$. Abbiamo quindi
$ \F(y, \th , \bar{\th}) = \f + \sqrt{2} \th \psi + \th^2 F$ o, nelle 
coordinate originali,
\be
\F( x, \th , \bar{\th}) = \f - i \th \s^\m \bar{\th} \part_\m \f
 - \frac{1}{4} \th^2 \bar{\th}^2 \Box \f + \sqrt{2} \th \psi 
+ \frac{i}{\sqrt{2}} \th^2( \part_\m \psi \s^\m \bar{\th} ) + \th^2 F     
\ .
\ee
Un supercampo che soddisfa il vincolo (\ref{chir}) \`e detto chirale;
in modo analogo il vincolo $D_\a \F = 0 $ definisce un supercampo 
antichirale. Un supercampo chirale contiene uno scalare complesso, 
uno spinore di Weyl e un campo complesso ausiliario $F$;
dalla definizione (\ref{variaz}) si ricavano le trasformazioni 
di supersimmetria di queste componenti
\beq
& & \d \f = \sqrt{2} \xi \psi  \ , \nonumber \\
& & \d \psi = -i \sqrt{2} \s^\m \bar{\xi} \part_\m \f + \sqrt{2} \xi F \ ,
\nonumber \\
& & \d F = -i \sqrt{2} \bar{\xi} \bar{\s}^\m \part_\m \psi \ .
\eeq
Per descrivere teorie di gauge supersimmetriche \`e essenziale introdurre 
supercampi vettoriali, definiti dal vincolo $V^{\dagger} = V$. Lo sviluppo in 
componenti \`e in questo caso
\beq
& & V = c - i \th \chi + i \bar{\th} \bar{\chi} + \frac{i}{2} \th^2 (M+iN)
- \frac{i}{2} \bar{\th}^2 (M-iN) - \th \s^\m \bar{\th} V_\m \nonumber \\
& & + i \th^2 \bar{\th}( \bar{\l} + \frac{i}{2} \bar{\s}^\m \part_\m \chi)
- i \bar{\th}^2 \th ( \l + \frac{i}{2} \s^\m \part_\m \bar{\chi})
+ \bar{\th^2 \bar{\th}^2}{2}( D - \frac{1}{2} \Box C )   \ .
\label{vetto}
\eeq
La trasformazione $V \mapsto V + \F + \F^{\dagger}$, con $\F$ un campo chirale, 
\`e una generalizzazione supersimmetrica dell'usuale trasformazione di gauge e 
implica infatti $ \l \mapsto \l$, $D \mapsto D$ e 
$V_\m \mapsto V_\m - \part_\m(2 Im \f)$. Utilizzando questa trasformazione \`e
possibile porre a zero i campi $C, \chi , M, N$ nella (\ref{vetto}); 
in questo gauge, detto gauge di Wess e Zumino, $V^n = 0$ per
$n \ge 3$. I campi $\l$ , $V_{\m\n} = \part_\m V_\n - \part_\n V_\m$ e $D$
formano da soli una rappresentazione dell'algebra di supersimmetria e le loro 
variazioni sono
\beq
& & \d V_{\m\n} = -i( \xi \s_\n \part_\m \bar{\l} 
+ \bar{\xi} \bar{\s}_\n \part_\m \l) 
-i( \xi \s_\m \part_\n \bar{\l} 
+ \bar{\xi} \bar{\s}_\m \part_\n \l)  \ , \nonumber \\
& & \d \l = i \xi D + i \s^{\m\n}V_{\m\n} \xi \ ,  \nonumber \\
& & \d D = \xi \s^\m \part_\m \bar{\l} - \bar{\xi} \bar{\s}^\m \part_\m \l 
\ .
\label{gaug}
\eeq

Osserviamo che la componente pi\`u alta di un supercampo trasforma in una 
derivata totale dei campi di dimensione pi\`u bassa, ed \`e quindi possibile
scrivere azioni supersimmetriche integrando in $dx^4$
i termini proporzionali a $\th^2$ 
dei campi chirali, o $\it{F-termini}$, e i termini proporzionali a
$\th^2 \bar{\th}^2$ dei campi vettoriali, o $\it{D-termini}$. Se definiamo gli 
elementi di misura $d^2 \th = - \frac{1}{4} \e_{\a\b} d \th^\a d \th^\b$,
$d^2 \bar{\th} = - \frac{1}{4} \e_{\a\b} d \bar{\th}^\a d \bar{\th}^\b$ e 
$d^4 \th = d^2 \th d^2 \bar{\th}$, gli F-termini si possono scrivere
$\int d^4x d^2 \th \F$, con $\F$ un supercampo chirale, e i D-termini
$\int d^4x d^4 \th V$, con $V$ un supercampo vettoriale.

Le trasformazioni di gauge dei campi chirali sono definite da
$ \F \mapsto e^{-g \L} \F$, dove $\L = \L^a T^a$, con $\L^a$ campi chirali e
$T^a$ generatori di un gruppo di Lie. L'interazione tra campi di gauge \`e 
materia \`e descritta dal D-termine del campo vettoriale
$\F^{\dagger}e^{gV} \F$, invariante se $V$ trasforma secondo
\be
e^{gV} \mapsto e^{-g \L^{\dagger}} e^{gV} e^{g \L} \ .
\ee
Il termine cinetico si ottiene definendo il supercampo spinoriale chirale
\be
W_\a = \frac{i}{4} \bar{D}^2(e^{-gV}D_\a e^{gV}) \ ,
\ee
che trasforma secondo $W_\a \mapsto e^{-ig \L} W_\a e^{ig \L}$. L'azione
per una teoria di supercampi chirali e vettoriali \`e data quindi da
\be
S = \int d^4x 
\left \{ d^2 \th ( \frac{1}{4g^2} TrW^\a W_\a + f(\F_i) + h.c.)
+  d^4 \th \sum_i ( \F^{\dagger}_i e^{gV} \F_i ) \right \} \ .
\label{sym}
\ee
La funzione $f(\F_i)$ \`e detta superpotenziale ed \`e una funzione olomorfa 
dei supercampi chirali; se si considerano solo teorie rinormalizzabili, $f$ \`e 
una funzione al pi\`u cubica
\be
f(\F_i) = c_i \F_i + \frac{m_{ij}}{2} \F_i \F_j 
+ \frac{\l_{ijk}}{3!} \F_i \F_j \F_k  \ ,
\ee
con coefficienti $c_i$, $m_{ij}$ e $\l_{ijk}$ tali da rispettare l'invarianza 
di gauge. La (\ref{sym}) pu\`o essere scritta in modo pi\`u esplicito nel gauge 
di Wess e Zumino, sviluppandola in componenti. Il risultato \`e
\beq
& & L = (D_\m \f_i   )^{\dagger} (D^\m \f_i) 
+( \frac{i}{2} \bar{\psi}_i \bar{\s}^\m D_\m \psi_i + h.c.)
- \frac{1}{4} (G^a_{\m\n})^2
+( \frac{i}{2} \bar{\l}^a \bar{\s}^\m D_\m \l^a + h.c.) \nonumber \\
& & - \left ( \frac{1}{2} \frac{\part^2 f( \f)}{\part \f_i \part \f_j}
+ i \sqrt{2} \frac{\part D^a}{ \part \f_i} \psi_i \l^a + h.c. \right )
- V(F, F^{\dagger}, D) \ .
\label{lsym}
\eeq
Nella (\ref{lsym}) $D_\m$ rappresenta l'usuale derivata covariante e
i campi $F$ e $D$ vanno espressi in termini degli altri 
campi della teoria risolvendo le loro equazioni del moto
\be
F^{\dagger}_i = - \frac{\part f( \f)}{\part \f_i} \ , \hspace{1cm}
D^a = - g \sum_i \f_i^{\dagger} T^a \f_i \ .
\ee
Il potenziale effettivo ad albero \`e
\be
V(F, F^{\dagger}, D) = \sum_i F^{\dagger}_i F_i + \frac{1}{2} \sum_a (D^a)^2
\ .
\label{pot}
\ee
Se il gruppo di gauge contiene dei fattori $U_Y(1)$ \`e possibile introdurre in
(\ref{sym}) il termine
\be
S_{FI} = 2 \xi_Y \int d^4x d^4 \th V_Y \ ,
\ee
detto termine di Fayet e Iliopoulos \cite{FI}; il D-termine per $V_Y$ \`e dato 
alora da
\be
D_Y = - g_Y \sum_i Y_i \f^{\dagger}_i \f_i - \xi_Y \ .
\ee
In assenza di termini di Fayet-Iliopoulos le lagrangiane supersimmetriche
sono prive di divergenze quadratiche. Esiste infatti un teorema di non 
rinormalizzazione \cite{GSR} in base al quale il superpotenziale non viene 
rinormalizzato; nella serie perturbativa si hanno solo divergenze logaritmiche 
che rinormalizzano la funzione d'onda dei campi vettoriali e chirali e la 
costante di accoppiamento. Grazie a queste buone propriet\`a ultraviolette, i 
parametri ad albero della teoria sono stabili ed una gerarchia di scale 
stabilita a livello ad albero non viene eliminata dalle correzioni 
quantistiche. In presenza di un termine di Fayet-Iliopoulos, si possono 
produrre termini della forma $\xi \int \d^4 \th V$ che divergono 
quadraticamente; in questo caso la divergenza \`e proporzionale, ad un loop, a
$TrY$ e si annulla se il gruppo $U_Y(1)$ \`e non anomalo.
I termini che possono essere aggiunti ad una lagrangiana supersimmetrica 
rompendo in modo esplicito la supersimmetria senza tuttavia introdurre 
divergenze quadratiche sono detti termini di rottura $\it{soffice}$ e sono 
stati classificati in \cite{GG}.

Osserviamo che il potenziale (\ref{pot}) \`e semidefinito positivo; questa \`e 
una propriet\`a generale delle teorie globalmente supersimmetriche. 
Dall'algebra di supersimmetria (\ref{susy}) segue infatti la relazione
\be
H = P^0 = \frac{1}{4} \sum_{i, \a} \{ Q^i_\a, \bar{Q}^i_{\dot{\a}} \} \ .
\label{envuot}
\ee
La supersimmetria \`e spontaneamente rotta quando lo stato di vuoto non \`e 
annullato da tutte le cariche di supersimmetria, ovvero quando
$Q^i_\a | 0 \ran \ne 0$. La (\ref{envuot}) 
implica allora che stati fondamentali 
con energia nulla sono supersimmetrici mentre stati fondamentali con energia 
positiva rompono la supersimmetria spontaneamente. Un modo equivalente di 
studiare la rottura spontanea della supersimmetria consiste nel considerare 
valori di vuoto per i campi della teoria non invarianti rispetto a 
trasformazioni supersimmetriche. In un campo chirale, l'unica componente che 
pu\`o acquistare un valore di vuoto che rompa la supersimmetria senza violare
l'invarianza di Lorentz \`e il campo ausiliario $F$; in questo caso si parla di 
rottura di tipo F o di O'Raifeartaigh \cite{OR}. In un campo vettoriale, 
l'unica componente utile \`e il campo ausiliario $D$ e si parla quindi di 
rottura di tipo D o di Fayet e Iliopoulos \cite{FI}. Quando 
$ \lan F_i \ran \ne 0$ o $ \lan D^a \ran \ne 0$, l'energia di vuoto \`e 
positiva, $V_{min} = f^2 >0$. Il fermione
\be
\psi_g = \frac{1}{f} [ - \lan F_i \ran \psi_i 
+ \frac{i}{\sqrt{2}} \lan D^a \ran \l^a ]  \ ,
\label{gold}
\ee
ha massa nulla e rappresenta l'analogo del bosone di Glodstone che compare 
quando una simmetria bosonica globale viene rotta spontaneamente. Il campo 
fermionico (\ref{gold}) viene pertanto detto goldstino e 
$ \lan \d \psi_g \ran = - \sqrt{2} f \xi$. Per il goldstino si possono 
ricavare relazioni del tutto analoghe a quelle che si ottengono per i bosoni di 
Goldstone e  in particolare per i pioni quando si studia la rottura spontanea 
della simmetria chirale. Ad esempio, se $S^\m_\a$ indica la supercorrente e
$|k, \l \ran$ un goldstino con momento $k$ e elicit\`a $\l$ si pu\`o dimostrare 
che
\be
\lan 0 | S^\m_\a |k, \l \ran = -i \sqrt{2}[\g^\m u(k, \l )]_\a f \ ,
\ee
dove $u(k, \l )$ \`e una soluzione d'onda piana dell'equazione di Dirac.
Il parametro d'ordine per la rottura della 
supersimmetria \`e quindi la costante di 
decadimento del goldstino; inoltre se $m_b$ e $m_f$ sono le masse di bosoni e 
fermioni in un dato supermultipletto e $e_g$ \`e
il loro accoppiamento di Yukawa al 
goldstino, si pu\`o dimostrare un 
analogo della relazione di Goldberger e Treiman
\be
m_b^2 - m_f^2 \sim f e_g \ .
\ee

Il pi\`u semplice modello supersimmetrico \`e il modello di Wess e Zumino che 
descrive un solo supercampo chirale con superpotenziale
\be
f(\F) = \frac{m}{2} \F^2 + \frac{\l}{3} \F^3 \ .
\ee
Questo modello ha due vuoti supersimmetrici, per $\f = 0$ e per
$ \f = - \frac{m}{\l}$. Per rompere spontaneamente la supersimmetria 
utilizzando gli F-termini \`e necessario che il superpotenziale contenga 
termini lineari nei campi chirali, altrimenti esiste sempre una soluzione 
supersimmetrica che si ottiene quando $\lan \f_i \ran = 0$, $ \forall i$.
Consideriamo ad esempio un modello con tre campi chirali $X,A,B$ e con 
il superpotenziale \cite{OR}
\be
W = \l X( A^2 - \m^2) + m AB \ .
\ee
Le equazioni 
\be
\frac{\part W}{\part X} = \l ( A^2 - \m^2 ) = 0 \ , \hspace{1cm}
\frac{\part W}{\part A} = 2 \l XA + mB = 0 \ , \hspace{1cm}
\frac{\part W}{\part B} = mA = 0 \ , 
\ee
non hanno soluzione e quindi la supersimmetria \`e spontaneamente rotta. Il 
minimo del potenziale si ha quando $\lan A \ran = \lan B \ran =0$;
si noti che 
$\lan X \ran $ \`e lasciato indeterminato. L'unico F-termine non nullo \`e
$\lan F^{\dagger}_X \ran = - \l \m^2$ 
e quindi $\m$ determina la scala di rottura 
della supersimmetria; il goldstino \`e $\psi_X$. 

Come esempio di rottura con D-termini consideriamo una teoria di gauge $U(1)$ 
accoppiata ad un multipletto chirale  
\be
L = \int d^2 \th \frac{1}{4 g^2} (W^\a W_\a) 
+ \int d^4 \th  ( \F^{\dagger} e^{gV} \F + \xi V )     \ .
\ee
Il superpotenziale \`e assente per invarianza di gauge, ma il D termine \`e 
permesso. La relazione
\be
D = - \xi - e \f^{\dagger} \f \ ,
\ee
implica che il minimo del potenziale si ha per 
$ \f^{\dagger} \f = - \frac{\xi}{e}$ se $ \frac{ \xi}{e} < 0$ e per
$ \f^{\dagger} \f = 0 $ se $ \frac{ \xi}{e} > 0$; 
nel primo vuoto la supersimmetria non 
\`e rotta ma \`e rotta la simmetria $U(1)$, nel secondo vuoto la simmetria di 
gauge \`e intatta mentre la supersimmetria \`e rotta spontaneamente,
poich\`e $V_{min} = \frac{\xi^2}{2}$.

\subsection{Supersimmetria locale}

Abbiamo visto che la rottura spontanea della supersimmetria nel caso globale 
\`e accompagnata dalla comparsa nello spettro di un fermione di massa nulla, il 
goldstino. Questo campo pu\`o essere eliminato con un meccanismo del tutto 
analogo a quello di Higgs se si considerano teorie nelle quali 
la supersimmetria \`e una simmetria locale, le teorie di supergravit\`a.
In supergravit\`a, come nelle altre teorie di gauge, quando la
supersimmetria si rompe spontaneamente, il gravitino acquista massa assorbendo 
gli stati di elicit\`a $\pm \frac{1}{2}$ del goldstino; questo fenomeno \`e 
detto fenomeno di super-Higgs. Per supergravit\`a con N=1 in quattro dimensioni
consideriamo 
la parte della Lagrangiana quadratica nei campi fermionici
\beq
e^{-1}L^{(2)} &=& -i   [ \frac{1}{2} \bar{\chi}_i Z_{ij} \g^\m \part_\m \chi_j
- \frac{1}{2e} \e^{\m\n\r\s} \bar{\psi}_\m \g_5 \g_\n D_\r \psi_\s  \nonumber \\
&+&  A_i \bar{\chi}_i \g \psi + M \bar{\psi}_\m \s^{\m\n} \psi_\n
+ \frac{1}{2} \bar{\chi}_i N_{ij} \chi_j ] - V(\f)   ,
\eeq
dove $Z_{ij}, A_i, M$ e $N_{ij}$ sono funzioni dei campi scalari. 
Dalle variazioni dei campi
\be
\d \chi_i = C_i \e  \ , \hspace{0.5cm}
\d \psi_\m = -D_\m \e + B \g_\m \e \   , \hspace{0.5cm}
\d \f_a = i\bar{\chi}_iL_{ia}\e    \   ,
\ee
dove $C_i( \f) = $,  seguono le equazioni basilari \cite{CJS}
\beq
& & 2C_iZ_{ij}C_j - 3M^2 = V   \  , \nonumber \\
& & N_{ij}C_j + 2Z_{ij}C_jM = - \frac{\part V}{\part \f_a}L_{ai}  \ .
\label{susyb}
\eeq
Osserviamo che in (\ref{susyb}) il potenziale scalare non \`e definito 
positivo come nel caso globale, 
a causa del contributo del gravitino. Tecnicamente il nuovo termine 
trae origine dai campi ausiliari $S$ e $P$ presenti nel multipletto 
gravitazionale.
Si ha supersimmetria non rotta se 
\be
\lan \d \chi \ran = 0 \rar C_i(\f_0) = 0   \   ,
\ee
dove $\f_0$ \`e il punto stazionario in cui 
$ \part V/ \part \f_a(\f_0) = 0$.
Se risulta $M(\f_0) = 0$ abbiamo come vuoto lo spazio di Minkowski; se
$M(\f_0) \neq 0$ abbiamo come vuoto lo spazio di anti-de Sitter o lo
spazio di de-Sitter.
La supersimmetria \`e rotta se $\lan \d \chi \ran \neq 0 $.
La costante cosmologica si annulla se
\be
3M^2(\f_0) = 2C_i(\f_0)Z_{ij}(\f_0)C_j(\f_0)    \  ,
\ee
da cui, reintroducendo la massa di Planck
\be
M(\f_0) = \frac{1}{m_p} \sqrt{\frac{8 \p}{3}} \sqrt{2C_iZ_{ij}C_j}  \ ,
\ee
dove le costanti $C_i$ hanno dimensioni di $[m]^2$. La relazione tra massa del 
gravitino $m_{3/2}$, massa di Planck $m_p$ e scala di rottura della 
supersimmetria $E$ \`e
\be
m_{3/2} = \sqrt{\frac{8 \p}{3}} \frac{E^2}{m_p}   \ .
\ee
Le (\ref{susyb}) si generalizzano al caso di teorie con supersimmetria estesa, 
per le quali \`e possibile una rottura parziale della supersimmetria. Per
$N = 1$, l'accoppiamento localmente supersimmetrico di campi di materia e di 
gauge \`e parametrizzato da due funzioni
\be
G(z,\bar{z}) = J(z, \bar{z}) + ln|f(z)|^2  , \hspace{1cm}
f_{\a\b}(z)  .
\ee
I termini cinetici dei campi scalari e dei vettori di gauge sono
\be
Z_i^j \partial \f^i \partial \f_j + Re(f_{\a\b}) F^{\a}_{\m\n}F^{\b\m\n}
\ ,
\ee
dove $Z^i_j = \frac{\part^2 G}{\part z_i \part \bar{z}^j} $
\`e una matrice hermitiana e $ f_{\a\b}$ \`e una matrice di funzioni olomorfe
la cui parte reale determina i termini cinetici e quella immaginaria i
$\th$ termini.
La massa del gravitino \`e determinata da $G$ ed \`e
\be
m_{3/2}(z, \bar{z}) = e^{G/2} = |f(z)|e^{J/2}   \ .
\ee
La rottura spontanea con costante cosmologica nulla si ha solo quando
nel punto stazionario  $\part V/ \part z = 0$, $f(z) \neq 0$ e
\be
e^G G_i (G^{-1})^i_j G^j + \frac{1}{2}(Ref^{-1}_{\a\b}D^\a D^\b) = 3e^G  .
\label{nocos}
\ee
Generalmente $D^\a =0$ e quindi la condizione (\ref{nocos}) si riduce a 
\be
G_i (G^{-1})^i_j G^j = 3 \  .
\label{nosc}
\ee
Esiste un`ampia classe di modelli, 
noti come supergravit\`a $\it{"no-scale"}$ \cite{noscale}
e ottenibili anche come limite di 
teorie di stringa, 
per i quali la (\ref{nosc}) \`e naturalmente soddisfatta senza 
aggiustare accuratamente i parametri della teoria.


\subsection{Teoria di stringhe}

Passiamo ora alla teoria di stringhe. In teoria di stringhe 
uno dei primi meccanismi proposti per la rottura spontanea della supersimetria
\`e la condensazione di gaugini \cite{glu}; 
il problema principale in questo caso \`e che il 
meccanismo si basa interamente 
sull'azione effettiva di bassa energia e non consente calcoli espliciti.
Consideriamo la teoria HE compattificata su di uno spazio di  Calabi-Yau
$K$; il gruppo di gauge \`e rotto naturalmente
a $E_6 \times E_8$ e linee di Wilson possono 
rompere ulteriormente $E_6$ ad un gruppo realistico per la fisica di bassa 
energia e $E_8$ ad un opportuno gruppo $Q$. In dieci dimensioni l'azione per i 
campi di gauge e per il campo gravitazionale \`e
\be
L = - \frac{m_p^8}{8 \p} R - \frac{1}{4 g_Y^2} \f^{-3/4} Tr F_{\m\n}F^{\m\n} \ ,
\label{10sug}
\ee
dove $m_p$ \`e la scala di Planck e $g_Y$ la costante di accoppiamento di 
gauge. Scegliamo su $K$ una metrica $g_{ij} = e^\s g_{ij}^0$ con
$\int_K \sqrt{g^0} = m_p^{-6}$. Integrando la (\ref{10sug}) su $K$ e riscalando 
la metrica quadridimensionale abbiamo per i campi di gauge
\be
L = - \frac{1}{4}\f^{-3/4}e^{3\s}(m_p^6g_Y^2)^{-1} Tr F^2_{\m\n}  \  .
\ee
Dato che il valore di $g_Y$ pu\`o essere assorbito nella definizione di $\f$, 
possiamo porre $m_p^6g_Y^2 = 1$. L'accoppiamento di gauge effettivo in quattro 
dimensioni \`e quindi $g_4^2 = \f^{3/4}e^{-3\s}$ e la teoria di gauge di $Q$ 
diviene fortemente accoppiata alla scala
\be
\m \sim m_pe^{-2\s}e^{-\frac{1}{2b_0}g_4^2}  \  ,
\ee
dove $b_0$ \`e il coefficiente della funzione $\b$ ad un loop di $Q$. Nella 
regione $\f, \m << m_p$, la gravit\`a rappresenta una piccola correzione alle
interazioni di gauge di $Q$. In una teoria di SYM con soli campi di gauge, la 
supersimmetria non viene spontaneamente rotta, ma si sviluppa un condensato di 
gaugini con $ \lan Tr \bar{\chi}\chi \ran \sim \m^3$; se la teoria viene 
accoppiata a multipletti scalari contenenti un assione, come avviene quando si 
compattifica la supergravit\`a in dieci dimensioni, il condensato di gaugini 
pu\`o invece
causare la rottura spontanea della supersimmetria. Per i gaugini di 
Majorana-Weyl in dieci dimensioni, l'unico bilineare gauge invariante \`e
$Tr \bar{\chi}\G_{\m\n\a}\chi$, e tenendo conto che un eventuale condensato
deve essere un singoletto 
del gruppo $SU(3)$ di olonomia, si vede che
$ \lan Tr\bar{\chi}\G_{ijk} \chi \ran = A \W_{ijk}$, dove $\W$ \`e la 3-forma 
olomorfa del Calabi-Yau e $A$ un numero complesso che va interpretato come
$ \lan Tr\bar{\chi} ( 1 - \g_5 ) \chi \ran $.

Studiando le trasformazioni dei campi, si pu\`o vedere che, 
in presenza di un condensato di gaugini, la supersimmetria \`e spontaneamente 
rotta; per quanto riguarda l'energia di vuoto, la lagrangiana contiene un 
termine
\be
\d L = - \frac{4 \p}{m_p^8} \frac{1}{384}(Tr \bar{\chi}\G_{\m\n\a}\chi)^2 \  ,
\label{pert}
\ee
che d\`a origine ad un potenziale per i campi $\s$ e $\f$ della forma
\be
V(\f, \s) \sim \frac{m_{Gut}^6}{m_p^2}e^{-\frac{3}{2b_0}g_4^2} \  ,
\ee
con $m_{Gut} = m_p e^{-2 \s}$.
Questo potenziale non si annulla in alcun punto a distanza finita nello spazio 
dei moduli e non ha estremi con $\f$ e $\s$ costanti al finito
( $\f = 0$ \`e infinitamente distante perch\`e il termine cinetico 
di $\f$ \`e $(\part_\m \f/ \f)^2$ ); di conseguenza predice un 
continuo cambiamento di $\f$ e $\s$ con $\s \rar \infty$ e $\f \rar 0$. Questo 
comportamento \`e accettabile se le costanti della natura cambiano abbastanza 
lentamente da essere in accordo con i limiti sperimentali e se l'energia di 
vuoto si annulla abbastanza rapidamente rispetto alla differenza di 
massa tra bosoni e fermioni. Vediamo quindi che la sola condensazione dei 
gaugini non \`e sufficiente ad indurre un potenziale scalare con minimi ad 
energia nulla;
uno scenario pi\`u soddisfacente si ottiene 
consentendo un valore d'aspettazione non nullo nello spazio interno 
per la field strenght 
$F_{\m\n\a}$ della 2-forma $a_{\m\n}$ che compare tra i campi della teoria.
In questo caso bisogna tenere conto di due nuovi termini che insieme a 
(\ref{pert}) formano un quadrato perfetto
\be
\d L = - \frac{3 \k^2}{4 g_Y^4} \f^{-3/2} ( F_{\m\n\r} 
- g_Y^2 \sqrt{2} \f^{3/4} Tr \bar{\chi} \G_{\m\n\r} \chi )^2  \  .
\ee
Possiamo allora cancellare l'energia di vuoto se 
\be
\lan  F_{\m\n\r}  \ran =
- g_Y^2 \sqrt{2} \f^{3/4} \lan ( Tr \bar{\chi} \G_{\m\n\r} \chi )^2 \ran  \ .
\ee
Si pu\`o verificare che anche in questo vuoto con 
$ \lan F_{\m\n\r} \ran \neq 0$ la supersimmetria \`e spontaneamente rotta. Posto
$ \lan F_{ijk} \ran = c m_p^3 \W_{ijk}$, si pu\`o verificare che la nuova forma 
del potenziale \`e
\be
V(\f,\s,\a) \sim \m_p^4 e^{-6 \a} \f^{-3/2}
| c - h e^{i \a} e^{ - \frac{3}{2b_0} g^2_4}|^2 \   ,
\label{pots}
\ee
dove $\a$ \`e la combinazione lineare dei due assioni, partners di $\s$ e $\f$,
che si accoppia al condensato dei gaugini. Entro 
il limite di validit\`a della teoria effettiva, non sono necessari particolari 
aggiustamenti dei parametri per ottenere una costante cosmologica nulla; per 
ogni valore di $c$ diverso da zero, esiste un vuoto con supersimmetria rotta e 
$\L = 0$ dato dal minimo di (\ref{pots}). 
Dal punto di vista della teoria quadridimensionale, si 
pu\`o verificare che la struttura del potenziale scalare \`e quella propria 
delle supergravit\`a no-scale.
Questa discussione \`e ovviamente solo qualitativa; inoltre il 
meccanismo di condensazione dei gaugini 
non \`e stato ancora riprodotto in un modello di stringhe consistente.

Per avere un migliore controllo quantitativo del fenomeno \`e essenziale 
costruire esplicitamente dei modelli di stringa che descrivano vuoti con 
supersimmetria spontaneamente rotta;
in questo capitolo discuteremo un metodo che consente di costruire modelli  
di questo tipo deformando la 
funzione di partizione di modelli supersimmetrici. Questo metodo \`e 
un'estensione alla teoria di stringhe del meccanismo introdotto da Scherk e 
Schwarz in teoria di campi \cite{ss} ed \`e stato descritto in 
\cite{sub1,sub2} per il caso di stringhe chiuse orientate. 
L'estensione di questo meccanismo a teorie di stringhe aperte \`e stata 
ottenuta solo recentemente \cite{ads}, e presenta caratteristiche peculiari 
legate alla struttura non perturbativa della teoria di stringhe
che ne rendono molto pi\`u 
interessanti le prospettive fenomenologiche. 

Osserviamo, prima di concludere, che in teoria di stringhe 
la costante cosmologica ad un loop coincide con l'integrale sullo spazio dei 
moduli della funzione di partizione; se la supersimmetria \`e non rotta 
$Z(\t)$ si anulla punto per punto nello spazio dei moduli e $\L = 0$; un metodo 
per ottenere una costante cosmologica nulla ad un loop quando lo spettro non 
\`e supersimmetrico \`e stato proposto in 
\cite{al} e consiste nel considerare modelli nei quali $\L$ \`e il 
prodotto di due 
forme modulari, ed \`e nullo se queste trasformano sotto due diverse 
rappresentazioni di una simmetria dello spazio dei moduli (simmetria di
Atkin-Lehner). Recentemente, motivati dalla corrispondenza AdS/CFT, sono stati 
costruiti modelli non supersimmetrici di stringa nei quali $\L$ si annulla ad 
uno e due loops \cite{ks}. Viene anche suggerito che la costante cosmologica 
continui ad annullarsi anche a loops superiori; un'indagine diretta di questa 
affermazione \`e resa 
particolarmente difficile dalle numerose sottigliezze che si incontrano nei 
calcoli di ampiezze per stringhe fermioniche, ma pu\`o essere condotta 
costruendo modelli duali per queste compattificazioni.


\section{Meccanismo di Scherk e Schwarz}
\markboth{}{}

\subsection{Teoria dei campi}
\markboth{}{}

La costruzione di modelli con supersimmetria spontaneamente rotta richiede una 
generalizzazione dell'usuale riduzione dimensionale \cite{ss,ss2}. 
Nella riduzione dimensionale ordinaria, le coordinate di una teoria in 
$D+E$ dimensioni vengono divise in $D$ coordinate spaziali $x^\m$ ed E
coordinate interne $y^\a$, che formano generalmente uno spazio
compatto. Nei casi pi\`u semplici, come le compattificazioni
toroidali, si considerano 
quindi solo  campi e leggi di trasformazione indipendenti dalle coordinate 
interne; questo in particolare implica che una simmetria della teoria in $D+E$ 
dimensioni corrisponde ad una simmetria in $D$ dimensioni; ad 
esempio, teorie con supersimmetria semplice in $D+E$ dimensioni danno
origine 
a teorie con supersimmetria estesa in $D$ dimensioni.
Possiamo generalizzare l'operazione di riduzione dimensionale in modo da 
rompere alcune simmetrie;
consideriamo come esempio una teoria 
in $D+1$ dimensioni, compattifichiamo su un cerchio ed espandiamo un campo 
scalare complesso in modi di Fourier come
\be
S(x,y) = e^{imy} \sum_{n = - \infty}^{\infty} S_n(x)e^{2 \p i y \frac{n}{L}}  \ ,
\ee
dove $L$ \`e il periodo della cordinata interna $y$ e $x$ rappresenta le 
coordinate non compatte. Questo campo non \`e ad un valore in $D+1$ dimensioni 
in quanto $S(x,y+L) = e^{imL}S(x,y)$ ma questo comportamento non \`e 
inconsistente se la teoria possiede una simmetria globale $U(1)$ tale che
$S \mapsto e^{i \a} S$. Il campo a pi\`u valori rappresenta allora
una sezione di un fibrato non banale, continua lungo il cerchio. 
Nel limite $L \rar 0$ i modi $S_n$ , $ n \neq 0$ divengono infinitamente 
massivi; l'unico campo che non pu\`o essere trascurato \`e $S_0$ che acquista 
una massa $m$. Per $m=0$ si ottiene l'usuale riduzione dimensionale.

La generalizzazione di questo meccanismo \`e chiara: si permette ai campi ed 
alle leggi di trasformazione di dipendere dalle coordinate $y$ 
in un modo ben definito utilizzando una simmetria della teoria, in modo che i 
campi definiscano sezioni continue di fibrati sullo spazio compatto 
E-dimensionale. In questo modo 
la dipendenza da $y$ si cancella essenzialmente dalla Lagrangiana e pu\`o 
essere fattorizzata fuori dalle leggi di trasformazione; la conseguenza pi\`u 
importante \`e che la dipendenza dalle coordinate $y$ d\`a origine a termini di 
massa e nuovi accoppiamenti in $D$ dimensioni. 

Si possono utilizzare due classi di simmetrie; simmetrie della
variet\`a interna
($\it{interne}$) 0 ulteriori simmetrie ($\it{esterne}$), come R
parit\`a. Ad esempio si pu\`o 
applicare questa tecnica alla supergravit\`a con $N=1$ in $d=4$  \cite{ss}
per ottenere la supergravit\`a con $N=2$ in $d=3$ in una fase spontaneamente 
rotta utilizzando la trasformazione di chiralit\`a $U(1)$ 
$\psi_\m \mapsto e^{ia \G_5} \psi_\m$ ponendo
\be
\psi_\m(x,y) = e^{im \G_5 y} \psi_\m(x)  \   , \hspace{1cm}
V^r_\m(x,y) = V^r_\m(x)  \   .
\ee
Lo stesso meccanismo pu\`o essere applicato anche a simmetrie locali; 
ad esempio 
per un campo di Yang-Mills si ha un'interessante realizzazione del meccanismo 
di Higgs realizzato con scalari nell'aggiunta. Se infatti poniamo
\be
A_\m(x,y) = e^{i \L y} A_\m e^{-1 \L y}   \  ,
\ee
e ridefiniamo $A_\m$ tramite una trasformazione di gauge, otteniamo il 
potenziale equivalente
\be
A^{'}_\m(x,y) = \frac{1}{ig}e^{i \L y} \partial_\m e^{-i \L y} 
+ A_\m(x)  \   .
\ee
Da questa formula segue che per $ \m = 0,...,4$ $~A_\m$ \`e immutato, mentre 
$A_4$, che in quattro dimensioni descrive uno scalare nell'aggiunta, viene 
traslato con una costante. Questo \`e un altro esempio di come lo shift in un 
campo scalare pu\`o dare origine a masse per i vettori di gauge.
Sfruttando l'invarianza rispetto a trasformazioni generali di coordinate, si 
pu\`o rompere la supersimmetria delle teorie di supergravit\`a con $N=4$ e $N=
8$ in $d=4$ realizzandole come riduzioni dimensionali generalizzate delle 
teorie di supergravit\`a con $N=1$ in $d=10$ e $d=11$ \cite{ss,ss2}.


\subsection{Teoria di stringhe}

L'estensione di questo meccanismo in teoria di stringhe \`e stata considerata 
in \cite{sub1,sub2}. In essenza si tratta di modificare  la 
funzione di partizione di un modello di stringhe chiuse, introducendo una 
dipendenza dal momento e dal winding lungo una dimensione compatta in modo 
compatibile con l'invarianza modulare. 
Applicando questo metodo a modelli fermionici in cinque 
dimensioni ulteriormente compattificati su di un cerchio,
si ottengono modelli con supersimmetria spontaneamente rotta il cui spettro di 
bassa energia riproduce quello descritto in \cite{ss,ss2} per le supergravit\`a 
con $N=4$ e $N=8$ in quattro dimensioni costruite come riduzioni dimensionali 
generalizzate. 

Consideriamo modelli di Tipo II ed eterotici in cinque dimensioni.
Descriviamo le 
coordinate interne introducendo 15 fermioni sinistri:
\be
\psi^\m \ , \psi^5 \ , \chi^i \ , \w^i \ , y^i \hspace{1cm} i = 2,...,6 \ ,
\ee
ed un insieme identico di fermioni destri per la tipo II oppure 42 fermioni 
destri $\f^a$, $a = 3,...,44$ per l'eterotica;
assumiamo inoltre che le condizioni al bordo consentano di formare delle coppie 
complesse $f_i$.
La funzione di partizione del modello si scrive:
\be
Z(\t,\bar{\t}) = \frac{1}{(\sqrt{\t_2}\h(\t)\h(\bar{\t}))^{3/2}}
\sum_{spin}C\pmatrix{a_L & a_R\cr b_L & b_R}Z_L\pmatrix{a_L\cr b_L}(0,\t)  
Z_R\pmatrix{a_R\cr b_R}(0,\bar{\t})  \    ,
\ee
dove
\be
Z_L\pmatrix{a_L\cr b_L}(v_L,\t)  = 
\prod_{i=1}^{N_L} \frac{\th\pmatrix{a^i_L\cr b^i_L}(v^i_L,\t)}{\h(\t)}  \  ,
\ee
e $Z_R$ si scrive in modo analogo; $a_L$, $b_L$ sono vettori di condizioni al 
bordo per gli $N_L$ fermioni complessi ($N_L = 9$ nel nostro caso), con
condizioni di quasiperiodicit\`a lungo le due direzioni del toro
\be
1: f_{i_{L(R)}} \mapsto - e^{2 \p i a_{i_{L(R)}}}f_{i_{L(R)}} \ , \hspace{1cm}
\t: f_{i_{L(R)}} \mapsto - e^{-2 \p i b_{i_{L(R)}}}f_{i_{L(R)}} \ .
\ee
Compattifichiamo ora la quinta coordinata su di un cerchio; la funzione di 
partizione diviene:
\be
Z(\t,\bar{\t}) = \frac{1}{(\sqrt{\t_2}\h(\t)\h(\bar{\t}))}
\sum_{m,n} Z_{m,n}(\t,\bar{\t})
\sum_{spin}C\pmatrix{a_L & a_R\cr b_L & b_R}Z_L\pmatrix{a_L\cr b_L}(0,\t)  
Z_R\pmatrix{a_R\cr b_R}(0,\bar{\t})  \    ,
\ee 
dove
\be
Z_{m,n}(\t,\bar{\t}) = \frac{R}{\sqrt{2\t_2}\h(\t)\h(\bar{\t})}
e^{-\frac{\p R^2}{2 \t_2}|m - n\t|^2} \   ,
\ee
descrive la coordinata bosonica compatta e trasforma, rispetto ad elementi del 
gruppo modulare, secondo:
\be
Z_{m,n}(\frac{a\t+b}{c\t+d}) = Z_{dm-bn,-cm+an}(\t) \    .
\ee
Per deformare il modello, definiamo
\beq
& & v_L = e_L(m - \t n) \  , \hspace{1cm} u_L = ne_L \cdot v_L \nonumber \\
& & v_L = e_R(m - \t n) \  , \hspace{1cm} u_R = ne_R \cdot v_R  \   .
\eeq
con $e_L$, $e_R$ vettori arbitrari con lo stesso numero di componenti dei
vettori di condizioni al bordo $a_{L,R}$, $b_{L,R}$. Possiamo ora costruire una 
nuova funzione di partizione invariante modulare:
\beq
Z(\t,\bar{\t}) &=& \frac{1}{(\sqrt{\t_2}\h(\t)\h(\bar{\t}))}
\sum_{m,n} Z_{m,n}(\t,\bar{\t}) \nonumber \\
& & \sum_{spin}C\pmatrix{a_L & a_R\cr b_L & b_R}
Z_L\pmatrix{a_L\cr b_L}(v_L,u_L,\t)  
Z_R\pmatrix{a_R\cr b_R}(v_R,u_R,\bar{\t})  \nonumber \\
&=& \frac{1}{\sqrt{\t_2}(\h(\t)\h(\bar{\t}))^{N_L}}
\sum_{m,n} Z_{m,n}(\t,\bar{\t}) \\
& & \sum_{spin}\tilde{C}\pmatrix{a_L & a_R\cr b_L & b_R}
\prod_{i_L=1}^{N_L}\th_L\pmatrix{a_{i_L} - ne_{i_L}\cr b_{i_L} + me_{i_L}}(\t)  
\prod_{i_R=1}^{N_R}\th_R\pmatrix{a_{i_R} - ne_{i_R}\cr b_{i_R} + me_{i_R}}
(\bar{\t})  \nonumber  ,
\label{defss}
\eeq
dove 
\beq
Z_L\pmatrix{a_L\cr b_L}(v_L,u_L,\t) &=& e^{-i\p u_L}
Z_L\pmatrix{a_L\cr b_L}(v_L,\t) \ , \nonumber \\
Z_R\pmatrix{a_R\cr b_R}(v_R,u_R,\bar{\t}) &=& e^{i\p u_R}
Z_R\pmatrix{a_R\cr b_R}(v_R,\bar{\t}) \  ,
\eeq
e dove, nella seconda espressione in (\ref{defss}) per la funzione di 
partizione,
\be
\tilde{C}\pmatrix{a\cr b} = e^{2 \p i n e \cdot(b + \frac{m}{2}e)}
C\pmatrix{a\cr b} \ ,
\ee
con prodotto scalare lorentziano : $a \cdot b = a_Lb_L - a_Rb_R$.
Che si tratti della corretta generalizzazione pu\`o essere stabilito nel modo 
seguente. Consideriamo un campo spaziotemporale $\F_i$ al quale corrisponde 
l'operatore vertice $V_i(z, \bar{z})$; l'operatore
$\F_i(k_\a)V_i(z,\bar{z})e^{ik^\a X_{\a}}$ deve avere dimensione 
conforme $(1,1)$ ed essere 
ad un valore. Se utilizziamo un background della forma
$e^{ieQX_5}\F_(X_\m,X_5)$, i vincoli soddisfatti in precedenza da $V_i$ 
devono essere ora 
soddisfatti da $\F_i(k)e^{ieQX_5}V_i(z,\bar{z})e^{ik^\m X_\m}$; 
questo significa 
che le condizioni al bordo per $V_i$ vanno modificate. Se $V_i$ \`e 
un autostato di $Q$ con carica $q_i$ e che la quinta coordinata \`e 
compattificata su di un cerchio, abbiamo, nel settore con momento e 
avvolgimento dati da $(n,m)$:
\be
V_i(\s+2 \p,\t)= e^{-ieq_inR}V_i(\s,\t)  \  ,  \hspace{1cm}
V_i(\s,\t+2 \p)= e^{-ieq_imR}V_i(\s,\t)  \   .
\ee
Si possono utilizzare anche delle simmetrie discrete realizzate da qualche 
operatore sul worldsheet $U$; in questo caso le condizioni al bordo sono
\be
V_i(\s+2 \p,\t)= U^n_{ij}V_j(\s,\t)  \  ,  \hspace{1cm}
V_i(\s,\t+2 \p)= U^m_{ij}V_j(\s,\t)  \   .
\ee
Nei modelli costruiti con fermioni liberi, utilizzando la carica
\be
Q = \oint dz e^i_L f^{*}_if_i + 
\oint d \bar{z} e^i_R \bar{f}^{*}_i \bar{f}_i   \   ,
\ee
si ottiene esattamente la funzione di partizione (\ref{defss}).
L'invarianza modulare si verifica facilmente utilizzando le propriet\`a di 
trasformazione delle funzioni $\th$. Quando $e_L=e_R=0$ la (\ref{defss}) 
coincide con l'usuale compattificazione toroidale da cinque a quattro 
dimensioni e d\`a origine a teorie con $N=8,6,4,2,0$ supersimmetrie; 
in generale descrive invece una nuova classe di modelli con supersimmetria 
spontaneamente rotta. Per vedere questo in modo esplicito, \`e conveniente 
riscrivere la (\ref{defss}) come una traccia sullo spazio degli stati; per fare 
questo basta sostituire alle funzioni $\th$ la loro rappresentazione in 
termini di traccia sullo spazio degli stati del fermione complesso 
corrispondente:
\beq
\frac{\th \pmatrix{a_L^i\cr b_L^i}(\t)}{\h(\t)} &=&
Tre^{2 \p i b^i_L Q^i_L}q^{H^i_L} \nonumber \\
&=& e^{2 \p i a^i_Lb^i_L}q^{\frac{a^{i2}_L}{2}-\frac{1}{24}}
\prod_{n=1}^{\infty}(1 + q^{n+a^i_L - \frac{1}{2}}e^{2 \p i b^i_L})
(1+q^{n - a^i_L - \frac{1}{2}}e^{2 \p i b^i_L}) \  ,
\eeq
dove $H^i_L$ \`e l'Hamiltoniana per il fermione $f^i$ twistato con $a^i_L$:
\be
H^i_L = \sum_{n= - \infty}^{\infty} \left (n + a^i_L - \frac{1}{2} \right )
:b^{\dagger}_{n+a^i_L - \frac{1}{2}} b_{n+a^i_L-\frac{1}{2}}:
+ \left ( \frac{a^2_{iL}}{2}-\frac{1}{24} \right ) \ ,
\ee
e dove $Q = a + F$, 
con $a$ vettore delle condizioni al bordo lungo la direzione 
$1$ del toro $(1, \t)$ e $F$ operatore numero fermionico per i fermioni 
complessi $f^i$.

Con questa sostituzione si ottiene:
\be
Z(\t, \bar{\t}) = \sum_{m,n=-\infty}^{\infty}Tr g q^{L_0} \bar{q}^{\bar{L_0}} \  ,
\label{ham}
\ee
dove $g$ \`e la proiezione GSO, indipendente dalle cariche $e$ e 
\beq
L_0 &=& \frac{(Q_L - ne_L)^2}{2} + 
\frac{  (\frac{m+e \cdot Q - \frac{n}{2}e^2}{R}+\frac{nR}{2}  )^2}{2} - 
\frac{1}{2} + osc.  \  ,
\nonumber \\
\bar{L}_0 &=& \frac{(Q_R - ne_R)^2}{2} + 
\frac{  (\frac{m+e \cdot Q - \frac{n}{2}e^2}{R}-\frac{nR}{2} )^2}{2} - 
\frac{1}{2} + osc.  \  .
\label{el}
\eeq
Da (\ref{el}) segue che
\beq
L_0 + \bar{L}_0 &=& (L_0 + \bar{L}_0)_S + \frac{1}{R^2}
 \left ( e \cdot Q - \frac{n}{2}e^2  \right )^2 - n(e_LQ_L + e_RQ_R) \nonumber \\
&+& \frac{n^2}{2}(e^2_L + e^2_R) - \frac{2m}{R^2} \left (e \cdot Q - \frac{n}{2}e^2 \right ) 
\  ,  \nonumber \\
L_0 - \bar{L}_0 &=& (L_0 - \bar{L}_0)_S \ ,
\label{ssspec}
\eeq
dove $(L_0)_S$ denota l'operatore di Virasoro per l'usuale compattificazione 
sul cerchio. Da (\ref{ssspec}) risulta evidente che i due modelli hanno gli
stessi stati fisici ma un diverso spettro di masse; in particolare, gli stati
con $m=n=0$, non massivi nel limite $e \rar 0$, hanno masse pari a 
\be
M = \frac{1}{R}|e \cdot Q| \  .
\ee
Questi modelli sono allora vuoti distinti di una stessa teoria, con rottura 
spontanea della simmetria quando si passa da un vuoto ad un altro meno 
simmetrico.
\`E utile, per avere una migliore comprensione di questo tipo di deformazione, 
formulare il modello utilizzando un reticolo bosonico piuttosto che fermioni 
interni. Si pu\`o passare dal modello fermionico a quello bosonico 
identificando il momento $p$ di uno stato bosonico con la carica $Q$ 
\cite{hig}; 
la funzione di partizione diviene allora:
\be
Z(\t, \bar{\t}) = \sum_{p \in \G_{N_L,N_R}} Tr g q^{L_0}\bar{q}^{\bar{L}_0}  \  ,
\ee
dove $g$ \`e lo stesso operatore presente in (\ref{ham}) e $\G_{N_L,N_R}$ \`e 
un reticolo lorentziano, pari e autoduale; inoltre
\be
L_0 = \frac{p_L^2}{2}+\frac{p_{0L}^2}{2} - \frac{1}{2} + osc. \   , \hspace{1cm}
\bar{L}_0 = \frac{p_R^2}{2}+\frac{p_{0R}^2}{2} - \frac{1}{2} + osc. \  ,
\label{ssbos}
\ee
dove
\be
p_{0L} = \left ( \frac{m}{R} + \frac{nR}{2} \right ) \ , \hspace{1cm}
p_{0R} = \left ( \frac{m}{R} - \frac{nR}{2} \right )  \  .
\ee
Lo spettro descritto da (\ref{ssspec}) si ottiene da (\ref{ssbos}) tramite la 
sostituzione
\be
p \mapsto p - ne \ , \hspace{1cm} m \mapsto m + e \cdot p - \frac{n}{2} e^2 \ .
\ee
Questa trasformazione coincide con la deformazione del reticolo
$\G_{N_L,N_R} \otimes \G_{1,1}$ data da
\beq
& & p_L \mapsto p_L - \xi_L(p^0_L-p^0_R) \ , \hspace{1cm}
p^0_L \mapsto p^0_L + \xi \cdot p - \frac{\xi^2}{2}(p^0_L-p^0_R) \   , 
\nonumber \\
& & p_R \mapsto p_R - \xi_R(p^0_L-p^0_R) \ , \hspace{1cm}
p^0_R \mapsto p^0_R + \xi \cdot p - \frac{\xi^2}{2}(p^0_L-p^0_R) \   ,
\eeq
con $\xi = \frac{e}{R}$. Possiamo scrivere questa deformazione in modo pi\`u 
compatto come
\be
P \mapsto Pe^T = P \left (1 + T + \frac{T^2}{2} \right )  \   ,
\ee
con $P = (p_L,p^0_L;p_R,p^0_R)$ e
\be
T = \pmatrix{0 & -\xi_L & 0 & \xi_L\cr \xi^T_L & 0 & -\xi_R^T & 0\cr
0 & -\xi_R & 0 & \xi_R\cr \xi^T_L & 0 & -\xi_R^T & 0}  \   .
\ee
La matrice $T$ \`e un elemento di $SO(N_L+1,N_R+1)$ e quindi il meccanismo di 
Scherk-Schwarz corrisponde nel linguaggio bosonico ad una deformazione 
di Narain del reticolo \cite{tor1}.

Quando si utilizza il meccanismo di Scherk-Schwarz per la rottura di simmetrie 
di gauge, le cariche $e$ possono essere arbitrarie e corrispondono a direzioni 
piatte del potenziale scalare della teoria effettiva. Un esempio ben noto di 
deformazioni continue \`e quello appena citato per le compattificazioni su 
reticoli lorentziani $\G_{p,q}$ che hanno uno spazio dei moduli 
$SO(p,q)/SO(p) \times SO(q)$. Possiamo dimostrare che queste deformazioni 
possono essere realizzate come compattificazioni $\it{a^{`}~la~}$ 
Scherk-Schwarz modificando una compattificazione toroidale sul reticolo
$\G_{1,1}^d$ ( $\G_{1,1}^d \otimes \G_{8k}$ nel caso eterotico ) utilizzando le 
$d$ simmetrie $Q^i$ che agiscono sui bosoni $X_i$ come traslazioni:
$ Q^j  : X_i \mapsto X_i + Q_i^j$. Detto $A_i^j = \d_i^j$ ed introducendo anche 
un campo $B_{ij}$ di background, si ottiene la deformazione
\beq
& & m_i \mapsto A^{-t}{}_i^j(m_j - Q_j^ap_a + \frac{1}{2} Q^a_jQ_a^ln_l
+ B_j^kn_k)  \  , \nonumber \\
& & n_i \mapsto A_i^j n_j  \ ,  \hspace{1cm}  p_a \mapsto p_a - Q_a^j n_j  \  ,
\eeq
che dipende proprio da $d(d+8k)$ parametri.

Il meccanismo di Scherk-Schwarz, oltre a riprodurre le deformazioni 
continue del vuoto di una teoria di stringhe, permette anche di realizzare 
deformazioni discrete. In particolare se la carica $Q$ non commuta con la 
supercorrente, non si pu\`o interpolare in modo continuo tra vuoti distinti, ma 
si ha solo un insieme discreto di vuoti consistenti. Osserviamo che la 
discussione fatta precedentemente per stabilire la corrispondenza
tra gli stati del modello originario e gli stati del modello 
deformato non dipende dalla 
natura continua o discreta delle cariche $e$. Sappiamo inoltre che le 
deformazioni continue sono associate a valori di aspettazione di correnti
$J_LJ_R$ con dimensione $(1,1)$, ovvero corrispondenti a stati fisici; d'altra 
parte i bilineari fermionici $\l^{*}\l$ non sono la componente alta di un 
supercampo primario e quindi non corrispondono a stati fisici; questo significa 
che la rottura della simmetria che avviene nei modelli con deformazioni 
discrete \`e analoga al fenomeno di super-Higgs, in cui acquista un valore 
d'aspettazione non un campo fisico, come nel caso del fenomeno di Higgs usuale, 
ma un campo ausiliario.
In definitiva, al meccanismo di Scherk-Schwarz per la
rottura di simmetrie di gauge e della supersimmetria 
corrispondono in teoria di stringhe deformazioni rispettivamente 
continue e discrete del reticolo Lorentziano che definisce la 
compattificazione.

Per rompere la supersimmetria dobbiamo quindi utilizzare una carica $Q$
che commuta con la supercorrente $T_F$ e che trasforma i gravitini in modo non 
banale. Consideriamo ad esempio il caso della Tipo II ed utilizziamo le cariche 
nella sottoalgebra di Cartan dei due gruppi $SO(4)$ che agiscono su
$ ( \chi^2  ,  \chi^3  ,  \chi^4   ,   \chi^5)_{L,R}$,
introducendo i parametri $e_L^1$, $e_L^2$, $e_R^1$ e $e_R^2$. Gli stati 
non massivi quando $e_L=e_R=0$ sono dati dal prodotto
\be
( |\m \ran + | 5 \ran + | i \ran + |4, \a \ran )_L \otimes
( |\n \ran + | 5 \ran + | j \ran + |4, \b \ran )_R   \  ,
\ee
dove $| \m \ran = \psi^\m_{-1/2}| 0 \ran$, 
$| 5 \ran = \psi^5_{-1/2}| 0 \ran$,
$| i \ran = \chi^i_{-1/2}| 0 \ran$  e
$|4, \a \ran $ sono gli stati del settore di Ramond dei fermioni
$\psi$ e $\chi$. 
Questo spettro riproduce quello della supergravit\`a con $N=8$. \`E facile 
vedere come le masse di questi stati cambiano quando $e_L$ e $e_R$ hanno un 
valore non nullo e lo spettro risultante coincide con quello calcolato in
\cite{ss2}.

La costante cosmologica ad un loop di questi modelli \`e fortemente soppressa 
\be
\L = \int_F \frac{d \t_1 d \t_2}{\t_2^2}  Z(e_L,e_R; \t)  \  .
\ee
L'assenza di tachioni assicura che si tratta di una quantit\`a finita; inoltre 
utilizzando la formula
\be
\frac{1}{2} \sum_{a,b=0}^1(-1)^{a+b+ab}\prod_{i=1}^4\th\pmatrix{a\cr b}(v_i) =
- \prod_{i=1}^4 \th_1(v_i^{'})   \   ,
\ee
dove
\beq
v_1^{'} &=& \frac{1}{2}(-v_1+v_2+v_3+v_4) \  , \hspace{1cm}
v_2^{'} = \frac{1}{2}(v_1-v_2+v_3+v_4)  \  ,  \nonumber \\
v_3^{'} &=& \frac{1}{2}(v_1+v_2-v_3+v_4)  \  ,  \hspace{1cm}
v_4^{'} = \frac{1}{2}(v_1+v_2+v_3-v_4) \ .
\eeq
si ottiene, per il modello con $N=8$,
\be
\L = \int_F \frac{d \t_1 d \t_2}{\t_2^2}  \sum_{m,n} 
[\th_1^2(v_L^1 + v_L^2;\t)\th_1^2(v_L^1 - v_L^2;\t)
\bar{\th}_1^2(v_R^1 + v_R^2;\t)\bar{\th}_1^2(v_R^1 - v_R^2;\t) 
\hat{Z}_{m,n}(\t)  \  ,
\ee
dove $\hat{Z}_{m,n}(\t)$ \`e la parte della funzione di partizione indipendente 
da $v_L$e $v_R$.
Ricordando che $\th_1(v;\t)$ \`e una funzione dispari di $v$, possiamo scrivere
\be
\L = cost. (\prod_{i=1}^4m^i_{3/2})^2  \  ,
\ee
dove le $m_{3/2}^i$ sono le masse dei gravitini
\beq
m^1_{3/2} &=& \frac{1}{2}|e_L^1 + e_L^2|  \ ,  \hspace{0.6cm}
m^2_{3/2} = \frac{1}{2}|e_L^1 - e_L^2|  \ ,  \nonumber \\
m^3_{3/2} &=& \frac{1}{2}|e_R^1 + e_R^2|  \ ,  \hspace{0.6cm}
m^4_{3/2} = \frac{1}{2}|e_R^1 - e_R^2|  \ .
\eeq
La costante cosmologica non riceve correzioni quadratiche o quartiche nei 
parametri di rottura della supersimetria, ma \`e sempre proporzionale a
$(m^1_{3/2}m^2_{3/2}m^3_{3/2}m^4_{3/2})^2$. Questo estende il risultato di pura 
supergravit\`a riguardo l'annullarsi delle correzioni a $\L$ della forma
$str M^{2n}$ per $n < 4$, all'intero spettro della teoria di stringhe.
Una trattazione del tutto analoga pu\`o farsi per teorie effettive con
N=4 compattificando un modello eterotico; in questo caso si verifica l'assenza 
di correzioni quadratiche.

Coma abbiamo visto,
una soluzione della teoria di stringhe con supersimmetria spontaneamente rotta, 
si ottiene agendo su un vuoto supersimmetrico
con un operatore definito sul worldsheet e consistente con 
l'invarianza di Lorentz e con la supercorrente; in 
questo modo si stabilisce una corrispondenza uno a uno tra stati nelle due 
fasi. Dato che la presenza di supersimmetria non rotta garantisce l'assenza di 
tachioni e la stabilit\`a del background piatto almeno a due loops 
nell'espansione perturbativa e probabilmente ad ogni ordine, ci si pu\`o 
aspettare che, almeno per un sottoinsieme dei modelli, questo livello di 
consistenza persista anche nella fase spontaneamente rotta.
Un altro vantaggio di questa costruzione consiste nel fatto che 
la conoscenza dello spettro del modello in entrambe le fasi, i vincoli 
imposti dalla supersimmetria e alcune caratteristiche generiche della teoria di 
stringhe, consentono di fissare in modo non ambiguo l'azione effettiva completa 
in presenza di rottura spontanea.

Quando si applica questa costruzione ad un orbifold $Z_2$, generato da un 
operatore $h$ che inverte la coordinata $X_5$, 
la carica usata per la deformazione e l'operatore $h$ 
devono soddisfare la condizione di consistenza
\be
\{ Q , h \} = 0 \ .
\label{suscons}
\ee
La (\ref{suscons}) deriva dal fatto che $h$ cambia segno al momento e 
all'avvolgimento lungo $X_5$ e deve commutare con $L_0$ e $\bar{L}_0$.
Come esempio consideriamo un modello fermionico eterotico in cinque dimensioni 
definito dai vettori di condizioni al bordo
\beq
& & F = \{ \frac{1}{2} : {\rm tutti~i~fermioni} \} \ ,    \nonumber \\
& & S = \{ \frac{1}{2} : \psi^\m, \psi^5, \chi^2, \chi^3, \chi^4, \chi^5, 
\chi^6; 0 : { \rm i~fermioni~restanti} \} \ ,    \\
& & b_1 = \{ \frac{1}{2} : \psi^\m, \psi^5, \chi^2, \w^3, \w^4, \w^5, 
\w^6, \f^3, \f^4, ..., \f^{17}, \f^{18}; 0 : { \rm i~fermioni~restanti} \} \  .
\nonumber
\eeq
Il modello ha gruppo di gauge $SO(16) \times SO(26)$. Possiamo costruire un 
orbifold $Z_2$ utilizzando la proiezione
\be
\hat{g} \f = - \f {\rm ~se~}  \f \in b_2 \ , \hspace{1cm}
\hat{g} \f = \f {\rm ~altrimenti}   \  ,
\ee
dove
\be
b_2 = \{ X^5, \psi^\m, \chi^3, \chi^5, y^2, y^4, y^6,
\f^1,..., \f^{10}, \f^{17}, ..., \f^{21+8k} \}  \  ,
\ee
con $k = 0,1$. 
Definiamo ora i fermioni complessi
\be
\chi_c^1 = \frac{\chi^3 + i \chi^4}{\sqrt{2}}  \ ,  \hspace{0.5cm}
\chi_c^2 = \frac{\chi^5 + i \chi^6}{\sqrt{2}}  \ ,  \hspace{0.5cm}
y_c^1 = \frac{y^3 + i y^4}{\sqrt{2}} \  ,  \hspace{0.5cm}
y_c^1 = \frac{y^3 + i y^4}{\sqrt{2}} \  .
\ee
Se introduciamo le cariche
\be
Q_1 = \oint \frac{dz}{2 \p i} [ \bar{\chi}_c^1 \chi_c^1 
+ \bar{y}_c^1 y_c^1 ]  \ ,   \hspace{1cm}
Q_2 = \oint \frac{dz}{2 \p i} [ \bar{\chi}_c^2 \chi_c^2 
+ \bar{y}_c^2 y_c^2 ]  \  ,
\ee
l' operatore
\be
T_3 = e^{2 \p i(e_1Q_1 + e_2Q_2)}  \    ,
\ee
con $e_1 = e_2 = \frac{1}{2}$ lascia invariata la supercorrente,
anticommuta con l'operazione di orbifold e pu\`o essere quindi utilizzato per
realizzare il meccanismo di Scherk-Schwarz.


\section{Stringhe aperte}
\markboth{}{}

In \cite{ads} si \`e osservato che l'estensione di questo meccanismo a modelli 
di stringhe aperte ha propriet\`a peculiari, legate alla struttura non 
perturbativa della teoria. Abbiamo visto che, 
rompendo la supersimmetria con il 
meccanismo di Scherk-Schwarz, si introduce una relazione tra la scala di 
rottura e la dimensione dello spazio interno; per modelli eterotici, si ha 
infatti ad albero
\be
m_{3/2} = m_{1/2} \sim R^{-1}  \  ,
\ee
con $m_{3/2}$ e $m_{1/2}$ masse del gravitino e del gaugino. Per avere masse 
fenomenologicamente accettabili dell'ordine del $TeV$, si deve quindi 
considerare un raggio molto grande, dell'ordine del $(TeV)^{-1}$. Dato che i 
bosoni di gauge del modello standard avvertono la dimensione aggiuntiva, solo 
in particolari modelli \`e possibile evitare pesanti correzioni agli 
accoppiamenti di gauge riproducendo la fenomenologia nota.

Quando si utilizza il meccanismo di Scherk-Schwarz per deformare un modello di 
stringhe aperte, la massa del gravitino \`e ancora legata al 
raggio interno, ma tutte le volte che i campi di gauge vivono su D-brane 
ortogonali alla direzione compatta la massa del gaugino se ne disaccoppia; 
abbiamo quindi a livello ad albero
\be
m_{3/2} \sim R^{-1}  \  ,  \hspace{1cm} m_{1/2} = 0 \ .
\ee
La rottura della supersimmetria sulla brana \`e mediata da interazioni 
gravitazionali e quindi soppressa da potenze della massa di Planck $m_p$, 
ovvero $m_{1/2} \sim O(m_{3/2}/m_p)$. \`E quindi possibile scegliere un raggio 
interno ad una scala intermedia: $R^{-1} \sim 10^{12} - 10^{14} ~GeV$. 

Abbiamo in effetti due meccanismi basilari di rottura; se le stringhe aperte 
hanno condizioni al bordo di Neumann lungo la direzione compatta, la
supersimmetria viene rotta dagli shifts
\be
n \mapsto n \ , \hspace{1cm} m \mapsto m + ep - \frac{n}{2}e^2 \ , \hspace{1cm}
p \mapsto p - ne \ ,
\label{sshift}
\ee 
con $n=0$, dato che stringhe aperte con condizioni al bordo di Neumann non 
hanno avvolgimento. La costruzione del discendente aperto mostra come in questo 
caso abbiamo la naturale estensione del meccanismo di Scherk-Schwarz, con 
supersimmetria rotta ad albero e masse proporzionali a $R^{-1}$.
Pi\`u interessante \`e il caso in cui le stringhe aperte hanno condizioni al 
bordo di Dirichlet lungo il cerchio, come avviene per la Tipo $I^{'}$ 
se si identifica il 
cerchio con la direzione non perturbativa della M teoria; 
effettuando una trasformazione di T dualit\`a, si ottiene un modello di Tipo I 
nel quale gli shifts introdotti nel settore chiuso alterano l'avvolgimento:
\be
m \mapsto m \ , \hspace{1cm} n \mapsto n + ep - \frac{m}{2}e^2 \ , \hspace{1cm}
p \mapsto p - me \ .
\label{mshift}
\ee 
In questo caso, la costruzione del discendente aperto mostra in modo esplicito 
come la supersimmetria non viene rotta ad albero per i settori non
massivi 
dello 
spettro aperto.

Per costruire questi modelli, partiamo da una teoria in $d=9$ con 
$e_L=e_R=(0,0,0,1)$ \cite{ads}. La 
funzione di partizione della $IIB$ riportata in (\ref{defss}), scritta in modo 
esplicito per questo caso \`e
\be
T = \frac{1}{(\sqrt{\t_2}\h\bar{\h})^{7/2}}\sum_{m,n}Z_{m,n}
\frac{1}{4(\h\bar{\h})^4}
|\th^4_3 - (-1)^m\th^4_2 - (-1)^n \th^4_4 - (-1)^{n+m}\th^4_1|^2  \  .
\label{momtor}
\ee
Il modello con shifts negli avvolgimenti si ottiene da (\ref{momtor}) 
scambiando $m$ ed $n$. Per semplicit\`a chiameremo il modello con shifts nei 
momenti modello di Scherk-Schwarz ed il modello con shifts negli avvolgimenti 
modello di M teoria. Possiamo riscrivere le (\ref{momtor}) in modo pi\`u adatto 
alla costruzione del discendente aperto utilizzando i caratteri di $SO(8)$ e le 
somme reticolari:
\beq
E^{'}_0 &=& \frac{1+(-1)^n}{2}Z_{m,n} \  , \hspace{1cm} 
O^{'}_0 = \frac{1-(-1)^n}{2}Z_{m,n}  \ , \nonumber \\
E^{'}_{1/2} &=& \frac{1+(-1)^n}{2}Z_{m+\frac{1}{2},n}  \ , \hspace{1cm}
O^{'}_{1/2} = \frac{1-(-1)^n}{2}Z_{m+ \frac{1}{2},n} \  .
\eeq
Per il modello di M teoria introduciamo in modo analogo
\beq
E_0 &=& \frac{1+(-1)^m}{2}Z_{m,n} \  , \hspace{1cm} 
O_0 = \frac{1-(-1)^m}{2}Z_{m,n} \  , \nonumber \\
E_{1/2} &=& \frac{1+(-1)^m}{2}Z_{m,n+\frac{1}{2}} \  , \hspace{1cm}
O_{1/2} = \frac{1-(-1)^m}{2}Z_{m,n+\frac{1}{2}}  \ .
\eeq
Risommando l'indice $m$ per il modello di Scherk-Schwarz e l'indice $n$ per il 
modello di M teoria, otteniamo le due funzioni di partizione:
\be
T_{SS} = E^{'}_0(V\bar{V}+S\bar{S}) - E^{'}_{1/2}(S\bar{V}+V\bar{S}) 
+ O^{'}_0(O\bar{O}+C\bar{C}) - O^{'}_{1/2}(O\bar{C}+C\bar{O})   \ ,
\label{torss}
\ee
\be
T_{M} = E_0(V\bar{V}+S\bar{S}) - E_{1/2}(S\bar{V}+V\bar{S}) 
+ O_0(O\bar{O}+C\bar{C}) - O_{1/2}(O\bar{C}+C\bar{O})  \  .
\label{torm}
\ee
Osserviamo che lo spettro del modello (\ref{torss}) contiene un tachione per
$R \leq \sqrt{\ap}$, associato a $O\bar{O}$ e che il modello di Tipo IIB si 
ottiene nel limite $R \rar \infty$; lo spettro del modello (\ref{torm}) 
contiene un tachione per $R \geq \sqrt{\ap}$ e il modello di Tipo IIB si 
ottiene nel limite $R \rar 0$.
La (\ref{torss}) si pu\`o ottenere come un orbifold asimmetrico del modello
OB utilizzando la simmetria $g = -(-1)^{G_L}(-1)^n$, dove $G_L$ \`e il numero 
fermionico sinistro sul worldsheet, cos\`\i \, che $-(-1)^{G_L}$ agisce come 
$1$ su $V$ e $S$ e come $-1$ su $O$ e $C$. Partendo infatti da
\be
T_{++} = \frac{1}{2}\sum_{m,n}Z_{m,n}(|O|^2+|V|^2+|S|^2+|C|^2)  \  ,
\ee
si costruiscono immediatamente
\beq
T_{+-} &=& \frac{1}{2}\sum_{m,n}Z_{m,n}(-1)^n(-O\bar{O}+V\bar{V}+S\bar{S}
-C\bar{C}) \   ,  \nonumber \\
T_{-+} &=& \frac{1}{2}\sum_{m,n}Z_{m+\frac{1}{2},n}
(-O\bar{C}-C\bar{O}-S\bar{V}-V\bar{S})  \  ,  \nonumber \\
T_{--} &=& \frac{1}{2}\sum_{m,n}Z_{m+\frac{1}{2},n}(-1)^n
(O\bar{C}+C\bar{O}-S\bar{V}-V\bar{S})  \  .
\eeq
Combinando i vari settori si ottiene (\ref{torss}); in modo analogo 
(\ref{torm}) si ottiene da OB con la simmetria discreta
$g = -(-1)^{G_L}(-1)^m$. Il modello di Scherk-Schwarz si pu\`o anche ottenere 
come orbifold simmetrico della IIB utilizzando $g = (-1)^F(-1)^m$, con 
$F = F_L+F_R$ numero fermionico spaziotemporale, e raddoppiando il raggio del 
modello risultante. Infatti
\beq
T_{++} &=& \frac{1}{2}\sum_{m,n}Z_{m,n}|V-S|^2 \  , \nonumber \\
T_{+-} &=& \frac{1}{2}\sum_{m,n}Z_{m,n}(-1)^m|V+S|^2  \ , \nonumber \\
T_{-+} &=& \frac{1}{2}\sum_{m,n}Z_{m,n+\frac{1}{2}}|O-C|^2 \  , \nonumber \\
T_{++} &=& \frac{1}{2}\sum_{m,n}Z_{m,n+\frac{1}{2}}(-1)^m|O+C|^2  \ ,
\eeq
e combinando i vari settori
\be
Z = E_0(V\bar{V}+S\bar{S})-O_0(V\bar{S}+S\bar{V})
+E_{1/2}(O\bar{O}+C\bar{C})-O_{1/2}(O\bar{C}+C\bar{O}) \ .
\ee
Utilizzando $E_0(2R) = E_0^{'}(R)$ , $O_0(2R) = E_{1/2}^{'}(R)$  , 
$E_{1/2}(2R) = O_0^{'}(R)$ , $O_{1/2}(2R) = O_{1/2}^{'}(R)$  , 
si vede subito che $Z(2R) = T_{SS}(R)$.

Prima di costruire il discendente aperto di questi due modelli
definiamo 
\be
Z_{m+a}(\t) = \frac{q^{\frac{1}{2}(\frac{m+a}{R})^2}}{\h(\t)} \  , \hspace{1cm}
\tilde{Z}_{n+b}(\t) = \frac{q^{\frac{1}{2}((n+b)\frac{R}{2})^2}}{\h(\t)}  \   .
\ee
Queste due funzioni sono legate, passando dal canale trasverso a quello 
diretto, da
\be
\sum_m e^{2i \p mb}Z_{m+a}(-\frac{1}{\t}) = Re^{-2i\p ab} \sum_n e^{-2i \p na}
\tilde{Z}_{2n+2b}(\t)  \   .
\ee
Cominciamo con il modello di Scherk-Schwarz. La parte dello spettro chiuso 
$\W$-invariante include il sottoreticolo $P_L=p_R$, $p^0_L=p^0_R$; per questi 
stati gli shifts (\ref{sshift}) si annullano e la bottiglia di Klein non viene 
alterata dalla rottura della supersimmetria. Abbiamo infatti
\be
K = \frac{1}{2}(V - S)\sum_m Z_m  \  ,
\ee
e nel canale trasverso
\be
\tilde{K} = \frac{2^{9/2}}{2} (V - S) R \sum_n \tilde{Z}_{2n}  \  .
\ee
Nell'anello trasverso pu\`o fluire la porzione diagonale dello spettro
di $T_{SS}$, ovvero $V$ ed $S$ con avvolgimento pari e $O$ e $C$ con 
avvolgimento dispari. Introducendo il numero minimo di coefficienti di 
riflessione, parametrizzati con $(n_1,n_2,n_3,n_4)$, possiamo scrivere
\beq
\tilde{A} &=& \frac{2^{-11/2}R}{2}
([(n_1+n_2+n_3+n_4)^2V - (n_1+n_2-n_3-n_4)^2S]\tilde{Z}_{2n} \nonumber \\
&+& [(n_1-n_2+n_3-n_4)^2O - (n_1-n_2-n_3+n_4)^2C]\tilde{Z}_{2n+1})  \  ,
\eeq
e nel canale diretto
\beq
A &=& \frac{n_1^2+n_2^2+n_3^2+n_4^2}{2}(VZ_{2m} - SZ_{2m+1})
+ (n_1n_2+n_3n_4)(VZ_{2m+1} - SZ_{2m})  \nonumber \\
&+& (n_1n_3+n_2n_4)(IZ_{2m} - CZ_{2m+1})    
+ (n_1n_4+n_2n_3)(IZ_{2m+1} - VZ_{2m})   \  ,
\eeq
che descrive uno spettro aperto con quattro tipi di cariche di Chan-Paton; 
ulteriori coefficienti di riflessione si ottengono introducendo 
linee di Wilson.

Infine, l'ampiezza di M\"obius trasversa \`e fissata dai caratteri comuni 
a $\tilde{K}$ e $\tilde{A}$, con coefficienti che sono la media geometrica dei 
coefficienti con cui compaiono in queste due ampiezze
\be
\tilde{M} = - \frac{R}{\sqrt{2}}((n_1+n_2+n_3+n_4)\hat{V}
(\tilde{Z}_{4n} + \tilde{Z}_{4n+2}) - (n_1+n_2-n_3-n_4)\hat{S}
(\tilde{Z}_{4n} - \tilde{Z}_{4n+2})   \   .
\ee
L'unica sottigliezza riguarda il segno relativo tra i termini proporzionali a 
$\tilde{Z}_{4n}$ e $\tilde{Z}_{4n+2}$, fissato dal vincolo
\be
M^i_a = A^i_{aa} \hspace{0.5cm} {\rm (mod2)}   \  .
\ee
Nel canale diretto
\be
M = - \frac{n_1+n_2+n_3+n_4}{2}\hat{V} Z_{2m} 
- \frac{n_1+n_2-n_3-n_4}{2}\hat{S}Z_{2m+1} \ .
\ee  
Annullando il coefficiente di riflessione totale per i modi non massivi, 
otteniamo le condizioni di tadpole
\be
n_1 + n_2 + n_3 + n_4 = 32  \  , \hspace{1cm}
n_1 + n_2 - n_3 - n_4 = 32  \  ,
\ee 
ovvero $n_3=n_4=0$ e $n_1+n_2 = 32$. Lo spettro aperto risultante
\beq
A &=& \frac{n_1^2+n_2^2}{2}(VZ_{2m} - SZ_{2m+1}) 
+ n_1n_2(VZ_{2m+1} - SZ_{2m})  \   ,    \nonumber \\
M &=& - \frac{n_1+n_2}{2}(\hat{V}Z_{2m} - \hat{S}Z_{2m+1})   \    ,
\label{sopen}
\eeq
corrisponde ad una famiglia di gruppi di gauge $SO(n_1) \times SO(n_2)$ con
$n_1+n_2 = 32$; per livelli con momento intero,  contiene 
vettori nella $(\bf{n_1(n_1-1)/2},\bf{1})$ + $(\bf{1},\bf{n_2(n_2-1)/2})$
e fermioni nella $(\bf{n_1},\bf{n_2})$,
mentre per livelli con momento semiintero, contiene
vettori nella $(\bf{n_1},\bf{n_2})$
e fermioni nella $(\bf{n_1(n_1-1)/2},\bf{1})$ + $(\bf{1},\bf{n_2(n_2-1)/2})$.

Questi modelli sono una deformazione discreta di compattificazioni toroidali
della Tipo I in presenza di 
linee di Wilson che rompono in parte la simmetria di gauge. Se infatti
compattifichiamo la Tipo I su un cerchio ed introduciamo una linea di Wilson 
$W = diag(e^{2 \p ia_1},e^{-2 \p ia_1}...e^{2 \p ia_{16}},e^{-2 \p ia_{16}})$
nella sottoalgebra di Cartan di $SO(32)$, otteniamo nel canale diretto
\beq
T &=& |V - S|^2Z_{m,n} \ , \hspace{1cm}
K = \frac{1}{2}(V - S)Z_m   \ ,\nonumber \\
A &=& \frac{1}{2}(V - S) \sum_{i,j=1}^{32}Z_{2(m+a_i+a_j)} \ , \hspace{1cm}
M = - \frac{1}{2}(\hat{V} - \hat{S})\sum_{i=1}^{32}Z_{2(m+2a_i)}  \ ,
\eeq
e in quello trasverso
\beq
\tilde{K} &=& \frac{2^{9/2}}{2}R(V-S)\tilde{Z}_{2n}  \  ,  \nonumber \\
\tilde{A} &=& \frac{2^{-11/2}}{2}R(V-S)\sum_n(TrW^n)^2\tilde{Z}_{n} \   ,
\nonumber \\
\tilde{M} &=& - \frac{R}{\sqrt{2}}(\hat{V}-\hat{S})
\sum_n TrW^{2n}\tilde{Z}_{2n}  \  .
\eeq
Se scegliamo la linea di Wilson $W = \pmatrix{-1_{n_1} & 0\cr 0 & 1_{32-n_1}}$,
in modo da rompere $SO(32)$ a $SO(n_1) \times SO(32-n_1)$, 
abbiamo, per il settore aperto
\beq
A &=& (V-S) \left ( \frac{n_1^2+n_2^2}{2}Z_{2m} + n_1n_2Z_{2(m+1/2)} \right ) \   , 
\nonumber  \\
M &=& - \frac{n_1+n_2}{2}(\hat{V} - \hat{S})Z_{2m} \  .
\label{susopen}
\eeq
Da queste espressioni \`e evidente che (\ref{sopen}) \`e una deformazione 
discreta, risultante in uno shift $m \mapsto m + 1/2$ delle masse fermioniche,
del modello supersimmetrico (\ref{susopen}). 

Possiamo anche introdurre linee di Wilson direttamente in (\ref{sopen}); 
partendo con $n_2=0$ e con 
tutti gli $a_i$ distinti, il gruppo di gauge viene rotto a $U(1)^{16}$ 
\beq
A &=& \frac{1}{2} \sum_{i,j=1}^{32}(VZ_{2(m+a_i+a_j)}-SZ_{2(m+1/2+a_i+a_j)})  \ 
,   \nonumber \\
M &=& - \frac{1}{2} \sum_{i=1}^{32}(\hat{V}Z_{2(m+2a_i)}
- \hat{S}Z_{2(m+1/2+2a_i)})  \   .
\eeq
Per particolari scelte degli $a_i$, si ristabilisce parzialmente  la simmetria 
originaria; ad esempio con $a_i= 1/2$ per $i =1,...,n_2/2$ e gli altri nulli  
si ottiene proprio il modello (\ref{sopen}) con $SO(n_1)\times SO(n_2)$.

Possiamo, infine, ottenere modelli con gruppi unitari, ovvero con un settore 
aperto orientato, rompendo $SO(32)$ a $U(16)$ con $a_i = 1/4,  i=1,...,16$.
Le ampiezze di anello e M\"obius sono
\beq
A &=& n \bar{n}(VZ_{2m} - SZ_{2m+1}) + \frac{n^2+\bar{n}^2}{2}
(VZ_{2m+1} - SZ_{2m})   \  , \nonumber \\
M &=& - \frac{n + \bar{n}}{2}(\hat{V}Z_{2m+1} - \hat{S}Z_{2m})   \  .
\eeq
Le condizioni di tadpole impongono $n = \bar{n}$ e $n = 16$; lo spettro 
consiste di un vettore nell'aggiunta e di uno spinore nell'antisimmetrica e 
nella sua coniugata per livelli con momento pari, e di un vettore 
nell'antisimmetrica e nella sua coniugata e di uno spinore nell'aggiunta per 
livelli con momenti dispari. Questo modello \`e stato descritto anche in 
\cite{blum}.

Consideriamo adesso il modello di M teoria; in questo caso la bottiglia di 
Klein risente della deformazione
\be
K = \frac{1}{2}(V - S)Z_{2m} = \frac{1}{2}(I - C) Z_{2m+1}  \   ,
\ee
\be
\tilde{K} = \frac{2^{9/2}}{2}R(V \tilde{Z}_{2n} - S\tilde{Z}_{2n+1})  \  .
\ee
Le ampiezze di anello e di M\"obius sono
\beq
 \tilde{A} &=& \frac{2^{-11/2}R}{2}
\Large ( \left [ (n_1+n_2+n_3+n_4)^2V - (n_1+n_2-n_3-n_4)^2S \right ]\tilde{Z}_{2n} \nonumber \\
&+& \left [ (n_1-n_2+n_3-n_4)^2V - (n_1-n_2-n_3+n_4)^2S \right ]\tilde{Z}_{2n+1} \Large ) \   ,
\eeq
\beq
A &=& \left ( \frac{n_1^2+n_2^2+n_3^2+n_4^2}{2}(V-S) + (n_1n_3+n_2n_4)(O - C) \right ) Z_{2m} 
\nonumber \\
&+& \left ( (n_1n_2+n_3n_4)(V - S) + (n_1n_4+n_2n_3)(O - C) \right ) Z_{2m+1}  \   ,
\eeq
\be
\tilde{M} = - \frac{R}{\sqrt{2}} \left ( (n_1+n_2+n_3+n_4)\hat{V}
\tilde{Z}_{2n} - (n_1-n_2-n_3+n_4)\hat{S}\tilde{Z}_{2n+1} \right )   \   ,
\ee
\be
M = - \frac{n_1+n_2+n_3+n_4}{2}\hat{V} Z_{2m} 
+ \frac{n_1-n_2-n_3+n_4}{2}\hat{S}(-1)^mZ_{2m} \ .
\ee  
Le condizioni di tadpole
\be
n_1+n_2+n_3+n_4 = 32  , \hspace{1cm} n_1+n_2-n_3-n_4 = 0  \ ,
\ee
non eliminano automaticamente il tachione presente nel settore $OZ_m$ in $A$.
Una soluzione interessante si ha per $n_2=n_3=0$; il gruppo di gauge \`e
$SO(16) \times SO(16)$ e il settore aperto
\beq
A &=& \frac{n_1^2+n_4^2}{2}(V - S)Z_{2m} 
+ n_1n_4(O - C)Z_{2m+1}   \  ,    \nonumber \\
M &=& - \frac{n_1+n_4}{2}\hat{V}Z_{2m} 
+ \frac{n_1+n_4}{2}\hat{S}(-1)^m Z_{2m}    \   ,
\eeq
\`e supersimmetrico nei livelli con momento pari, ed in particolare nel settore 
non massivo, e descrive un vettore ed uno spinore nell'aggiunta di 
$SO(16) \times SO(16)$; per livelli con momento dispari, il vettore \`e 
nell'aggiunta e lo spinore nella simmetrica
$(\bf{135},\bf{1})$+2$(\bf{1},\bf{1})$ + $(\bf{1},\bf{135})$; infine 
livelli con momento semiintero sono scalari e spinori nella 
$(\bf{16},\bf{16})$.

Un'altra soluzione delle condizioni di tadpole \`e $n_1=n_4=0$ e $n_2=n_3=16$;
anche questo modello ha come gruppo di gauge $SO(16) \times SO(16)$,
ma la supersimmetria \`e rotta gi\`a a livello ad albero in quanto
le rappresentazioni dei fermioni in livelli con momento pari 
e in livelli con momento dispari vengono scambiate.
Osserviamo tuttavia che l'annullamento dei 
tadpoles dei campi che divengono non massivi 
nel limite $R \rar 0$ selezionano la soluzione $n_2 = n_3 =0$.

Il contributo ad un loop del settore aperto all'energia di vuoto e alle masse 
degli scalari \`e soppresso esponenzialmente nel limite $R \rar 0$, che dal 
punto di vista della Tipo $I^{'}$ \`e il limite di decompattificazione.


\section{Modelli in sei dimensioni}
\markboth{}{}

Modelli di stringhe aperte in sei dimensioni contengono tipicamente, a meno 
di T dualit\`a, sia D9 sia D5 brane; in presenza di un solo tipo di brana, 
l'estensione del meccanismo di Scherk-Schwarz \`e immediata. Quando entrambe le 
brane sono presenti, ci si aspetta una combinazione dei 
due modi basilari di rottura della 
supersimmetria illustrati nel paragrafo precedente, con masse proporzionali a 
$1/R$ ed $R$. In particolare, nel settore 99 
la supersimmetria viene rotta ad 
albero mentre i settori 55 e 59, risentono della rottura della supersimmetria 
solo attraverso correzioni radiative.
Come sappiamo, vuoti della Tipo I in sei dimensioni contengono nello spettro un 
numero variabile di multipletti tensoriali; consideriamo il 
modello con un unico multipletto e con gruppo di gauge 
$U(16)_9 \times U(16)_5$
\cite{od3,or2}. Questo modello si ottiene compattificando la teoria su
$T^4/I$, dove $I$ \`e l'inversione delle coordinate del toro e localizzando 
tutte le D5 brane nello stesso punto fisso. Lo spettro aperto contiene, 
nei settori 99 e 55,
multipletti vettoriali ed ipermultipletti nelle rappresentazioni
$(\bf{120+ \bar{120}},\bf{1})$ e $(\bf{1},\bf{120+ \bar{120}})$ 
e nel settore 59 un ipermultipletto nella $(\bf{16},\bf{16})$.

Se fermionizziamo i modi bosonici compatti secondo
\be
4\frac{\h^2}{\th_2^2} = (OO - VV)_B \  , \hspace{1cm}
4\frac{\h^2}{\th_4^2} = (Q_S + Q_C)_B  \  ,  \hspace{1cm}
4\frac{\h^2}{\th_3^2} = (Q_S - Q_C)_B  \  .
\ee
possiamo scrivere la funzione di partizione della IIB su $T^4/I$ nel modo 
seguente
\beq
T &=& \frac{1}{2}\L_{4,4}|V - S|^2 + \frac{1}{2}|Q_O - Q_V|^2|OO - VV|^2_B
\nonumber \\
&+& \frac{1}{2} \{ |Q_S + Q_C|^2|Q_S + Q_C|^2_B 
+ |Q_S - Q_C|^2|Q_S - Q_C|^2_B \}  \    ,
\label{k3b}
\eeq
dove $Q_O,  Q_V,  Q_S,  Q_C$ sono le combinazioni di caratteri di $SO(4)$ 
introdotte nel terzo capitolo e $\L_{4,4}$ \`e il contributo del reticolo di 
Narain $\G_{4,4}$.
Per rompere la supersimmetria, scegliamo quest'ultimo del tipo
$\G_{1,1} \otimes \G_{3,3}$, deformiamo il primo termine in (\ref{k3b}) 
con $e = (0,0,0,1)$ e 
quozientiamo con la simmetria $Z_2$. Le due operazioni sono
consistenti 
perch\`e
la carica utilizzata per il meccanismo di Scherk-Schwarz anticommuta con 
l'operazione di orbifold. Il risultato \`e
\beq
T &=& \frac{\L_{3,3}}{2}\{ E_0^{'}(|V|^2 + |S|^2) + 
O_0^{'}(|O|^2 + |C|^2) - E_{1/2}^{'}(V\bar{S} + S\bar{V}) 
- O_{1/2}^{'}(O\bar{C} + C\bar{O}) \}   \nonumber \\
&+& \frac{1}{4}(|Q_O - Q_V|^2 + |Q^{'}_O - Q^{'}_V|^2)|OO - VV|^2_B  \nonumber  \\
&+& \frac{1}{4}\{ (|Q_S + Q_C|^2 + |Q^{'}_S + Q^{'}_C|^2)|Q_S + Q_C|^2_B 
\nonumber \\
&+& (|Q_S - Q_C|^2 + |Q^{'}_S - Q^{'}_C|^2)|Q_S - Q_C|^2_B \}  \   .  
\label{6dt}
\eeq
I caratteri primati si ottengono da quelli usuali scambiando $S$ e $C$
\beq
& & Q_O^{'} = VO - SS    \    , \hspace{1cm} Q_V^{'} = OV - CC  \   , \nonumber \\
& & Q_S^{'} = OS - CO    \    , \hspace{1cm} Q_C^{'} = VC - SV  \   .
\eeq
Vediamo quindi che la deformazione di Scherk-Schwarz \`e accompagnata da un 
cambiamento di chiralit\`a per i fermioni twistati gi\`a nel settore chiuso.
Costruiamo l'ampiezza di bottiglia di Klein; indicando con $\L_3$ e 
$\tilde{\L}_3$ rispettivamente la somma solo sui momenti e la somma solo sugli 
avvolgimenti e con $\L_{3e}$ e $\tilde{\L}_{3e}$ le stesse somme ma ristrette a 
valori pari, abbiamo
\beq
K &=& \frac{1}{4}[(V - S)(Z_m \L_3 + \tilde{Z}_{2n} \tilde{\L})
+ (O - C)\tilde{Z}_{2n+1}\tilde{\L}_3 \nonumber \\
&+& (Q_S + Q_C + Q^{'}_S + Q^{'}_C )
(Q_S + Q_C)_B]   \    ,   \nonumber \\
\tilde{K} &=& \frac{2^5}{4}[v_4(V - S)\tilde{Z}_{2n}\tilde{\L}_{3e}
+ \frac{1}{v_4}(VZ_{2m} - SZ_{2m+1})\L_{3e}]  \nonumber \\
&+& \frac{2^5}{4}(Q_O - Q_V + Q_O^{'} - Q_V^{'})|OO - VV|_B   \    ,
\label{6dk} 
\eeq
dove $v_4$ \`e il volume dello spazio compatto. Possiamo verificare 
esplicitamente che $\tilde{K}$ si presenta come una somma di caratteri con 
coefficienti che sono 
quadrati perfetti; questo \`e evidente per il contributo dei punti 
del reticolo fuori dal'origine; per quanto riguarda il contributo dell'origine 
abbiamo
\beq
\tilde{K}_0 &=& \frac{2^5}{4}(\sqrt{v_4} + \frac{1}{\sqrt{v_4}})^2
[V_4O_4(O_4O_4)_B + O_4V_4(V_4V_4)_B]  \nonumber \\
&+& \frac{2^5}{4}(\sqrt{v_4} - \frac{1}{\sqrt{v_4}})^2
[V_4O_4(V_4V_4)_B + O_4V_4(O_4O_4)_B] \nonumber \\
&-& \frac{2^5}{4}(\sqrt{v_4})^2
(S_4S_4 + C_4C_4)(O_4O_4 + V_4V_4)_B   \    .
\eeq
L'ampiezza di anello trasverso contiene contributi dai vari settori dello 
spettro chiuso, pesati con corrispondenti coefficienti 
di riflessione; termini con 
momento nullo corrispondono a condizioni al bordo di Neumann mentre termini con 
avvolgimento nullo corrispondono a condizioni al bordo di Dirichlet. 
Introduciamo tre cariche di Chan-Paton, $n_N$ , $n_{D_1}$ e $n_{D_2}$ per i
settori non twistati e i tre coefficienti di riflessione corrispondenti, 
$R_N$ , $R_{D_1}$ e $R_{D_2}$ per i settori twistati. Abbiamo quindi
\beq
2^7 \tilde{A} &=& v_4n_N^2[(V - S)\tilde{Z}_{2n} + (O - C)\tilde{Z}_{2n+1}]
\tilde{\L}_3   \nonumber \\
&+& \frac{1}{v_4}[(n_{D_1}+n_{D_2})^2(VZ_{2m} - SZ_{2m+1}) 
+ (n_{D_1}-n_{D_2})^2(VZ_{2m+1} - SZ_{2m})]\L_3   \nonumber \\
&+& [2n_Nn_{D_1}(Q_O^{'} - Q_V^{'}) + 2n_Nn_{D_2}(Q_O - Q_V)]
(O_4O_4 - V_4V_4)_B  \nonumber \\
&+& [2R^2_N(Q_S + Q_C + Q_S^{'} + Q_C^{'}) 
+ 4R^2_{D_2}(Q_S + Q_C) + 4R^2_{D_1}(Q_S^{'} + Q_C^{'})](Q_S + Q_C)_B 
\nonumber \\
&+& [2R_NR_{D_1}(Q_S^{'} - Q_C^{'}) + 2R_NR_{D_2}(Q_S - Q_C)](Q_S - Q_C)_B \ .
\eeq
Anche per $\tilde{A}$ possiamo mostrare che i coefficienti di riflessione sono 
quadrati perfetti; in particolare i contributi derivanti dall'origine 
del reticolo si scrivono
\beq
2^7\tilde{A}_0 &=& 
\left (\sqrt{v_4}n_N + \frac{n_{D_1} + n_{D_2}}{\sqrt{v_4}} \right )^2
[V_4O_4(O_4O_4)_B+O_4V_4(V_4V_4)_B]   \nonumber \\
&+& \left ( \sqrt{v_4}n_N - \frac{n_{D_1} + n_{D_2}}{\sqrt{v_4}} \right )^2
[V_4O_4(V_4V_4)_B+O_4V_4(O_4O_4)_B]  \nonumber \\
&-&\left (\sqrt{v_4}n_N + \frac{n_{D_1} - n_{D_2}}{\sqrt{v_4}} \right )^2
[S_4S_4(O_4O_4)_B+C_4C_4(V_4V_4)_B]  \nonumber \\
&-& \left(\sqrt{v_4}n_N - \frac{n_{D_1} - n_{D_2}}{\sqrt{v_4}} \right )^2
[S_4S_4(V_4V_4)_B+C_4C_4(O_4O_4)_B]  \nonumber \\
&+&\left (\frac{(R_N + 4R_{D_2})^2}{4} + \frac{7}{4}R^2_N \right )(Q_SQ_{SB}
+Q_CQ_{CB}  )
\nonumber \\
&+&\left (\frac{(R_N - 4R_{D_2})^2}{4} + \frac{7}{4}R^2_N \right )(Q_SQ_{CB}
+Q_CQ_{SB})
\nonumber \\
&+&\left (\frac{(R_N + 4R_{D_2})^2}{4} + \frac{7}{4}R^2_N \right )(Q_S^{'}Q_{SB}
+ Q_C^{'}Q_{CB})
\nonumber \\
&+&\left (\frac{(R_N - 4R_{D_2})^2}{4} + \frac{7}{4}R^2_N \right )(Q_S^{'}Q_{CB}
+ Q_C^{'}Q_{SB}) \  .
\label{posd}
\eeq
I coefficienti di riflessione dei settori twistati sono molto interessanti in 
quanto contengono informazioni sulla posizione delle D brane; 
nel nostro caso, stiamo 
utilizzando due tipi di caratteri diversi $(Q_S,Q_C)$ e $(Q_S^{'},Q_C^{'})$ per 
descrivere il settore twistato e possiamo quindi 
distinguere due tipi di punti fissi.
Da (\ref{posd}) risulta che dei $16$ punti fissi di $T^4/I$, 14 
sono vuoti mentre dei restanti due, uno contiene 8 D5 brane del primo tipo 
e l'altro 8 D5 brane del secondo tipo.

Nel canale diretto abbiamo
\beq
A &=& \frac{n^2_N}{4}(VZ_{2m}-SZ_{2m+1})\L_{3e} + \frac{1}{4} 
[(n_{D_1}^2+n_{D_2}^2)(V - S)\tilde{Z}_{2n} 
+ 2n_{D_1}n_{D_2}(O - C)\tilde{Z}_{2n+1} ] \tilde{\L}_{3e} \nonumber \\
&+& \frac{1}{4}[2n_Nn_{D_1}(Q_S^{'} + Q_C^{'}) 
+ 2n_Nn_{D_2}(Q_S + Q_C)]
\left ( \frac{Q_S+Q_C}{4} \right )_B  \nonumber \\
&+& \frac{1}{4} \left [\frac{1}{2}R^2_N(Q_O - Q_V + Q_O^{'} - Q_V^{'}) 
+ R^2_{D_2}(Q_O - Q_V) + R^2_{D_1}(Q_O^{'} - Q_V^{'}) \right ](O_4O_4 - V_4V_4)_B 
\nonumber \\
&+& \frac{1}{4}[2R_NR_{D_1}(Q_S^{'} - Q_C^{'}) 
+ 2R_NR_{D_2}(Q_S - Q_C)] \left ( \frac{Q_S - Q_C}{4} \right )_B \ .
\eeq
L'ampiezza di M\"obius trasversa \`e
\beq
-2 \tilde{M} &=& n_Nv_4[\hat{V}(\tilde{Z}_{4n} + \tilde{Z}_{4n+2})
- \hat{S}(\tilde{Z}_{4n} - \tilde{Z}_{4n+2})] \tilde{\L}_{3e} \nonumber \\
&+& \frac{n_{D_1}+n_{D_2}}{v_4}(\hat{V}Z_{2m} - \hat{S}Z_{2m+1})\L_{3e} 
\nonumber \\
&+& [n_N(\hat{V}_4\hat{O}_4 - \hat{O}_4\hat{V}_4) 
+ n_{D_1}(\hat{Q}_O^{'} - \hat{Q}_V^{'}) \nonumber \\
&+& n_{D_2}(\hat{Q}_O - \hat{Q}_V)](\hat{O}_4\hat{O}_4 - \hat{V}_4\hat{V}_4)_B  \ . 
\eeq
e i termini nell'origine si scrivono
\beq
-2 \tilde{M}_0 &=& 
\left (\sqrt{v_4} + \frac{1}{\sqrt{v_4}} \right ) \left (\sqrt{v_4}n_N 
+ \frac{n_{D_1}+n_{D_2}}{\sqrt{v_4}} \right )
[\hat{V}_4\hat{O}_4(\hat{O}_4\hat{O}_4)_B
+ \hat{O}_4\hat{V}_4(\hat{V}_4\hat{V}_4)_B]   \nonumber \\
&+& \left (\sqrt{v_4} - \frac{1}{\sqrt{v_4}} \right ) \left (\sqrt{v_4}n_N 
- \frac{n_{D_1}+n_{D_2}}{\sqrt{v_4}} \right )
[\hat{V}_4\hat{O}_4(\hat{V}_4\hat{V}_4)_B
+ \hat{O}_4\hat{V}_4(\hat{O}_4\hat{O}_4)_B]   \nonumber \\
&-& \sqrt{v_4} \left ( \sqrt{v_4}n_N 
+ \frac{n_{D_1}+n_{D_2}}{\sqrt{v_4}} \right )
[\hat{S}_4\hat{S}_4(\hat{O}_4\hat{O}_4)_B
+ \hat{C}_4\hat{C}_4(\hat{V}_4\hat{V}_4)_B]   \nonumber \\
&-& \sqrt{v_4} \left (\sqrt{v_4}n_N 
- \frac{n_{D_1}+n_{D_2}}{\sqrt{v_4}} \right )
[\hat{S}_4\hat{S}_4(\hat{V}_4\hat{V}_4)_B
+ \hat{C}_4\hat{C}_4(\hat{O}_4\hat{O}_4)_B] \    .
\eeq
Infine nel canale diretto abbiamo
\beq
M &=& -\frac{n_N}{4}[\hat{V}Z_{2m} - \hat{S}Z_{2m+1}]\L_{3}
- \frac{n_{D_1}+n_{D_2}}{4}(\hat{V}\tilde{Z}_{2n} - 
\hat{S}(-1)^n\tilde{Z}_{2n})
\tilde{\L}_{3}  \\
&+& \frac{1}{4}[n_N(\hat{V}_4\hat{O}_4 - \hat{O}_4\hat{V}_4) 
+ n_{D_1}(\hat{Q}_O^{'} - \hat{Q}_V^{'}) 
+ n_{D_2}(\hat{Q}_O - \hat{Q}_V)](\hat{O}_4\hat{O}_4 - \hat{V}_4\hat{V}_4)_B 
\nonumber \ . 
\eeq
Dalle condizioni di tadpole dei settori twistati si ottiene
\be
R_N = 0  \  , \hspace{1cm} R_{D_1} = 0  \  ,\hspace{1cm} R_{D_2} = 0  \  ,
\ee
mentre dai settori non twistati
\be
n_N = 32  \ , \hspace{1cm} n_{D_1} = 16  \ , \hspace{1cm}  n_{D_2} = 16 \ .
\ee
Possiamo introdurre cariche di Chan-Paton complesse; se poniamo 
\beq
n_N &=& n + \bar{n} \ , \hspace{1.7cm} R_N = i(n - \bar{n}) \ , \nonumber \\
n_{D_1} &=& m_1 + \bar{m_1}  \ , \hspace{1cm} 
R_{D_1} = i(m_1 - \bar{m_1})  \ , \nonumber \\
n_{D_2} &=& m_2 + \bar{m_2} \ , \hspace{1cm} 
R_{D_2} = i(m_2 - \bar{m_2}) \ ,
\eeq
otteniamo $n = 16$, $m_1 = 8$, $m_2 =8$.
Lo spettro aperto non massivo \`e dato quindi da
\beq
A &=& 
n\bar{n} V_4O_4 + m_1\bar{m}_1(V_4O_4 - S_4S_4) + m_2\bar{m}_2(V_4O_4 - C_4C_4)
\nonumber \\
&+& \frac{n^2 + \bar{n}^2}{2}O_4V_4  
+ \frac{m_1^2 + \bar{m}_1^2}{2}(O_4V_4 - C_4C_4) 
+ \frac{m_2^2 + \bar{m}_2^2}{2}(O_4V_4 - S_4S_4) \nonumber \\
&+& (n\bar{m}_1 + \bar{n}m_1)Q_S^{'} + (n\bar{m}_2 + \bar{n}m_2)Q_S  \  , \\
M &=& - \frac{n + \bar{n}}{2}\hat{O}_4\hat{V}_4
- \frac{m_1 + \bar{m}_1}{2}(\hat{O}_4\hat{V}_4 - \hat{C}_4\hat{C}_4) \nonumber \\
&-& \frac{m_2 + \bar{n}}{2}(\hat{O}_4\hat{V}_4 - \hat{C}_4\hat{C}_4)  \  .
\eeq
Il gruppo di gauge \`e $U(8)_5 \times U(8)_{5}^{'} \times U(16)_9$; 
Il settore 99 contiene i bosoni di gauge e quartetti di scalari nella 
$\bf{120} + \bf{\bar{120}}$; i fermioni corrispondenti sono massivi come 
conseguenza della rottura dela supersimmetria. Nel settore 55 abbiamo i bosoni 
di gauge di $U(8)_5 \times U(8)_{5}^{'}$, i gaugini nell'aggiunta,
$(\bf{64},\bf{1})_R$ e $(\bf{1},\bf{64})_L$, quartetti di scalari e loro 
partners fermionici nelle rappresentazioni 
$(\bf{28},\bf{1})_L$ , $(\bf{\bar{28}},\bf{1})_L$ , 
$(\bf{1},\bf{28})_R$ e $(\bf{1},\bf{\bar{28}})_R$. 
Il settore 59 contiene infine coppie di scalari e 
semifermioni nelle rappresentazioni
$(\bf{\bar{8}},\bf{1},\bf{16})_R$ e $(\bf{1},\bf{8},\bf{\bar{16}})_L$. 
Come ci si aspettava, gli stati 
non massivi provenienti da D brane ortogonali alla direzione utilizzata per 
la compattificazione di Scherk-Schwarz non risentono a livello ad albero 
della rottura della supersimmetria.

Il polinomio di anomalia non contiene termini irriducibili in $R^4$, 
dato che non \`e presente un numero netto di fermioni o forme chirali in 
seguito al cambiamento di chiralit\`a prodotto dalla deformazione; 
le condizioni di 
tadpole eliminano i termini $F^4$ e il polinomio residuo \`e
\be
A = \frac{1}{4}(trF^2_5 - trF^2_{5^{'}}) \left ( trF^2_9 - \frac{1}{2}trR^2
\right )  \  .
\ee


\section{Modelli in quattro dimensioni}
\markboth{}{}

Dal punto di vista fenomenologico i modelli pi\`u interessanti sono quelli che 
presentano rottura spontanea $N=1 \rar N=0$ con uno spettro chirale 
\cite{worpro}. 
Questi possono essere costruiti come discendenti di un orbifold 
$Z_6 = Z_2 \times Z_3$ nel quale la supersimmetria viene rotta utilizzando 
l'operatore $e^{i \p \oint J}$, dove
\be
J = \psi_{56} \psi_{78} + \bar{\psi_{56}} \bar{\psi_{78}} 
+ X_{56} \part X_{78} 
+ X^{*}_{56} \part X^{*}_{78}  \ ,
\ee
e $\psi_{56}$ indica la combinazione $\frac{1}{\sqrt{2}}( \psi_5 +  i \psi_6)$. 
In questa Tesi riportiamo solo la costruzione dei discendenti della 
compattificazione toroidale deformata dall'azione di $e^{i \p \oint J}$
\cite{worpro}. 
La 
funzione di partizione chiusa descrive la rottura spontanea $N=4 \rar N=2$
ed \`e data da \cite{Kir}
\be
T = \frac{1}{(\h \bar{\h})^4} \sum_{m,n} | \sum_{a,b=0,1/2} 
C \pmatrix{a\cr b} \frac{\th^2\pmatrix{a\cr b} 
\th\pmatrix{a - n/2\cr b + m/2} 
\th\pmatrix{a + n/2\cr b - m/2}}{\h^2 \th\pmatrix{1/2 - n/2\cr 1/2 + m/2}
\th\pmatrix{1/2 + n/2\cr 1/2 - m/2}}|^2 \L_{4,4} \L_{1,1}Z_{m,n} \ ,
\ee
Questa funzione di partizione coincide con quella di un orbifold il cui gruppo 
agisce senza punti fissi \cite{Kir}. Ponendo infatti
$n = 2N + k$ e $m = 2M + l$ con $l,k = 0,1$, si ottiene
\beq
T &=& \frac{1}{2} 
\{ |Q_O^{'} + Q_V^{'}|^2 \frac{\L_{4,4}}{(\h \bar{\h})^4} Z_{2m,2n} 
+ |Q_O^{'} - Q_V^{'}|^2 |I_4I_4 - V_4V_4|^2 Z_{2m+1,2n} \nonumber \\
&+& |Q_S^{'} + Q_C^{'}|^2 |Q_S + Q_C|_B^2 Z_{2m,2n+1} \nonumber \\
&+& |Q_S^{'} - Q_C^{'}|^2 |Q_S - Q_C|_B^2 Z_{2m+1,2n+1} \} \L_{1,1}  \ ,
\eeq
e risommando l'indice $m$
\beq
T &=& \frac{1}{2} 
 \{ |Q_O^{'} + Q_V^{'}|^2 \frac{\L_{4,4}}{(\h \bar{\h})^4} (Z_{m,2n} +
Z_{m+1/2,2n})  \nonumber \\ 
&+& |Q_O^{'} - Q_V^{'}|^2 |I_4I_4 - V_4V_4|^2 (Z_{m,2n} -
Z_{m+1/2,2n})   \nonumber \\
&+& |Q_S^{'} + Q_C^{'}|^2 |Q_S + Q_C|_B^2 (Z_{m,2n+1} +
Z_{m+1/2,2n+1})   \nonumber \\ 
&+& |Q_S^{'} - Q_C^{'}|^2 |Q_S - Q_C|_B^2 (Z_{m,2n+1} -
Z_{m+1/2,2n+1})   \} \L_{1,1}  \ .
\label{tor12}
\eeq
Se raddoppiamo il raggio del cerchio nella (\ref{tor12}) otteniamo infine
\beq
T &=& \frac{1}{2} 
\{ |Q_O^{'} + Q_V^{'}|^2 \frac{\L_{4,4}}{(\h \bar{\h})^4} Z_{m,n} 
+ |Q_O^{'} - Q_V^{'}|^2 |I_4I_4 - V_4V_4|^2 (-1)^m Z_{m,n} \nonumber \\
&+& |Q_S^{'} + Q_C^{'}|^2 |Q_S + Q_C|_B^2 Z_{m,n+1/2} \nonumber \\
&+& |Q_S^{'} - Q_C^{'}|^2 |Q_S - Q_C|_B^2 (-1)^m Z_{m,n+1/2} \} 
\L_{1,1}  \ ,
\eeq
ovvero la funzione di partizione dell'orbifold $IIB/(-1)^m I$, dove 
$I$ \`e l'inversione delle coordinate del toro 
$I X_{5,6,7,8} = - X_{5,6,7,8}$ e $(-1)^m$ \`e la traslazione di $\p R$ lungo 
il cerchio $X_9 \mapsto X_9 + \p R$.
Il modello $IIB/(-1)^n I$, con shift sugli avvolgimenti, si ottiene scambiando 
$m$ ed $n$, come in nove dimensioni.

Costruiamo il discendente aperto del modello $IIB/(-1)^m I$, che chiamiamo 
Modello 1. Il reticolo $\L_{1,1}$ \`e inerte rispetto all'azione dell'orbifold 
e verr\`a quindi sottinteso nelle equazioni seguenti.
Il contenuto operatoriale della teoria chiusa fissa la bottiglia di 
Klein
\be
K = \frac{1}{4}(Q_O + Q_V) [ (P + W) Z_{2m} + (P - W)Z_{2m+1} ] \ ,
\label{m1dk}
\ee
dove $P$ e $W$ sono rispettivamente la somma sui momenti e la somma sugli 
avvolgimenti relative a $T^4$. Nel canale trasverso
\be
\tilde{K} = \frac{2^5}{4}R(Q_O + Q_V) \left [ v_4 W^e \tilde{Z}_{2n} +
\frac{P^e}{v_4} \tilde{Z}_{2n+1} \right ]  \ ,
\label{m1tk}
\ee
dove $P^e$ e $W^e$ sono somme ristrette a valori pari del momento e 
dell'avvolgimento e $v_4$ \`e il volume di $T^4$. Dal toro ricaviamo il 
contenuto dell'anello trasverso
\beq
\tilde{A} &=& \frac{2^{-5}}{4} R \{ 
[ (Q_O + Q_V) \left ( \a^2 v_4 W + \b^2 \frac{P}{v_4} \right )
+ (Q_O - Q_V)(O_4O_4 - V_4V_4)_B 2 \a \b ] \tilde{Z_m} \nonumber \\
&+& [ \g^2(Q_O - Q_V)(O_4O_4 - V_4V_4)_B
+ \d^2(Q_O - Q_V)(O_4O_4 - V_4V_4)_B  \nonumber \\
&+& 2 \g \d (Q_S - Q_C)(Q_S - Q_C)_B] (-1)^m \tilde{Z}_{n + 1/2} \} \ .
\label{m1ta}
\eeq
Dalla struttura delle ampiezze (\ref{m1tk} - \ref{m1ta})
nell'origine del reticolo di $T^4$ si pu\`o quindi costruire l'ampiezza di 
M\"obius trasversa
\beq
\tilde{M} &=& - \frac{R}{2} [ (Q_O + Q_V) \left ( \a v_4 W^e \tilde{Z}_{2n}
+ \b \frac{P^e}{v_4}\tilde{Z}_{2n+1} \right ) \nonumber \\
&+& (Q_O - Q_V) (O_4O_4 - V_4V_4)_B ( \b \tilde{Z}_{2n} + \a \tilde{Z}_{2n+1} )]
\ .
\label{m1tm}
\eeq
Nel canale diretto le ampiezze che determinano lo spettro aperto sono quindi
\beq
A &=& \frac{1}{4} \{ [ (Q_O + Q_V) (\a^2 P + \b^2 W) +
2 \a \b (Q_S + Q_C)(Q_S + Q_C)_B]Z_m \nonumber \\
&+& [ \g^2 (Q_O - Q_V)(O_4O_4-V_4V_4)_B +
\d^2 (Q_O - Q_V)(O_4O_4-V_4V_4)_B   \nonumber \\
&+& 2 \g \d (Q_S - Q_C)(Q_S - Q_C)_B](-1)^mZ_m \} \ ,
\label{m2da}
\eeq
\beq
M &=& - \frac{1}{4} \{ (Q_O + Q_V)(\a P Z_m + \b W (-1)^m Z_m) \nonumber \\
&+&(Q_O - Q_V)(O_4O_4-V_4V_4)_B ( \b Z_m + \a (-1)^m Z_m ) \} \ .
\label{m2dm}
\eeq
Le condizioni di tadpole impongono $\a = 32$ e $\b = 0$; il settore di D-brana 
viene quindi rimosso dallo spettro a causa dello shift nei momenti. 
Parametrizzando nel modo solito con $\a = n_1 + n_2$, $\g = n_1 - n_2$
otteniamo modelli con gruppi ortogonali, in particolare $SO(32)$.
Se sostituiamo $n_1 + n_2$ con $m + \bar{m}$ e  $n_1 - n_2$ con 
$i(m - \bar{m})$, passiamo da gruppi ortogonali a gruppi unitari, 
e in particolare otteniamo un modello con gruppo di gauge $U(16)$. 

Molto pi\`u 
interessante \`e il modello $IIB/(-1)^n I$, che chiameremo Modello 2. 
La bottiglia di Klein \`e in questo caso
\be
K = \frac{1}{4}[(Q_O + Q_V)(P+W) Z_n + 2 \cdot 16
(Q_S + Q_C)(Q_S + Q_C) Z_{n+1/2}]
\ ,
\label{m2dk}
\ee
e nel canale trasverso
\be
\tilde{K} = \frac{2^5}{4}R [ (Q_O + Q_V) \left ( v_4W^e + \frac{P^e}{v_4} \right ) 
\tilde{Z}_{2n}
+ 2(Q_O - Q_V) (O_4O_4-V_4V_4)_B(-1)^n\tilde{Z}_{2n}] \ .
\label{m2tk}
\ee
All'anello trasverso danno un contributo non nullo solo stati con avvolgimento 
pari lungo il cerchio $S_1$ e quindi
\beq
\tilde{A} &=& \frac{2^{-5}}{4}R \{ 
[(Q_O + Q_V) \left ( \a_1^2Wv_4 + \b_1^2 \frac{P}{v_4} \right )  \nonumber \\
&+& 2 \a_1 \b_1(Q_O - Q_V) (O_4O_4-V_4V_4)_B] \tilde{Z}_{4n} \nonumber \\
&+& [(Q_O + Q_V) \left ( \a_2^2Wv_4 + \b_2^2 \frac{P}{v_4} \right )  \nonumber \\
&+& 2 \a_2 \b_2(Q_O - Q_V) (O_4O_4-V_4V_4)_B] \tilde{Z}_{4n+2} \} \ .
\label{m2ta}
\eeq
In (\ref{m2ta}) abbiamo separato $\tilde{Z}_{4m}$ e $\tilde{Z}_{4m+2}$ in modo 
analogo a quanto fatto per il modello di M teoria.
L'ampiezza di M\"obius trasversa \`e in questo caso
\beq
\tilde{M} &=& -\frac{R}{2} \{ 
[(Q_O + Q_V) \left ( \a_1 W^ev_4 + \b_1 \frac{P^e}{v_4} \right ) + 
(\a_1 + \b_1) (Q_O - Q_V) (O_4O_4-V_4V_4)_B] \tilde{Z}_{4n} \nonumber \\
&+& [(Q_O + Q_V) \left ( \a_2 W^ev_4 - \b_2 \frac{P^e}{v_4} \right )  
\nonumber \\
&+& ( \b_2 - \a_2) (Q_O - Q_V) (O_4O_4-V_4V_4)_B] \tilde{Z}_{4n+2} \} \ .
\label{m2tm}
\eeq
Nel canale diretto abbiamo
\beq
A &=& \frac{1}{4}\{ 
[(Q_O + Q_V)(\a_1^2 P + \b_1^2 W) + \nonumber \\
&+& 2 \a_1 \b_1(Q_S + Q_C) (Q_S + Q_C)_B] 
\frac{1}{4}( Z_m + Z_{m+1/2} + Z_{m+1/4} + Z_{m+3/4} ) \nonumber \\
&+& [(Q_O + Q_V)(\a_2^2Wv_4 + \b_2^2 \frac{P}{v_4})  \\
&+& 2 \a_2 \b_2(Q_S + Q_C) (Q_S + Q_C)_B] 
\frac{1}{4}( Z_m + Z_{m+1/2} - Z_{m+1/4} - Z_{m+3/4} ) 
\} \nonumber \ ,
\label{m2da}
\eeq
\beq
M &=& -\frac{1}{4} \{ 
[(Q_O + Q_V)(\a_1 P + \b_1 W)  \nonumber \\
&-& (\a_1 + \b_1) (Q_O - Q_V) (O_4O_4-V_4V_4)_B]
\frac{1}{2}(Z_m + Z_{m+1/2}) \nonumber \\
&+& [(Q_O + Q_V)(\a_2 P - \b_2 W)  \nonumber \\
&-& ( \b_2 - \a_2) (Q_O - Q_V) (O_4O_4-V_4V_4)_B] 
\frac{1}{2}(Z_m - Z_{m+1/2})  \} \ .
\label{m2dm}
\eeq
Le condizioni di tadpole sono $\a_1 = 32$ e $\b_1 = 32$. Per avere 
un'interpretazione fisica dello spettro usiamo la  
parametrizzazione seguente
\be
\a_1 = 2(n_1+n_2)  \ , \hspace{0.5cm}
\a_2 = 2(n_1-n_2)  \ , \hspace{0.5cm}
\b_1 = 2(d_1+d_2)  \ , \hspace{0.5cm}
\b_2 = 2(d_1-d_2)  \ .
\ee
Dalle (\ref{m2da}- \ref{m2dm}) segue che lo spettro aperto ha gruppo di gauge
\be
SO(n_1) \times Sp(16 - n_1) \times Sp(d_1) \times SO(16 - d_1) \ . 
\ee
L'annullamento 
dei tadpoles per gli stati che divengono non massivi nel limite $R \rar 0$ 
seleziona anche questa volta il gruppo di gauge $SO(16) \times SO(16)$.



\begin{thebibliography}{99}
\addcontentsline{toc}{chapter}{Bibliografia}
\markboth{} {}



\bibitem{Pol} A.~M.~Polyakov,
``Quantum Geometry Of Bosonic Strings,''
Phys.\ Lett.\ B {\bf 103} (1981) 207;
``Quantum Geometry Of Fermionic Strings,''
Phys.\ Lett.\ B {\bf 103} (1981) 211.
\bibitem{BDH} 
L.~Brink, P.~Di Vecchia and P.~Howe,
``A Locally Supersymmetric And Reparametrization Invariant Action For The Spinning String,''
Phys.\ Lett.\ B {\bf 65} (1976) 471.
\bibitem{GSO} F.~Gliozzi, J.~Scherk and D.~I.~Olive,
``Supersymmetry, Supergravity Theories And The Dual Spinor Model,''
Nucl.\ Phys.\ B {\bf 122} (1977) 253.
\bibitem{BPZ} 
A.~A.~Belavin, A.~M.~Polyakov and A.~B.~Zamolodchikov,
``Infinite Conformal Symmetry In Two-Dimensional Quantum Field Theory,''
Nucl.\ Phys.\ B {\bf 241} (1984) 333.
\bibitem{FMS} D.~Friedan, E.~J.~Martinec and S.~H.~Shenker,
``Conformal Invariance, Supersymmetry And String Theory,''
Nucl.\ Phys.\ B {\bf 271} (1986) 93.
\bibitem{D'Ho} E.~D'Hoker and S.~B.~Giddings,
``Unitary Of The Closed Bosonic Polyakov String,''
Nucl.\ Phys.\ B {\bf 291} (1987) 90.
\bibitem{cb1} E.~S.~Fradkin and A.~A.~Tseytlin,
``Effective Field Theory From Quantized Strings,''
Phys.\ Lett.\ B {\bf 158} (1985) 316,
``Quantum String Theory Effective Action,''
Nucl.\ Phys.\ B {\bf 261} (1985) 1;\\
C.~G.~Callan, E.~J.~Martinec, M.~J.~Perry and D.~Friedan,
``Strings In Background Fields,''
Nucl.\ Phys.\ B {\bf 262} (1985) 593.
\bibitem{ob1} 
E.~S.~Fradkin and A.~A.~Tseytlin,
``Nonlinear Electrodynamics From Quantized Strings,''
Phys.\ Lett.\ B {\bf 163} (1985) 123;\\
A.~Abouelsaood, C.~G.~Callan, C.~R.~Nappi and S.~A.~Yost,
``Open Strings In Background Gauge Fields,''
Nucl.\ Phys.\ B {\bf 280} (1987) 599.
\bibitem{Mar} 
E.~J.~Martinec,
``Nonrenormalization Theorems And Fermionic String Finiteness,''
Phys.\ Lett.\ B {\bf 171} (1986) 189.
\bibitem{An1} 
L.~Alvarez-Gaume and E.~Witten,
``Gravitational Anomalies,''
Nucl.\ Phys.\ B {\bf 234} (1984) 269.
\bibitem{an2} M.~B.~Green and J.~H.~Schwarz,
``Anomaly Cancellation In Supersymmetric D=10 Gauge Theory 
And Superstring Theory,''
Phys.\ Lett.\ B {\bf 149} (1984) 117,
``Infinity Cancellations In SO(32) Superstring Theory,''
Phys.\ Lett.\ B {\bf 151} (1985) 21.
\bibitem{het1} 
D.~J.~Gross, J.~A.~Harvey, E.~J.~Martinec and R.~Rohm,
``Heterotic String Theory. 1. The Free Heterotic String,''
Nucl.\ Phys.\ B {\bf 256} (1985) 253,
``Heterotic String Theory. 2. The Interacting Heterotic String,''
Nucl.\ Phys.\ B {\bf 267} (1986) 75.
\bibitem{sugy1} 
J.~H.~Schwarz,
``Covariant Field Equations Of Chiral N=2 D = 10 Supergravity,''
Nucl.\ Phys.\ B {\bf 226} (1983) 269.
\bibitem{Tonin} 
P.~Pasti, D.~Sorokin and M.~Tonin,
``On Lorentz invariant actions for chiral p-forms,''
Phys.\ Rev.\ D {\bf 55} (1997) 6292
[arXiv:hep-th/9611100].
\bibitem{Rom} L.~J.~Romans,
``Massive N=2a Supergravity In Ten-Dimensions,''
Phys.\ Lett.\ B {\bf 169} (1986) 374.
\bibitem{n2a} 
T.~Banks, L.~J.~Dixon, D.~Friedan and E.~J.~Martinec,
``Phenomenology And Conformal Field Theory Or Can String Theory Predict The Weak Mixing Angle?,''
Nucl.\ Phys.\ B {\bf 299} (1988) 613.
\bibitem{n2b} 
T.~Banks and L.~J.~Dixon,
``Constraints On String Vacua With Space-Time Supersymmetry,''
Nucl.\ Phys.\ B {\bf 307} (1988) 93.
\bibitem{n2c} 
T.~Banks and N.~Seiberg,
``Nonperturbative Infinities,''
Nucl.\ Phys.\ B {\bf 273} (1986) 157.
\bibitem{seib} 
N.~Seiberg,
``Observations On The Moduli Space Of Superconformal Field Theories,''
Nucl.\ Phys.\ B {\bf 303} (1988) 286.
\bibitem{war} A.~N.~Schellekens and N.~P.~Warner,
``Anomalies, Characters And Strings,''
Nucl.\ Phys.\ B {\bf 287} (1987) 317.
\bibitem{g1}  
D.~J.~Gross and E.~Witten,
``Superstring Modifications Of Einstein's Equations,''
Nucl.\ Phys.\ B {\bf 277} (1986) 1.
\bibitem{g2}  D.~J.~Gross and J.~H.~Sloan,
``The Quartic Effective Action For The Heterotic String,''
Nucl.\ Phys.\ B {\bf 291} (1987) 41.
\bibitem{tc1} 
V.~S.~Kaplunovsky,
``One Loop Threshold Effects In String Unification,''
Nucl.\ Phys.\ B {\bf 307} (1988) 145
[Erratum-ibid.\ B {\bf 382} (1988) 436]
[arXiv:hep-th/9205068];
L.~J.~Dixon, V.~Kaplunovsky and J.~Louis,
``Moduli dependence of string loop corrections to gauge coupling constants,''
Nucl.\ Phys.\ B {\bf 355} (1991) 649;
I.~Antoniadis, E.~Gava and K.~S.~Narain,
``Moduli corrections to gauge and gravitational couplings 
in four-dimensional superstrings,''
Nucl.\ Phys.\ B {\bf 383} (1992) 93
[arXiv:hep-th/9204030];
I.~Antoniadis, E.~Gava, K.~S.~Narain and T.~R.~Taylor,
``Superstring threshold corrections to Yukawa couplings,''
Nucl.\ Phys.\ B {\bf 407} (1993) 706
[arXiv:hep-th/9212045].
\bibitem{tc2} 
E.~Kiritsis and C.~Kounnas,
``Infrared regularization of superstring theory and the one loop calculation of coupling constants,''
Nucl.\ Phys.\ B {\bf 442} (1995) 472
[arXiv:hep-th/9501020].
\bibitem{tc3} 
C.~Bachas and C.~Fabre,
``Threshold Effects in Open-String Theory,''
Nucl.\ Phys.\ B {\bf 476} (1996) 418
[arXiv:hep-th/9605028].


\bibitem{tor1} 
K.~S.~Narain,
``New Heterotic String Theories In Uncompactified Dimensions < 10,''
Phys.\ Lett.\ B {\bf 169} (1986) 41;
K.~S.~Narain, M.~H.~Sarmadi and E.~Witten,
``A Note On Toroidal Compactification Of Heterotic String Theory,''
Nucl.\ Phys.\ B {\bf 279} (1987) 369.
\bibitem{tor2} 
A.~Giveon, M.~Porrati and E.~Rabinovici,
``Target space duality in string theory,''
Phys.\ Rept.\  {\bf 244} (1994) 77
[arXiv:hep-th/9401139].
\bibitem{tor3} 
K.~S.~Narain, M.~H.~Sarmadi and C.~Vafa,
``Asymmetric Orbifolds,''
Nucl.\ Phys.\ B {\bf 288} (1987) 551.
\bibitem{spin1} 
N.~Seiberg and E.~Witten,
``Spin Structures In String Theory,''
Nucl.\ Phys.\ B {\bf 276} (1986) 272.
\bibitem{spin2} 
L.~Alvarez-Gaume, G.~W.~Moore and C.~Vafa,
``Theta Functions, Modular Invariance, And Strings,''
Commun.\ Math.\ Phys.\  {\bf 106} (1986) 1.
\bibitem{spin3} 
E.~Witten,
``Global Anomalies In String Theory,''
Print-85-0620 (PRINCETON)
{\it To appear in Proc. of Argonne Symp. on Geometry, Anomalies and Topology, Argonne, IL, Mar 28-30, 1985}.
\bibitem{tor4} L.~J.~Dixon and J.~A.~Harvey,
``String Theories In Ten-Dimensions Without Space-Time Supersymmetry,''
Nucl.\ Phys.\ B {\bf 274} (1986) 93;\\
L.~Alvarez-Gaume, P.~Ginsparg, G.~W.~Moore and C.~Vafa,
``An O(16) X O(16) Heterotic String,''
Phys.\ Lett.\ B {\bf 171} (1986) 155;
P.~Ginsparg and C.~Vafa,
``Toroidal Compactification Of Nonsupersymmetric Heterotic Strings,''
Nucl.\ Phys.\ B {\bf 289} (1987) 414.
\bibitem{tor6} 
M.~Dine, P.~Huet and N.~Seiberg,
``Large And Small Radius In String Theory,''
Nucl.\ Phys.\ B {\bf 322} (1989) 301.
\bibitem{tor7} 
P.~Ginsparg,
``Comment On Toroidal Compactification Of Heterotic Superstrings,''
Phys.\ Rev.\ D {\bf 35} (1987) 648.
\bibitem{tor5} V.~P.~Nair, A.~D.~Shapere, A.~Strominger and F.~Wilczek,
``Compactification Of The Twisted Heterotic String,''
Nucl.\ Phys.\ B {\bf 287} (1987) 402.
\bibitem{orb0} Satake,J. {\it Proc.~Nat.~Acad.~Sci.~U.S.A.} $\bf{42}$
        (1956) 359.
\bibitem{orb1} 
L.~J.~Dixon, J.~A.~Harvey, C.~Vafa and E.~Witten,
``Strings On Orbifolds,''
Nucl.\ Phys.\ B {\bf 261} (1985) 678,
``Strings On Orbifolds. 2,''
Nucl.\ Phys.\ B {\bf 274} (1986) 285.
\bibitem{orb2} 
L.~J.~Dixon, D.~Friedan, E.~J.~Martinec and S.~H.~Shenker,
``The Conformal Field Theory Of Orbifolds,''
Nucl.\ Phys.\ B {\bf 282} (1987) 13.
\bibitem{orb3} 
S.~Hamidi and C.~Vafa,
``Interactions On Orbifolds,''
Nucl.\ Phys.\ B {\bf 279} (1987) 465.
\bibitem{orb4} C.~Vafa,
``Modular Invariance And Discrete Torsion On Orbifolds,''
Nucl.\ Phys.\ B {\bf 273} (1986) 592;
D.~S.~Freed and C.~Vafa,
``Global Anomalies On Orbifolds,''
Commun.\ Math.\ Phys.\  {\bf 110} (1987) 349
[ {\bf 117} (1987) 349].
\bibitem{orb4b} 
C.~Vafa and E.~Witten,
``On orbifolds with discrete torsion,''
J.\ Geom.\ Phys.\  {\bf 15} (1995) 189
[arXiv:hep-th/9409188].
\bibitem{orb6} 
L.~E.~Ibanez, J.~Mas, H.~Nilles and F.~Quevedo,
``Heterotic Strings In Symmetric And Asymmetric Orbifold Backgrounds,''
Nucl.\ Phys.\ B {\bf 301} (1988) 157.
\bibitem{orb8} E. Zaslow, \CMP{156}{93}{301}.
E.~Zaslow,
``Topological orbifold models and quantum cohomology rings,''
Commun.\ Math.\ Phys.\  {\bf 156} (1993) 301
[arXiv:hep-th/9211119].
\bibitem{ff1} 
H.~Kawai, D.~C.~Lewellen and S.~H.~Tye,
``Construction Of Fermionic String Models In Four-Dimensions,''
Nucl.\ Phys.\ B {\bf 288} (1987) 1;
I.~Antoniadis, C.~P.~Bachas and C.~Kounnas,
``Four-Dimensional Superstrings,''
Nucl.\ Phys.\ B {\bf 289} (1987) 87;
H.~Kawai, D.~C.~Lewellen, J.~A.~Schwartz and S.~H.~Tye,
``The Spin Structure Construction Of String Models And Multiloop Modular Invariance,''
Nucl.\ Phys.\ B {\bf 299} (1988) 431;
\bibitem{ff2} I.~Antoniadis, C.~Bachas, C.~Kounnas and P.~Windey,
``Supersymmetry Among Free Fermions And Superstrings,''
Phys.\ Lett.\ B {\bf 171} (1986) 51.
\bibitem{ff3} I.~Antoniadis and C.~Bachas,
``4-D Fermionic Superstrings With Arbitrary Twists,''
Nucl.\ Phys.\ B {\bf 298} (1988) 586.
\bibitem{ff4} I.~Antoniadis, J.~R.~Ellis, J.~S.~Hagelin and D.~V.~Nanopoulos,
``The Flipped SU(5) X U(1) String Model Revamped,''
Phys.\ Lett.\ B {\bf 231} (1989) 65.
\bibitem{cy1}  
P.~Candelas, G.~T.~Horowitz, A.~Strominger and E.~Witten,
``Vacuum Configurations For Superstrings,''
Nucl.\ Phys.\ B {\bf 258} (1985) 46.
\bibitem{cy2}E.~Witten,
``New Issues In Manifolds Of SU(3) Holonomy,''
Nucl.\ Phys.\ B {\bf 268} (1986) 79.
\bibitem{cy3} 
A.~Strominger,
``Superstrings With Torsion,''
Nucl.\ Phys.\ B {\bf 274} (1986) 253.
\bibitem{cy4} 
M.~Dine, N.~Seiberg, X.~G.~Wen and E.~Witten,
``Nonperturbative Effects On The String World Sheet,''
Nucl.\ Phys.\ B {\bf 278} (1986) 769,
``Nonperturbative Effects On The String World Sheet. 2,''
Nucl.\ Phys.\ B {\bf 289} (1987) 319.
\bibitem{cy5} 
A.~Strominger and E.~Witten,
``New Manifolds For Superstring Compactification,''
Commun.\ Math.\ Phys.\  {\bf 101} (1985) 341.
\bibitem{cy6} M.~T.~Grisaru, A.~E.~van de Ven and D.~Zanon,
``Four Loop Beta Function For The N=1 And N=2 Supersymmetric Nonlinear Sigma Model In Two-Dimensions,''
Phys.\ Lett.\ B {\bf 173} (1986) 423,
``Two-Dimensional Supersymmetric Sigma Models On Ricci Flat Kahler Manifolds Are Not Finite,''
Nucl.\ Phys.\ B {\bf 277} (1986) 388,
``Four Loop Divergences For The N=1 Supersymmetric Nonlinear Sigma Model In Two-Dimensions,''
Nucl.\ Phys.\ B {\bf 277} (1986) 409.
\bibitem{mm} B.~R.~Greene and M.~R.~Plesser,
``Duality In Calabi-Yau Moduli Space,''
Nucl.\ Phys.\ B {\bf 338} (1990) 15.
\bibitem{gep1} 
D.~Gepner,
``Exactly Solvable String Compactifications On Manifolds Of SU(N) Holonomy,''
Phys.\ Lett.\ B {\bf 199} (1987) 380,
``Space-Time Supersymmetry In Compactified String Theory And Superconformal Models,''
Nucl.\ Phys.\ B {\bf 296} (1988) 757.
\bibitem{gep2} 
D.~Gepner,
``Yukawa Couplings For Calabi-Yau String Compactification,''
Nucl.\ Phys.\ B {\bf 311} (1988) 191.
\bibitem{mmlg} W.~Lerche, C.~Vafa and N.~P.~Warner,
``Chiral Rings In N=2 Superconformal Theories,''
Nucl.\ Phys.\ B {\bf 324} (1989) 427;
E.~Witten,
``On the Landau-Ginzburg description of N=2 minimal models,''
Int.\ J.\ Mod.\ Phys.\ A {\bf 9} (1994) 4783
[arXiv:hep-th/9304026].

\bibitem{mmcy}
B.~R.~Greene, C.~Vafa and N.~P.~Warner,
``Calabi-Yau Manifolds And Renormalization Group Flows,''
Nucl.\ Phys.\ B {\bf 324} (1989) 371;
E.~J.~Martinec,
``Algebraic Geometry And Effective Lagrangians,''
Phys.\ Lett.\ B {\bf 217} (1989) 431;
E.~Witten,
``Phases of N = 2 theories in two dimensions,''
Nucl.\ Phys.\ B {\bf 403} (1993) 159
[arXiv:hep-th/9301042].


\bibitem{CP} 
J.~E.~Paton and H.~Chan,
``Generalized Veneziano Model With Isospin,''
Nucl.\ Phys.\ B {\bf 10} (1969) 516.
\bibitem{grap1}  J.~H.~Schwarz,
``Gauge Groups For Type I Superstrings,''
CALT-68-906-REV
{\it Presented at 6th Johns Hopkins Workshop on Current Problems in High-Energy Particle Theory, Florence, Italy, Jun 2-4, 1982}.
\bibitem{grap2} N.~Marcus and A.~Sagnotti,
``Tree Level Constraints On Gauge Groups For Type I Superstrings,''
Phys.\ Lett.\ B {\bf 119} (1982) 97.
\bibitem{grap3} N.~Marcus and A.~Sagnotti,
``Group Theory From 'Quarks' At The Ends Of Strings,''
Phys.\ Lett.\ B {\bf 188} (1987) 58.
\bibitem{aa} V.~Alessandrini,
``A General Approach To Dual Multiloop Diagrams,''
Nuovo Cim.\ A {\bf 2} (1971) 321;
V.~Alessandrini and D.~Amati,
``Properties Of Dual Multiloop Amplitudes,''
Nuovo Cim.\ A {\bf 4} (1971) 793.
\bibitem{od1} A.~Sagnotti,
``Open Strings And Their Symmetry Groups,''
ROM2F-87-25
{\it Talk presented at the Cargese Summer Institute on Non-Perturbative Methods in Field Theory, Cargese, France, Jul 16-30, 1987}.
\bibitem{od2} G.~Pradisi and A.~Sagnotti,
``Open String Orbifolds,''
Phys.\ Lett.\ B {\bf 216} (1989) 59.
\bibitem{od3} M.~Bianchi and A.~Sagnotti,
``On The Systematics Of Open String Theories,''
Phys.\ Lett.\ B {\bf 247} (1990) 517,
``Twist symmetry and open string Wilson lines,''
Nucl.\ Phys.\ B {\bf 361} (1991) 519.
\bibitem{od4} 
M.~Bianchi, G.~Pradisi and A.~Sagnotti,
``Toroidal compactification and symmetry breaking in open string theories,''
Nucl.\ Phys.\ B {\bf 376} (1992) 365.
\bibitem{od5} J.~L.~Cardy,
``Boundary Conditions, Fusion Rules And The Verlinde Formula,''
Nucl.\ Phys.\ B {\bf 324} (1989) 581.
\bibitem{od6} E.~Verlinde,
``Fusion Rules And Modular Transformations In 2-D Conformal Field Theory,''
Nucl.\ Phys.\ B {\bf 300} (1988) 360.
\bibitem{od7} J.~Polchinski and Y.~Cai,
``Consistency Of Open Superstring Theories,''
Nucl.\ Phys.\ B {\bf 296} (1988) 91.
\bibitem{ocon} 
M.~Bianchi, G.~Pradisi and A.~Sagnotti,
``Planar duality in the discrete series,''
Phys.\ Lett.\ B {\bf 273} (1991) 389;
D.~Fioravanti, G.~Pradisi and A.~Sagnotti,
``Sewing constraints and nonorientable open strings,''
Phys.\ Lett.\ B {\bf 321} (1994) 349
[arXiv:hep-th/9311183];
G.~Pradisi, A.~Sagnotti and Y.~S.~Stanev,
``Planar duality in SU(2) WZW models,''
Phys.\ Lett.\ B {\bf 354} (1995) 279
[arXiv:hep-th/9503207],
``The Open descendants of nondiagonal SU(2) WZW models,''
Phys.\ Lett.\ B {\bf 356} (1995) 230
[arXiv:hep-th/9506014],
``Completeness Conditions for Boundary Operators in 2D Conformal Field Theory,''
Phys.\ Lett.\ B {\bf 381} (1996) 97
[arXiv:hep-th/9603097].
\bibitem{bs} M.~Bianchi and A.~Sagnotti,
``The Partition Function Of The SO(8192) Bosonic String,''
Phys.\ Lett.\ B {\bf 211} (1988) 407.
\bibitem{4d}C.~Angelantonj, M.~Bianchi, G.~Pradisi, A.~Sagnotti and Y.~S.~Stanev,
``Chiral asymmetry in four-dimensional open- string vacua,''
Phys.\ Lett.\ B {\bf 385} (1996) 96
[arXiv:hep-th/9606169];
M.~Bianchi, S.~Ferrara, G.~Pradisi, A.~Sagnotti and Y.~S.~Stanev,
``Twelve-dimensional aspects of four-dimensional N = 1 type I vacua,''
Phys.\ Lett.\ B {\bf 387} (1996) 64
[arXiv:hep-th/9607105].
\bibitem{og} 
C.~Angelantonj, M.~Bianchi, G.~Pradisi, A.~Sagnotti and Y.~S.~Stanev,
``Comments on Gepner models and type I vacua in string theory,''
Phys.\ Lett.\ B {\bf 387} (1996) 743
[arXiv:hep-th/9607229].
\bibitem{an3} J. Wess e B. Zumino, \PLB{73}{71}{95}.
\bibitem{6d1} L.~J.~Romans,
``Selfduality For Interacting Fields: Covariant Field Equations For Six-Dimensional Chiral Supergravities,''
Nucl.\ Phys.\ B {\bf 276} (1986) 71.
\bibitem{6d2} H.~Nishino and E.~Sezgin,
``The Complete N=2, D = 6 Supergravity With Matter And Yang-Mills Couplings,''
Nucl.\ Phys.\ B {\bf 278} (1986) 353.
\bibitem{6d3} A.~Sagnotti,
``A Note on the Green-Schwarz mechanism in open string theories,''
Phys.\ Lett.\ B {\bf 294} (1992) 196
[arXiv:hep-th/9210127].
\bibitem{6d4} S.~Ferrara, R.~Minasian and A.~Sagnotti,
``Low-Energy Analysis of $M$ and $F$ Theories on Calabi-Yau Threefolds,''
Nucl.\ Phys.\ B {\bf 474} (1996) 323
[arXiv:hep-th/9604097].
\bibitem{6d5} N.~Seiberg and E.~Witten,
``Comments on String Dynamics in Six Dimensions,''
Nucl.\ Phys.\ B {\bf 471} (1996) 121
[arXiv:hep-th/9603003];
M.~J.~Duff, H.~Lu and C.~N.~Pope,
``Heterotic phase transitions and singularities of the gauge dyonic string,''
Phys.\ Lett.\ B {\bf 378} (1996) 101
[arXiv:hep-th/9603037].
\bibitem{6d6} M.~J.~Duff, R.~Minasian and E.~Witten,
``Evidence for Heterotic/Heterotic Duality,''
Nucl.\ Phys.\ B {\bf 465} (1996) 413
[arXiv:hep-th/9601036].
\bibitem{6d7} P.~Pasti, D.~Sorokin and M.~Tonin,
``On Lorentz invariant actions for chiral p-forms,''
Phys.\ Rev.\ D {\bf 55} (1997) 6292
[arXiv:hep-th/9611100].
\bibitem{6df} S.~Ferrara, F.~Riccioni and A.~Sagnotti,
``Tensor and vector multiplets in six-dimensional supergravity,''
Nucl.\ Phys.\ B {\bf 519} (1998) 115
[arXiv:hep-th/9711059].
\bibitem{vettor} M.~Bianchi,
``A note on toroidal compactifications of the type I superstring and  other superstring vacuum configurations with 16 supercharges,''
Nucl.\ Phys.\ B {\bf 528} (1998) 73
[arXiv:hep-th/9711201];
E.~Witten,
``Toroidal compactification without vector structure,''
JHEP {\bf 9802} (1998) 006
[arXiv:hep-th/9712028].
\bibitem{orienti} 
E.~G.~Gimon and C.~V.~Johnson,
``K3 Orientifolds,''
Nucl.\ Phys.\ B {\bf 477} (1996) 715
[arXiv:hep-th/9604129];
A.~Dabholkar and J.~Park,
``An Orientifold of Type-IIB Theory on $K3$,''
Nucl.\ Phys.\ B {\bf 472} (1996) 207
[arXiv:hep-th/9602030],
``Strings on Orientifolds,''
Nucl.\ Phys.\ B {\bf 477} (1996) 701
[arXiv:hep-th/9604178],
``A note on orientifolds and F-theory,''
Phys.\ Lett.\ B {\bf 394} (1997) 302
[arXiv:hep-th/9607041];
R.~Gopakumar and S.~Mukhi,
``Orbifold and orientifold compactifications of F-theory and M-theory  to six and four dimensions,''
Nucl.\ Phys.\ B {\bf 479} (1996) 260
[arXiv:hep-th/9607057];
J.~D.~Blum and A.~Zaffaroni,
``An orientifold from F theory,''
Phys.\ Lett.\ B {\bf 387} (1996) 71
[arXiv:hep-th/9607019];
J.~D.~Blum,
``F-theory orientifolds, M-theory orientifolds, and twisted strings,''
Nucl.\ Phys.\ B {\bf 486} (1997) 34
[arXiv:hep-th/9608053].
\bibitem{ob2} 
C.~Bachas and M.~Porrati,
``Pair creation of open strings in an electric field,''
Phys.\ Lett.\ B {\bf 296} (1992) 77
[arXiv:hep-th/9209032].
\bibitem{db1} J.~Polchinski,
``Dirichlet-Branes and Ramond-Ramond Charges,''
Phys.\ Rev.\ Lett.\  {\bf 75} (1995) 4724
[arXiv:hep-th/9510017].
\bibitem{db2} J.~Dai, R.~G.~Leigh and J.~Polchinski,
``New Connections Between String Theories,''
Mod.\ Phys.\ Lett.\ A {\bf 4} (1989) 2073;
R.~G.~Leigh,
``Dirac-Born-Infeld Action From Dirichlet Sigma Model,''
Mod.\ Phys.\ Lett.\ A {\bf 4} (1989) 2767.
\bibitem{db2a} 
M.~B.~Green,
``Space-time duality and Dirichlet string theory,''
Phys.\ Lett.\ B {\bf 266} (1991) 325.
\bibitem{db3} C.~Bachas,
``D-brane dynamics,''
Phys.\ Lett.\ B {\bf 374} (1996) 37
[arXiv:hep-th/9511043].
\bibitem{db4} 
C.~P.~Bachas, M.~R.~Douglas and M.~B.~Green,
``Anomalous creation of branes,''
JHEP {\bf 9707} (1997) 002
[arXiv:hep-th/9705074].
\bibitem{db5} D.~Kabat and P.~Pouliot,
``A Comment on Zero-brane Quantum Mechanics,''
Phys.\ Rev.\ Lett.\  {\bf 77} (1996) 1004
[arXiv:hep-th/9603127];
U.~H.~Danielsson, G.~Ferretti and B.~Sundborg,
``D-particle Dynamics and Bound States,''
Int.\ J.\ Mod.\ Phys.\ A {\bf 11} (1996) 5463
[arXiv:hep-th/9603081];
M.~R.~Douglas, D.~Kabat, P.~Pouliot and S.~H.~Shenker,
``D-branes and short distances in string theory,''
Nucl.\ Phys.\ B {\bf 485} (1997) 85
[arXiv:hep-th/9608024].
\bibitem{mm1} T.~Banks, W.~Fischler, S.~H.~Shenker and L.~Susskind,
``M theory as a matrix model: A conjecture,''
Phys.\ Rev.\ D {\bf 55} (1997) 5112
[arXiv:hep-th/9610043].
\bibitem{db6} M.~B.~Green, J.~A.~Harvey and G.~W.~Moore,
``I-brane inflow and anomalous couplings on D-branes,''
Class.\ Quant.\ Grav.\  {\bf 14} (1997) 47
[arXiv:hep-th/9605033].
\bibitem{bh1} A.~Strominger and C.~Vafa,
``Microscopic Origin of the Bekenstein-Hawking Entropy,''
Phys.\ Lett.\ B {\bf 379} (1996) 99
[arXiv:hep-th/9601029].
\bibitem{bft1} A.~Hanany and E.~Witten,
``Type IIB superstrings, BPS monopoles, and three-dimensional gauge  dynamics,''
Nucl.\ Phys.\ B {\bf 492} (1997) 152
[arXiv:hep-th/9611230].
\bibitem{elit} S.~Elitzur, A.~Giveon and D.~Kutasov,
``Branes and N = 1 duality in string theory,''
Phys.\ Lett.\ B {\bf 400} (1997) 269
[arXiv:hep-th/9702014];
S.~Elitzur, A.~Giveon, D.~Kutasov, E.~Rabinovici and A.~Schwimmer,
``Brane dynamics and N = 1 supersymmetric gauge theory,''
Nucl.\ Phys.\ B {\bf 505} (1997) 202
[arXiv:hep-th/9704104].
\bibitem{bqcd} E.~Witten,
``Solutions of four-dimensional field theories via M-theory,''
Nucl.\ Phys.\ B {\bf 500} (1997) 3
[arXiv:hep-th/9703166],
``Branes and the dynamics of {QCD},''
Nucl.\ Phys.\ B {\bf 507} (1997) 658
[arXiv:hep-th/9706109].
\bibitem{hod3}
E.~Witten,
``Small Instantons in String Theory,''
Nucl.\ Phys.\ B {\bf 460} (1996) 541
[arXiv:hep-th/9511030].
\bibitem{ncg} A.~Connes, M.~R.~Douglas and A.~Schwarz,
``Noncommutative geometry and matrix theory: Compactification on tori,''
JHEP {\bf 9802} (1998) 003
[arXiv:hep-th/9711162];
M.~R.~Douglas and C.~M.~Hull,
``D-branes and the noncommutative torus,''
JHEP {\bf 9802} (1998) 008
[arXiv:hep-th/9711165].
\bibitem{subs} M.~R.~Douglas and G.~W.~Moore,
``D-branes, Quivers, and ALE Instantons,''
arXiv:hep-th/9603167;
M.~R.~Douglas, B.~R.~Greene and D.~R.~Morrison,
``Orbifold resolution by D-branes,''
Nucl.\ Phys.\ B {\bf 506} (1997) 84
[arXiv:hep-th/9704151];
S.~Mukhopadhyay and K.~Ray,
``Conifolds from D-branes,''
Phys.\ Lett.\ B {\bf 423} (1998) 247
[arXiv:hep-th/9711131];
B.~R.~Greene,
``D-brane topology changing transitions,''
Nucl.\ Phys.\ B {\bf 525} (1998) 284
[arXiv:hep-th/9711124].
\bibitem{KN}P.~B.~Kronheimer,
``The Construction Of Ale Spaces As Hyperkahler Quotients,''
J.\ Diff.\ Geom.\  {\bf 29} (1989) 665,
``A Torelli Type Theorem For Gravitational Instantons,''
J.\ Diff.\ Geom.\  {\bf 29} (1989) 685.
P. Kronheimer e H. Nakajima, Math. Ann. $\bf{288}$ (1990) 263




\bibitem{Bog}E.~B.~Bogomolny,
``Stability Of Classical Solutions,''
Sov.\ J.\ Nucl.\ Phys.\  {\bf 24} (1976) 449
[Yad.\ Fiz.\  {\bf 24} (1976) 861].
\bibitem{BPS} M.~K.~Prasad and C.~M.~Sommerfield,
``An Exact Classical Solution For The 'T Hooft Monopole And The Julia-Zee Dyon,''
Phys.\ Rev.\ Lett.\  {\bf 35} (1975) 760.
\bibitem{MO}  C.~Montonen and D.~I.~Olive,
``Magnetic Monopoles As Gauge Particles?,''
Phys.\ Lett.\ B {\bf 72} (1977) 117.
\bibitem{towcen} P.~K.~Townsend,
``P-brane democracy,''
arXiv:hep-th/9507048.
\bibitem{Zwi} M.~R.~Gaberdiel and B.~Zwiebach,
``Exceptional groups from open strings,''
Nucl.\ Phys.\ B {\bf 518} (1998) 151
[arXiv:hep-th/9709013].
\bibitem{hs}J.~A.~Harvey and A.~Strominger,
``The heterotic string is a soliton,''
Nucl.\ Phys.\ B {\bf 449} (1995) 535
[Erratum-ibid.\ B {\bf 458} (1995) 456]
[arXiv:hep-th/9504047].
\bibitem{hid1} E.~Cremmer, J.~Scherk and S.~Ferrara,
``SU(4) Invariant Supergravity Theory,''
Phys.\ Lett.\ B {\bf 74} (1978) 61.
\bibitem{11d} E.~Cremmer and B.~Julia,
``The N=8 Supergravity Theory. 1. The Lagrangian,''
Phys.\ Lett.\ B {\bf 80} (1978) 48,
``The SO(8) Supergravity,''
Nucl.\ Phys.\ B {\bf 159} (1979) 141.
\bibitem{thp} G.~'t Hooft,
``Magnetic Monopoles In Unified Gauge Theories,''
Nucl.\ Phys.\ B {\bf 79} (1974) 276;
A.~M.~Polyakov,
``Particle Spectrum In Quantum Field Theory,''
JETP Lett.\  {\bf 20} (1974) 194
[Pisma Zh.\ Eksp.\ Teor.\ Fiz.\  {\bf 20} (1974) 430].
\bibitem{jz}  B.~Julia and A.~Zee,
``Poles With Both Magnetic And Electric Charges In Nonabelian Gauge Theory,''
Phys.\ Rev.\ D {\bf 11} (1975) 2227.
\bibitem{OW} E.~Witten and D.~I.~Olive,
``Supersymmetry Algebras That Include Topological Charges,''
Phys.\ Lett.\ B {\bf 78} (1978) 97.
\bibitem{wit} E.~Witten,
``Dyons Of Charge E Theta / 2 Pi,''
Phys.\ Lett.\ B {\bf 86} (1979) 283.
\bibitem{pq1} A.~Dabholkar and J.~A.~Harvey,
``Nonrenormalization Of The Superstring Tension,''
Phys.\ Rev.\ Lett.\  {\bf 63} (1989) 478;
A.~Dabholkar, G.~W.~Gibbons, J.~A.~Harvey and F.~Ruiz Ruiz,
``Superstrings And Solitons,''
Nucl.\ Phys.\ B {\bf 340} (1990) 33.
\bibitem{pq2} J.~H.~Schwarz,
``An SL(2,Z) multiplet of type IIB superstrings,''
Phys.\ Lett.\ B {\bf 360} (1995) 13
[Erratum-ibid.\ B {\bf 364} (1995) 252]
[arXiv:hep-th/9508143],
``The power of M theory,''
Phys.\ Lett.\ B {\bf 367} (1996) 97
[arXiv:hep-th/9510086].
\bibitem{br1}C.~G.~Callan, J.~A.~Harvey and A.~Strominger,
``World sheet approach to heterotic instantons and solitons,''
Nucl.\ Phys.\ B {\bf 359} (1991) 611.
\bibitem{br2} C.~G.~Callan, J.~A.~Harvey and A.~Strominger,
``Worldbrane actions for string solitons,''
Nucl.\ Phys.\ B {\bf 367} (1991) 60.
\bibitem{br3} M.~J.~Duff, R.~R.~Khuri and J.~X.~Lu,
``String solitons,''
Phys.\ Rept.\  {\bf 259} (1995) 213
[arXiv:hep-th/9412184].
\bibitem{br4} R.~I.~Nepomechie,
``Magnetic Monopoles From Antisymmetric Tensor Gauge Fields,''
Phys.\ Rev.\ D {\bf 31} (1985) 1921;
C.~Teitelboim,
``Monopoles Of Higher Rank,''
Phys.\ Lett.\ B {\bf 167} (1986) 69.
\bibitem{br5} 
A.~Strominger,
``Open p-branes,''
Phys.\ Lett.\ B {\bf 383} (1996) 44
[arXiv:hep-th/9512059];
P.~K.~Townsend,
``D-branes from M-branes,''
Phys.\ Lett.\ B {\bf 373} (1996) 68
[arXiv:hep-th/9512062].
\bibitem{br7} E.~Witten,
``Bound States Of Strings And p-Branes,''
Nucl.\ Phys.\ B {\bf 460} (1996) 335
[arXiv:hep-th/9510135].
\bibitem{bb} G.~T.~Horowitz and A.~Strominger,
``Black strings and P-branes,''
Nucl.\ Phys.\ B {\bf 360} (1991) 197.
\bibitem{con} A.~Strominger,
``Massless black holes and conifolds in string theory,''
Nucl.\ Phys.\ B {\bf 451} (1995) 96
[arXiv:hep-th/9504090].
\bibitem{con2} B.~R.~Greene, D.~R.~Morrison and A.~Strominger,
``Black hole condensation and the unification of string vacua,''
Nucl.\ Phys.\ B {\bf 451} (1995) 109
[arXiv:hep-th/9504145].
\bibitem{spg1} B.~de Wit, P.~G.~Lauwers and A.~Van Proeyen,
``Lagrangians Of N=2 Supergravity - Matter Systems,''
Nucl.\ Phys.\ B {\bf 255} (1985) 569.
\bibitem{sw} N.~Seiberg and E.~Witten,
``Electric - magnetic duality, monopole condensation, and confinement in N=2 supersymmetric Yang-Mills theory,''
Nucl.\ Phys.\ B {\bf 426} (1994) 19
[Erratum-ibid.\ B {\bf 430} (1994) 485]
[arXiv:hep-th/9407087].
\bibitem{dualp} 
S.~Ferrara, J.~A.~Harvey, A.~Strominger and C.~Vafa,
``Second quantized mirror symmetry,''
Phys.\ Lett.\ B {\bf 361} (1995) 59
[arXiv:hep-th/9505162];
C.~Vafa and E.~Witten,
``Dual string pairs with N = 1 and N = 2 supersymmetry in four  dimensions,''
Nucl.\ Phys.\ Proc.\ Suppl.\  {\bf 46} (1996) 225
[arXiv:hep-th/9507050];
A.~Sen and C.~Vafa,
``Dual pairs of type II string compactification,''
Nucl.\ Phys.\ B {\bf 455} (1995) 165
[arXiv:hep-th/9508064].

\bibitem{du8} S.~Kachru and C.~Vafa,
``Exact results for N=2 compactifications of heterotic strings,''
Nucl.\ Phys.\ B {\bf 450} (1995) 69
[arXiv:hep-th/9505105].

\bibitem{or2} E.~G.~Gimon and J.~Polchinski,
``Consistency Conditions for Orientifolds and D-Manifolds,''
Phys.\ Rev.\ D {\bf 54} (1996) 1667
[arXiv:hep-th/9601038].

\bibitem{sen1}  A.~Sen,
``Strong - weak coupling duality in four-dimensional string theory,''
Int.\ J.\ Mod.\ Phys.\ A {\bf 9} (1994) 3707
[arXiv:hep-th/9402002];
A.~Sen,
``Dyon - monopole bound states, selfdual harmonic forms on the multi - monopole moduli space, and SL(2,Z) invariance in string theory,''
Phys.\ Lett.\ B {\bf 329} (1994) 217
[arXiv:hep-th/9402032].

\bibitem{sen2} A.~Sen,
``Unification of string dualities,''
Nucl.\ Phys.\ Proc.\ Suppl.\  {\bf 58} (1997) 5
[arXiv:hep-th/9609176].

\bibitem{du1} C.~M.~Hull and P.~K.~Townsend,
``Unity of superstring dualities,''
Nucl.\ Phys.\ B {\bf 438} (1995) 109
[arXiv:hep-th/9410167].

\bibitem{du2} E.~Witten,
``String theory dynamics in various dimensions,''
Nucl.\ Phys.\ B {\bf 443} (1995) 85
[arXiv:hep-th/9503124].

\bibitem{tw1} P.~K.~Townsend,
``The eleven-dimensional supermembrane revisited,''
Phys.\ Lett.\ B {\bf 350} (1995) 184
[arXiv:hep-th/9501068].

\bibitem{du3} P.~Horava and E.~Witten,
``Heterotic and type I string dynamics from eleven dimensions,''
Nucl.\ Phys.\ B {\bf 460} (1996) 506
[arXiv:hep-th/9510209].

\bibitem{du3b} P.~Horava and E.~Witten,
``Eleven-Dimensional Supergravity on a Manifold with Boundary,''
Nucl.\ Phys.\ B {\bf 475} (1996) 94
[arXiv:hep-th/9603142].


\bibitem{du4} J.~Polchinski and E.~Witten,
``Evidence for Heterotic - Type I String Duality,''
Nucl.\ Phys.\ B {\bf 460} (1996) 525
[arXiv:hep-th/9510169].

\bibitem{hod2} A.~A.~Tseytlin,
``On SO(32) heterotic - type I superstring duality in ten dimensions,''
Phys.\ Lett.\ B {\bf 367} (1996) 84
[arXiv:hep-th/9510173],
``Heterotic - type I superstring duality and low-energy effective actions,''
Nucl.\ Phys.\ B {\bf 467} (1996) 383
[arXiv:hep-th/9512081].

\bibitem{ft1} C.~Bachas and E.~Kiritsis,
``F**4 terms in N = 4 string vacua,''
Nucl.\ Phys.\ Proc.\ Suppl.\  {\bf 55B} (1997) 194
[arXiv:hep-th/9611205].

\bibitem{ohdu} I.~Antoniadis, C.~Bachas, C.~Fabre, H.~Partouche and T.~R.~Taylor,
``Aspects of type I - type II - heterotic triality in four dimensions,''
Nucl.\ Phys.\ B {\bf 489} (1997) 160
[arXiv:hep-th/9608012].

\bibitem{hod1} W.~Lerche,
``Elliptic Index And Superstring Effective Actions,''
Nucl.\ Phys.\ B {\bf 308} (1988) 102.

\bibitem{Fth} C.~Vafa,
``Evidence for F-Theory,''
Nucl.\ Phys.\ B {\bf 469} (1996) 403
[arXiv:hep-th/9602022];
D.~R.~Morrison and C.~Vafa,
``Compactifications of F-Theory on Calabi--Yau Threefolds -- I,''
Nucl.\ Phys.\ B {\bf 473} (1996) 74
[arXiv:hep-th/9602114],
``Compactifications of F-Theory on Calabi--Yau Threefolds -- II,''
Nucl.\ Phys.\ B {\bf 476} (1996) 437
[arXiv:hep-th/9603161].

\bibitem{For} A.~Sen,
``F-theory and Orientifolds,''
Nucl.\ Phys.\ B {\bf 475} (1996) 562
[arXiv:hep-th/9605150],
``A non-perturbative description of the Gimon-Polchinski orientifold,''
Nucl.\ Phys.\ B {\bf 489} (1997) 139
[arXiv:hep-th/9611186],
``Orientifold limit of F-theory vacua,''
Phys.\ Rev.\ D {\bf 55} (1997) 7345
[arXiv:hep-th/9702165].

\bibitem{winx} M.~B.~Green and M.~Gutperle,
``D-particle bound states and the D-instanton measure,''
JHEP {\bf 9801} (1998) 005
[arXiv:hep-th/9711107],
``D-instanton partition functions,''
Phys.\ Rev.\ D {\bf 58} (1998) 046007
[arXiv:hep-th/9804123].

\bibitem{dist} M.~B.~Green and M.~Gutperle,
``Effects of D-instantons,''
Nucl.\ Phys.\ B {\bf 498} (1997) 195
[arXiv:hep-th/9701093];
M.~B.~Green, M.~Gutperle and P.~Vanhove,
``One loop in eleven dimensions,''
Phys.\ Lett.\ B {\bf 409} (1997) 177
[arXiv:hep-th/9706175].

\bibitem{sus} L.~Susskind,
``Another conjecture about M(atrix) theory,''
arXiv:hep-th/9704080.

\bibitem{olo} C.~R.~Stephens, G.~'t Hooft and B.~F.~Whiting,
``Black hole evaporation without information loss,''
Class.\ Quant.\ Grav.\  {\bf 11} (1994) 621
[arXiv:gr-qc/9310006].

\bibitem{mal} J.~Maldacena,
``The large $N$ limit of superconformal field theories and supergravity,''
Adv.\ Theor.\ Math.\ Phys.\  {\bf 2} (1998) 231
[Int.\ J.\ Theor.\ Phys.\  {\bf 38} (1998) 1113]
[arXiv:hep-th/9711200].



\bibitem{FI} P.~Fayet and J.~Iliopoulos,
``Spontaneously Broken Supergauge Symmetries And Goldstone Spinors,''
Phys.\ Lett.\ B {\bf 51} (1974) 461.

\bibitem{OR} L. O'Raifeartaigh, \NPB{96}{75}{331}.

\bibitem{GG} L.~Girardello and M.~T.~Grisaru,
``Soft Breaking Of Supersymmetry,''
Nucl.\ Phys.\ B {\bf 194} (1982) 65.

\bibitem{GSR} M.T. Grisaru, M. Rocek e W. Siegel, \NPB{194}{82}{65}.
\bibitem{CJS} E.~Cremmer, B.~Julia, J.~Scherk, S.~Ferrara, L.~Girardello and P.~van Nieuwenhuizen,
``Spontaneous Symmetry Breaking And Higgs Effect In Supergravity Without Cosmological Constant,''
Nucl.\ Phys.\ B {\bf 147} (1979) 105;
E.~Cremmer, S.~Ferrara, L.~Girardello and A.~Van Proeyen,
``Yang-Mills Theories With Local Supersymmetry: Lagrangian, Transformation Laws And Superhiggs Effect,''
Nucl.\ Phys.\ B {\bf 212} (1983) 413;
E.~Cremmer, S.~Ferrara, C.~Kounnas and D.~V.~Nanopoulos,
``Naturally Vanishing Cosmological Constant In N=1 Supergravity,''
Phys.\ Lett.\ B {\bf 133} (1983) 61.

\bibitem{ss} J.~Scherk and J.~H.~Schwarz,
``Spontaneous Breaking Of Supersymmetry Through Dimensional Reduction,''
Phys.\ Lett.\ B {\bf 82} (1979) 60,
``How To Get Masses From Extra Dimensions,''
Nucl.\ Phys.\ B {\bf 153} (1979) 61.

\bibitem{ss2} E.~Cremmer, J.~Scherk and J.~H.~Schwarz,
``Spontaneously Broken N=8 Supergravity,''
Phys.\ Lett.\ B {\bf 84} (1979) 83.

\bibitem{ss3} P.~Fayet,
``Supersymmetric Grand Unification In A Six-Dimensional Space-Time,''
Phys.\ Lett.\ B {\bf 159} (1985) 121,
``Six-Dimensional Supersymmetric QED, R Invariance And N=2 Supersymmetry Breaking By Dimensional Reduction,''
Nucl.\ Phys.\ B {\bf 263} (1986) 649.

\bibitem{sub1} S.~Ferrara, C.~Kounnas and M.~Porrati,
``Superstring Solutions With Spontaneously Broken Four-Dimensional Supersymmetry,''
Nucl.\ Phys.\ B {\bf 304} (1988) 500.

\bibitem{sub2}C.~Kounnas and M.~Porrati,
``Spontaneous Supersymmetry Breaking In String Theory,''
Nucl.\ Phys.\ B {\bf 310} (1988) 355;
S.~Ferrara, C.~Kounnas and M.~Porrati,
``Multiloop Modular Invariance In Spontaneously Broken Superstrings,''
Phys.\ Lett.\ B {\bf 197} (1987) 135,
``N=1 Superstrings With Spontaneously Broken Symmetries,''
Phys.\ Lett.\ B {\bf 206} (1988) 25.
S.~Ferrara, C.~Kounnas, M.~Porrati and F.~Zwirner,
``Superstrings With Spontaneously Broken Supersymmetry And Their Effective Theories,''
Nucl.\ Phys.\ B {\bf 318} (1989) 75;
C.~Kounnas and B.~Rostand,
``Coordinate Dependent Compactifications And Discrete Symmetries,''
Nucl.\ Phys.\ B {\bf 341} (1990) 641;
M.~Porrati,
``Off-Shell Ward Identities And Gauge Symmetries In String Theory,''
Phys.\ Lett.\ B {\bf 231} (1989) 403;
I.~Antoniadis,
``A Possible New Dimension At A Few Tev,''
Phys.\ Lett.\ B {\bf 246} (1990) 377;
I.~Antoniadis and C.~Kounnas,
``Superstring phase transition at high temperature,''
Phys.\ Lett.\ B {\bf 261} (1991) 369.

\bibitem{Kir} E.~Kiritsis and C.~Kounnas,
``Perturbative and non-perturbative partial supersymmetry breaking:  N = 4 $\to$ N = 2 $\to$ N = 1,''
Nucl.\ Phys.\ B {\bf 503} (1997) 117
[arXiv:hep-th/9703059].

\bibitem{ks} S.~Kachru, J.~Kumar and E.~Silverstein,
``Vacuum energy cancellation in a non-supersymmetric string,''
Phys.\ Rev.\ D {\bf 59} (1999) 106004
[arXiv:hep-th/9807076];
J.~A.~Harvey,
``String duality and non-supersymmetric strings,''
Phys.\ Rev.\ D {\bf 59} (1999) 026002
[arXiv:hep-th/9807213];
S.~Kachru and E.~Silverstein,
``Self-dual nonsupersymmetric type II string compactifications,''
JHEP {\bf 9811} (1998) 001
[arXiv:hep-th/9808056];
G.~Shiu and S.~H.~Tye,
``Bose-Fermi degeneracy and duality in non-supersymmetric strings,''
Nucl.\ Phys.\ B {\bf 542} (1999) 45
[arXiv:hep-th/9808095].

\bibitem{ads} I.~Antoniadis, E.~Dudas and A.~Sagnotti,
``Supersymmetry breaking, open strings and M-theory,''
Nucl.\ Phys.\ B {\bf 544} (1999) 469
[arXiv:hep-th/9807011].

\bibitem{hig} I.~Antoniadis, C.~Bachas and C.~Kounnas,
``Higgs Phenomenon In String Theories,''
Phys.\ Lett.\ B {\bf 200} (1988) 297.

\bibitem{blum} J.~D.~Blum and K.~R.~Dienes,
``Duality without supersymmetry: The case of the SO(16) x SO(16) string,''
Phys.\ Lett.\ B {\bf 414} (1997) 260
[arXiv:hep-th/9707148],
``Strong/weak coupling duality relations for non-supersymmetric string  
theories,''
Nucl.\ Phys.\ B {\bf 516} (1998) 83
[arXiv:hep-th/9707160].

\bibitem{glu} M.~Dine, R.~Rohm, N.~Seiberg and E.~Witten,
``Gluino Condensation In Superstring Models,''
Phys.\ Lett.\ B {\bf 156} (1985) 55.

\bibitem{al} G. Moore, G.~W.~Moore,
``Atkin-Lehner Symmetry,''
Nucl.\ Phys.\ B {\bf 293} (1987) 139
[Erratum-ibid.\ B {\bf 299} (1987) 847].

\bibitem{worpro} I.~Antoniadis, G.~D'Appollonio, E.~Dudas and A.~Sagnotti,
``Partial breaking of supersymmetry, open strings and M-theory,''
Nucl.\ Phys.\ B {\bf 553} (1999) 133
[arXiv:hep-th/9812118].


\end{thebibliography}
\end{document}